\RequirePackage{fix-cm}
\documentclass[smallextended,final,natbib]{svjour3}       
\smartqed  
\usepackage{graphicx}
\usepackage{amsmath}%
\usepackage{mathptmx}      
\usepackage{txfonts}
\usepackage{booktabs}
\usepackage{rotating}
\usepackage{colortbl}
\usepackage{color}
\usepackage[usenames,dvipsnames,svgnames,table]{xcolor}
\usepackage{hyperref}
\hypersetup{%
  pdftitle    = {Gamma-ray emission from binaries},
  pdfsubject  = {review},
  pdfkeywords = {gamma rays},
  pdfauthor   = {Guillaume Dubus},
}
\def\gd#1{{#1}}

\def\degr{\hbox{$^\circ$}}

\def\ergs{\hbox{erg\,s$^{-1}$}}
\def\la{\mathrel{\hbox{\rlap{\hbox{\lower4pt\hbox{$\sim$}}}\hbox{$<$}}}}
\def\ga{\mathrel{\hbox{\rlap{\hbox{\lower4pt\hbox{$\sim$}}}\hbox{$>$}}}}

\def\arcmin{\hbox{$^\prime$}}
\def\arcsec{\hbox{$^{\prime\prime}$}}

\def\fdg{\hbox{$.\!\!^\circ$}}

\newcommand{\ra}[1]{\renewcommand{\arraystretch}{#1}}
\newcommand{\msol}{${\rm M}_{\odot}$}
\newcommand{\eps}{erg\,s$^{-1}$}
\def\fermi{{\em Fermi}/LAT}
\def\ls{LS~5039}
\def\hessj{HESS~J0632+057}
\def\fgl{1FGL~J1018.6-5856}
\def\lsi{LS~I~+61\degr303}
\def\psrb{PSR~B1259-63}
\def\cyg{Cyg~X-3}

\begin{document}

\title{Gamma-ray binaries and related systems
}


\author{Guillaume Dubus
}


\institute{G. Dubus \at
               Institut de Plan\'etologie et d'Astrophysique de Grenoble (IPAG) UMR 5274,\\
               UJF-Grenoble 1 / CNRS-INSU, Grenoble, F-38041, France\\
              \email{Guillaume.Dubus@obs.ujf-grenoble.fr}           
}

\date{Received: date / Accepted: date}

\maketitle

\begin{abstract}
{After initial claims and a long hiatus, it is now established that several binary stars emit high (0.1$-$100 GeV) and very high energy ($>$100 GeV) gamma rays. A new class has emerged called ``gamma-ray binaries'', since most of their radiated power is emitted beyond 1 MeV. Accreting X-ray binaries, novae and a colliding wind binary ($\eta$ Car) have also been detected --- ``related systems'' that confirm the ubiquity of particle acceleration in astrophysical sources. Do these systems have anything in common ? What drives their high-energy emission ? How do the processes involved compare to those in other sources of gamma rays: pulsars, active galactic nuclei, supernova remnants ? I review the wealth of observational and theoretical work that have followed these detections, with an emphasis on gamma-ray binaries. I present the current evidence that gamma-ray binaries are driven by rotation-powered pulsars.  Binaries are laboratories giving access to different vantage points or physical conditions on a regular timescale as the components revolve on their orbit. I explain the basic ingredients that models of gamma-ray binaries use, the challenges that they currently face, and how they can bring insights into the physics of pulsars. I discuss how gamma-ray emission from microquasars  provides a window into the connection between accretion--ejection and acceleration, while $\eta$ Car and novae raise new questions on the physics of these objects --- or on the theory of diffusive shock acceleration. Indeed, explaining the gamma-ray emission from binaries strains our theories of high-energy astrophysical processes, by testing them on scales and in environments that were generally not foreseen, and this is  how these detections are most valuable.}
\end{abstract}

\section{Introduction\label{introduction}}
\label{intro}
 The advent of a new generation of observatories in the mid-2000s enabled the discovery of binary systems emitting high energy (HE, 0.1-100 GeV) or very high energy (VHE, $>$100 GeV) gamma rays.  A new class has emerged called ``gamma-ray binaries'', composed of a compact object and a massive star, distinguished by their radiative output with a peak in $\nu F_\nu$ beyond 1 MeV. At the time of writing, gamma-ray emission has also been detected from microquasars (Cyg X-3, possibly Cyg X-1), a colliding wind binary ($\eta$ Car), three novae (including the symbiotic binary V407 Cyg) and dozens of millisecond pulsars in binaries. 
 How singular are the binaries at this extreme end of the electromagnetic spectrum ? What physical processes are involved in the production of gamma rays ? How do they compare to those at work in other astrophysical sources ?
 
Binary stars have a unique property compared to other objects: that of giving access to different vantage points or physical conditions on a regular timescale as the components revolve on their orbit. The ensuing geometrical and dynamical constraints are precious. In the past, observations of binaries allowed accurate measurements of the masses and radii of stars, setting the stage for stellar physics; closer to us, binary radio pulsars have provided tests of general relativity ; X-ray binaries have provided the first dynamical evidence for black holes and spurred accretion theory. Binaries detected in gamma rays provide new opportunities for the study  of particle acceleration, magnetised relativistic outflows, and accretion-ejection physics. 

\subsection{The checkered history of binaries in gamma rays\label{checkered}}
Detecting gamma-ray emission from binaries is an effort that traces back decades. In the late 1970s, {\em Cos B} observations led to the discovery of the first HE gamma-ray source (2CG 135+01) where the search for a counterpart revealed a binary (\citealt{Gregory:1978or}, Fig.~\ref{fig:maps}). The binary is composed of a Be star (\lsi) and an unidentified compact object. Many such Be binary systems had already been detected in X-rays: what made this one highly unusual was that it was also a radio emitter, bursting at periodic intervals \citep{Gregory:1979ng}. A firm identification of the gamma-ray source with the binary could not be made because of the limited angular resolution in gamma rays, even using data collected with {\em CGRO}/EGRET in the 1990s  \citep{Hartman:1999mn,Reimer:2003nx}. The limited sensitivity also did not allow detailed timing studies, although there were hints of gamma-ray variability \citep{Tavani:1998qi}. Several other EGRET sources were tentatively associated to binaries over the years based on positional coincidence: Cyg X-3 \citep{1997ApJ...476..842M}, GRO J1838-04 \citep{1997ApJ...479L.109T}, GX 304-1 \citep{Nolan:2003sn}, LS~5039 \citep{2000Sci...288.2340P}, GX 339-4 \citep{Reimer:2003nx}, Cen X-3 \citep{1997ApJ...483L..49V}, SAX J0635+0533 \citep{1999ApJ...523..197K}. Most associations remain unconfirmed to this day.

At very high energies, the 1970s saw pioneering efforts to detect the flash
of Cherenkov light emitted by the electromagnetic shower created by the arrival of VHE gamma rays in the upper
atmosphere. In the 1980s, various instruments reported constant, flaring, pulsing or episodic VHE emission from
binaries. The binary Cyg X-3 played a notorious role in this bubble. Discovered by Giacconi in 1966, this X-ray binary fired up the imagination of theorists and observers alike, with dozens of papers published in {\em Nature} and {\em Science}, after it was found by \citet{1972Natur.239..440G} that the system is the site of major radio flares during which it becomes one of the brightest sources in the radio sky. Reports poured in claiming HE and VHE detections of Cyg X-3 --- including a periodic muon signal that challenged conventional physics (see \citealt{Chardin:1989ga} for a critical assessment). Confirmation remained elusive and by the end of the 1980s the situation had become confused if not controversial \citep{Weekes:1992lf}. However, these claims motivated many of the instrumental developments that were to bear fruit later.

\subsection{Discovery of gamma-ray binaries}

The story unfolded in the mid-2000s when the HESS, MAGIC and VERITAS collaborations secured the first VHE detections of binaries.  The latest generation of instruments combine stereoscopic imaging, large collecting areas, high resolution pixelation and improved analysis techniques to reject background particle triggers, lower the energy threshold and identify the imprint of an incoming gamma ray. These imaging arrays of Cherenkov telescopes (IACTs) have increased the number of known sources from a handful in 2004 to more than a hundred today\footnote{The \href{http://tevcat.uchicago.edu/}{\color{violet}{TeVCat}} online catalog keeps an up-to-date list}. EGRET was followed in 2007 by {\em AGILE} and in 2008 by \fermi, enabling the discovery of nearly 2000 sources of HE gamma rays \citep{2012ApJS..199...31N}.

\begin{figure}
\centering\includegraphics[height=4.75cm]{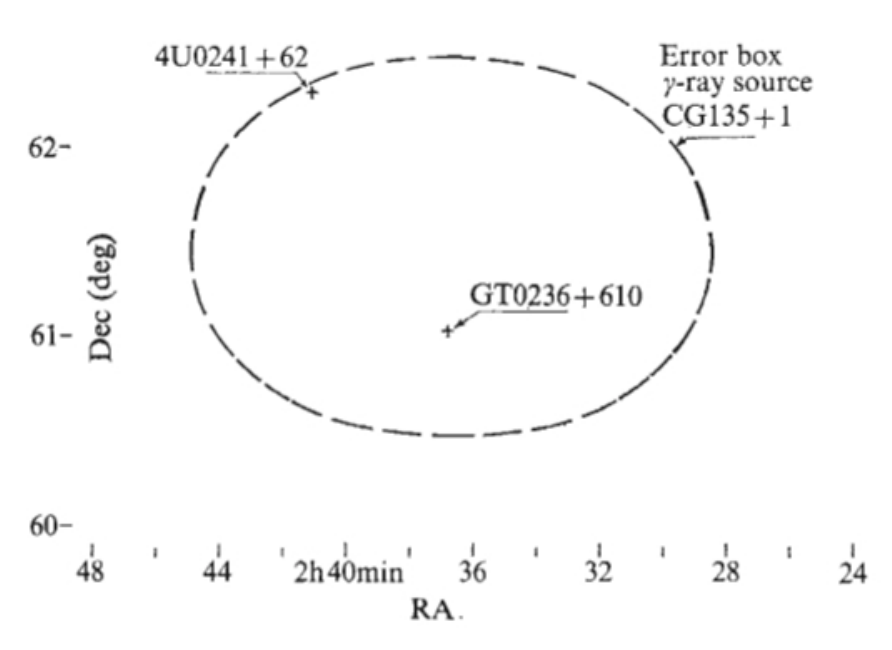}\includegraphics[height=4.75cm]{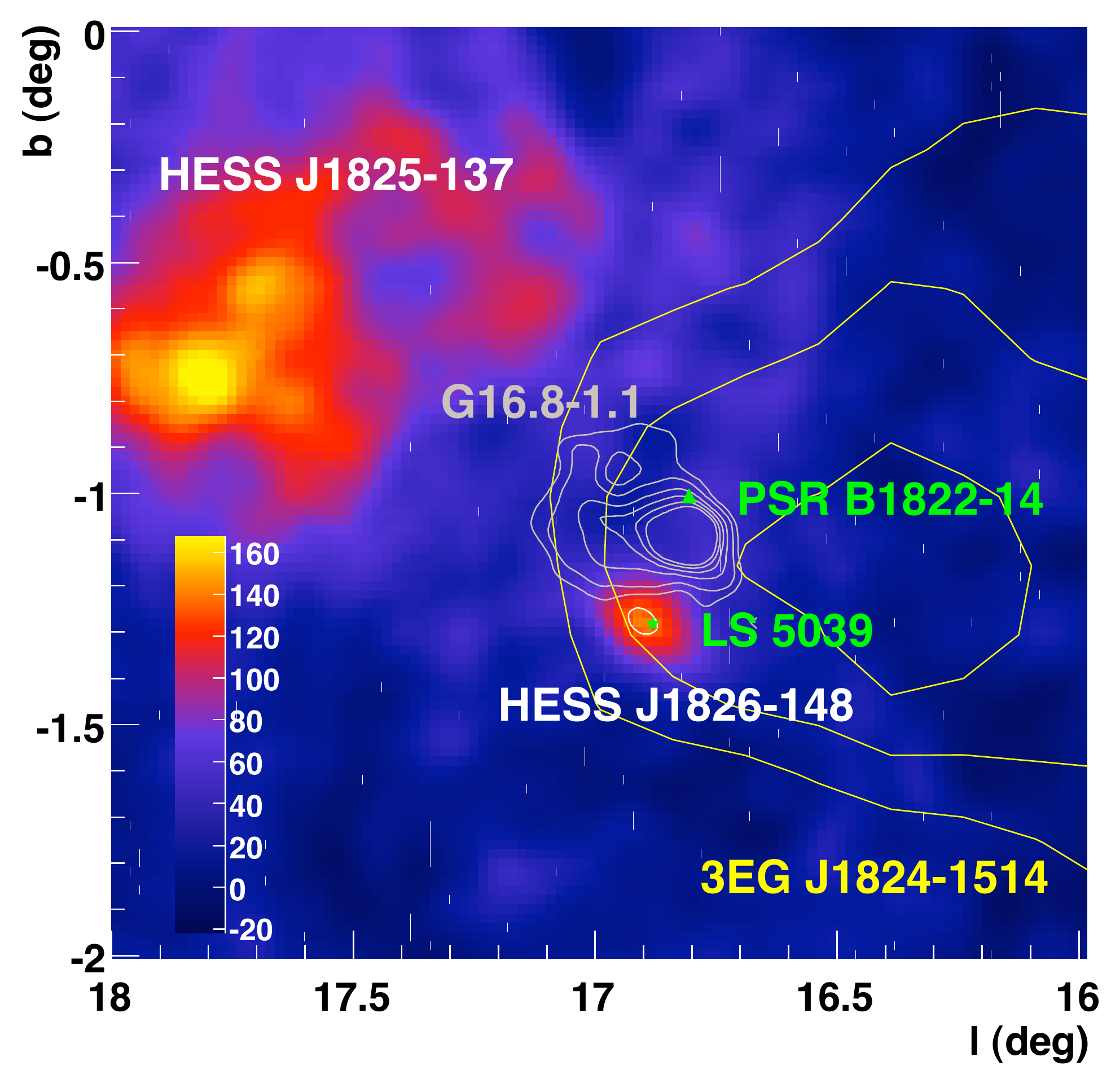}
\caption{Left: The error circle of the {\em Cos B} source 2CG 135+01 showing the possible association with the binary \lsi\ (radio source GT0236+610, reprinted from \citealt{Gregory:1978or}, by permission from Nature \copyright Macmillan Publishers Ltd 1978). Right: VHE gamma-ray map illustrating the detection of LS 5039. The nearby supernova remnant (grey contours) and pulsar are excluded by the high angular resolution of the VHE observations.  The yellow contours show the positional uncertainty of the possible EGRET counterpart in HE gamma rays (reprinted from \citealt{Aharonian:2005nj} with permission from AAAS).}
\label{fig:maps}
\end{figure}

Binaries have now firmly been identified as gamma ray sources. The latest observatories have much improved sensitivity and angular resolution (Fig.~\ref{fig:maps}). The gamma-ray sources are point-like, with any extension constrained to less than an arcminute, and localised to within 20\arcsec\ of their stellar counterpart. They have been consistently detected by various groups. Crucially, all of them show gamma-ray variability on the orbital period timescale.

At the time of writing, all five  binaries\footnote{\psrb, \ls, \lsi, \hessj, \fgl} with secure detections in VHE gamma-rays are ``gamma-ray binaries": bright in gamma-rays yet easily overlooked at other wavelengths. The second part of this review (\S2) summarises the wealth of observational data that has been gathered on these sources. What makes them so bright in gamma rays ? The current prevailing idea is that their non-thermal emission is due to particles accelerated at the shock between the wind of the massive star and the wind of a pulsar. Hence, gamma-ray emission is ultimately powered by the spindown of a rotating neutron star with a strong magnetic field $\sim 10^{11}-10^{13}$\,G. They are pulsar wind nebulae in a binary environment. The third part (\S3) discusses the evidence in favour of this interpretation and its main alternative: gamma-ray emission from a relativistic jet. The fourth part (\S4) presents the theoretical work that has been pursued to understand the observations -- especially the HE and VHE orbital modulations -- and derive new insights into the physics of pulsars, concluding on the current problems faced by models. 

This is quite a different picture from the accretion/ejection scenario that had been previously thought to hold the most promise for gamma-ray emission. Microquasars have proven elusive in the gamma-ray domain even with the latest instrumentation.  A new chapter in the long history of the microquasar Cyg X-3 opened up in 2010 when gamma-ray emission from the system was conclusively and simultaneously detected with both {\em AGILE} and \fermi.  Cyg X-3 is a superb window into how non-thermal processes connect the release of accretion power with the launch of relativistic jets. Recent years have seen surprises with the discovery of gamma-ray emission associated with nova eruptions in binaries, and from the colliding wind binary $\eta$ Car. Many of the  tools developed for gamma-ray binaries are relevant to the interpretation of these systems. The observations, interpretation and the understanding derived thereof are described in the last part (\S5). Observational and theoretical prospects of research on gamma-ray emission from binary systems are discussed in the conclusion.

\section{Observations of gamma-ray binaries\label{parti}}

Gamma-ray binaries are systems composed of a massive star and a compact object where the non-thermal emission peaks above 1 MeV in a $\nu F_{\nu}$ spectral luminosity diagram. There is also a strong contribution to the spectral energy distribution from the luminous massive star, but this is easily separated from the non-thermal continuum as it is thermal with a maximum temperature of a few eV. The definition is simple and based on observational features. The discovery and orbits of these binaries are described in \S\ref{systems}, the salient multi-wavelength facts are listed in \S\ref{mwl}, by wavelength domain instead of by object to highlight common properties. The non-specialist reader will find the main properties briefly summarised in Tab.~\ref{mwltable} (\S\ref{summwl}). The conclusion is that these binaries do indeed appear to be members of a common class. Hence, the definition also unites binaries where the same basic mechanism is at work, distinguishing them from the other binaries that emit in gamma rays.

\subsection{Binary systems\label{systems}}

\subsubsection{Initial discovery in gamma rays}
Three of the binaries (\lsi, \hessj, \fgl) were unknown before gamma-ray sources brought attention to them. The other two had been singled out by surveys with different objectives than finding sources of gamma-ray radiation. All are within $\approx$1\degr\ of the Galactic Plane. 

\paragraph{\psrb} was the first binary to be conclusively detected in VHE gamma rays, in 2004 by the \citet{Aharonian:2005br}. \psrb\ had initially been found in a search for radio pulsars \citep{1992ApJ...387L..37J} and predicted to be a gamma-ray source \citep{Tavani:1997wv,Kirk:1999hr}.

\paragraph{\ls} followed in 2005 (Fig.~\ref{fig:maps}, \citealt{Aharonian:2005nj}). \citet{Motch:1997qk} were the first to identify \ls\ as a high-mass X-ray binary, from a cross-correlation of  unidentified {\em ROSAT} X-ray sources with OB star catalogues.  \citet{2000Sci...288.2340P} pointed out that LS 5039 was within the 0.5\degr\ error box of the EGRET source 3EG 1824-1314, earmarking it as a possible gamma-ray source (Fig.~\ref{fig:maps}). 

\paragraph{\lsi} was the third gamma-ray binary detected in VHE gamma rays (\citealt{Albert:2006wi}). The 8.7$\sigma$ detection, 2\arcmin\ localisation and variability of the source (later associated with the orbit, \citealt{2009ApJ...693..303A}) confirmed \lsi\ is a gamma-ray emitter, more than 25 years after this had been proposed based on the {\em Cos B} detection (see \S\ref{introduction} and Fig.~\ref{fig:maps}).

\paragraph{\hessj} was found serendipitously as a point-like source in the HESS survey of the Galactic Plane \citep{2007A&A...469L...1A}. Part of the HESS data had actually been collected to search (unsuccessfully) for gamma-ray emission from the nearby massive X-ray binary SAX J0635+0533, which had a proposed EGRET counterpart \citep{1999ApJ...523..197K}. The VHE source is coincident with a variable radio source associated with a Be star (MWC 148), much like \lsi\ \citep{2007A&A...469L...1A,Hinton:2008eg, 2009MNRAS.399..317S}. The binary orbital period was found later from X-ray monitoring and tied to the VHE variability observed with HESS, MAGIC and VERITAS  \citep{2011ApJ...737L..11B}.

\paragraph{\fgl} was found by \citet{2011ATel.3221....1C} in a search for periodic flux variations from \fermi\ sources. Follow-up work proved that the HE gamma-ray source is associated with a radio, X-ray source and a O massive star  \citep{2012Sci...335..189F}. \citet{2012A&A...541A...5H} independently discovered a VHE source positionally coincident with the \fermi\ source.\\

The Be star MWC 656 has been proposed as a counterpart to the transient gamma-ray source  AGL J2241+4454 \citep{2010ApJ...723L..93W,2012MNRAS.421.1103C}. However, no additional information has been provided on the {\em AGILE} detection besides the ATel \citep{2010ATel.2761....1L}, and the source was not detected with \fermi\ (Fermi \href{http://fermisky.blogspot.fr/2010_07_01_archive.html}{\color{violet}{blog}}), so it is not discussed further here.

\subsubsection{The massive star companion\label{wind}}
All five binary systems harbour a massive O or Be star, with a mass of 10-20 \msol, a radius of $\approx$ 10 R$_{\odot}$, and a photosphere temperature of 20\,000 -- 40\,000 K (Tab.~\ref{parameters}). The luminosity of the star reaches 0.5--5$\times10^{38}$ erg s$^{-1}$,  a significant fraction of its Eddington luminosity. Hence, massive stars drive strong radiation-driven winds, with supersonic speeds of 1500--2500 km s$^{-1}$ and mass loss rates from 10$^{-8}$ up to 10$^{-5}$ \msol\,yr$^{-1}$ in the case of Wolf-Rayet stars. This wind is, to first approximation, isotropic. In addition to this radiation-driven wind, Be stars have a dense equatorial outflow responsible for line emission in optical and excess continuum emission in infrared \citep{2003PASP..115.1153P}. Interferometric observations have established that the Be discs are thin and that they are keplerian. The infrared excess is modelled as thermal bremsstrahlung emission from material distributed as a power-law function of radius ($\rho\propto r^{-n}$, $n\approx 2-3$) with a subsonic radial outflow speed of a few $\sim$1--10 km s$^{-1}$ \citep{1986A&A...162..121W,1988A&A...198..200W}. The results imply mass outflow rates of order $10^{-6}$\,\msol\,yr$^{-1}$. The formation of Be  discs is thought to be linked to the near breakup rotation velocity of the star (88\% of $\Omega_{\rm break}$ for the Be star companion of \psrb, \citealt{2011ApJ...732L..11N}), with material possibly lifted up by stellar pulsations. Accurate measurements of the stellar and wind parameters rely on models of massive stars. Some parameters, such as the stellar temperature $T_{\star}$ or wind speed $v_{\rm w}$ are reasonably well determined using optical/UV spectra. Considerable uncertainties remain on many others, notably the wind mass loss rate $\dot{M}_{\rm w}$.

\begin{table}
\ra{1.27}
\begin{tabular}{@{}llllll@{}}
\toprule 
 & \psrb$^{\star}$ & \ls$^{\dagger}$ & \lsi$^{\bullet}$ & \hessj$^{\diamond}$ & \fgl$^{\ddagger}$ \\
 \midrule
 P$_{\rm orb}$ (days) & 1236.72432(2) 	&  3.90603(8) 		& 26.496(3) 	& 315(5) 		& 16.58(2) \\
 $e$ 				& 0.8698872(9) 	&  0.35(3) 		& 0.54(3) 		& 0.83(8) 		& - \\
 $\omega$ (\degr)	& 138.6659(1)$^\sharp$ 		&  212(5) 		& 41(6) 		& 129(17)  & - \\
 $i$ (\degr)		& 19--31 			& 13--64 			& 10--60 		& 47--80 		& -\\
 $d$ (kpc)			& 2.3(4) 			& 2.9(8) 			& 2.0(2) 		& 1.6(2)  		& 5.4\\
 \midrule
spectral type 			& O9.5Ve 			& O6.5V((f))		& B0Ve 		& B0Vpe 		& O6V((f)) \\
 $M_{\star}$ (M$_{\odot}$)  	& 31 	& 23 			& 12 		& 16 		& 31  \\
$R_{\star}$ (R$_{\odot}$) 	& 9.2 	& 9.3 			& 10 		& 8 			& 10.1 \\
$T_{\star}$ (K) 				&  33500 & 39000 			& 22500 		& 30000 		& 38900 \\
 \midrule
$d_{\rm periastron}$ (AU) 	& 0.94		& 0.09  	& 0.19 & 0.40 & (0.41)\\
$d_{\rm apastron}$ (AU) 		& 13.4		& 0.19 	& 0.64 & 4.35 & (0.41)\\
$\phi_{\rm periastron}$ 		& 0		& 0		& 0.23	& 0.967 & - \\
$\phi_{\rm sup.\ conj.}$ 		& 0.995	& 0.080	& 0.036	& 0.063 & -\\
$\phi_{\rm inf.\ conj.}$ 		& 0.048	& 0.769	& 0.267	& 0.961 & -\\
 \bottomrule 
\multicolumn{6}{l}{ $\star$ \citet{Wang:2004gt,2011ApJ...732L..10M,2011ApJ...732L..11N}}\\
\multicolumn{6}{l}{ $\dagger$ \citet{2004ApJ...600..927M,Casares:2005gg,2011heep.conf..559C}}\\ 
\multicolumn{6}{l}{ $\bullet$ \citet{1983MNRAS.203..801H,1991AJ....101.2126F,1995AA...298..151M,2002ApJ...575..427G,2009ApJ...698..514A}}\\
\multicolumn{6}{l}{ $\diamond$ \citet{2010ApJ...724..306A,2012MNRAS.421.1103C,2012arXiv1212.0350B}}\\
\multicolumn{6}{l}{$\ddagger$ \citet{2012Sci...335..189F,2011PASP..123.1262N}}\\
\multicolumn{6}{l}{$\sharp$ argument of periastron of the pulsar orbit (massive star for the others systems)}
\end{tabular}
\caption{System parameters of gamma-ray binaries\label{parameters}}
\end{table}

\begin{figure*}
\center
\includegraphics[width=5cm]{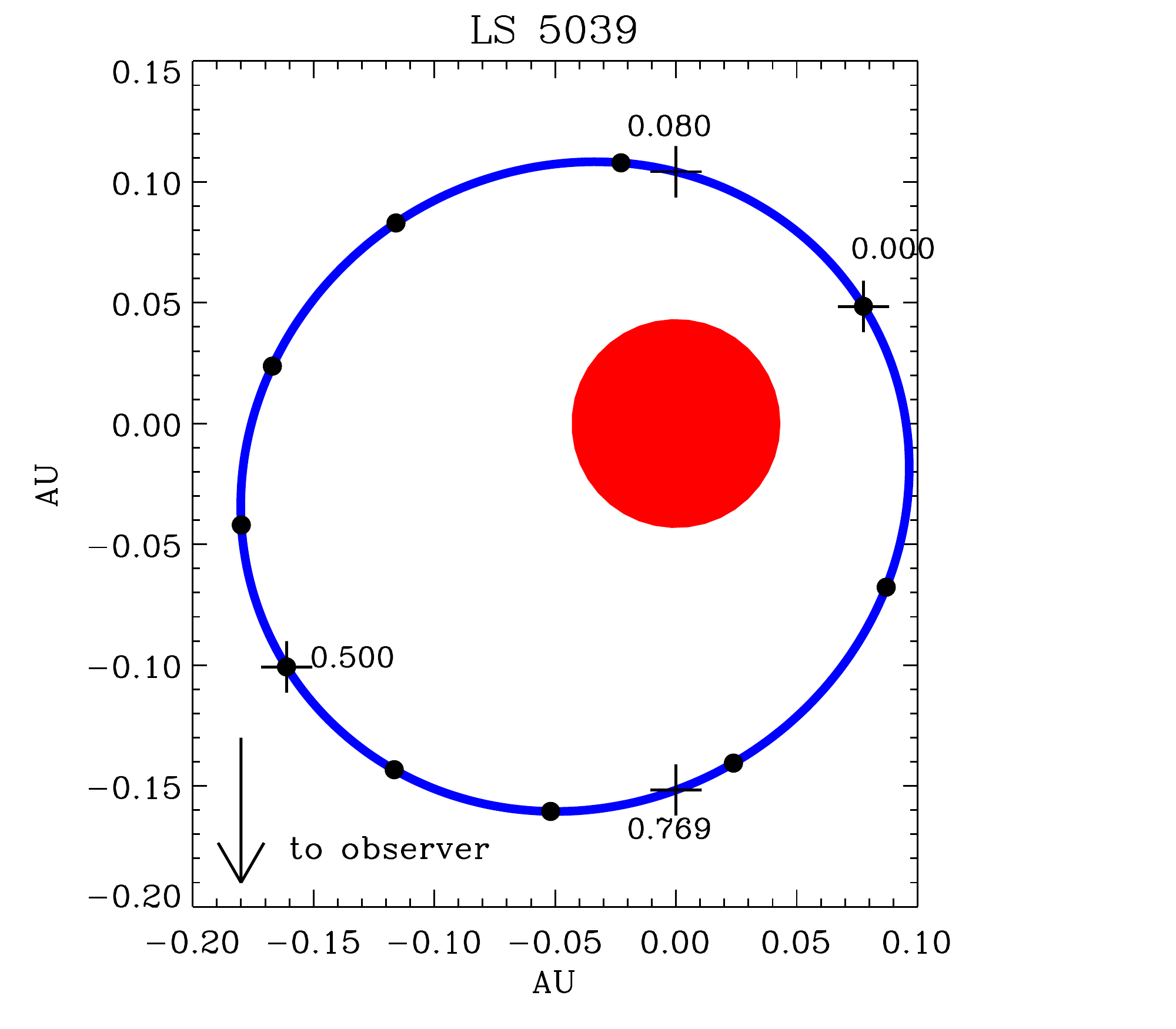}
\includegraphics[width=5cm]{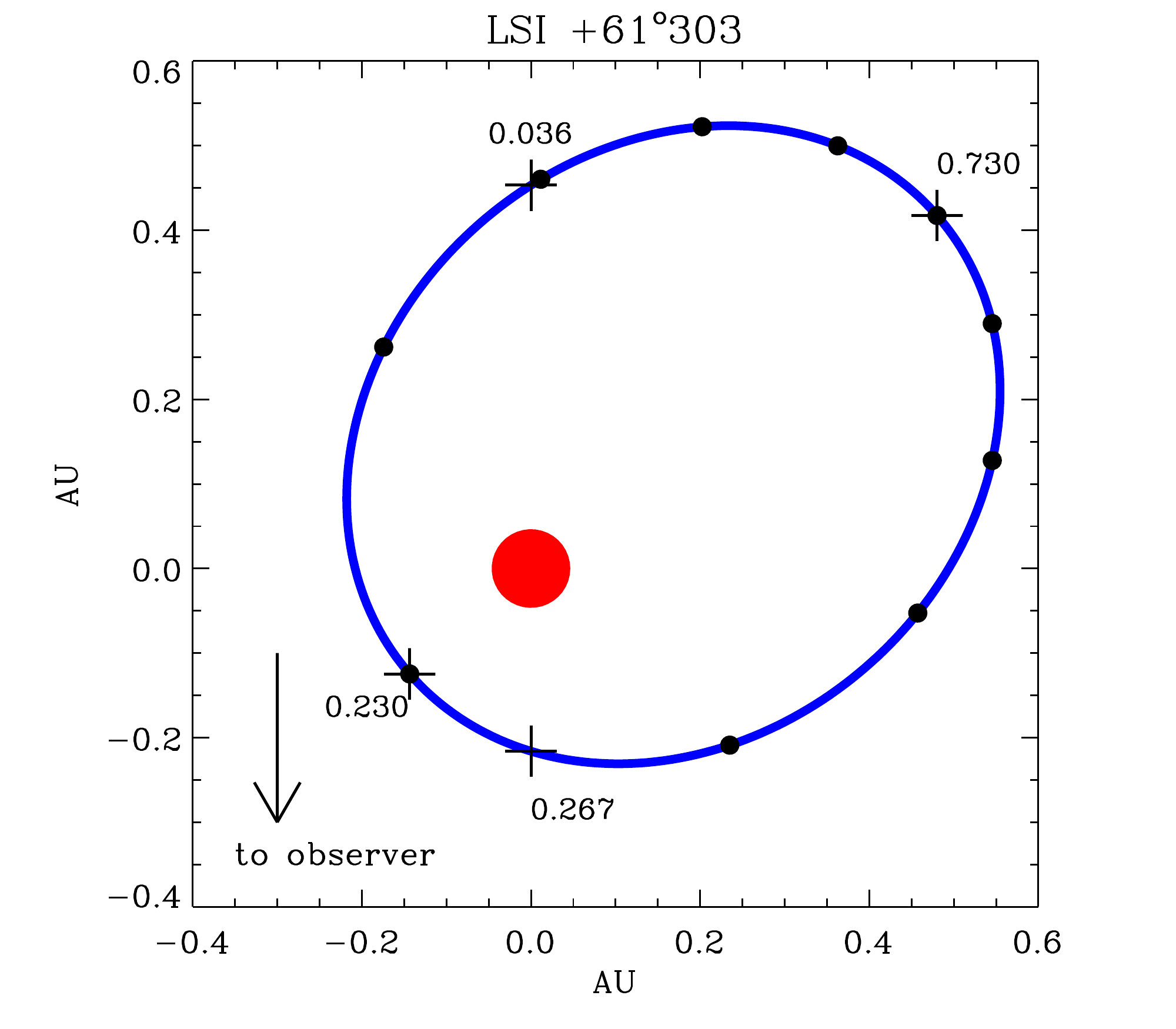}
\includegraphics[width=5cm]{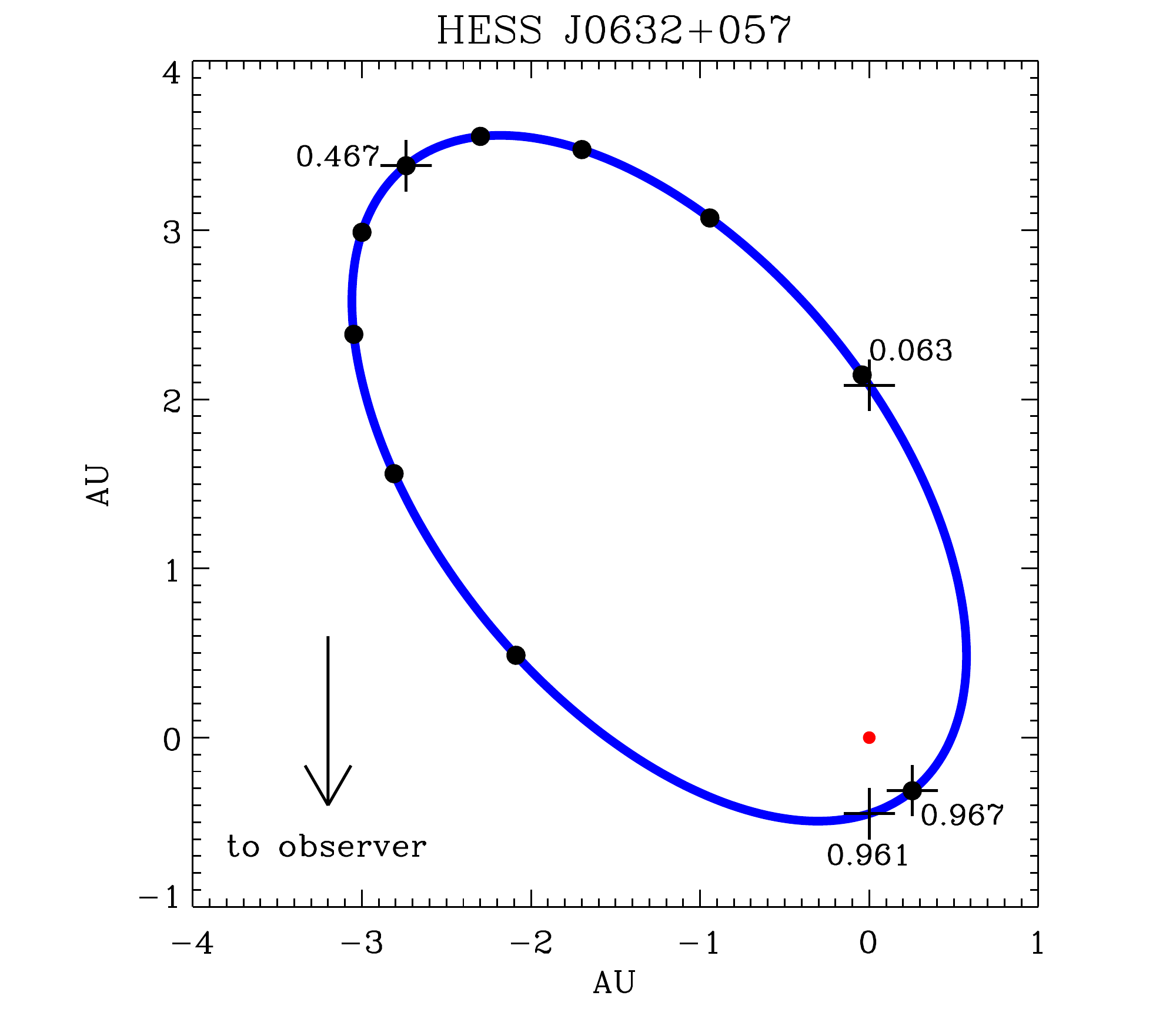}
\includegraphics[width=5cm]{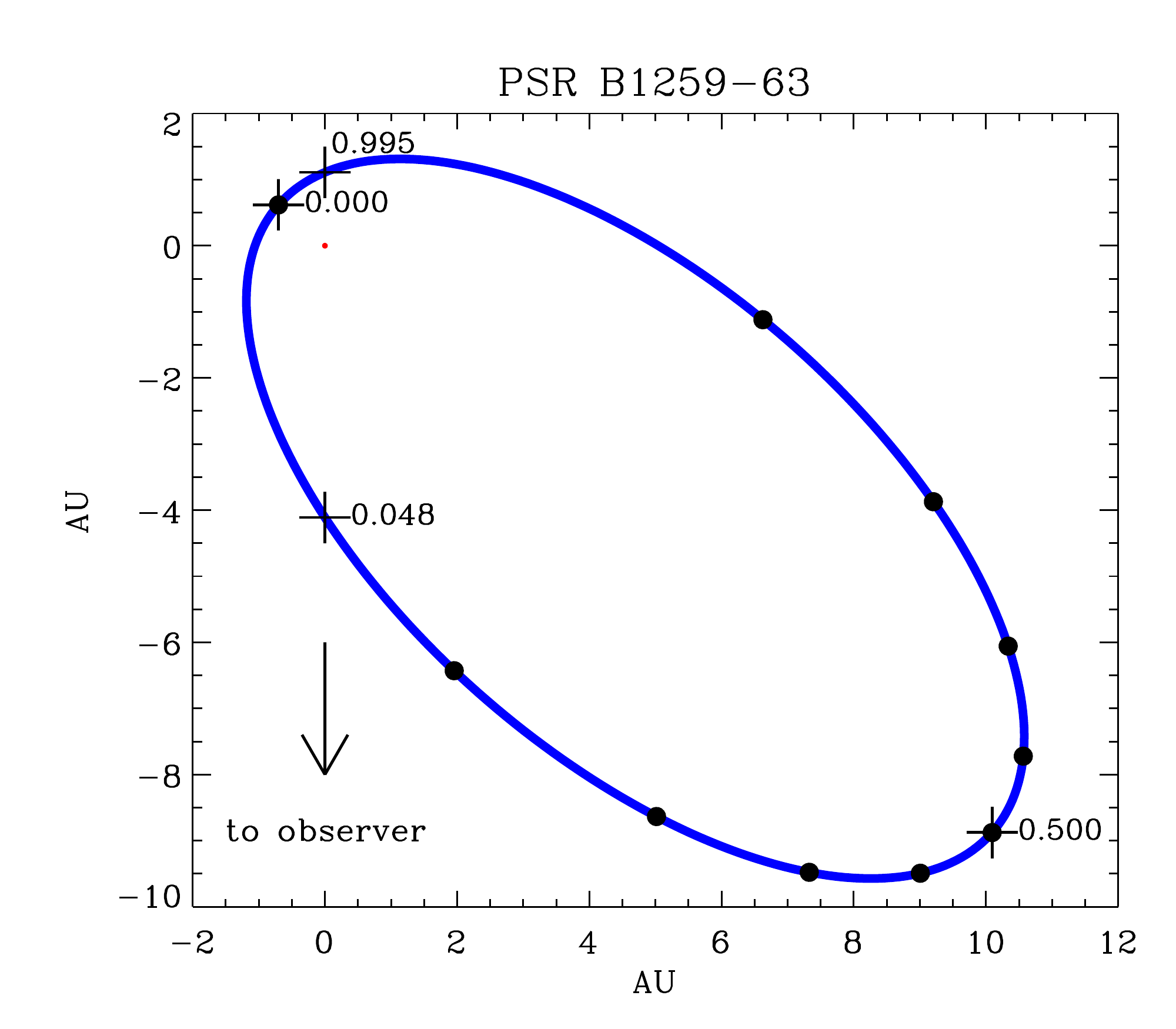}
\caption{Orbits of gamma-ray binaries. The star size is to scale. Crosses mark the phases of apastron/periastron and conjunctions (superior when the compact object is seen behind the massive star, inferior when in front, with the observer towards bottom of the figure, see Tab.~\ref{parameters}). Dots mark intervals of 0.1 in orbital phase, starting from periastron.\label{fig:orbit}}
\end{figure*}

\subsubsection{Orbits\label{orbit}}
The system parameters of gamma-ray binaries are grouped in Table~\ref{parameters} and illustrated in Figure~\ref{fig:orbit}. The orbital sizes have been calculated assuming a neutron star of 1.4\,\msol. \psrb\ has the best determined orbital parameters, thanks to timing of its 47.76 ms radio pulsar. Analysis of the pulse time-of-arrival is fitted to an orbital solution, giving the orbital period $P_{\rm orb}$, eccentricity $e$, projected semi-major axis $a \sin i$, argument of periastron $\omega$ and time of periastron passage $t_{0}$ to very high accuracy.  Orbital solutions are also obtained using measurements of the radial velocity of absorption lines originating at or close to the massive star photosphere, albeit with much poorer accuracy. The O/Be star is clearly the most massive component so the amplitude of the variations are small and easily contaminated by fluctuations from the intervening stellar wind. Long orbital periods and high eccentricities also make the detection of the orbital reflex motion difficult. In \lsi, \hessj, and \fgl, the binary orbital period is best determined by the modulation observed in radio, X-ray, or gamma rays.

The accurate determination of the orbital parameters is of fundamental importance to models. The errors on the last significant digit are indicated in parenthesis in Table~\ref{parameters} --- these are only indicative since more important systematic errors are possible. For instance, the orbital elements for \hessj\ remain tentative given the limited phase coverage and large scatter in measurements (Casares, priv. com.). As will be discussed later (\S\ref{modulations}), knowing the phases  of periastron/apastron passage and of conjunctions is essential to interpret the orbital modulations. Superior conjunction is when the compact object passes behind the massive star as seen by the observer, inferior is when the compact object passes in front of the O/Be star. The phase of periastron passage in Table~\ref{parameters} is different from 0 for \lsi\ (resp. \hessj) because the reference time has traditionally been set by using the radio (resp. X-ray) lightcurve. The orbital phase $\phi$ is defined in the interval $[0,1]$.

The non-detection of pulsed emission from gamma-ray binaries (\psrb\ excepted, see \S\ref{pulsations}) leaves open the question of the nature of the compact object. Determining the compact object mass could distinguish between a black hole candidate and a neutron star. The mass function derived from the radial velocity of the O/Be star gives a lower limit on the mass of the compact object, which is too small to be constraining by itself. A 1.4\,\msol\ neutron star fits the mass function within the loose constraints on the mass of the companion star and the orbit inclination. A $>$3\,\msol\ compact star (black hole) requires  a low binary inclination (face-on), typically $i<30$\degr. However, low inclinations are statistically disfavoured, assuming the systems that we see have random inclinations. The lack of eclipses can be used to place an upper limit on the inclination, typically $i<70$\degr. At the other end, the  rotational broadening of the stellar lines yields a lower limit on the inclination, typically $i>10$\degr, assuming the star rotates at less than breakup speed, spin-orbit alignment and pseudo-synchronisation. Table~\ref{parameters} lists the range of possible $i$ derived by various means, as summarised by \citet{2012MNRAS.421.1103C}.

\subsection{The multi-wavelength picture\label{mwl}}
\begin{figure}
\center
\includegraphics[width=6.7cm]{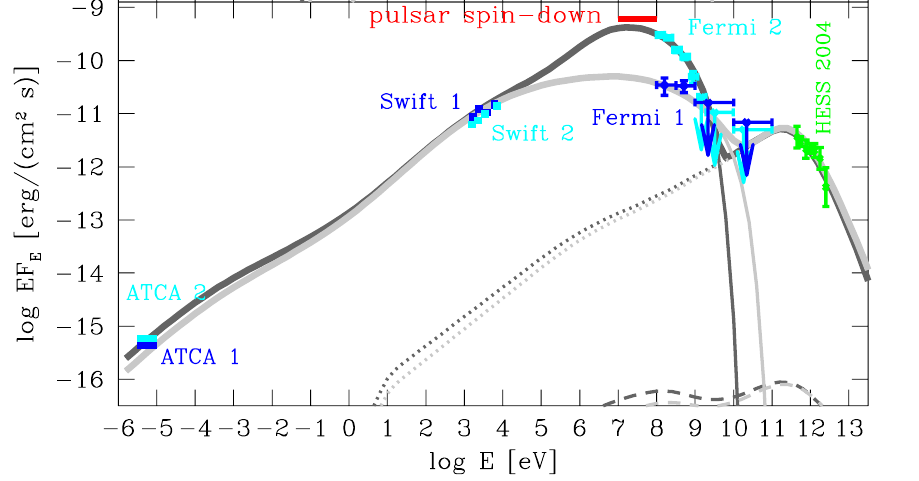}
\includegraphics[width=6.7cm]{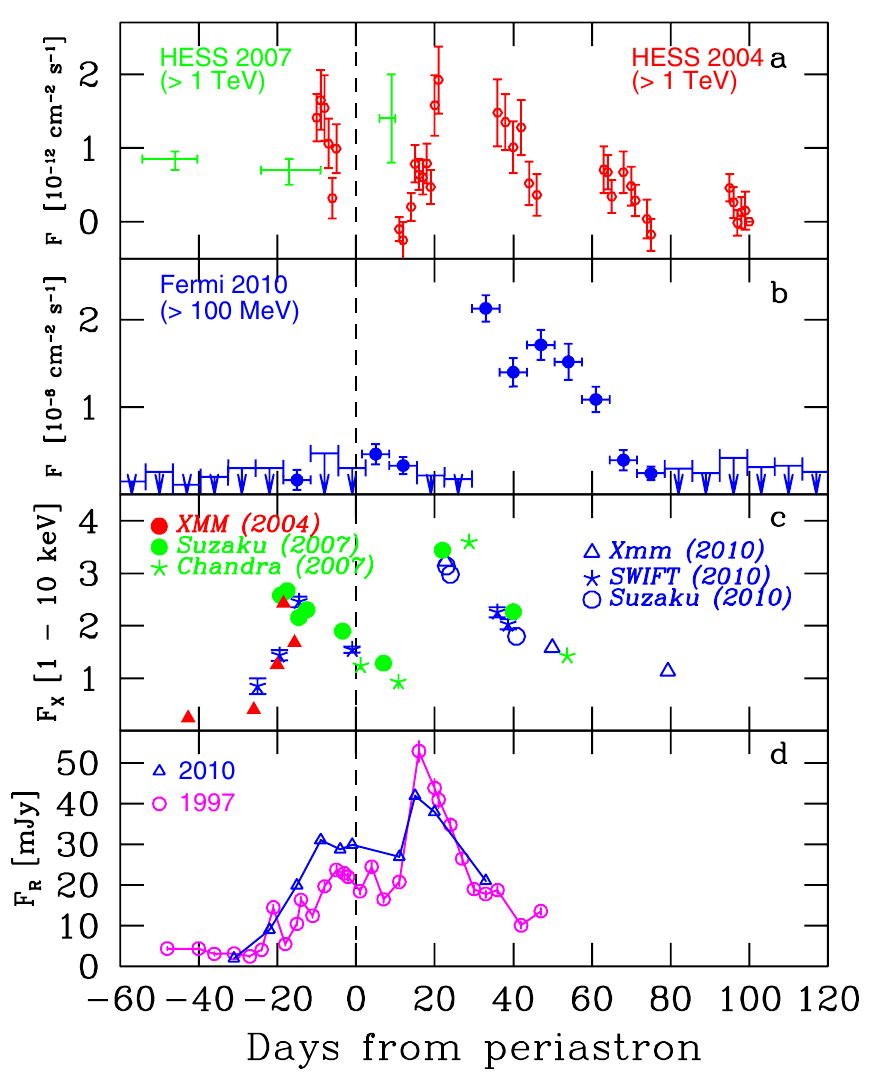}
\caption{Observations of \psrb. From top to bottom: spectral energy distribution, TeV, GeV, X-ray, and radio lightcurves \citep[reproduced by permission of the AAS from][]{2011ApJ...736L..11A}.\label{fig:psrb}}
\end{figure}
\begin{figure}
\centering
\includegraphics[height=4.25cm]{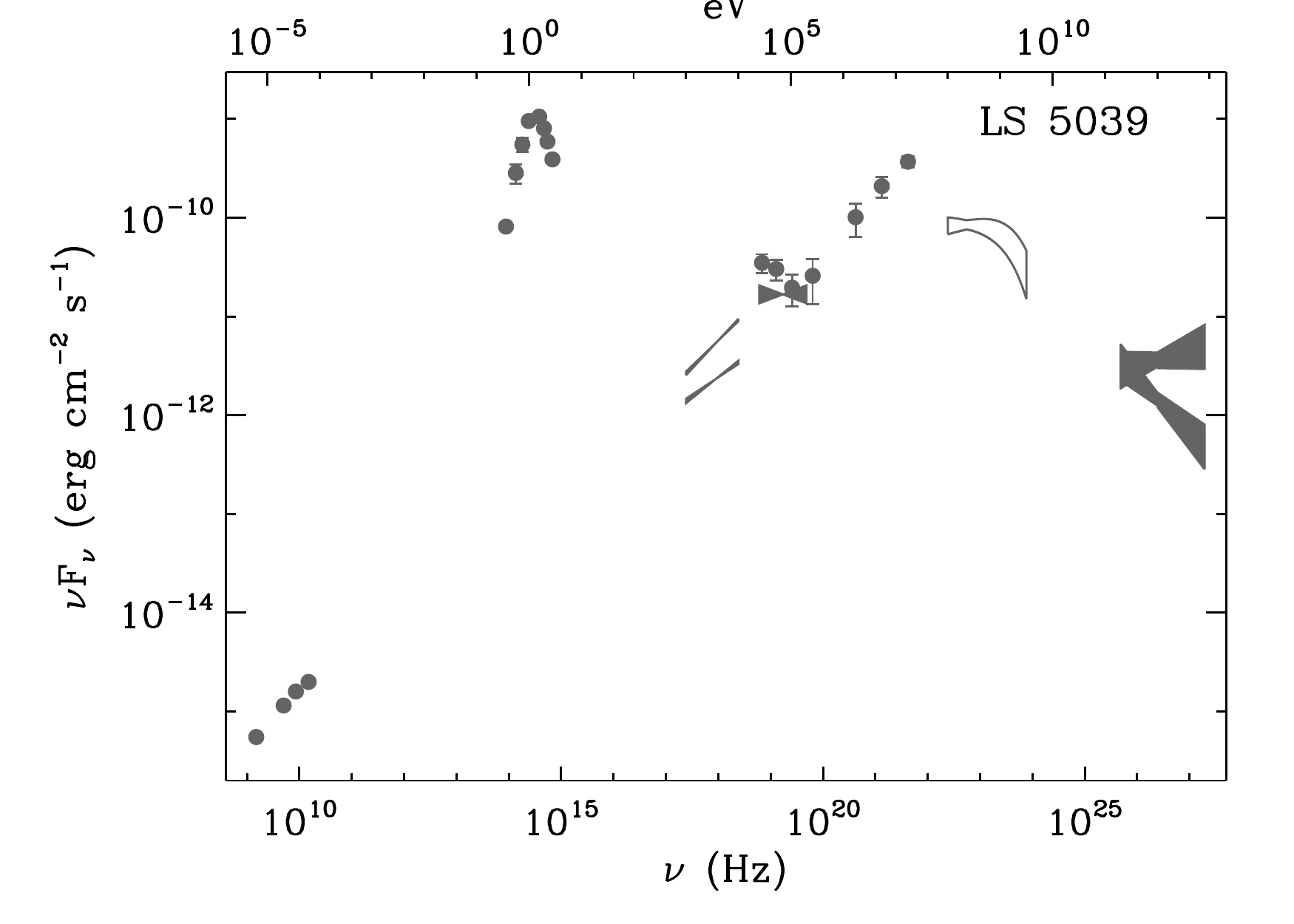}
\includegraphics[height=3.2cm]{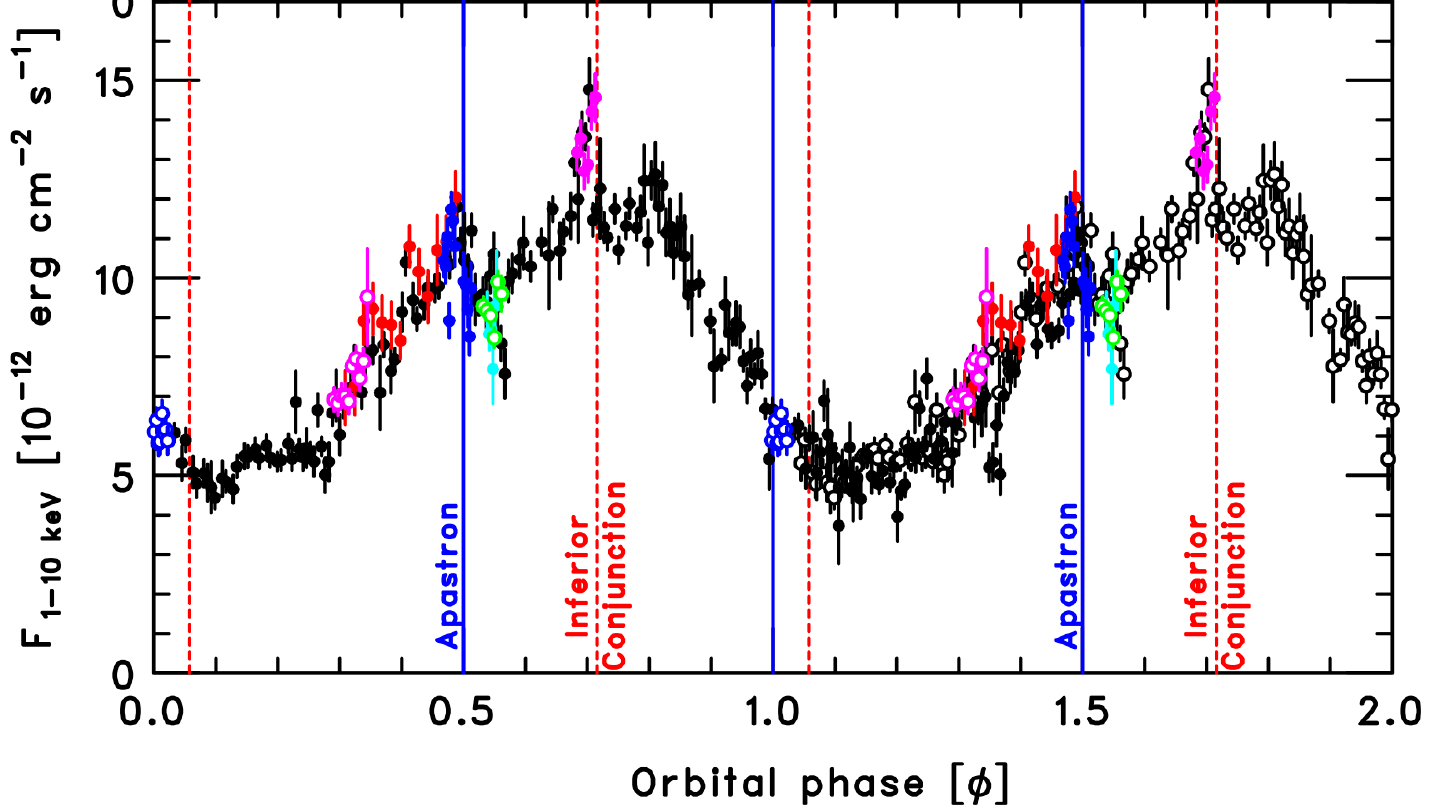}\\
\includegraphics[height=4.5cm]{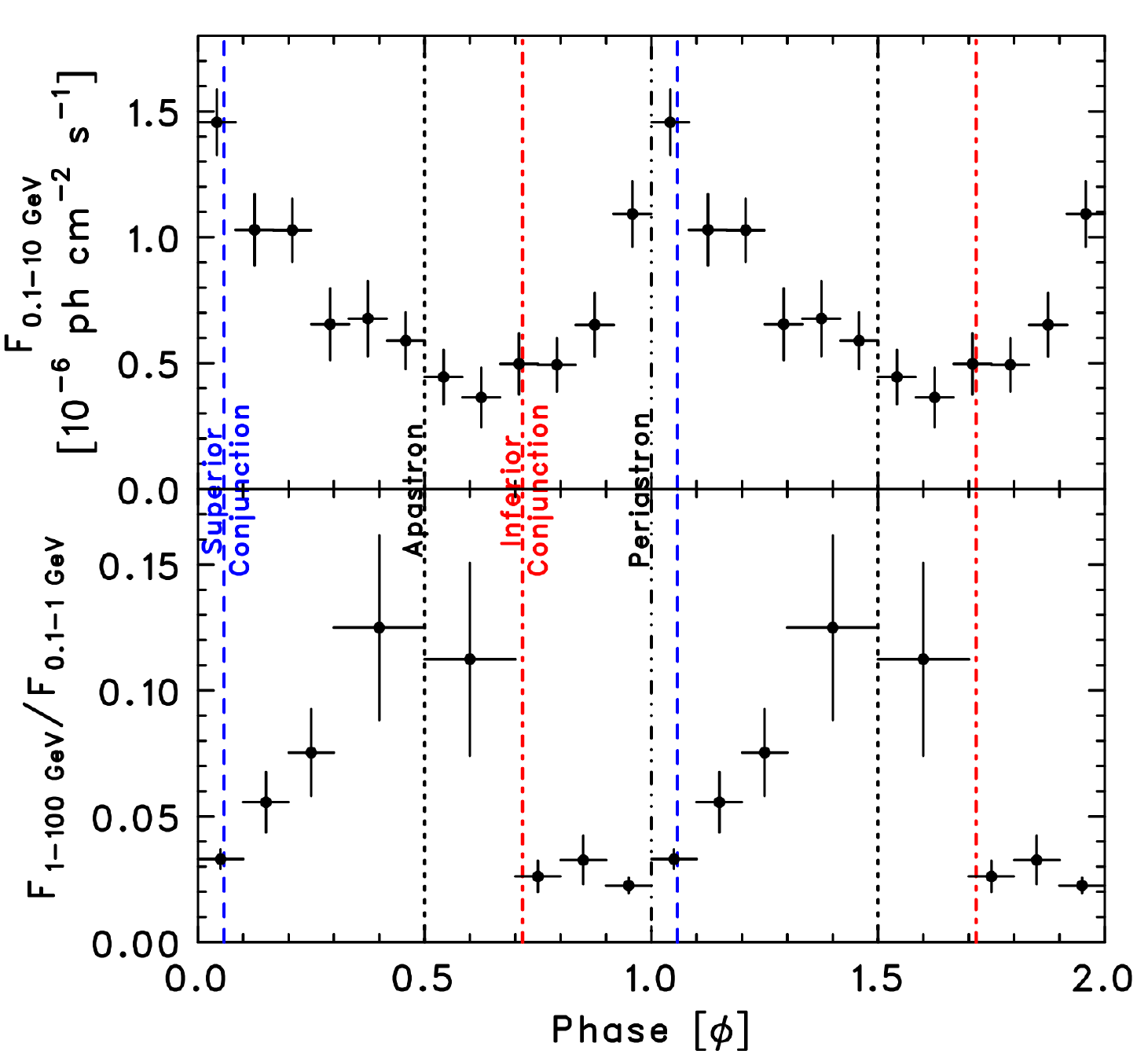}
\includegraphics[height=4.5cm]{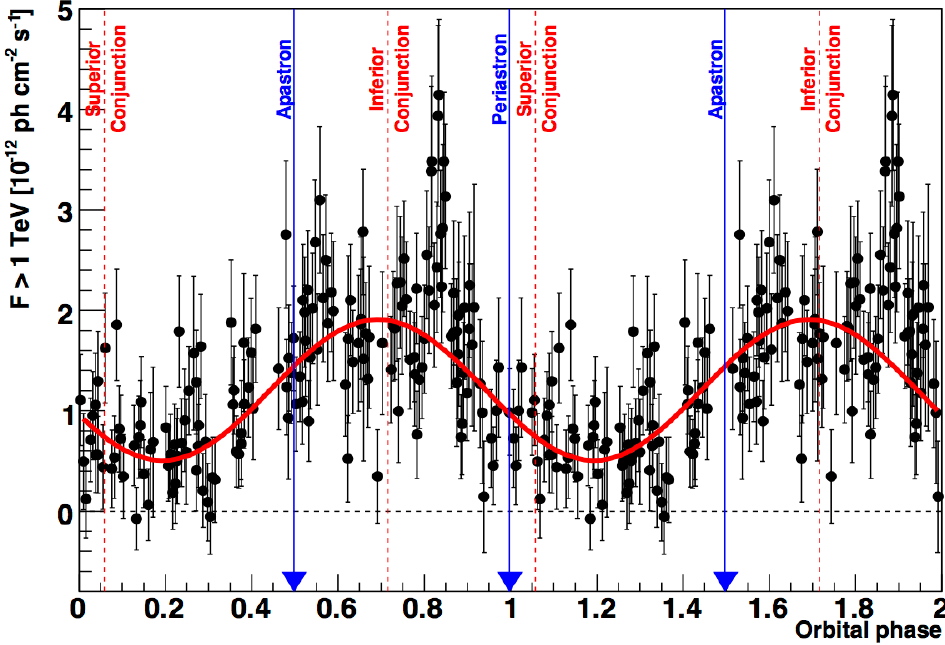}
\caption{Observations of \ls. From top to bottom: spectral energy distribution adapted from \citet{Dubus:2006lc}; X-ray modulation \citep[reproduced by permission of the AAS from][]{2009ApJ...697L...1K}; GeV modulation \citep[reproduced by permission of the AAS from][]{2009ApJ...706L..56A}; TeV modulation (reproduced with permission from \citealt{Aharonian:2005nj} \copyright ESO).\label{fig:ls5}}
\end{figure}
\begin{figure}
\center
\includegraphics[height=4.25cm]{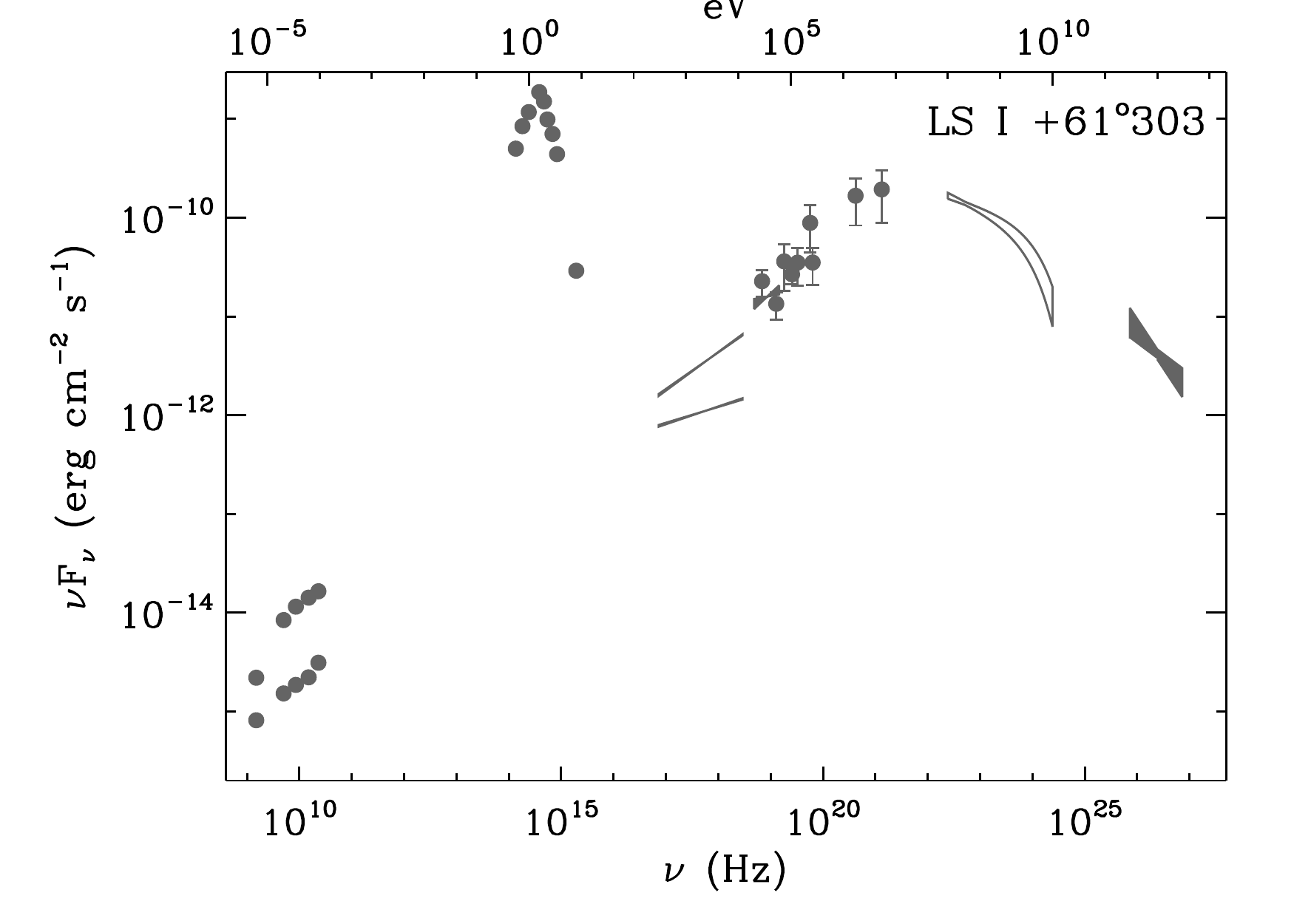}
\includegraphics[height=4.5cm]{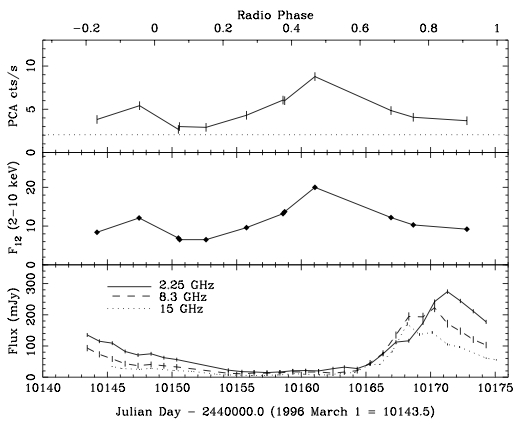}\\
\includegraphics[height=4.5cm]{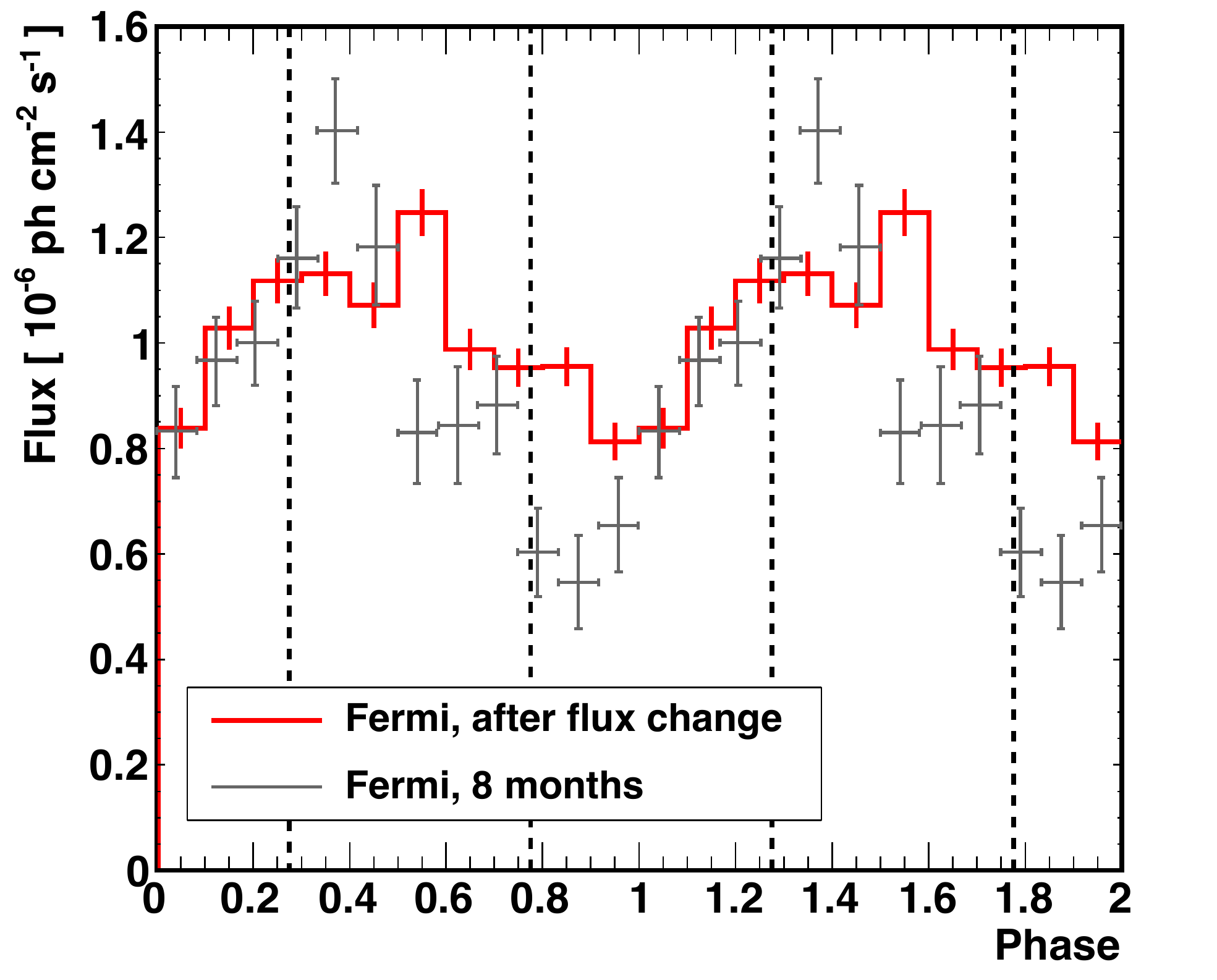}
\includegraphics[height=4.5cm]{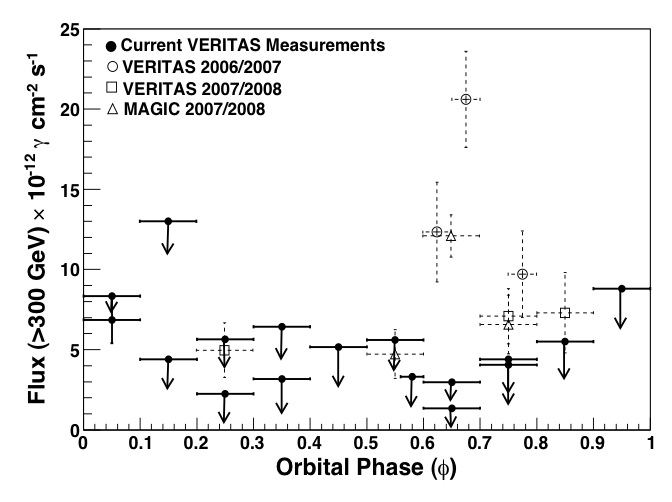}
\caption{Observations of \lsi. From top to bottom: spectral energy distribution (this work); one example of the periodic radio and X-ray outburst that occur once per orbit \citep[reproduced by permission of the AAS from][]{Harrison:2000bk} ; GeV modulation \citep[reproduced by permission of the AAS from][]{2012ApJ...749...54H}; TeV modulation \citep[reproduced by permission of the AAS from][]{2011ApJ...738....3A}. Periastron passage is at $\phi=0.23$.\label{fig:lsi}}
\end{figure}
\begin{figure}
\center
\includegraphics[height=4cm]{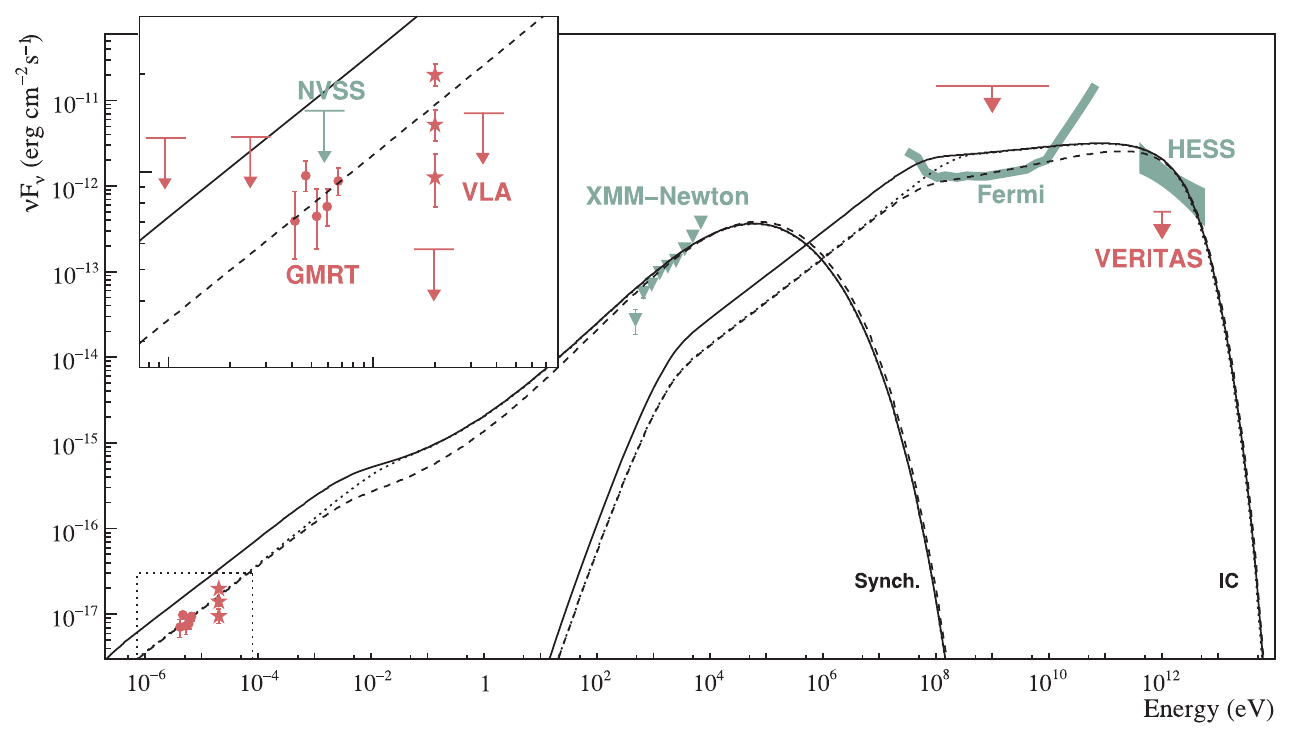}
\includegraphics[height=3.75cm]{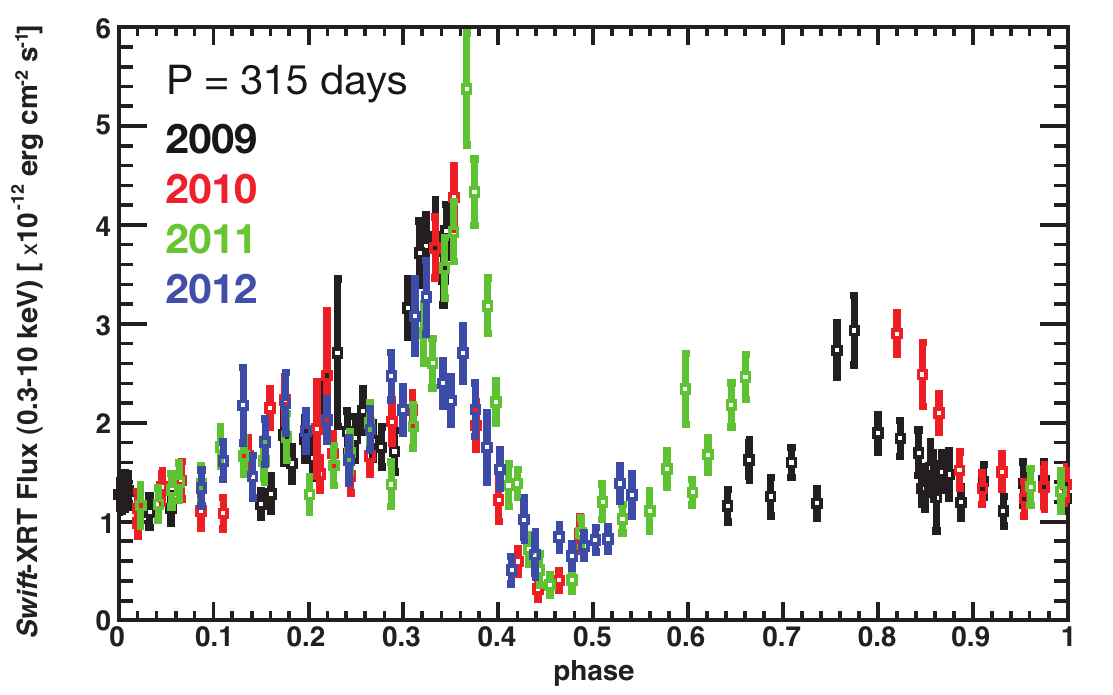}
\includegraphics[height=3.75cm]{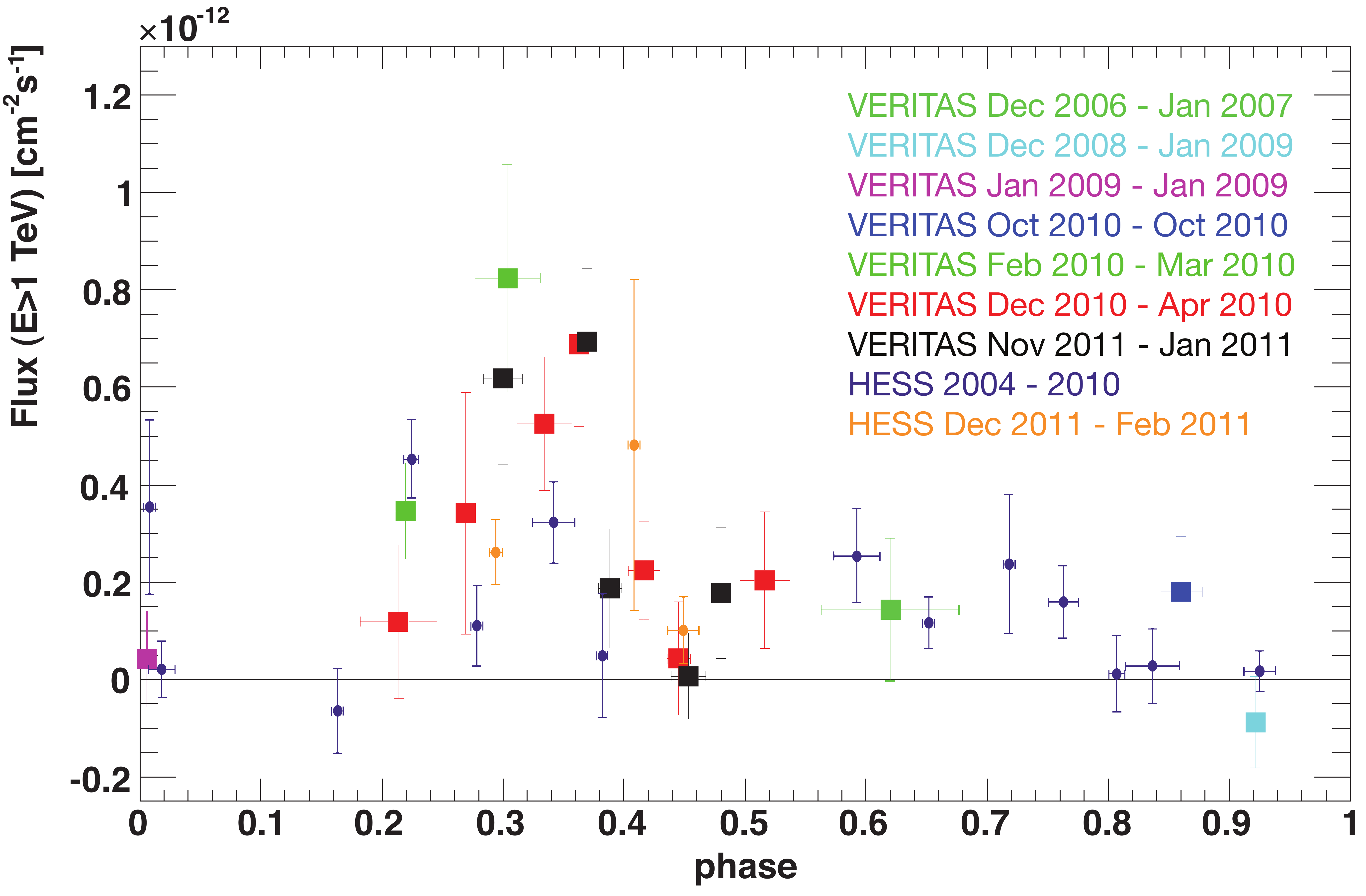}
\caption{Observations of \hessj. From top to bottom: spectral energy distribution \citep[figure reprinted from][by permission of Oxford University Press, on behalf of the RAS]{2009MNRAS.399..317S};  X-ray and TeV lightcurve folded on the 315 day period (reproduced with permission from \citealt{2012arXiv1212.0350B}, \copyright 2012, American Institute of Physics).\label{fig:hessj}}
\end{figure}

\begin{figure}
\center
\includegraphics[height=4.25cm]{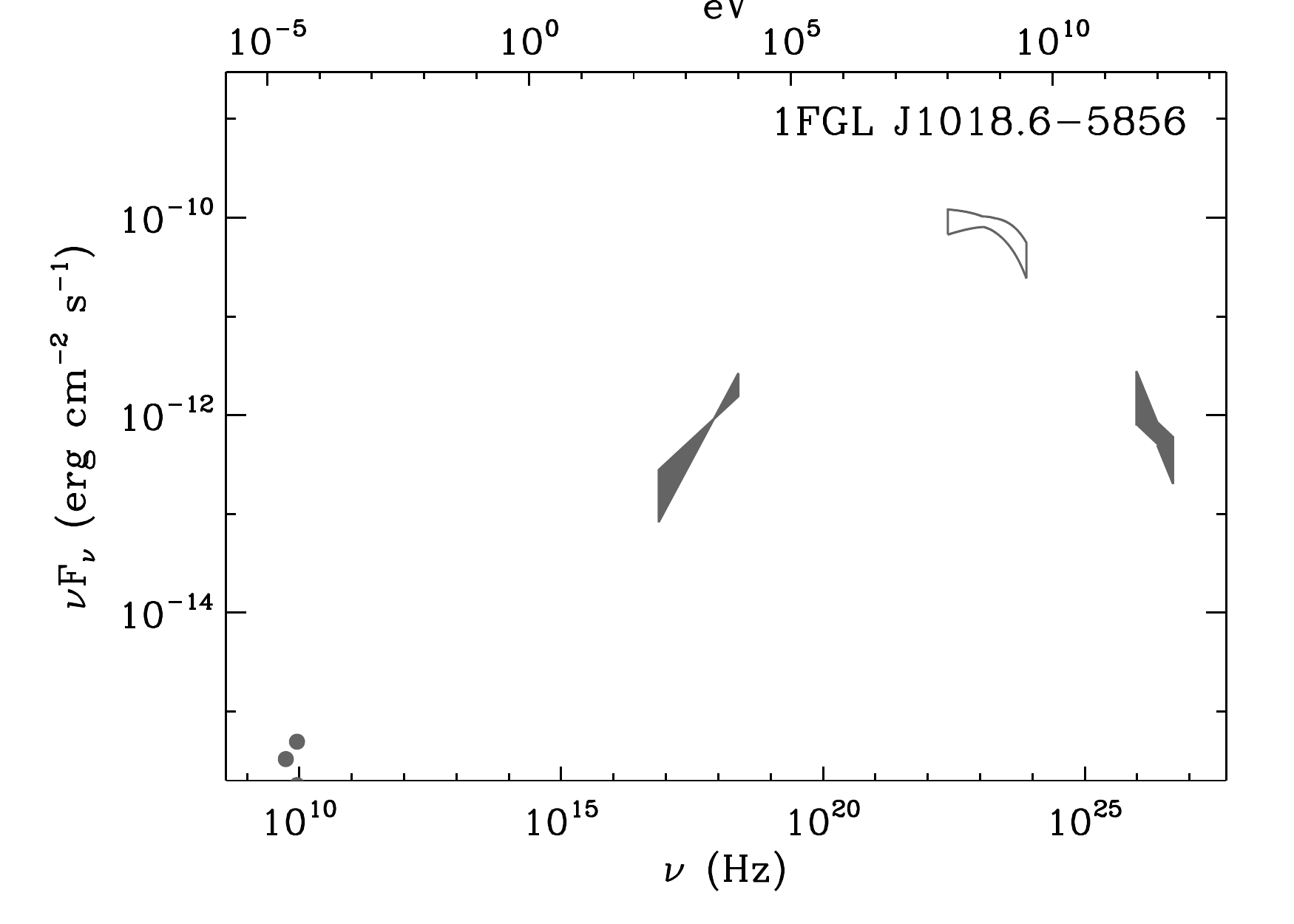}\\
\includegraphics[height=4.5cm]{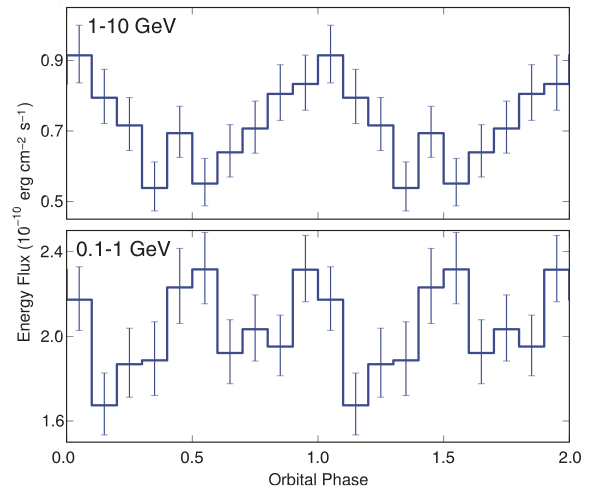}
\includegraphics[height=4.5cm]{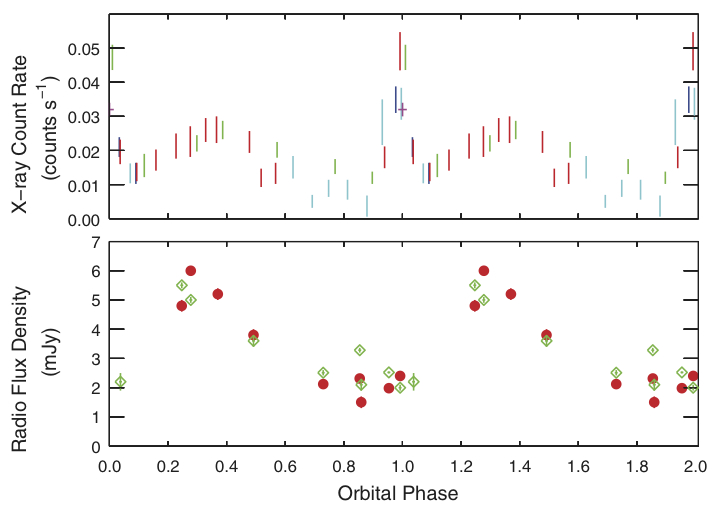}
\caption{Observations of \fgl. From top to bottom: spectral energy distribution (this work), GeV, X-ray, and radio orbital modulations (reprinted from \citealt{2012Sci...335..189F} with permission from AAAS).\label{fig:1fgl}}
\end{figure}

A summary of the main spectral characteristics in the various bands is given below in Table~\ref{mwltable}. Figures~\ref{fig:psrb}-\ref{fig:1fgl} present the spectral energy distribution and lightcurves for the five gamma-ray binaries.
\subsubsection{VHE gamma rays (TeV)\label{vhe}}
All five gamma-ray binaries are detected by IACTs above 100 GeV. The VHE counterparts are point-like, with a typical limit on extended emission $\la  0\fdg1$. Nearly all other VHE sources in the Galactic Plane ($|b|\la 3$\degr) are extended. The HESS Galactic Plane survey led to the discovery of only three point-like sources besides the \gd{Crab and the} VHE source at the Galactic Center: \hessj, \fgl, and HESS J1943+213. The first two are binaries, the last one is likely a blazar \citep{2011A&A...529A..49H}. 

\paragraph{\psrb} has now been detected on three occasions using HESS when the pulsar was in the vicinity of periastron during its 3.5 year orbit (Fig.~\ref{fig:psrb}). The repeatability of the detection at this orbital phase is proof of the association of the VHE source with the binary. The earliest detection occurred 55 days before and the latest detection occurred 100 days after periastron. The VHE lighcurve also shows variability on timescales of days, but sampling has been limited due to the observing constraints of IACTs (moonless nights). Observations away from periastron have only yielded upper limits. The three different epochs have yielded consistent results for the average spectrum: a power-law with a photon index $\Gamma_{\rm VHE}\approx 2.7$ and a normalisation at 1 TeV of $1.3 \times 10^{-12}\,\mathrm{TeV}^{-1} \mathrm{cm}^{-2}\,\mathrm{s}^{-1}$ \citep{Aharonian:2005br,2009A&A...507..389A,H.E.S.S.Collaboration:2013fk}

\paragraph{\ls} was detected in the HESS Galactic Plane survey \citep{Aharonian:2005nj}. A Lomb-Scargle periodogram of the VHE lightcurve gives a period of 3.90678$\pm$0.0015 days that  corresponds to the orbital period determined independently using radial velocity measurements \citep{Aharonian:2006qw}. The minimum is close to superior conjunction or to periastron (the two phases are separated by only $\Delta\phi=0.057$). Maximum flux occurs around inferior conjunction (Fig.~\ref{fig:ls5}). Spectral variability is detected between superior and inferior conjunction (``SUPC'' $\phi \leq 0.45$ and $\phi \geq 0.9$, ``INFC'' $0.45\leq \phi\leq0.9$). At INFC, the best fit spectrum is a power law with $\Gamma_{\rm VHE}\approx 1.8$ and an exponential cutoff at $E_{c}\approx 8.7$\,TeV.  At SUPC, the source is fainter and best described by a single power-law with a softer index $\Gamma_{\rm VHE}\approx 2.5$. The average normalisation at 1 TeV is $1.8 \times 10^{-12}\,\mathrm{TeV}^{-1} \mathrm{cm}^{-2}\,\mathrm{s}^{-1}$ (see Fig.~\ref{fig:exp}). There is no report of long term changes in the orbit-averaged flux. 

\paragraph{\lsi} was detected by the MAGIC and VERITAS collaborations. Again, the VHE emission is tied to the orbital motion. However, unlike \ls, the orbital phases of VHE detections have varied considerably with epoch (Fig.~\ref{fig:lsi}). Early observations (Oct. 2005-Jan 2008) indicated that VHE emission was confined to phases $0.5 \leq \phi \leq 0.9$, peaking at 0.6-0.8 \citep{2009ApJ...693..303A,Albert:2006wi,Acciari:2008vf,2011ApJ...738....3A}. Later observations then failed to detect the source, until late 2010, when the source was detected at $0.0\leq \phi \leq 0.1$ \citep{2011ApJ...738....3A}. The 26.5 day orbital period is long enough and close enough to the lunar cycle to make repeated, homogeneous sampling of the orbital lightcurve difficult compared to \ls. The average spectra reported by both collaborations are compatible within systematic errors. The best fit is a power-law with a photon index $\Gamma_{\rm VHE}\approx 2.4$ to 2.7, and a normalisation at 1 TeV from 2.4 to $2.9 \times 10^{-12}\,\mathrm{TeV}^{-1} \mathrm{cm}^{-2}\,\mathrm{s}^{-1}$ \citep[][Fig.~\ref{fig:exp}]{Albert:2008zs,2012ApJ...746...80A,2009ApJ...700.1034A,2011ApJ...738....3A}. 

\paragraph{\hessj} was detected by the HESS Galactic Plane survey \citep{2007A&A...469L...1A}. The VHE detections cluster around $0.2\leq \phi\leq0.4$  when folded on the 315$\pm$5\,d period (Fig.~\ref{fig:hessj}, \citealt{2011ApJ...737L..11B,2012arXiv1212.0350B}). In hindsight, the initial VHE detection was lucky since follow-up observations by IACTs failed repeatedly to re-detect the source before the long orbital period was identified in the X-ray lightcurve \citep{2009ApJ...698L..94A,2012ApJ...754L..10A,2011ATel.3153....1O}. This the only gamma-ray binary that can be observed by all three major IACTs. The spectrum is compatible with a power-law of photon index $\Gamma_{\rm VHE}\approx 2.5$ and a normalisation at 1 TeV of $9.1\times 10^{-13}\,\mathrm{TeV}^{-1} \mathrm{cm}^{-2}\,\mathrm{s}^{-1}$ \citep{2007A&A...469L...1A}.

\paragraph{\fgl} is associated with HESS J1018-589 \citep{2012A&A...541A...5H}. The VHE source is decomposed into a point source (A) and a source (B) with an extension of 0\fdg15 $\pm$ 0\fdg03. The position of the VHE point source (A) is compatible with the GeV, X-ray, optical and radio sources associated with the binary. The best fit VHE spectrum is a power-law with a photon index $\Gamma_{\rm VHE}\approx 2.4$ and a normalisation at 1 TeV of $3.2 \times 10^{-13}\,\mathrm{TeV}^{-1} \mathrm{cm}^{-2}\,\mathrm{s}^{-1}$. Significant variability is detected in the (sparsely distributed) VHE observations, formally associating the VHE source with the binary. When folded and rebinned on the known 16.58\,d period, the VHE lightcurve is modulated in phase with the 1-10 GeV folded lightcurve measured with the \fermi\ (H.E.S.S. collaboration, 2013, in prep.).

\subsubsection{HE gamma rays (GeV)\label{he}}
\lsi\ and \ls\ had tentative associations with HE sources long before their VHE detections. These were confirmed thanks to the detection of orbital modulations directly from the \fermi\ data. The typical HE spectrum of a gamma-ray binary is a power-law with an exponential cutoff around a GeV.

\paragraph{\psrb} had been searched for in EGRET data, without success \citep{1996A&AS..120C.243T}. Weak emission was detected using \fermi\  starting a month before and ending a couple of weeks after periastron \citep{2011ApJ...736L..11A,2011ApJ...736L..10T}. The \fermi\ data does not confirm the earlier {\em AGILE} detection reported by \citet{2010ATel.2772....1T}. The average flux over this part of the orbit is $F_{\rm HE}$(0.1-1\,GeV)$\approx2.5\times 10^{-7}\,\mathrm{cm}^{-2}\,\mathrm{s}^{-1}$ (below the EGRET sensitivity) using a power-law index $\Gamma_{\rm HE}\approx 2.4$. \psrb\  brightened dramatically starting about 1 month after periastron passage (Fig.~\ref{fig:psrb}). This unexpected flare lasted seven weeks, with a peak luminosity $\approx 7\times10^{35}$ \ergs\ close to the pulsar spindown power $\dot{E}\approx 8\times10^{35}$ \ergs. Such a flare could have been detected with EGRET but the instrument had stopped pointing to the source two weeks after periastron passage. The average spectrum during the flare is a power law of photon index $\Gamma_{\rm HE}\approx1.4$ with an exponential cutoff at $E_{\rm c}\approx 0.3$ GeV, for an  average flux of $F_{\rm HE}$($>$0.1\,GeV)$\approx 1.3\times 10^{-6}\,\mathrm{cm}^{-2}\,\mathrm{s}^{-1}$. The spectrum softens with increasing flux, from $\Gamma_{\rm HE}\approx2.2$ to 3.2 when fitting by a power law and no cutoff. There is no HE detection at other times, with a collective upper limit of $F_{\rm HE}$($>$0.1\,GeV)$\la 9\times 10^{-9}\,\mathrm{cm}^{-2}\,\mathrm{s}^{-1}$. 

\paragraph{\lsi} The HE orbital modulation was easily detected within the first few months of operation of \fermi\,\citep{2009ApJ...701L.123A} settling the long-standing issue of the identification with the {\em Cos B} source 2CG 135+01. \lsi\ is the 12$^{\rm th}$ brightest source of the HE sky (in $\nu F_{\nu}$, \citealt{2012ApJS..199...31N}). The Lomb-Scargle periodogram of the HE lightcurve gives a period of 26.71 $\pm$ 0.05 days, consistent with the orbital period \citep{2012ApJ...749...54H}. The folded lightcurve peaks slightly after periastron passage ; the minimum occurs after apastron passage (Fig.~\ref{fig:lsi}). The phase-resolved spectra softens with increasing HE flux. The best fit spectrum is a power law $\Gamma_{\rm HE}\approx 2.1$ with an exponential cutoff at $E_{\rm c}\approx 3.9$ GeV \citep{2012ApJ...749...54H}, for a flux $F_{\rm HE}(>0.1\,\mathrm{GeV})\approx9.5 \times 10^{-7} \mathrm{cm}^{-2}\,\mathrm{s}^{-1}$. The \fermi\ spectral point at 30 GeV deviates from the exponential cutoff, signaling the emergence of the VHE component above this energy (Fig.~\ref{fig:exp}). The orbit-averaged flux increased by 40\% around March 2009, accompanied by a decrease in the modulation amplitude with no spectral change \citep{2012ApJ...749...54H}. This long-term variability has been related to the 1667\,day superorbital period observed in radio (\citealt{2013arXiv1307.6384T}, \S\ref{radio}).
   
\paragraph{\ls} The HE orbital modulation was detected soon after \lsi\ \citet{2009ApJ...706L..56A}. The Lomb-Scargle periodogram gives $3.90532 \pm 0.0008$ days \citep{2012ApJ...749...54H}. The HE modulation is in anti-phase with the VHE modulation (Fig.~\ref{fig:ls5}). \ls\ has been stable during the first 2.5 years of \fermi\ operations. Like \lsi, the spectrum is a power law $\Gamma_{\rm HE}\approx 2.1$ with an exponential cutoff at $E_{\rm c}\approx 2.2$ GeV, for a flux of $F_{\rm HE}(>0.1\,{\rm GeV})\approx 6.1 \times 10^{-7} \mathrm{cm}^{-2}\,\mathrm{s}^{-1}$ (Fig.~\ref{fig:exp}). The spectrum is softer when brighter \citep{2009ApJ...706L..56A} yet there is no statistical difference in spectra when binned into INFC and SUPC \citep{2012ApJ...749...54H}. The accumulated dataset shows evidence for a hard spectral component emerging above 10 GeV with $\Gamma\approx 1.6$ and $F(>10\rm\,GeV)\approx1.6\times 10^{-10} \mathrm{cm}^{-2}\,\mathrm{s}^{-1}$, compatible with the low-energy tail of the VHE emission \citep{2012ApJ...749...54H}. 

\paragraph{\hessj}\ The source remains undetected using the \fermi\ with an upper limit at the 95\% confidence level of $F_{\rm HE}(>0.1\,\mathrm{GeV})\la 3.0\times10^{-8}$\,ph\,cm$^{-2}$\,s$^{-1}$ \cite{2013arXiv1308.5234C}.

\paragraph{\fgl} was discovered thanks to its 16.58$\pm$0.02 day modulation in the  \fermi\ data (Fig.~\ref{fig:1fgl}, \citealt{2011ATel.3221....1C,2012Sci...335..189F}). The HE spectrum can be fitted by a power law $\Gamma_{\rm HE}\approx 1.9$ with an exponential cutoff $E_{\rm c}\approx 2.5$ GeV. However, the preferred spectral fit is a broken power-law with $\Gamma$(0.1--1\,GeV)$\approx2.0$ and $\Gamma$(1--10\,GeV)$\approx$3.1, with $F_{\rm HE}(>0.1\,\mathrm{GeV})\approx5.3\times 10^{-7}\,\mathrm{cm}^{-2}\,\mathrm{s}^{-1}$.  The spectral curvature varies significantly with orbital phase: the $\sim 1$ GeV peak in  $\nu F_{\nu}$ apparently shifts below the \fermi\ range at orbital phases $0.2\leq \phi\leq 0.6$.

\subsubsection{X-rays and LE gamma rays (MeV)\label{xray}}
All of the systems are detected in X-rays and all are modulated on the orbital period. The spectra are hard power-laws with $\Gamma_{\rm X}=1.5-2$ with no measured cutoff in X-ray, nor any Fe line or Compton reflection component.\footnote{The Fe line in {\em RXTE} observations of \ls\ is Galactic Ridge emission \citep{Bosch-Ramon:2005zc}.} X-ray fluxes are given in the 1-10\,keV band, except where noted otherwise.

\paragraph{\psrb} has a well-documented regular behaviour in X-rays, with two peaks in the lightcurve bracketing periastron at roughly $\pm 20$ days (\citealt{2009MNRAS.397.2123C} and references therein, Fig~\ref{fig:psrb}). From apastron to peak, the X-ray flux varies from $F_{X}\approx (1\ \mathrm{to}\ 35)\times10^{-12}\rm\,erg\,cm^{-2}\,s^{-1}$. The absorption column density increases from $N_H\approx$ (3 to 6)$\times 10^{21}\,\mathrm{cm}^{-2}$ \citep{2006MNRAS.367.1201C}. The variations are consistent with the Be disc crossing times (\S\ref{radio}). There is a noticeable hardening just before the first X-ray peak with $\Gamma_{\rm X}\approx 1.2$ compared with 1.6-1.8 at other times. The X-ray spectrum breaks to $\Gamma_{\rm X}\approx 2$ at 5 keV \citep{2009ApJ...698..911U} and thereafter continues at least  up to 200 keV \citep{1996A&AS..120C.221T}. The source is weakly variable (10-30\%) on timescales of 3-10 ks \citep{2009MNRAS.397.2123C,2009ApJ...698..911U}. Emission extending up to 4\arcsec\ away southward is detected in {\em Chandra} observations taken near apastron. This extended component has a luminosity $L_{\rm X}\sim10^{32}$\,\ergs (about 8 times fainter than the point source) and might be accompanied by a very faint ($L_{\rm X}\sim10^{31}$\,\ergs) jet-like feature extending up to 15\arcsec\ away southwest \citep{2011ApJ...730....2P}.
 
\paragraph{\ls} The X-ray flux is modulated on the orbital period (Fig.~\ref{fig:ls5}). The flux also varies at the 10--20\% level on timescales of 0.1-1\,ks \citep{Martocchia:2005kq}. Nevertheless, the orbital X-ray lightcurve is very regular with a peak at $\phi\approx0.65\pm0.05$ and a minimum at $\phi\approx 0.15\pm0.05$ for an overall amplitude of $\approx 40$\% \citep{Takahashi:2008vu,2009ApJ...697L...1K}. The photon index varies with orbital phase from $\Gamma_{X}\approx 1.45$ (at $\phi=0.5$, apastron) to 1.61 (at $\phi=0.1$, close to $\phi_{\rm sup}$). The X-ray flux varies from $F_{X}\approx (5.2\ \mathrm{to}\ 12.1) \times10^{-12}\rm\,erg\,cm^{-2}\,s^{-1}$. The column density  stays constant at a value compatible with the ISM value  $N_{\rm H}\approx7.0\times10^{21}\,\mathrm{cm}^{-2}$ \citep{Martocchia:2005kq}. The spectrum is a single power-law up to at least 70 keV. The source is detected with INTEGRAL at 200 keV with a photon index of 2$\pm0.2$ \citep{Hoffmann:2008ys}. LS 5039 is the counterpart to the ``$l$=18\degr'' COMPTEL source (one of the 10 brightest sources of the MeV sky), which has a spectrum  $\Gamma_{\rm LE}\approx1.5$ compatible with the extrapolation of the X-ray spectrum up to 10 MeV (\citealt{2001AIPC..587...21S}). A reanalysis of COMPTEL data found the source to be periodic, clinching the identification (Collmar et al., in prep.). The MeV orbital lightcurve is in phase with the X-ray lightcurve. \citet{2011ApJ...735...58D} found X-ray emission extending up to 2\arcmin\ in {\em Chandra} data ; it was not confirmed by  \citet{2011MNRAS.416.1514R} using a different {\em Chandra} dataset.
 \begin{figure}\sidecaption
\resizebox{0.53\hsize}{!}{\includegraphics*{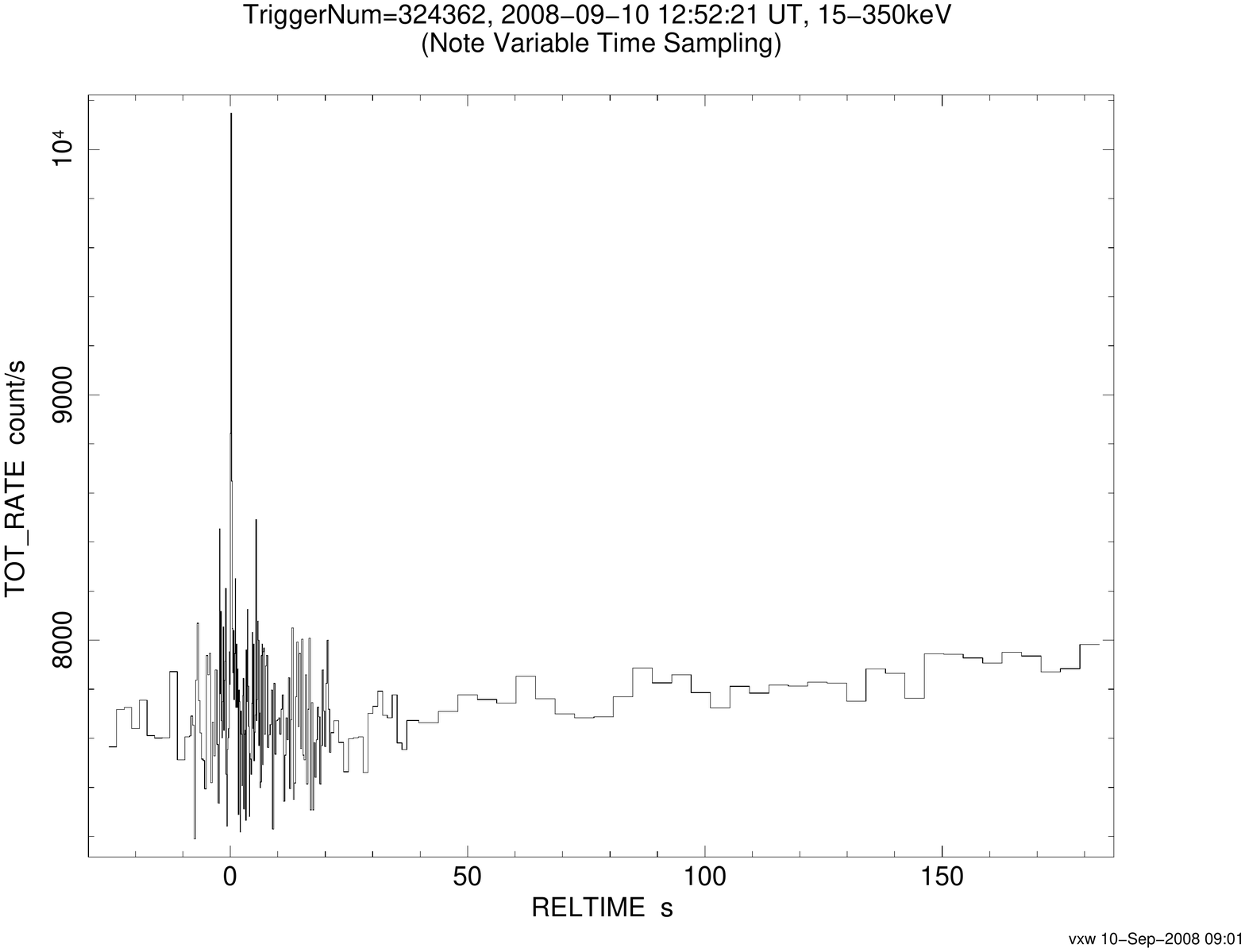}}
\caption{The 2008 burst of X-rays in the direction of \lsi\  that triggered the {\em Swift}/BAT \citep{2008GCN..8215....1B}. The lightcurve and spectrum are characteristic of magnetar bursts. Figure taken from the \href{http://gcn.gsfc.nasa.gov/swift_grbs.html}{\color{violet}{online data products}} provided by the {\em Swift} team. \label{fig:magburst}}
\end{figure}

\paragraph{\lsi} The X-ray flux is modulated on its 26.5 day orbital period, albeit with important long-term changes that have been related to the 4.5 year superorbital cycle (\S\ref{radio}): for instance, the X-ray peak varies from $\phi\approx0.4$ to $\phi\approx 0.8$ depending on cycle \citep{1997A&A...320L..25P,2010ApJ...719L.104T,2012ApJ...744L..13L,2012ApJ...747L..29C}. The X-ray flux varies from $F_{X} \approx (0.5\ \mathrm{to}\ 3)\times10^{-11}$ \ergs cm$^{-2}$, with some indication of a correlation with the VHE fluxes measured during one orbit \citep{2009ApJ...706L..27A}. The \citet{2011ApJ...738....3A} did not confirm the correlation using data from multiple orbits. The X-ray spectrum is well-fitted by an absorbed power-law. There is no sign of intrinsic absorption with $N_{\rm H}$ typically fixed at the ISM value  (estimated to be in the range $(4.9~{\rm to}~7.5)\times 10^{21}\,\mathrm{cm}^{-2}$, \citealt{2007ApJ...664L..39P}). The spectrum hardens from  $\Gamma_{\rm X}=$1.9 to 1.5 with increasing X-ray flux \citep{Albert:2008zs,2009ApJ...700.1034A}. There is no cutoff or break up to 60 keV (INTEGRAL, \citealt{Chernyakova:2006cu}) or 300 keV (OSSE, \citealt{1998ApJ...497..419S}). A COMPTEL source is detected in the 3-10 MeV range with a flat spectrum $\Gamma_{\rm LE}\approx 2$ so the X-ray spectrum must break around 1 MeV \citep{1996A&AS..120C.243T,1996A&A...315..485V}. The X-ray lightcurve fluctuates by 25\% on timescales of 1-10 ks \citep{Sidoli:2006an,2007A&A...474..575E}. In addition, several flares during which the flux increases by a factor 3-6  were observed with {\em Swift} and {\em RXTE}. The flares last tens to hundreds of seconds with doubling timescales as fast as a few seconds \citep{2009ApJ...693.1621S, 2011ApJ...733...89L}\footnote{The report of an X-ray quasi-periodic oscillation was disproved by \citet{2011ApJ...733...89L}.}. The {\em Swift}/BAT also triggered on two bursts of X-ray emission lasting less than 230 ms in the direction of \lsi\ (Fig.~\ref{fig:magburst}, \citealt{2008GCN..8215....1B,2012GCN..12914...1B}). The total energy involved is 10$^{37} (d/{\rm 2\,kpc})^{2}$ erg~s$^{-1}$ and the radius of the 7.5 keV blackbody that best fits the spectrum is 100 m.  The lightcurve, duration, fluence and spectrum have all the right characteristics of magnetar bursts \citep{2008ATel.1715....1D}. The only likely source within the 2.2\arcmin\ error box is \lsi\ \citep{2009A&A...497..457M,2010ASPC..422...23D,2012ApJ...744..106T}. \citet{2007ApJ...664L..39P} found evidence in {\em Chandra} data, at the 3$\sigma$ level, for  X-ray emission extending beyond 5\arcsec\ N of \lsi. This weak emission ($\approx 2\times 10^{31}$\,\eps) was not confirmed by \citet{2011MNRAS.416.1514R} using a different {\em Chandra} dataset. 
 
\paragraph{\hessj} The orbital period of 315$\pm$5 days was discovered thanks to X-ray monitoring observations (Fig.~\ref{fig:hessj}, \citealt{2011ApJ...737L..11B,2012arXiv1212.0350B}). The folded lightcurve is reminiscent of the X-ray lightcurve of the colliding wind binary $\eta$ Car (\S\ref{cwb}), with an X-ray flare lasting 20-30 days ($0.25\leq\phi\leq 0.35$) followed by a dip below the mean flux, again lasting 20-30 days ($0.35\leq\phi\leq 0.45$). The spectrum hardens during the dip. The flux has also been observed to change by 40\% on a timescale of 10 ks \citep{Hinton:2008eg}. The average photon index is $\Gamma_{\rm X}\approx1.6$ with  $N_{\rm H}=3.1\times10^{21}\,\mathrm{cm}^{-2}$ and $F_{\rm X}\approx (0.3\ \mathrm{to}\ 4.1)\times 10^{-12}\,\mathrm{erg}\,\mathrm{cm}^{-2}\,\mathrm{s}^{-1}$.  \citet{2011ApJ...737L..12R} found that the column density increases from $N_{H}\approx (2\ \mathrm{to}\ 4)\times 10^{21}\,\mathrm{cm}^{-2}$ and that the spectrum softens with $\Gamma_{\rm X}\approx 1.2$ to 1.65 when the flux increases.

\paragraph{\fgl} has a double-peaked orbital modulation, with a sharp peak superposed on a sine-like modulation shifted by $\Delta\phi=0.4$ (Fig.~\ref{fig:1fgl}, \citealt{2012Sci...335..189F}). The XMM photon index is $\Gamma_{\rm X}\approx 1.7$ for a column density of $N_{\rm H}\approx 6.6\times10^{21}\,\mathrm{cm}^{-2}$ \citep{2012A&A...541A...5H}. The average  X-ray flux with {\em Swift} is $F_{\rm X}\approx 2.3\times 10^{-12}\,\mathrm{erg}\,\mathrm{cm}^{-2}\,\mathrm{s}^{-1}$ with the spectrum fitted by a harder mean photon index $\Gamma_{\rm X}\approx 1.3$ \citep{2012Sci...335..189F}. There is a weak INTEGRAL detection in the 18--40 keV band \citep{2011ApJ...738L..31L}.
 
\subsubsection{IR, optical, and UV\label{optical}}
The massive star dominates the near-infrared, optical, and ultraviolet emission in all gamma-ray binaries. The disc in the Be systems is manifest through its emission lines and its infrared emission. Optical observations bring information on the distance, star, orbital parameters, or proper motion of the system (\S\ref{orbit}). Variability in the optical associated with the stellar wind can also shed light on processes at other wavebands.

\paragraph{\psrb} \citet{2011ApJ...732L..11N}  revised some of the initial measurements from  \citet{1994MNRAS.268..430J} that identified the companion as a Be star. The fast rotation of the star (LS 2883) induces a 6500 K temperature gradient from pole to equator. The system is likely a member of the Cen OB1 association, at a distance of 2.3$\pm0.4$ kpc,  further away than the previous best value of 1.5 kpc. The large equivalent width of H$\alpha$ ($W_{{\rm H}\alpha}\approx -54$\,\AA) implies a Be disc extending to 15-20 $R_\star$ \citep{Grundstrom:2006tc}. No changes in optical spectra have been reported around the time of periastron, when the pulsar is closest to the disc. There are no published radial velocity measurements from the optical lines.

\paragraph{\ls} Non-radial oscillations of the O star  with an amplitude $\approx$ 7 km\,s$^{-1}$ appear to be present in the radial velocity data  \citep{2011heep.conf..559C}. The system has a high peculiar velocity of 70--140\,km\,s$^{-1}$ \citep{McSwain:2002hp}: the Ser OB2 association is a possible birthplace, at a distance of 1.5-2.0 kpc, implying the system has an age of 1.0-1.2 Myr \citep{2002A&A...384..954R,2012A&A...543A..26M}. CNO-processed gas in the stellar atmosphere also suggests the system is young \citep{2004ApJ...600..927M}. The line equivalent widths are modulated on the orbital period ; the optical continuum is not, down to an amplitude $\approx$ 0.002 mag \citep{2011MNRAS.411.1293S}.

\paragraph{\lsi} has been associated with the open cluster IC 1805 at 2.3\,kpc \citep{Gregory:1979ng}. H{\sc i} radio observations place the system at 2.0$\pm0.2$\,kpc \citep{1991AJ....101.2126F}. \citet{Grundstrom:2007kv} observed a weakening of H$\alpha$ emission within a few days, to $W_{{\rm H}\alpha}\approx$-1\,\AA\ instead of the usual $W_{{\rm H}\alpha}\approx$-12\,\AA\ to -18\,\AA, perhaps because of sudden ionisation of the Be disc. They estimated the disc size to be $\approx 4-6 R_{\star}$. Changes in the approaching component of the double-peaked H$\alpha$ line  indicate the presence of a spiral wave due to tidal forces but also of more complex velocity structures \citep{2010ApJ...724..379M}. The spectral lines properties vary together with the 4.5\,yr superorbital period \citep{Zamanov:1999of}. The optical and infrared fluxes are modulated on the orbital period by 5-10\% \citep{1989MNRAS.239..733M,1994A&A...288..519P,2003AstL...29..188Z}. \citet{1995AA...298..151M} interpreted this modulation as attenuation by the Be disc of light from the vicinity of the compact object.

\paragraph{\hessj} The star (HD 259440 $\equiv$ MWC148) was identified early on as a possible counterpart \citep{2007A&A...469L...1A,Hinton:2008eg}. \citet{2010ApJ...724..306A} proposed that the system originates from  the cluster NGC 2244 at 1.6$\pm$0.2 kpc. \citet{2010ApJ...724..306A} and  \citet{2012MNRAS.421.1103C} reported on their search for radial velocities.  The  large equivalent width of H$\alpha$ (W$_{{\rm H}\alpha}\approx -50$\AA) points towards a large size for the Be disc size $\geq 10\,R_\star$, compatible with the long orbital period, and favours a low inclination \citep{Grundstrom:2006tc}. The hydrogen line profiles are asymmetric, as with \lsi. Shallow He {\sc i} and He {\sc ii} lines in absorption as well as numerous double-peaked Fe {\sc ii} lines are also detected.

\paragraph{\fgl} The star was unknown prior to the \fermi\ discovery. Optical observations have determined the spectral type, estimated the distance and searched (unsuccessfully) for modulations \citep{2012Sci...335..189F}. Follow-up studies of this bright ($B=12.8$) star with a 16.6 day orbital period should bring more information on the orbit and stellar parameters.

\subsubsection{Radio and mm\label{radio}}
All of the gamma-ray binaries are radio sources -- something unusual amongst the wider population of high-mass X-ray binaries\footnote{The only radio sources amongst the 117 HMXBs of \citet{Liu:2006kk} are \ls, \lsi, $\gamma$ Cas, IGR J17091-3624, V4641 Sgr, SS 433, Cyg X-1, Cyg X-3 and CI Cam (likely a nova).}. Their variable radio spectra is consistent with synchrotron emission. The sources are resolved on milliarsecond scales.
\begin{figure}
\centering
\includegraphics[height=8cm]{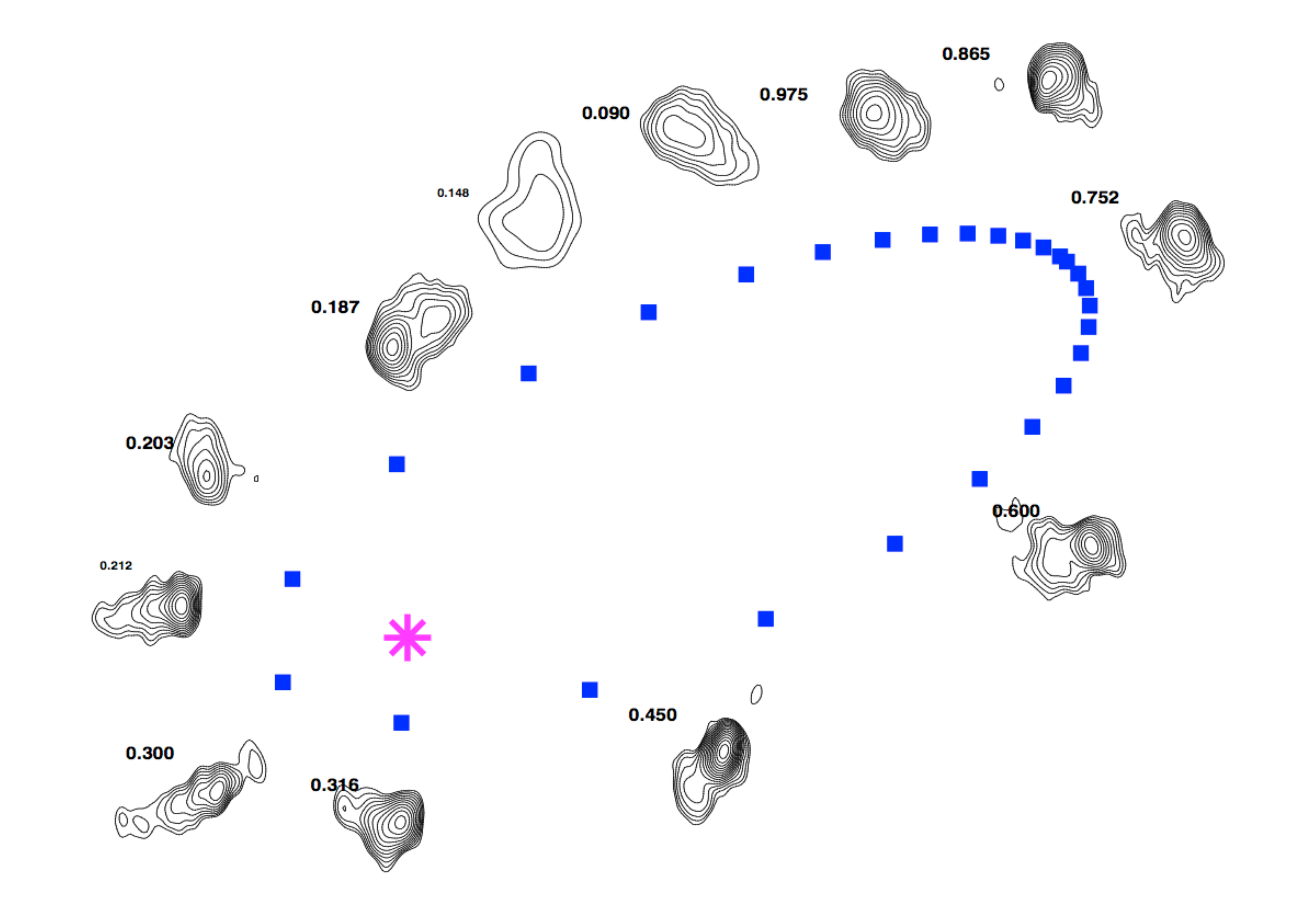}
\caption{Maps of the 8 GHz radio emission from \lsi\ obtained in consecutive VLBA observations covering an orbit of the binary. The above is a montage with the maps (resolution 1.1$\times$1.5 mas) organized by phase around an exaggerated orbit (real size 0.3 mas). Figure reprinted with permission from \citet{Dhawan:2006kr}. \label{lsiradio}}
\end{figure}

\paragraph{\psrb} has pulsed radio emission with an amplitude of 20 mJy at 1.4 GHz for a mean flux of a few mJy with a flat spectral index \citep{Manchester:1995ck,2011MNRAS.418L.114K}. The pulse lightcurve has two peaks separated by 140\degr, interpreted as a wide cone of emission from a single polar region \citep{Manchester:1995ck}.  Spindown of the 47.7\,ms pulse provides an estimated magnetic field $B\approx 3.3\times10^{11}$\,G and  age $\tau\approx 3.3\times10^{5}$\,yr\ for the pulsar (\S\ref{pw1}). The pulsation disappears completely for $\approx$30-40 days around periastron, as a result of free-free absorption by the circumstellar material. The absorption and the varying rotation and dispersion measures are best modelled by an intervening disc inclined by 10\degr\ from the orbital plane \citep{1995MNRAS.275..381M}. Higher inclinations have been suggested \citep{1998MNRAS.298..997W}. Non-pulsed emission, related to the interaction of the pulsar with the Be disc, appears $\approx$ 20 days before periastron, lasting up to 100 days after. The radio spectrum is a power-law $F_{\nu}\propto\nu^\alpha$ with a spectral index $\alpha\approx-0.6$,  consistent with synchrotron emission, with signs of absorption below 2 GHz before periastron passage \citep{Johnston:2005jq}. The lightcurve peaks once $\approx$10 days before periastron and, again, $\approx 20$ days after, reaching flux densities of 50 mJy at 2 GHz (Fig.~\ref{fig:psrb}). The changes in lightcurve shape from orbit to orbit probably reflect changes in the circumstellar environment probed by the pulsar. \citet{2011ApJ...732L..10M}  resolved the radio emission during the 2007 passage: its size is $\approx 50$\,mas (120\,AU at 2.3\,kpc) and it peaks at a distance 10 to 20\,mas  from the system (25--45 AU). 

\paragraph{\ls} has persistent radio emission with $\la 30$\% variability around a mean level $\approx 30$ mJy at 2 GHz \citep{Ribo:1999qo}. The radio flux is not known to vary on the orbital period. The spectrum is a power-law of spectral index $\alpha\approx{-0.5}$ breaking to an absorbed, optically thick spectrum below 1 GHz with $\alpha\approx {0.75}$ \citep{2008MNRAS.390L..43G,2012MNRAS.421L...1B}. The radio emission was first resolved by \citet{2000Sci...288.2340P} on scales of 1--6\,mas ($\geq$2.5\,AU). Further observations found that the radio morphology changes with orbital phase and is stable from orbit to orbit (\citealt{2008A&A...481...17R,2012arXiv1209.6073M}). Emission on larger scales (60--300\,mas) has also been reported \citep{2002A&A...393L..99P}.

\paragraph{\lsi}'s 26.5 day orbital period is best determined from its regular radio outbursts (Fig.~\ref{fig:lsi}). An additional  period at 1667$\pm8$ days appears in the radio lightcurve accumulated since the 1980s \citep{2002ApJ...575..427G}. This superorbital period is also present in HE gamma-rays, X-rays, and optical. The peak phase ($\phi\approx 0.4$ to 1.0), maximum intensity (50--300 mJy), and amplitude (factor 2-10) of the radio outburst depend on the superorbital cycle. Taking this into account, \citet{2012ApJ...747L..29C} noticed that the radio peak follows the X-ray peak after a delay of $\Delta\phi\approx0.2$ (see also \citealt{Harrison:2000bk}). The spectrum varies from optically thick to thin indices during the outburst ($\alpha\approx{0.4}$ to $\alpha\approx{-0.5}$, \citealt{1997ApJ...491..381R,2009ApJ...702.1179M}). \citet{Dhawan:2006kr} presented VLBI observations throughout an orbit (Fig.~\ref{lsiradio}). The radio source is resolved on mas scales (2 AU), elongated by $\approx 7$ mas, the position angle changing with $\phi$, with lower frequency emission occurring further away along the radio ``tail''. Identical morphologies are seen in maps obtained at the same orbital phase but at different times \citep{Albert:2008zs}. Large scale radio emission surrounds \lsi, probably an H{\sc ii} region \citep{1987AJ.....93.1506F,Marti:1998mf}. If \lsi\ were to be at the center of its supernova remnant then this would constrain the system age to be $\la 10^{4}$ yr.

\paragraph{\hessj} has a variable radio counterpart with an average spectral index $\alpha=-0.6$ \citep{2009MNRAS.399..317S}. The source is weak $\approx 0.2-0.7$ mJy at 1--5 GHz. \citet{2011A&A...533L...7M} obtained two VLBI maps, one during decline from the X-ray peak and another during the X-ray dip. The emission in the latter is fainter and extended by $\approx 50$ mas with the peak displaced by 14 mas from the earlier observation (more than the orbital size). 

\paragraph{\fgl} is modulated in radio on the orbital period with a sine wave pattern from 2 to 6 mJy, close in phase with the sine wave X-ray modulation \citep{2012Sci...335..189F}. The spectral index is variable around $\alpha\approx 0$. If due to absorption, the modulation should disappear at higher frequencies when the emission becomes optically thin. Finding the transition frequency constrains the location of radio emitting electrons. For instance, in LS 5039, the 9 GHz emission is optically thin, unmodulated, and must come from regions beyond a few AU based on the free free opacity (\S\ref{pulsations}). \citet{2012A&A...541A...5H} noted that \fgl\ is located in the centre of SNR G284.3--1.8: a physical association would imply a young age for the system.

\begin{sidewaystable}
\vspace{12cm}
\ra{1.3}
\begin{tabular}{llllll}
\toprule 
 & \psrb & \ls & \lsi & \hessj & \fgl \\
 \midrule
 norm @1\,TeV ($10^{-12}\,\mathrm{ph}\,\mathrm{cm}^{-2}\,\mathrm{s}^{-1}\,\mathrm{TeV}^{-1})$  & $<$0.4--2.9 	& 0.5-3		& $<$0.5--5	& $<$0.3--1.2		& 0.1-0.9 \\
$\Gamma_{\rm VHE}$ &  2.7 		& 1.8$_{\rm cut}$--3.1	& 2.6		& 2.5			& 2.4\\
\rowcolor[gray]{.9}$L_{\rm VHE}$\,(10$^{35}$\,erg\,s$^{-1}$)  & 0.09 & 0.14 &  0.13 & 0.02 & 0.09\\
\midrule
 $F_{\rm HE}(>0.1\,\mathrm{GeV})\, (10^{-7}\,\mathrm{ph}\,\mathrm{cm}^{-2}\,\mathrm{s}^{-1})$  & $<$0.09--35 	& 4--15		& 6--14	& $<$0.3 	& 5.0-5.6 \\
 $\Gamma_{\rm HE}$ &  1.4$_{\rm var}$		& 2.1  		& 2.1		& 	(2.9)		& 1.9$_{\rm var}$\\
 $E_{\rm cut}$ (GeV) &  0.3 		& 2.2		& 3.9 		& 	-		& 2.5\\
\rowcolor[gray]{.9}$L_{\rm HE}$\,(10$^{35}$\,erg\,s$^{-1}$) & 2.8 & 2.8 & 2.3 & $<$0.03 & 9.7 \\
\midrule
$F_{\rm X}$(1-10\,keV) ($10^{-12}$\,erg\,s$^{-1}$cm$^{-2}$) &  1--37 & 5--12 & 5--30 & 0.3--4.1 &  0.5--5\\
 $\Gamma_{\rm X}$ &  1.2--2.0		& 1.4--1.6  		& 1.5--1.9		& 1.2--1.7			& 1.3--1.7\\
\rowcolor[gray]{.9}$L_{\rm X}$\,(10$^{35}$\,erg\,s$^{-1}$) & 0.23 & 0.12 & 0.14 & 0.01 & 0.17 \\
\midrule
$F_{\rm radio}$(2\,GHz) (mJy) &  2--50 & 30 & 20--300 & 0.2--0.7  &  1.5--6\\
\rowcolor[gray]{.9}$L_{\rm radio}$\,(10$^{29}$\,erg\,s$^{-1}$) & 6.3 & 6.0 & 28.7 & 0.04 & 4.2 \\
\bottomrule
\multicolumn{6}{l}{\small VHE luminosity above 100 GeV, HE luminosity from 0.1--10 GeV, derived from the values in the text \& distances from Tab.~1}\\
\multicolumn{6}{l}{\small  X-ray flux modulation and (peak) luminosity in the 1--10 keV range, radio flux and (peak) luminosity at $\approx$ 2 GHz.}\\
\multicolumn{6}{l}{\small The HE spectra marked {\tiny var} show more complex variability with orbital phase than is summarised here.}
\end{tabular}\\
\caption{Summary of multiwavelength observations of gamma-ray binaries\label{mwltable}.}
\end{sidewaystable}
\subsection{The multi-wavelength picture: a summary\label{summwl}}
Figures~\ref{fig:psrb}-\ref{fig:1fgl} show the lightcurves and spectral energy distributions of the five binaries. Table~\ref{mwltable} summarises the multiwavelength picture of gamma-ray binaries. The range in fluxes for each band indicates the minimum and maximum values of the orbit-related variability  For VHE gamma rays, the normalisation at 1 TeV gives the easiest comparisons. For HE gamma rays, the parameters correspond to spectral fits assuming a power-law with an exponential cutoff, although there is substantial spectral variability for \psrb\ and \fgl\ (\S\ref{he}). $L_{\rm HE}$ for \psrb\ is computed from the average spectrum during the flare. The unabsorbed X-ray flux is given in the 1-10 keV band, with the range indicating the amplitude of the modulation (ignoring the flares or bursts from \lsi, \S{\ref{xray}). The isotropic luminosities use the distances in Tab.~\ref{parameters}. The VHE luminosities are derived from the average spectral fit mentioned in the text (\S\ref{vhe}) (typically corresponding to a high VHE state). The HE luminosities are derived from the average fluxes given in the text (\S\ref{he}), except for \psrb\ where the average flux during the ``flare" was used. The X-ray and radio luminosities use the maximum flux value given in the table.

All systems have $L_{\rm X}\sim L_{\rm VHE}$, hard X-ray slopes and soft VHE slopes, demonstrating that the dominant radiative output occurs in the 1 MeV -- 100 GeV range. Indeed, all systems have $L_{\rm HE}\ga 20 L_{\rm VHE}$ except \hessj, which clearly stands out with $L_{\rm HE}\la L_{\rm VHE}$. Even though \hessj\ is intrinsically fainter by a factor  $\approx 10$ than the other binaries, this peculiar dearth of HE emission is a challenge to models (\S\ref{hespec}). The comparable VHE fluxes from system to system reflects the sensitivity of IACTs. The X-ray sources are not exceptionally bright: there are $\ga 100$ X-ray sources in the Galactic Plane, typically accreting X-ray binaries, that are brighter than \lsi\ (compare the mean count rate of 0.2 counts s$^{-1}$ reported in \citealt{1997A&A...320L..25P} to the log$N$-log $S$ derived from the RXTE/ASM  by \citealt{Grimm:2002xx}). The radio sources, at flux minimum, are near the completeness limit of radio surveys (2.5 mJy at 1.4 GHz for the NRAO VLA Sky Survey). Hence, VHE and HE gamma-ray observations remain the best way to identify them at present. A point-like gamma-ray source in the Galactic plane coincident with a radio, X-ray and a massive star counterpart makes for a very good  gamma-ray binary candidate.

\section{Accretion-powered microquasars or rotation-powered pulsars ?\label{disguise}}

Gamma-ray binaries share many spectral and variability properties, notably the ubiquity of orbital modulations at all wavelengths. All have a massive stellar companion, radio emission, modest X-ray fluxes, hard X-ray spectra up to high energies. Taken together, these characteristics indicate that gamma-ray binaries form a distinct class of systems from high-mass X-ray binaries (HMXBs). HMXBs also have compact objects in orbit around massive stars, but reach higher X-ray luminosities, rarely show radio emission (\S\ref{radio}), have curved X-ray spectra and X-ray pulsations \citep{White:1995ld}. Nearly all HMXBs harbour neutron stars, with only a handful containing black hole candidates. 
 HMXBs are driven by accretion of material from the stellar wind or circumstellar disc of the O or Be companion. The plasma follows the magnetic field lines close to the neutron star, accreting onto the magnetic poles, producing X-ray pulses as the poles cross the line-of-sight.  HMXBs are not radio pulsars because the density of infalling material close to the poles shorts out the electric field and coherent emission responsible for radio pulsations. 

Radio pulsars do not accrete because the pulsar wind pressure is large enough to prevent material from falling into the gravitational potential of the neutron star (\S\ref{containment}). In \psrb, a shock forms between the pulsar wind and the circumstellar material. The non-thermal emission is thought to be due to high energy particles that are scattered and accelerated at the shock, as in pulsar wind nebulae \citep[PWN, see e.g.][for a review]{Gaensler:2006qi}. \psrb\ is at present the only binary system with a massive star where this  is proven to occur. This scenario, initially sketched by \citet{Shvartsman:1971rw}, was discussed by \citet{Illarionov:1975yk}, soon after the importance of  pulsar winds in shaping the nebulae around pulsars was realized (notably the Crab nebula). \citet{Basko:1974zv} and \citet{1977A&A....55..155B} put forward the idea to explain the high-energy emission from Cyg X-3. This was abandoned when the high luminosity, variability, radio ejections indicated Cyg X-3 is an accreting system (see \S\ref{cygx3}). Later, \citet{Maraschi:1981nj} proposed this scenario for \lsi\ and \citet{Tavani:1994qu} applied it to \psrb. 

However, by the end of the 1990s, radio observations of X-ray binaries had established accreting binaries as sources of non-thermal radiation and relativistic jets  \citep{Fender:2006ww}. The discovery of elongated radio emission on milliarcsecond scales in both \ls\ \citep{2000Sci...288.2340P} and \lsi\ \citep{Massi:2001jg} were thus interpreted as relativistic jets. Relativistic jets are the most striking analogy between accretion onto the stellar-mass compact objects in X-ray binaries and onto the supermassive black holes in Active Galactic Nuclei (AGN), hence the name ``microquasars'' for X-ray binaries with relativistic jets \citep{1992Natur.358..215M}. Similarities also exist in timing and spectral characteristics. The analogy prompted speculation that, like AGNs, some X-ray binaries could be gamma-ray emitters with particles accelerated in the jet responsible for the non-thermal emission. \ls\ and \lsi\ would be examples of such gamma-ray microquasars.

The detection of gamma-ray binaries initiated a debate as to whether the high-energy emission was ultimately due to accretion energy released in the form of a relativistic jet ({\em microquasar scenario}) or due to rotational energy released as a pulsar wind ({\em pulsar scenario}), with the cometary tail of shocked pulsar wind material mimicking a microquasar jet  (\citealt{Dubus:2006lc,2006Sci...312.1759M,Romero:2007ly}, Fig.\ref{fig:mirabel}). A minimum mass $>3 M_\odot$ would imply a black hole compact object and rule out a pulsar (\S\ref{orbit}). Inversely, pulsations would confirm the pulsar wind interpretation (\S\ref{pulsations}). Both have proven elusive to obtain, so distinguishing between the two scenario relies on indirect evidence.

\begin{figure}
\centering\includegraphics[height=4.4cm]{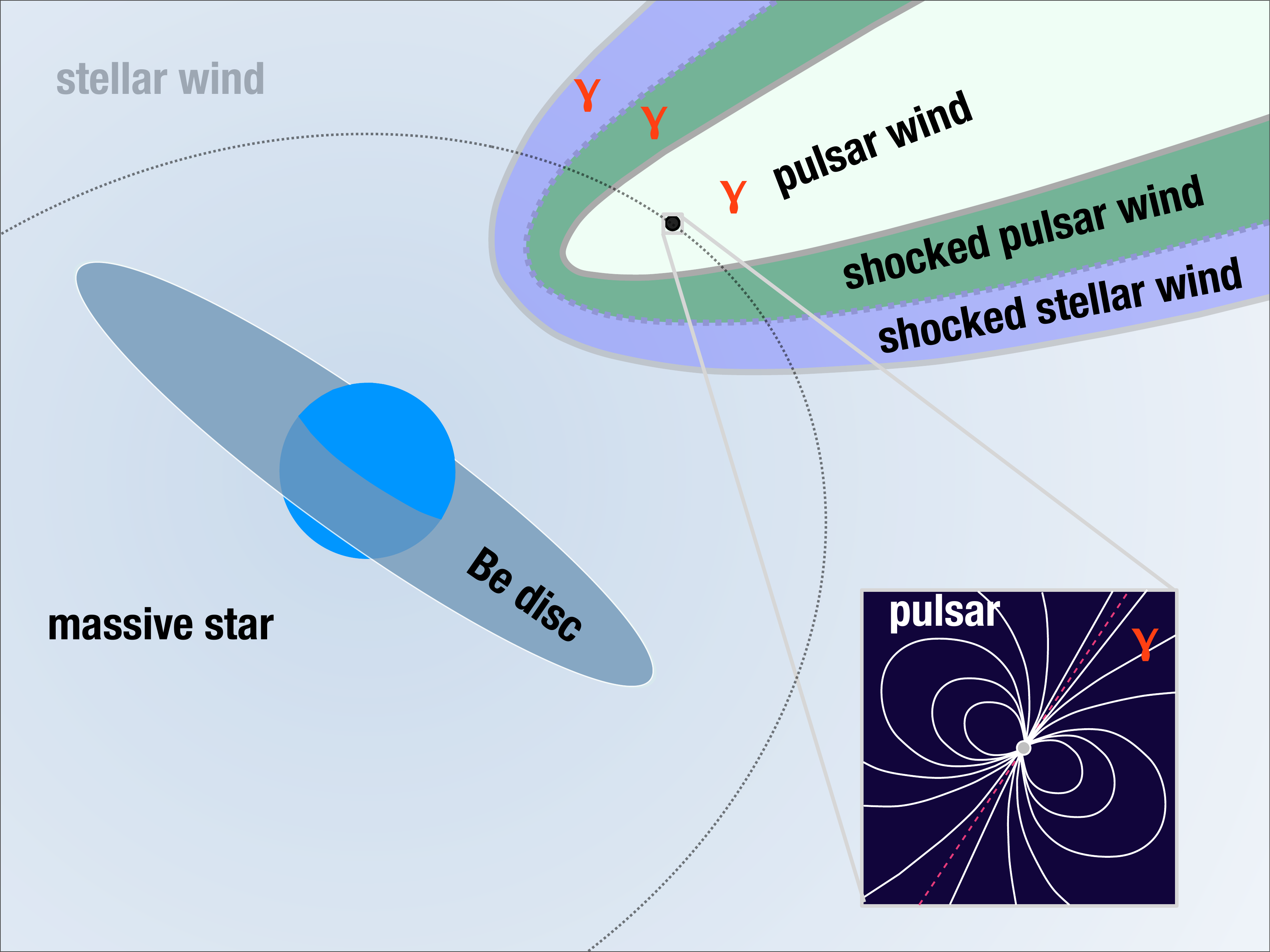}
\centering\includegraphics[height=4.4cm]{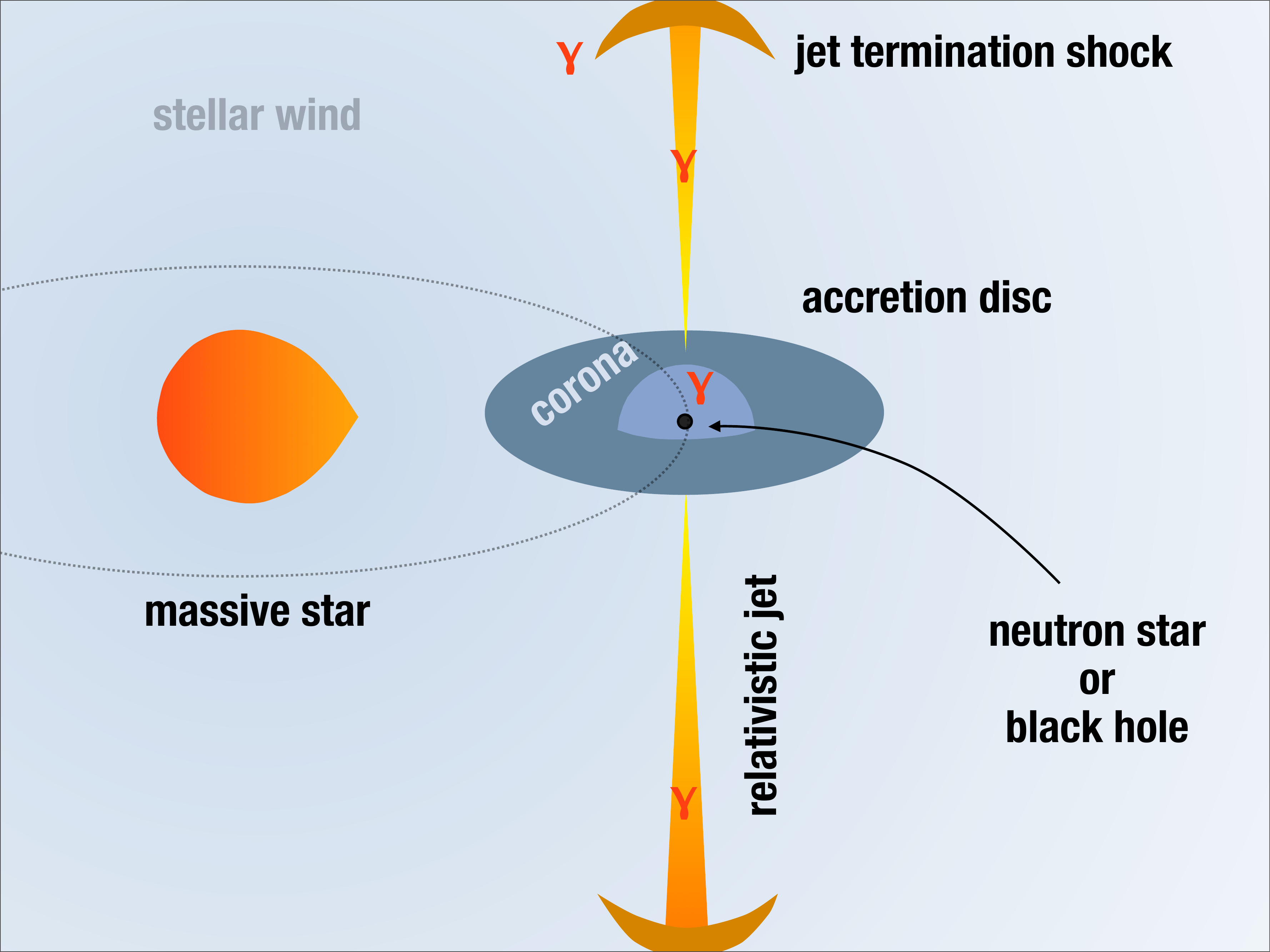}
\caption{Scenarios for gamma-ray emission from binaries. Left: the relativistic wind from a rotation-powered pulsar interacts with the stellar wind (and Be disc if present) of its massive companion star.  Pulsar inset shows the magnetic field lines within the light cylinder. Gamma-ray emission can occur near the pulsar, within the pulsar wind, or at the shocks terminating the pulsar and stellar wind. Right: the compact object accretes matter from the stellar wind or Be disc. Part of the  energy released in the accretion disc is used to launch a relativistic jet. Gamma-ray emission can arise from the corona of the accretion disc, within the jet, or at the termination shock of the jet with the ISM.}
\label{fig:mirabel}
\end{figure}

\subsection{Indirect evidence for the pulsar scenario \label{indirect}}

The current array of indirect evidence favours the pulsar interpretation over the accreting microquasar scenario. The main arguments are given below, more or less in order of decreasing weight, placed in the context of isolated pulsars and pulsar wind nebulae (PWN), along with caveats.
\begin{itemize}
\item The spectral and timing characteristics are similar in all five systems suggesting that, like \psrb, all are powered by the spindown of a pulsar. High gamma-ray luminosities, as seen in gamma-ray binaries, are also a generic property of pulsars. Pulsars and their nebulae are the most common class of galactic sources in the HE and VHE domain. Compared with isolated PWN, gamma-ray binaries have steeper VHE spectral indices by about $\Delta\Gamma\approx 0.5$ and a lower ratio $L_{\rm VHE}/L_{\rm X}\approx 1$. Taken at face value, the latter is compatible with young, $\la 10^4$\,yr old, PWN \citep{2009ApJ...694...12M,2010AIPC.1248...25K}.
\item The detection of two magnetar bursts from the direction of \lsi\  with no other obvious counterpart (\S\ref{xray}). Magnetar behaviour would imply a high magnetic field for the pulsar (\S\ref{containment}) and that magnetic energy is also tapped (at least sporadically) in addition to the rotational energy.
\item Radio emission elongated along one direction with position angle dependent on orbital phase and a repeatable morphology from orbit to orbit (\S\ref{radio}). Such ``comet tail'' behaviour is natural in a binary PWN scenario. The alternative is that the one-sided radio emission is due to Doppler boosting of the approaching jet in a microquasar scenario. However, the jet direction is not expected to change with orbital phase. The constraints derived from the flux ratio between approaching and receding jet point require precession of the jet direction on the orbital timescale \citep{Massi:2004pf,2008A&A...481...17R}, which would be unique. 
\item The HE spectrum with an exponential cutoff is a hallmark of pulsars, used as a criterion to identify candidate pulsars amongst \fermi\ sources \citep{2012ApJ...753...83A}. The photon index and cutoff energy are in the range of values observed from gamma-ray pulsars \citep{2010ApJS..187..460A}. However, the orbital variability in HE gamma rays is not expected from the standard pulsar magnetospheric emission models (\S\ref{hespec}).
\item X-ray indices are in the range of PWN \citep{2010AIPC.1248...25K} and show no cutoffs up to MeV energies. Based on isolated PWN, the X-ray luminosity of gamma-ray binaries suggests $\dot{E}\approx10^{37}$\,\eps\ \citep{2009ASSL..357...91B,2009ApJ...694...12M,2010AIPC.1248...25K}. HMXB spectra turn over around 30 keV. Low-mass X-ray binary (LMXB) spectra cut off around 100\,keV. Hard photon indices in accreting binaries are always associated with strong, red noise variability up to kHz frequencies \citep{Remillard:2006kg}, not seen in gamma-ray binaries despite their hard spectra. The X-ray spectral changes commonly seen in accreting X-ray binaries have  not been observed in gamma-ray binaries. However, wind accretion leads to small discs \citep{Beloborodov:2001qx}. In extreme cases no accretion disc is formed, yet a jet can still be launched by the Blandford-Znajek process \citep{2012MNRAS.421.1351B}.
\item Radio emission is expected from a PWN but rare in a HMXB. The radio indices in the $\approx$ 1--10 GHz band are mostly optically thin, as expected from a PWN, whereas an optically thick index (from the self-absorbed jet emission) is always found together with a hard X-ray index in LMXB \citep{Remillard:2006kg}. The optically thick radio indices found at lower frequencies in gamma-ray binaries can be explained by absorption in the wind. They have also been discussed in the  microquasar scenario \citep{2009ApJ...702.1179M}.
\item The repeatable orbital modulations at nearly all wavelengths suggest a steady injection of energy, as expected from a pulsar wind. However, accretion can also lead to modulations since the Bondi rate changes along the eccentric orbit \citep[e.g.][]{1995AA...298..151M}. 
\end{itemize}

Some of the observations remain puzzling within the usual pulsar/PWN phenomenology. They might be related to the binary nature of the source. The orbital variability of the ``pulsar-like'' HE spectrum is probably the thorniest (\S\ref{hespec}). The X-ray flaring behaviour on timescales of 1--1000\,s (\S\ref{xray}) is also unexpected in a standard PWN scenario. However, the timescales are compatible with the scale of the termination shock and the changes could be due to clumps in the stellar wind  \citep{2009ApJ...693.1621S,2010MNRAS.403.1873Z,2010MNRAS.405.2206R}. The detection of X-ray and gamma-ray flares from the Crab nebula  \citep{2011ApJ...727L..40W,2012ApJ...749...26B} proves phenomenologically that PWN can be variable on much shorter timescales than previously thought possible. X-ray emission on scales of arcseconds to arcminutes, if confirmed, is also at odds with an intra-binary termination shock: perhaps this emission arises from a larger shock with the ISM \citep{2011A&A...535A..20B}. The long-term changes in flux in \lsi\ (\S\ref{vhe}-\ref{he}), the periodic radio flares of \lsi\ and \psrb\ (\S\ref{radio}) might be linked to changes in circumstellar material  probed by the pulsar along its orbit.

\subsection{Detecting pulsations\label{pulsations}}

The detection of pulsed non-thermal emission would prove the pulsar scenario. This section presents the current upper limits and prospects to detect pulsations in gamma-ray binaries (besides \psrb).

\paragraph{Radio} In radio the most constraining upper limits are from \citet{2011ApJ...738..105M}, who report limits of 4-15 $\mu$Jy at 1.6 GHz to 9.3 GHz for \ls\ and \lsi. The orbital phases were chosen to be close to apastron and to have minimal Doppler shifts due to orbital motion. The upper limits are smaller than the amplitude of the radio pulsation in \psrb. However, they note that their upper limit is still compatible with the presence of a radio pulsar because of free-free absorption by the stellar wind in those systems. 

For an isotropic, isothermal stellar wind with constant mass-loss rate $\dot{M}_{\rm w}$ and speed $v_{\rm w}$, if the pulsed emission is at a distance $d$ from the massive star then, integrating radially outward for simplicity, the free-free opacity is:
\begin{equation}
\tau_{\rm ff}=\int_{d}^{\infty} 0.018 T^{-3/2} Z^2 n_e n_i g_{\rm ff} \nu^{-2} dr\approx 0.018\ g_{\rm ff} \nu^{-2} T_{\rm w}^{-3/2} Z^2  \int_{d}^{\infty}  \frac{\dot{M}_{\rm w}^2}{(4\pi r^2 v_{\rm w})^2\mu_i \mu_e m_p^2} dr
\end{equation}
\begin{equation}
\tau_{\rm ff}\approx 14.7\ g_{\rm ff} \left(\frac{\nu}{10^9\,\mathrm{Hz}}\right)^{-2} \left(\frac{\dot{M}_{\rm w}}{10^{-7}\,\mathrm{M}_\odot\,\mathrm{yr}^{-1}}\right)^2\left(\frac{{v}_{\rm w}}{1000\,\mathrm{km}\,\mathrm{s}^{-1}}\right)^{-2}\left(\frac{{T}_{\rm w}}{10\,000\,\mathrm{K}}\right)^{-3/2}\left(\frac{d}{1\,\mathrm{AU}}\right)^{-3}
\label{eq:freeabs}
\end{equation}
using the absorption coefficient for $h\nu\ll kT_{\rm w}$ from \citet{Rybicki:1979za}, with  $Z^2\approx 1.4$, $\mu_i\approx 1.30$, $\mu_e\approx 1.18$ for an ionised solar composition and typical parameters for the stellar wind (\S\ref{wind}). In \psrb, the stellar wind mass loss rate is $\dot{M}_{\rm w}\sim 10^{-8}$\,\msol\,yr$^{-1}$ \citep{1995MNRAS.275..381M}. The free-free opacity becomes optically thick, $\tau_{\rm ff}\ga1$, at distances $d \la 1.9$\,AU using Eq.~\ref{eq:freeabs}, roughly consistent with the disappearance of radio pulsations $\approx$20 days before and after periastron passage ($d\la 1.5$\,AU). For \ls\ and \lsi\ the free-free absorption is too high  to detect radio pulsations because of the tight orbits \citep{1982ApJ...255..210T,Dubus:2006lc,2010MNRAS.403.1873Z,2011ApJ...738..105M,2012A&A...543A.122C}. Higher frequencies would be preferable to reduce absorption, to be balanced against the typical pulsar radio spectrum $F_\nu\propto \nu^{-1.5}$ \citep{1977puls.book.....M}. The orbit of \hessj\ leaves more hope for detection.

\paragraph{X-ray} Radio pulsars can also have X-ray pulsations (although in this case the mechanism for X-ray emission is different from that of X-ray pulsars in HMXBs). Upper limits on the pulsed fraction of  8\%--15\% for \psrb\ \citep{1999ApJ...521..718H,2010MNRAS.405.2206R}, 10\%--15\%  for \ls\ and \lsi\ \citep{Martocchia:2005kq,2010MNRAS.405.2206R,2011MNRAS.416.1514R}, 35\% for \hessj\ \citep{2011ApJ...737L..12R} have been obtained over a typical frequency range of 0.001--175 Hz. These limits are compatible with the X-ray pulsed fractions of rotation-powered pulsars, which range from 2\% to 99\% \citep{2011MNRAS.416.1514R}. The pulsed fraction is also diluted because the contribution to the X-ray flux from the (unpulsed) emission of the pulsar wind nebula is likely to dominate in gamma-ray binaries (\S\ref{pwn}).

\paragraph{High-energy gamma rays} The last window where pulsations might be found is in HE gamma rays. Pulsations from \psrb\ were searched for, without success, despite the precise radio pulsar ephemeris \citep{2011ApJ...736L..11A}. However, \psrb\ is not an isolated case: 30 pulsars with $\dot{E}\geq 10^{36}$\,\eps\ and available pulse ephemeris remain undetected in gamma rays \citep{2013arXiv1305.4385T}. They could be intrinsically faint or the geometry could be unfavourable for detecting gamma rays. A direct detection of GeV pulsations from the other gamma-ray binaries appears difficult. While dozens of isolated pulsars have been found directly from blind pulsation searches of the gamma-ray data, the detection of pulsars in binaries requires knowledge of the orbital parameters to correct for orbital motion. The gamma rays cannot be folded coherently on the pulse period if this correction is not applied with high precision.  The search becomes computationally prohibitive when the orbit is unknown. Hence, new millisecond pulsars in binaries are usually detected by following-up candidate gamma-ray sources in radio, where the shorter integration times make it easier to correct for orbital motion. The first detection of a pulsar in a binary directly from a (semi-)blind search of the gamma-ray photons was reported only recently (the search used a circular orbit with a period determined from optical observations to a precision of $\approx 0.1$\,s, \citealt{Pletsch25102012}). The current precision on the orbital parameters is insufficient to make such a search feasible in gamma-ray binaries \citep{2012arXiv1209.2034C}.

\subsection{The population of gamma-ray binaries\label{population}}
\psrb\ is not the only known pulsar with a massive companion. The pulsar \href{http://www.atnf.csiro.au/people/pulsar/psrcat/}{\color{violet}{catalog}} \citep{2005AJ....129.1993M} lists three systems where the minimum mass of the companion, derived from pulsar timing, is greater than the minimum mass of the companion of \psrb\ (Tab.~\ref{radiopulsar}). The pulsars are not eclipsed in PSR J1740-3052 and PSR J0045-7319. The very small dispersion measured throughout the orbit restricts the stellar wind mass loss rate in these systems to (respectively) $\dot{M}_{\rm w}\approx 10^{-9}$\,\msol\,yr$^{-1}$ \citep{2012MNRAS.425.2378M} and $10^{-11}$\,\msol\,yr$^{-1}$ \citep{Kaspi:1996id}. The very low $\dot{M}_{\rm w}$ in PSR J0045-7319 has been interpreted as evidence for the impact of the low SMC metallicity on radiatively-driven winds. PSR J1638-4725 is known to be radio-quiet over a fifth of its orbit \citep{2004xmm..prop..229M}. The $\dot{M}_{\rm w}$ in  PSR J1740-3052 is on the low side for a massive companion, although similarly low values have been observed in other stars \citep{2012MNRAS.425.2378M}. None of these systems are known gamma-ray sources, which is no surprise given their much lower $\dot{E}$ and larger distances compared with \psrb.
\begin{table}
\centering
\ra{1.3}
\begin{tabular}{@{}lllll@{}}
\toprule 
						&  {\small \psrb }&  {\small PSR J1740-3052$^\star$} &  {\small PSR J1638-4725$^\diamond$} & {\small PSR J0045-7319$^ \dagger $} \\
 \midrule
 $M_{\star}$ (M$_{\odot}$) 	& 31 ($>3$) 		& $>$11 & $>6$ & $>4$ \\
spectral type 			& O9.5Ve 			& B?V	& ?		& B1V \\
 P$_{\rm orb}$ (days) 		& 1237 		&  231 & 1941 & 51$^\ddagger$\\
 P$_{\rm pulse}$ (s)			& 0.048    		&  0.570 & 0.764 & 0.926 \\
 $\dot{E}$ ($10^{34}$\,\eps)	& 80			&  0.5 & 0.04 & 0.02 \\
$\tau_{\rm sd}$ ($10^5$\,yr) & 3.3 & 3.5 & 52.8 & 32.9\\
$e$						& 0.87 		& 0.58 & 0.96 & 0.81 \\
 $d$ (kpc) 					& 2.3 & 11 & (6-7)$^\oplus$ & 61 (SMC)\\
 $d_{\rm periastron}$ (AU)  & 0.9 & 0.7  & 0.2 & 0.1\\
 $d_{\rm apastron}$ (AU) 		& 13 & 2.7 &  11.6 & 0.9 \\
 \bottomrule 
\multicolumn{5}{l}{\small $\star$ \citet{2001MNRAS.325..979S,2011MNRAS.412L..63B,2012MNRAS.425.2378M}}\\
\multicolumn{5}{l}{\small $\diamond$ \citet{2006MNRAS.372..777L}}\\
\multicolumn{5}{l}{\small $\oplus$ scaling the $K$ mag of coincident source 2MASS J16381302-4725335 with that of \lsi,}\\
\multicolumn{5}{l}{\small ~~~gives a distance consistent with the dispersion measure distance in Lorimer et al. 2006.}\\
\multicolumn{5}{l}{\small $\dagger$ \citet{1994ApJ...423L..43K,1995ApJ...447L.117B,1996Natur.381..584K,Kaspi:1996id}}\\
\multicolumn{5}{l}{\small $\ddagger$ 7.3 year superorbital modulation \citep{2011MNRAS.413.1600R}}
\end{tabular}\\
\caption{Radio pulsars in binaries with massive companions\label{radiopulsar} (pulsar \href{http://www.atnf.csiro.au/people/pulsar/psrcat/}{\color{violet}{catalog}}, \citealt{2005AJ....129.1993M}).}
\end{table}

Together with gamma-ray binaries (assuming these are pulsars), these systems constitute the progenitors of HMXBs \citep{Shvartsman:1971rw} and, consequently, also of the merging neutron star -- stellar-mass black hole or neutron star systems that are targeted by ground-based searches for gravitational waves \citep{2006LRR.....9....6P}. Magnetic flux and angular momentum conservation imply that neutron stars are born spinning rapidly and with a high magnetic field {\em i.e.} with a high $\dot{E}$. The power in the pulsar wind decreases as the pulsar spins down until pressure from the wind becomes too weak to hold off Bondi-Hoyle accretion of the stellar wind (\S\ref{containment}). Accretion starts at this point, turning off the radio pulses, the pulsar wind, and the associated non-thermal emission. The gamma-ray binary becomes a HMXB. The lifetime of a wind-fed HMXB is at most that of its massive companion, a few $10^6$\,yr. In comparison, the lifetime of the gamma-ray binary phase is at most the spindown timescale 
\begin{equation}
\tau_{\rm sd}=\frac{{P}}{(n-1)\dot{P}}\left[1-\left(\frac{P_{\rm i}}{P}\right)^{n-1}\right]\approx\frac{P}{2\dot{P}}\approx\frac{2\pi^2 I}{P^2\dot{E}}\approx 6\times10^5\ \left(\frac{0.1\rm\,s}{P}\right)^{2} \left(\frac{10^{35}\rm\, erg\,s^{-1}}{\dot{E}}\right)\,\mathrm{yr}
\end{equation}
where $n\approx 3$ for dipole radiation, $I=10^{45}$\,g\,cm$^2$ is the moment of inertia of the neutron star, $P_{\rm i}\ll P$ is the birth period, and the timescale is scaled for a pulse period of 0.1\,s and a spindown luminosity of $10^{35}$\,\eps, typical values for the young pulsars detected by \fermi. High spindown luminosities are necessary to have a strong pulsar wind pressure ($p\propto \dot{E}$) and a bright gamma-ray source. Hence, the gamma-ray binary phase is short in the evolution of the binary system. Indeed, there are only 5 known gamma-ray binaries (plus a couple more from Tab.~\ref{radiopulsar}) compared with $>100$ known HMXBs ($\ga$80\% of them with Be companions).

The number of such systems in our Galaxy may be estimated from models of the HMXB population. These models depend on assumptions for the initial binary parameters and various key processes in its evolution (winds, common envelope,  supernova, etc). Constraints on  mass loss and neutron star kick during the supernova are actually derived from the orbital eccentricity of gamma-ray binaries and the orientation of the star with respect to the orbital plane \citep{2002A&A...384..954R,McSwain:2002hp,2004ApJ...600..927M}. The HMXB birth rate is typically $\sim 10^{-3}$\,yr$^{-1}$  \citep{Meurs:1989zp,Iben:1995vu,Portegies-Zwart:1996ky,Portegies-Zwart:1998ic}. Hence, the number of gamma-ray binaries  is $\sim$100 with a $10^5$\,yr lifetime. Alternatively, the three brightest systems  (\psrb, \lsi, \ls) are within 3 kpc of the Sun, in a volume corresponding to $\sim$ 10\% of the volume of our Galaxy. The total number in the Galaxy of similarly luminous systems, if they are distributed uniformly, would be $\sim$ 30.

The HE luminosity of \psrb, \lsi, and \ls\ is $\ga 10^{35}$\,\eps\ (Tab.~\ref{mwltable}). Any equivalently-luminous gamma-ray binary within 9 kpc would be bright enough to appear in the second \fermi\ source catalog ($F_{\rm HE}\ga 10^{-11}$\,\eps\,cm$^{-2}$, see Fig. 7 in \citealt{2012ApJS..199...31N}), meaning a  dozen could be lurking amongst the 254 \fermi\ sources with $|b|\leq 2$\degr. Some could be found in searches for variable sources in the Galactic Plane \citep{2012A&A...543L...9N,2013ApJ...771...57A}. In VHE, the HESS Galactic Plane survey is thought to be complete down to 8.5\% of the flux from the Crab nebula \citep{2009arXiv0905.1287R}. The corresponding normalisation at 1 TeV is $K\approx 3\times10^{-12}\,\mathrm{ph}\,\mathrm{cm}^{-2}\,\mathrm{s}^{-1}\,\mathrm{TeV}^{-1}$,  comparable to those of detected gamma-ray binaries\footnote{The Crab normalised flux at 1 TeV is $K\approx3.5\times10^{-11}\,\mathrm{ph}\,\mathrm{cm}^{-2}\,\mathrm{s}^{-1}\,\mathrm{TeV}^{-1}$ for a single power-law fit with $\Gamma_{\rm VHE}\approx 2.6$ \citep{2006A&A...457..899A}. The VHE slope of gamma-ray binaries is comparable.}. However, the faintest sources in the survey are about 4 times fainter so there is also scope for a few more detections using the current generation of IACTs. With a ten-fold improvement in sensitivity, the {\em Cherenkov Telescope Array} (CTA) can be expected to detect a couple dozen gamma-ray binaries \citep{2012arXiv1210.3215P}. These numbers are subject to caution. Detectability is dependent on the duty cycle of gamma-ray emission. \psrb\ and \hessj\ illustrate how VHE detectability strongly depends on orbital phase, and the orbital periods can be long. \hessj\ (and the radio pulsars in Tab.~\ref{radiopulsar}) proves that much less gamma-ray luminous sources exist. Surveys could reveal many more than extrapolated here, depending on the  $\log N- \log S$ distribution of gamma-ray binaries. 

\subsection{Low-mass gamma-ray binaries ?\label{lmgb}}
The radio pulsar catalog also lists more than a hundred radio pulsars in low-mass binaries with companions ranging from 1.7\,\msol\ down to 0.02\,\msol, orbital periods from $0.07$ days up to hundreds of days, and $\dot{E}$ as high as 10$^{35}$\,\eps. Dozens of those have also been detected in HE gamma rays by \fermi\ \citep{2013arXiv1305.4385T}. The processes at work in some of these systems are similar to those in  gamma-ray binaries.

Gamma-ray pulsars in low-mass binaries are old pulsars that have been spun-up (``recycled'') to millisecond rotation periods by the accretion of mass from their companion \citep{Lorimer:2005ic}. Accretion is stopped once the neutron star rotates fast enough, followed by the radio pulsar -- and pulsar wind -- switching on (see \S\ref{containment}). Hence, these systems are the rotation-powered descendants of accreting LMXB, mirroring the passage of a gamma-ray binary to an accreting HMXB. A couple of systems have been observed to transit from the accreting, LMXB stage to the radio pulsar stage (PSR J1023+0038, \citealt{2009Sci...324.1411A}; IGR J18245-2452, \citealt{2013arXiv1305.3884P}). PSR J1023+0038 is also a weak \fermi\ source \citep{2010ApJ...724L.207T}.  The low-mass X-ray binary XSS J12270--4859, also associated with a \fermi\ source and observationally similar to PSR J10123+0038, may be another example of a source undergoing this transition \citep{2010A&A...515A..25D,2013A&A...550A..89D,2011MNRAS.415..235H}.

In black widow pulsars, the pulsar wind ablates the surface of its low-mass companion, producing a bow shock \citep{1988Natur.333..832P}. Another shock occurs where the pulsar wind interacts with the ISM. Redbacks are similar systems with 0.2-0.4 $M_\odot$ companions. \fermi\ has detected gamma rays from more than a dozen black widows and redbacks \citep{Roberts:2013aa}. Non-thermal emission from the intra-binary shock, detected in X-rays in these systems  (e.g. \citealt{Bogdanov:2005dt}), could contribute unpulsed gamma-ray emission. In nearly all systems, the gamma-ray emission is identical to that of isolated pulsars and shock emission is undetected. However, \cite{2012ApJ...761..181W} found evidence for an orbital modulation of the HE gamma-ray emission in the original black widow pulsar (PSR B1957+20). The pulsar spectrum with a cutoff at $E_c\approx 1\,\rm GeV$ is not variable ; the modulation is restricted to the second spectral  component that appears above 2.7\,GeV, which they attribute to pulsar wind emission. The system behaves like a  ``low-mass'' gamma-ray binary.


\section{Models of gamma-ray binaries}
If gamma-ray binaries harbour rotation-powered pulsars, how do the observations described above fit in our broad understanding of the way pulsars function and what are the consequences ? \S\ref{pw} presents pulsar winds and the conditions for their presence in binaries. \S\ref{shock} describes what happens at the shock. \S\ref{modulations} addresses the generic radiative models that have been put forward to explain the modulated gamma-ray emission in binaries. \S\ref{probe} discusses challenges to models and how gamma-ray binaries probe pulsar winds in a novel way. 

\subsection{Gamma-ray binaries as pulsars in binaries\label{pw}}

\subsubsection{Pulsar winds\label{pw1}}
Conservation of angular momentum and magnetic flux implies that neutron stars are born from their massive star progenitors as fast-rotating, intense dipoles. Dipolar field lines open up beyond the light cylinder radius
\begin{equation}
R_L=\frac{c}{\Omega}=\frac{cP}{2\pi}=5\times 10^9\, \left(\frac{P}{\rm 1\,s}\right)\, \mathrm{~cm}
\end{equation}
where co-rotation with the neutron star would require superluminal speeds. Particles moving along open field lines escape, forming a pulsar wind carrying away energy at a rate  \citep[][and references therein]{Spitkovsky:2006mq,2012MNRAS.424..605P}
\begin{equation}
\dot{E}\approx\frac{B_{\rm ns}^2 R_{\rm ns}^6 \Omega^4}{c^3}\left(1+\sin^2 \chi \right)
\label{eq1}
\end{equation}
where $\chi$ is the obliquity (inclination of the magnetic field axis with respect to the spin axis, $\chi=0$\degr\ when aligned), $B_{\rm ns}$ the magnetic field intensity at the neutron star surface, and $\Omega=2\pi/P$. The rotation of the neutron star provides the energy reservoir that is being tapped at a rate
\begin{equation}
\dot{E}=-I\Omega\frac{d\Omega}{dt}=-\frac{2}{5}GMR_{\rm ns}^{2}\Omega\dot{\Omega}.
\label{eq2}
\end{equation}
with $I$ the moment of inertia of the neutron star. The surface magnetic field of  pulsars $B_{\rm ns}$ can be estimated from the observation of $\Omega$ and $\dot{\Omega}$ by equating Eqs.~\ref{eq1}-\ref{eq2}. Old, ``recycled'' millisecond pulsars typically have $B_{\rm ns}\sim 10^{9}$\,G. Young pulsars have $B_{\rm ns}\sim 10^{12}$\,G. Magnetars have magnetic fields typically above the critical ``Schwinger'' field $B_{\rm crit}=(m_e^2c^3)/(e\hbar)\approx 4.4\times10^{13}\,{\rm G}$, up to $10^{15}$\,G.  

The emission from pulsars represents only a fraction of the measured spindown power $\dot{E}$ (with a caveat on beaming corrections). Most of the pulsar's spindown power is thought to be carried by the pulsar wind. This energy is released when the wind interacts with the surrounding medium (ISM, supernova ejecta, stellar wind) to create a pulsar wind nebula. The wind  initially has a high magnetisation
\begin{equation}
\sigma=\frac{B^{2}}{4\pi\Gamma n_{e}m_{e}c^{2}}
\end{equation}
where $\Gamma$ is the Lorentz factor of the particles with number density $n_{e}$. A high $\sigma$ means that the energy is carried by the magnetic field. However, assuming the termination shock behaves like an MHD shock, high $\sigma$ winds do not give enough energy to the particles at the shock to explain PWN emission, suggesting a transition occurs from a high $\sigma$ close to the pulsar to a low $\sigma$ at the shock  \citep{Kennel:1984pd,Kennel:1984gu}. 

The ``$\sigma$ problem'' is a long-standing issue of pulsar physics, also applicable to relativistic jets where a similar transition may occur. In pulsars, the open field lines originating from opposed magnetic poles have a different polarity and are thus separated by a current sheet that oscillates around the equator as the neutron star rotates. The current sheet is ``striped''. Reconnection in the current sheet may redistribute energy from the magnetic field  to the particles. The compression of the stripes at the termination shock can also trigger reconnection. Standard pulsar models assume that a transition from high to low $\sigma$ has occurred and that the pulsar wind consists of cold $e^+e^-$ pairs, frozen in with the magnetic field, carried outwards with  a high bulk Lorentz factor $\Gamma\sim 10^{4}-10^{6}$. Therefore, the particles do not radiate synchrotron emission (consistent with the dearth of emission between the pulsar and the termination shock in images of the Crab pulsar wind nebula) so little information is available on the processes occurring in pulsar winds (see \citealt{2009ASSL..357..373A} and \citealt{Kirk:2007wh} for reviews on all of the above). 

Gamma-ray binaries offer new opportunities to understand pulsar winds, their composition and their interaction, because the termination shocks are much closer to the pulsar than in isolated PWNs because of the high stellar wind densities, because orbital motion enables us to see the interaction from a regularly changing viewpoint, and because inverse Compton scattering off the companion light can reveal the particles in the pulsar wind (\S\ref{probe}).

\subsubsection{Colliding wind geometry\label{cwgeometry}}
\begin{figure}\sidecaption 
\resizebox{0.6\hsize}{!}{\includegraphics*{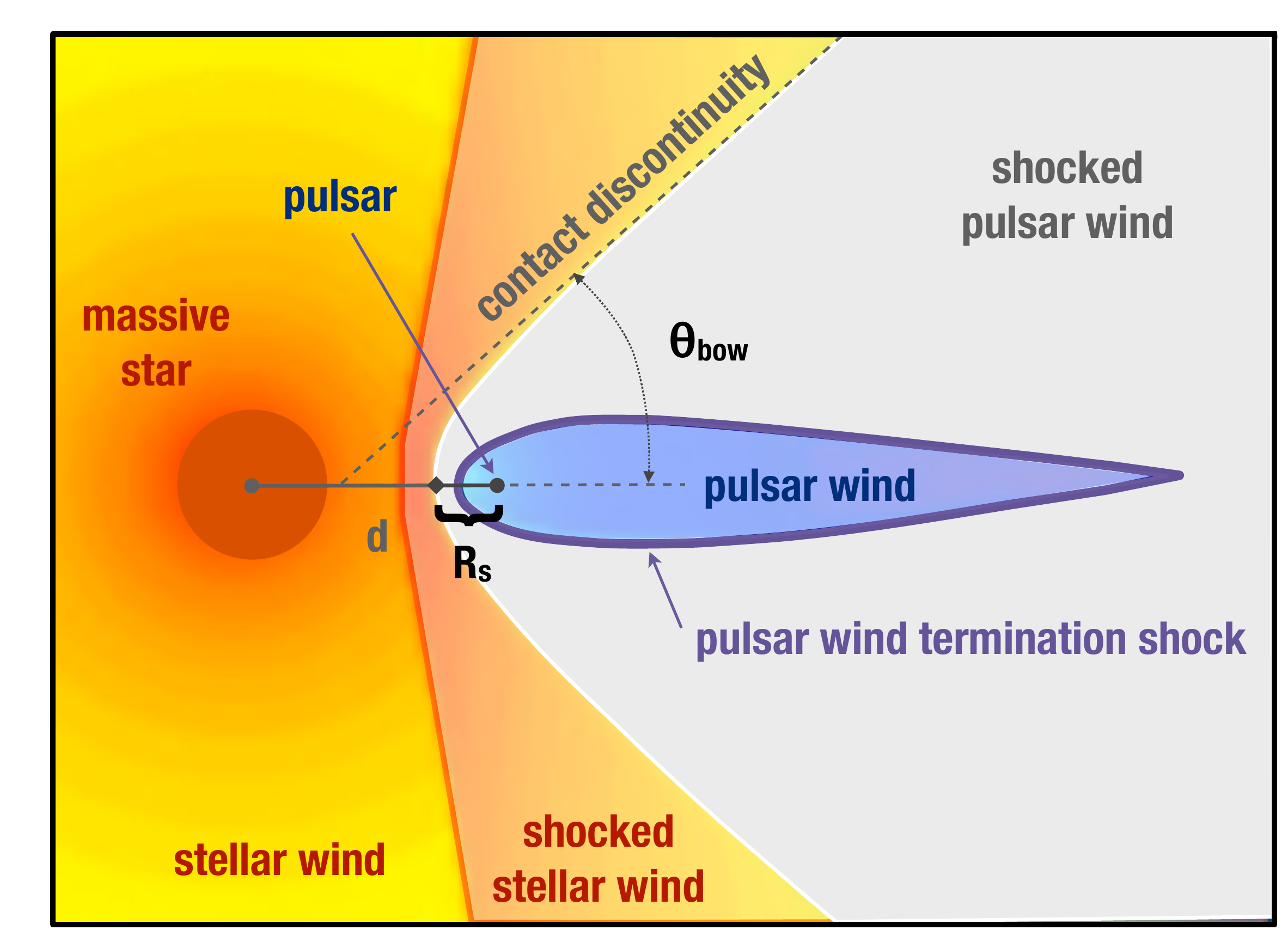}} \caption{Geometry of the interaction region between a pulsar wind and a stellar wind. Here, the pulsar wind is completely confined. Sketch based on a relativistic hydro simulation (courtesy of A. Lamberts).\label{fig:geom}}
   \end{figure}
The pulsar wind creates a dynamical (ram) pressure
\begin{equation}
p_{\rm pw}=\frac{\dot{E}}{4\pi R^2 c}
\label{ppw}
\end{equation}
that pushes on the surrounding environment at a distance $R$ from the pulsar. A steady situation is  reached if a stable pressure balance can be established. In the  case of binaries, the counteracting pressure can be the pressure from infalling material or the pressure of the companion's stellar wind or magnetic field \citep{Illarionov:1975yk,Harding:1990gb}. Whichever one dominates depends on the parameters of the pulsar and the stellar wind. Let us assume here that the pulsar wind is stopped by the stellar wind, leaving to the next section (\S\ref{containment}) the question of what happens when material is accreting towards the neutron star.

The interaction is supersonic for both mediums, resulting in a double shock structure with a contact discontinuity separating the shocked material from each region  (Fig.~\ref{fig:geom}). The location of this discontinuity on the binary axis (the stagnation point or standoff distance) can be derived by matching the ram pressures. In the case of a pulsar wind interacting with a coasting, supersonic, isotropic stellar wind 
\begin{equation}
p_{\rm pw}=\frac{\dot{E}}{4\pi R_s^2 c}=p_{\rm w}=\rho_{\rm w}v_{\rm w}^{2}=\frac{\dot{M}}{4\pi (d-R_s)^{2} v_{\rm w}}
\end{equation}
where $d$ is the orbital separation and $R_s$ is the distance of the discontinuity from the pulsar. The shock is stationary. The standoff distance is therefore
\begin{equation}
\frac{R_s}{d}=\frac{\eta^{1/2}}{1+\eta^{1/2}}
\end{equation}
with
\begin{equation}
\eta=\frac{\dot{E}/c}{\dot{M}_{\rm w}v_{\rm w}}\approx 0.05 \left(\frac{\dot{E}}{10^{36}\,\mathrm{erg}\,\mathrm{s}^{-1}}\right)\left(\frac{10^{-7}\,\mathrm{M}_\odot\,\mathrm{yr}^{-1}}{\dot{M}_{\rm w}}\right) \left(\frac{1000\,\mathrm{km}\,\mathrm{s}^{-1}}{{v}_{\rm w}}\right) 
\label{eq:eta}
\end{equation}
This dimensionless parameter, representing the ratio of momentum flux, is a key parameter in colliding wind binaries \citep{1990FlDy...25..629L}.

The usual (relativistic)(magneto) hydrodynamical fluid equations with jump conditions must be solved to get the  detailed structure (\S\ref{shock}). Semi-analytical estimates of the location of the interaction region are obtained by assuming the shock is thin and by balancing the  ram pressure normal to the discontinuity \citep{Stevens:1992on} or the linear and angular momentum  across the discontinuity \citep{Canto:1996jj}. The interaction region has a bow shape, starting at $R_s$ and reaching a constant opening angle $\theta_{\rm bow}$ at infinity, which depends only on $\eta$. \cite{2008MNRAS.387...63B} found that the formula first proposed by \cite{Eichler:1993wt}
\begin{equation}
\theta_{\rm bow}\approx 28.6\degr\,(4-\eta^{2/5}) \eta^{1/3}
\label{eq:opening}
\end{equation}
provides a good fit to the asymptotic angle of the contact discontinuity in their relativistic hydro simulations of a pulsar wind colliding with a stellar wind (Fig.~\ref{fig:bow}).
\begin{figure}
\sidecaption
\resizebox{0.68\hsize}{!}{\includegraphics*{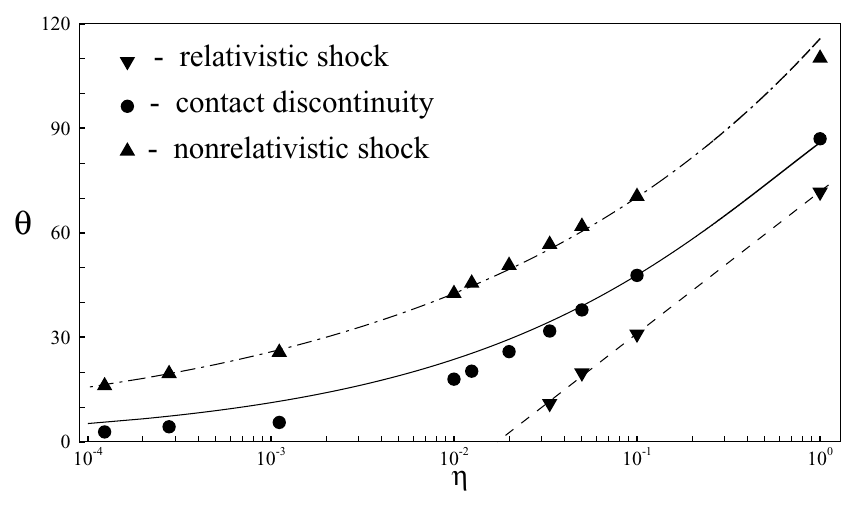}}
\caption{Asymptotic angle of the different discontinuities obtained from numerical simulations. The opening angle of the contact discontinuity (circles)  corresponds to $\theta_{\rm bow}$ in Fig.~\ref{fig:geom}. In these simulations the pulsar wind is completely confined when $\eta\la 0.02$. Figure reprinted from \citet{2008MNRAS.387...63B} by permission of Oxford University Press, on behalf of the RAS.\label{fig:bow}}
\end{figure}

Some consequences of the bow-shock geometry for gamma-ray binaries are:
\begin{itemize}
\item If $d-R_s\leq R_\star$ the wind collides with the surface of the companion star. This is what happens in black widow systems because the stellar wind of the low-mass star is weak so $\eta\gg 1$. If that were to happen in gamma-ray binaries, this would prevent the formation of a line-driven stellar wind over the hemisphere impacted by the pulsar wind. The available UV data does not show wind lines disappearing at some orbital phases, implying $\eta\la0.6$ for \ls\ \citep{2011MNRAS.411..193S}. Even without direct impact, the presence of a pulsar wind with a sizeable solid angle as seen from the massive star can lead to detectable modulations in the profile of stellar wind lines  \citep{2012MNRAS.420.3521S}.
\item If $\eta\la 1$ the stellar wind dominates and the bow shock wraps around the pulsar, collimating the shocked flow as suggested by the radio maps (Fig.~\ref{lsiradio}). However, in \lsi, the fast polar stellar wind is not strong enough to achieve this \citep{Romero:2007ly}. The value of $\eta$ is $\approx 5$ when taking  $\dot{M}_{\rm w}\approx 10^{-8}\rm\, M_\odot\,yr^{-1}$ (appropriate for the Be star) and $\dot{E}\approx 10^{37}$\,\eps\ (assuming the correlation observed in PWN between $\dot{E}$ and $F_{\rm VHE}/F_X$ also holds here, \citealt{2009ApJ...694...12M}). Although the estimate is uncertain, a low value of $\eta$ seems unlikely unless the dense Be disc plays a role. Material in the disc is in keplerian rotation with a density distribution $\rho_{\rm disc}\sim 10^{-11} (R/R_\star)^{-3}$\,g\,cm$^{-3}$ \citep{1988A&A...198..200W}. The pulsar wind pushes against nearly static material in the co-rotating frame, leading to the formation of a transient, anisotropic, PWN bubble, which could be responsible for the radio outbursts of \psrb\ and \lsi\ \citep{Dubus:2006lc} or the GeV flare of \psrb\ \citep{2012ApJ...752L..17K}. For \lsi\ at periastron ($v_{\rm orb}\approx 300$\,km\,s$^{-1}$, $R\approx 4R_\star$, $\rho_{\rm disc}\approx 10^{-13}$\,g\,cm$^{-3}$), equating the ram pressures gives $R_s\approx 2\times 10^{11} \dot{E}_{36}^{1/2}$\, cm or an ``equivalent'' $\eta\approx 0.3$ (see discussion in \citealt{Sierpowska-Bartosik:2008wt}). This time-dependent interaction has only begun to be addressed by numerical simulations using SPH codes (\citealt{2011PASJ...63..893O,2012ApJ...750...70T}, see also \citealt{1993ApJ...406..638K} for an early attempt).
\item Gamma-ray binaries are likely to radiate  the spindown energy efficiently if they are to collimate the shocked pulsar wind. The reason is that collimation requires a low $\eta$, hence the smallest value of $\dot{E}$ possible. Since $\dot{E}$ is bounded by the gamma-ray luminosity, this is equivalent to saying that the radiative efficiency must be as high as possible.This is supported by the high radiative efficiency observed in \psrb, where the peak HE gamma-ray luminosity was nearly equal to the spindown luminosity (\S\ref{he}).
\end{itemize}

\subsubsection{Turning accretion on and off\label{containment}}
Under what conditions can the pulsar wind hold off the pressure from accretion ? The simplest approach is to assume Bondi-Hoyle accretion of the stellar wind  \citep{Illarionov:1975yk}. Assuming a capture radius $R_{\rm ac}\approx 2GM_{\rm ns}/v^{2}_{\rm w}$ and a relative velocity of the accreted material $\approx v_{\rm w}$, the Bondi-Hoyle accretion rate is 
\begin{eqnarray}
\dot{M}_{\rm ac}&\approx& \pi R^2_{\rm ac} \rho_{\rm ac} v_{\rm w}= \frac{1}{4} \left(\frac{R_{\rm ac}}{d}\right)^2 \dot{M}_{\rm w}\\
&\approx& 3\times 10^{14}\,\left(\frac{\dot{M}_{\rm w}}{10^{-7}\,\mathrm{M}_\odot\,\mathrm{yr}^{-1}}\right)\left(\frac{1000\,\mathrm{km}\,\mathrm{s}^{-1}}{{v}_{\rm w}}\right)^{4}\left(\frac{0.1\,\mathrm{AU}}{d}\right)^{2}\,\mathrm{g}\,\mathrm{s}^{-1}
\label{eq:bondi}
\end{eqnarray}
for a 1.4\,\msol\ neutron star with an isotropic, coasting, radiatively-driven wind.  Here, $\rho_{\rm ac}= \rho_{\rm w}(R_{\rm ac})\approx \dot{M}_{\rm w}/(4\pi d^2 v_{\rm w})$ with $d$ the orbital separation.  The ram pressure from accretion is
\begin{equation}
p_{\rm ac}\approx \rho_{\rm ac} v_{\rm ac}^2 \approx \frac{\dot{M}_{\rm ac}}{4\pi R^2} \left(\frac{2GM_{\rm ns}}{R}\right)^{1/2}
\end{equation}
where taking the escape velocity for $v_{\rm ac}$ gives the correct order-of-magnitude. Equating $p_{\rm ac}$ with the pulsar wind pressure  $p_{\rm pw}$ (Eq.~\ref{ppw}) gives the equilibrium radius. 

Such an equilibrium is unstable. Consider a  perturbation that increases $p_{\rm ac}$ at the equilibrium radius while $p_{\rm pw}$ stays constant. As the equilibrium radius is pushed closer to the pulsar, the accretion pressure $p_{\rm ac}\propto R^{-2.5}$ increases faster than the pulsar wind pressure $p_{\rm pw}\propto R^{-2}$. The accretion flow pushes in unimpeded until  the light cylinder is reached and the fast rising  dipole magnetic pressure $\propto R^{-6}$ enables a new equilibrium. Inversely, a decrease in $p_{\rm ac}$ will lead to the pulsar wind blowing away the accretion flow until the pressure from the ISM or the stellar wind holds it off beyond the capture radius. 

In a gamma-ray binary, the system is thought to have a high $\dot{E}$ pulsar in equilibrium with the stellar wind at $R_{s}>R_{\rm ac}$. As $\dot{E}$ decreases with spindown, the stagnation point $R_{s}$ draws closer to the pulsar until it reaches $R_{\rm ac}$. Given the above, the equilibrium becomes unstable and a small inflow of matter will be enough to swamp the pulsar wind down to $R_L$ and turn it off. The pulsar wind holds off accretion as long as $p_{\rm pw}>p_{\rm w}$ at $R_{\rm ac}$ ($p_{\rm w}$ is also $>p_{\rm ac}$ at $R_{\rm ac}$). The minimum $\dot{E}$ above which the pulsar can prevent accretion is thus
\begin{eqnarray}
\dot{E}>4\dot{M}_{\rm ac} v_{\rm w} c&\approx& 10^{35}\,  \left(\frac{\dot{M}_{\rm ac}}{10^{16}\,\mathrm{g\,s}^{-1}}\right)\left(\frac{{v}_{\rm w}}{1000\,\mathrm{km}\,\mathrm{s}^{-1}}\right)\,\mathrm{erg\,s}^{-1}\\
&\approx& 4\times 10^{33}\, \left(\frac{\dot{M}_{\rm w}}{10^{-7}\,\mathrm{M}_\odot\,\mathrm{yr}^{-1}}\right)\left(\frac{1000\,\mathrm{km}\,\mathrm{s}^{-1}}{{v}_{\rm w}}\right)^{3}\left(\frac{0.1\,\mathrm{AU}}{d}\right)^{2}\,\mathrm{erg\,s}^{-1}
\label{eq:holdoff1}
\end{eqnarray}
The condition is satisfied by gamma-ray binaries since the luminosities imply at least $\dot{E}\ga L_{\gamma}\approx 10^{35}$\,\eps\ and the Bondi accretion rates are $< 10^{15}$\,g\,s$^{-1}$ (Eq.~\ref{eq:bondi}). It is also satisfied by the radio pulsars with stellar companions listed in Tab.~\ref{radiopulsar}. Alternatively, for a given $\dot{E}$, Eq.~\ref{eq:holdoff1} gives the minimum $\dot{M}_{\rm ac}$ above which accretion starts.

Assume now that the pulsar wind is turned off because accretion occurs within the light cylinder \citep{Illarionov:1975yk}. The accretion flow is truncated at the Alfv\'en radius where the neutron star magnetic dipole pressure is comparable to the accretion pressure \citep {Pringle:1972kx,1973ApJ...179..585D}. If the accretion rate is dropping, the Alfv\'en radius moves outward and the pulsar wind can turn on again when $p_{\rm pw}>p_{\rm ac}$ at $R_{\rm L}$. The critical $\dot{M}_{\rm ac}$ below which the pulsar turns back on, for fixed pulsar parameters, is:
\begin{eqnarray}
\dot{M}_{\rm ac}<\frac{\dot{E}}{c}\left(\frac{cP}{4\pi GM_{\rm ns}}\right)^{1/2}&\approx&4\times 10^{16}\,\left(\frac{\dot{E}}{10^{36}\,\mathrm{erg}\,\mathrm{s}^{-1}}\right) \left(\frac{P}{0.1\,\mathrm{s}}\right)^{1/2}\,{\rm g\,s}^{-1}\\
&\approx& 1.4\times10^{16}\,\left(\frac{B_{\rm ns}}{10^{12}\,{\rm G}}\right)^2 \left(\frac{0.1\,\mathrm{s}}{P}\right)^{7/2}\,{\rm g\,s}^{-1}
\label{eq:holdoff2}
\end{eqnarray}
High accretion rates must be sustained throughout the orbit in order to prevent the pulsar wind from re-establishing itself. This condition also means that recycled pulsars in LMXBs ($B_{\rm ns}\approx 10^9$\,G, $P\approx $ a few ms) can turn back on when the mass transfer rate from the companion becomes very low \citep{Burderi:2001nx}.

Accretion of material from the Be disc can lead to much higher mass inflow rates because of the high densities and low speeds. Bondi-Hoyle capture provides only very rough guidance to estimate $\dot{M}_{\rm ac}$. In the case of \lsi\ at periastron, $v_{\rm orb}\approx 300$\,km\,s$^{-1}$, $R\approx 4R_\star$ so $\rho_{w}\approx 10^{-13}$\,g\,cm$^{-3}$, $\dot{M}_{\rm ac}\approx 10^{18}$\,g\,s$^{-1}$, and the minimum spindown power to hold off accretion becomes quite high $\dot{E}\ga 6\times 10^{36}$\,\eps. Hence, \citet{2012ApJ...744..106T} proposed that  the pulsar in  \lsi\ is not turned on throughout the orbit, with a hysteresis since the turn on and turn off occur for different accretion rates.

The pulsar parameters of \lsi\ can be constrained if there is no accretion. Combining the minimum $\dot{E}$ found above with the magnetic field $B_{\rm ns}\ga 10^{13}$\,G suggested by low-level magnetar activity in \lsi\ gives 
\begin{equation}
P\la 0.3\,\left(\frac{B}{10^{13}\rm\,G}\right)^{1/2}\left(\frac{\dot{E}}{10^{36}\rm\,erg\,s^{-1}}\right)^{-1/4} \ {\rm s}
\end{equation}
The pulsar would be placed between magnetars and radio-loud pulsars in the $(P,\dot{P})$ diagram \citep{2010ASPC..422...23D}. There are pulsars with such parameters: PSR J1846-0258  (associated with SNR Kes 75, also showing magnetar bursts, $B\approx 4.9\times 10^{13}$\,G, $P\approx0.33$\,s, $\dot{E}\approx 8\times10^{36}$\,\eps) and PSR J1119-6127 (SNR G292.2-00.5, $B\approx 4.1\times 10^{13}$\,G, $P\approx 0.41$\,s, $\dot{E}\approx 2\times10^{36}$\,\eps). These are amongst the youngest pulsars in our Galaxy, with a spindown timescale $\tau\approx 10^3$\,yr and an associated supernova remnant. \lsi\ could be a similarly young system.

\subsubsection{The transition to high-mass X-ray binaries: propellers}
Once accretion starts, the flow penetrates inside the light cylinder and is stopped by magnetic dipole pressure at the Alfv\'en radius. However, an X-ray pulsar does not turn on straightaway because when accretion starts the Alfv\'en radius is just inside the light cylinder, so that the neutron star magnetic field necessarily has a higher specific angular momentum $R^2\Omega$ than the incoming material.  The interaction propels matter out of the system, torquing down the star, with little accretion onto the poles  \citep[e.g.][]{Stella:1986qd}. Pole accretion proceeds once the neutron star has spun down enough that the Alfv\'en radius becomes smaller than the corotation radius, completing the transition from gamma-ray binary to HMXB.

\citet{2009MNRAS.397.1420B} and \citet{2012ApJ...744..106T} proposed that \lsi\ becomes a propeller over part of its orbit, when the pulsar wind cannot hold off accretion. The consequences of a propeller on gamma-ray emission are uncertain. There is at present only one well-established propeller system from which to derive insights: the intermediate polar AE Aqr, composed of a 33\,s rapidly-spinning white dwarf orbiting a low-mass star filling its Roche lobe. The strong spindown is compatible with the propeller torque. The system shows non-thermal emission in the form of strong radio flares, which is unique amongst cataclysmic variables, attributed to material being ejected from the system \citep{1988ApJ...324..431B}.  The binary is on the notorious list of past claimed VHE emitters which have not been confirmed by present IACTs (\citealt{Lang:1998rz}, \citealt{2008ICRC....2..715S}).

\subsection{Modeling the pulsar wind -- stellar wind interaction\label{shock}}

\subsubsection{Conditions at the pulsar wind shock\label{conditions}}
Following the standard picture of pulsar wind nebulae \citep{Rees:1974xr,Kennel:1984pd,Kennel:1984gu}, the pulsar wind arrives at the shock with a bulk radial ultra-relativistic motion ($\Gamma\sim 10^6$) and a low magnetisation ($\sigma\sim 10^{-3}$). The properties of the shocked region can be studied by using the equations of relativistic magnetohydrodynamics, assuming the fluid description is adequate to describe the large-scale behaviour of the collisionless plasma. If the upstream velocity normal to the shock is ultra-relativistic then the downstream post-shock velocity converges to $c/3$ (when the fluid is a perfect gas composed of relativistic particles i.e. with a polytropic index $\hat{\gamma}=4/3$), the lab frame magnetic field and density are amplified by a factor 3, while the energy per particle downstream tends to $p/\rho=\Gamma c^2/\sqrt{18}$. The bulk kinetic energy is transferred to the random motion of the $e^{+}e^{-}$ pairs, which then radiate downstream of the termination shock. If the magnetic field is perpendicular to the shock normal (reasonable for a pulsar wind since the field is expected to be toroidal), its post-shock amplitude for a low-$\sigma$ wind is given by \citep{Kennel:1984pd}
\begin{equation}
B= 3(1-4\sigma) \left(\frac{\dot{E}/c}{R_s^2}\frac{\sigma}{1+\sigma}\right)^{1/2}\approx 1.7~ \left(\frac{\dot{E}}{10^{37}\,\mathrm{erg}\,\mathrm{s}^{-1}}\right)^{1/2}\left(\frac{\sigma}{10^{-3}}\right)^{1/2} \left(\frac{10^{12}\mathrm{\,cm}}{R_s}\right)~\mathrm{G}
\label{magnetic}
\end{equation}
The typical magnetic field expected at the termination shock of gamma-ray binaries is of order 1 G, much larger than the magnetic field at the shock in isolated pulsars where $R_s\approx 10^{17}$ cm. 

The fluid description erases all information of the particle momentum distribution. The usual assumption is that the wind particles will be accelerated to a power-law distribution in energy, an assumption that is required by observations but remains challenging to prove theoretically \citep{2011CRPhy..12..234L}. Numerical simulations find this occurs in unmagnetised pair plasmas \citep{2008ApJ...682L...5S} or for nearly parallel magnetised shocks \citep{2009ApJ...698.1523S}. Assuming a downstream power-law distribution in energy with a  given slope between $E_{\rm min}$ and $E_{\rm max}\gg E_{\rm min}$, the normalisation and $E_{\rm min}$ of the electron distribution can be derived by matching the shocked wind density and pressure, with $E_{\rm min}$ typically $\approx 0.1 \Gamma m_e c^2$. 

The maximum energy of the accelerated pairs is limited by radiation. The timescale for diffusive shock acceleration $\tau_{\rm acc}$ at a relativistic shock is no less than the timescale for gyration in the magnetic field
\begin{equation}
\tau_{\rm acc}\ga\xi \frac{R_L}{c}=\xi \frac{E}{eBc}\approx 0.1\, \xi \left(\frac{E}{1\,{\rm TeV}} \right) \left(\frac{1\, {\rm G}}{B}\right)\ {\rm s}
\end{equation}
where $\xi>1$. The Larmor radius becomes comparable to the characteristic size of the shock (say $R_{s}\approx 10^{12}$\,cm) for 300 TeV electrons. Such particle energies are not reached. The timescale for synchrotron radiative losses decreases with $E$ as \citep{Blumenthal:1970op}
\begin{equation}
\tau_{\rm sync}=\frac{3}{4}\frac{m_e^2 c^3}{\sigma_T}\frac{8\pi}{B^2}\frac{1}{E}\approx 400\, \left(\frac{1\,{\rm TeV}}{E} \right) \left(\frac{1\, {\rm G}}{B}\right)^2\ {\rm s}
\label{eq:tsync}
\end{equation}
where $\sigma_{T}$ is the Thompson cross-section. Since $\tau_{\rm acc}$ must be smaller than $\tau_{\rm sync}$, the maximum electron energy is 
\begin{equation}
E_{\rm max} \approx 60\ \xi^{-1/2} \left(\frac{1\, {\rm G}}{B}\right)^{1/2}\rm\ TeV
\end{equation}
This is enough to emit TeV radiation by upscattering seed photons. Inversely, the detection of VHE photons means that the magnetic field cannot be too high. 

Taking the $B$ estimated above (Eq.~\ref{magnetic}), the detection of photons with energies up to 10\,TeV implies efficient particle acceleration is ongoing ($\xi\sim1$), close to the theoretical limits for diffusive shock acceleration \citep{Khangulyan:2007me}.  Detecting $>10$ TeV photons would point to a different acceleration mechanism. One possibility is   acceleration at reconnection sites, already invoked to explain the gamma-ray flares from the Crab PWN \citep{2011ApJ...737L..40U}. The termination shock provides such reconnection sites if the pulsar wind remains striped: compression of the magnetic field at the shock can then trigger reconnection \citep{2003MNRAS.345..153L}. The pulsar wind is more likely to stay striped in gamma-ray binaries than in isolated pulsars because the termination shock is closer to the pulsar, leaving less time for reconnection to dissipate the structure before the shock. The  dissipation time is at most the light crossing time of the stripe wavelength, to be compared with the time to reach the shock in the wind frame. Combining the two gives an upper limit on the Lorentz factor of the wind $\Gamma\la (R_{\rm s}/R_{\rm L})^{1/2}$ (see \citealt{2009ASSL..357..373A}) i.e. $\Gamma\la 100$ in gamma-ray binaries. Since pulsar winds are thought to have much higher values of $\Gamma$, the stripes remain until the shock is reached. Simulations of shock-driven reconnection by \citet{2011ApJ...741...39S} find that particles initially accelerated by reconnection in the equatorial plane are further accelerated by a ``Fermi-like'' process, yielding a power-law spectrum, with the magnetic energy efficiently transferred to the particles. They also find that obtaining power-laws requires $4\pi \kappa (R_{\rm L}/R_{\rm s})\ga 10$, with $\kappa$ the pair multiplicity (with $\kappa$ typically $\approx 10^{4}$ to $10^{6}$). This should be substantially easier to achieve for gamma-ray binaries where $(R_{\rm L}/R_{\rm s})\sim 10^{-4}$ than in PWN where the termination shock is a factor 10$^4$ further away. Hence, there is some theoretical support for assuming a power-law in a low-$\sigma$ plasma (at least in the equatorial plane: at high latitudes a substantial  post-shock magnetic field remains and the distribution of particles is Maxwellian).

\subsubsection{Simulating the interaction on large scales\label{pwn}}
The shock region has a comet shape directed away from the stronger wind, when both winds are radial and when neglecting orbital motion (\S\ref{cwgeometry}). The shocked pulsar wind radiates synchrotron and inverse Compton emission (see \S\ref{modulations}). Particles  cool as they are advected in the comet-shaped flow, emitting at progressively longer wavelength (see example in Fig.~\ref{evol}). Radio emission becomes visible far from the binary once it emerges from the fog due to free-free absorption by the stellar wind. 

The comet tail turns with the movement of the pulsar along its orbit, creating a spiral pattern with a step size $ v P_{\rm orb}\approx 700 (v/c)$ AU (for \ls). The speed $v$ at which the pattern propagates outward is either the shocked pulsar wind speed, with asymptotic velocity $v=\sigma c$ \citep{Kennel:1984gu}, or the stellar wind speed depending on which is dominant \citep{2012A&A...546A..60L}. Kinematic models of the VLBI radio structures give $v/c\sim 0.01$, consistent with stellar wind speeds of a few 1000\,km\,s$^{-1}$ and/or $\sigma\sim 0.01$ \citep{Dubus:2006lc,2011ApJ...732L..10M,2012arXiv1209.6073M}. The radio properties of gamma-ray binaries (extended source, non-thermal emission at progressively longer wavelengths away from the compact object, repeatable orbital phase-dependent morphology, \S\ref{radio}) are thus consistent with a comet-like PWN.

\begin{figure}
\resizebox{!}{4.05cm}{\includegraphics{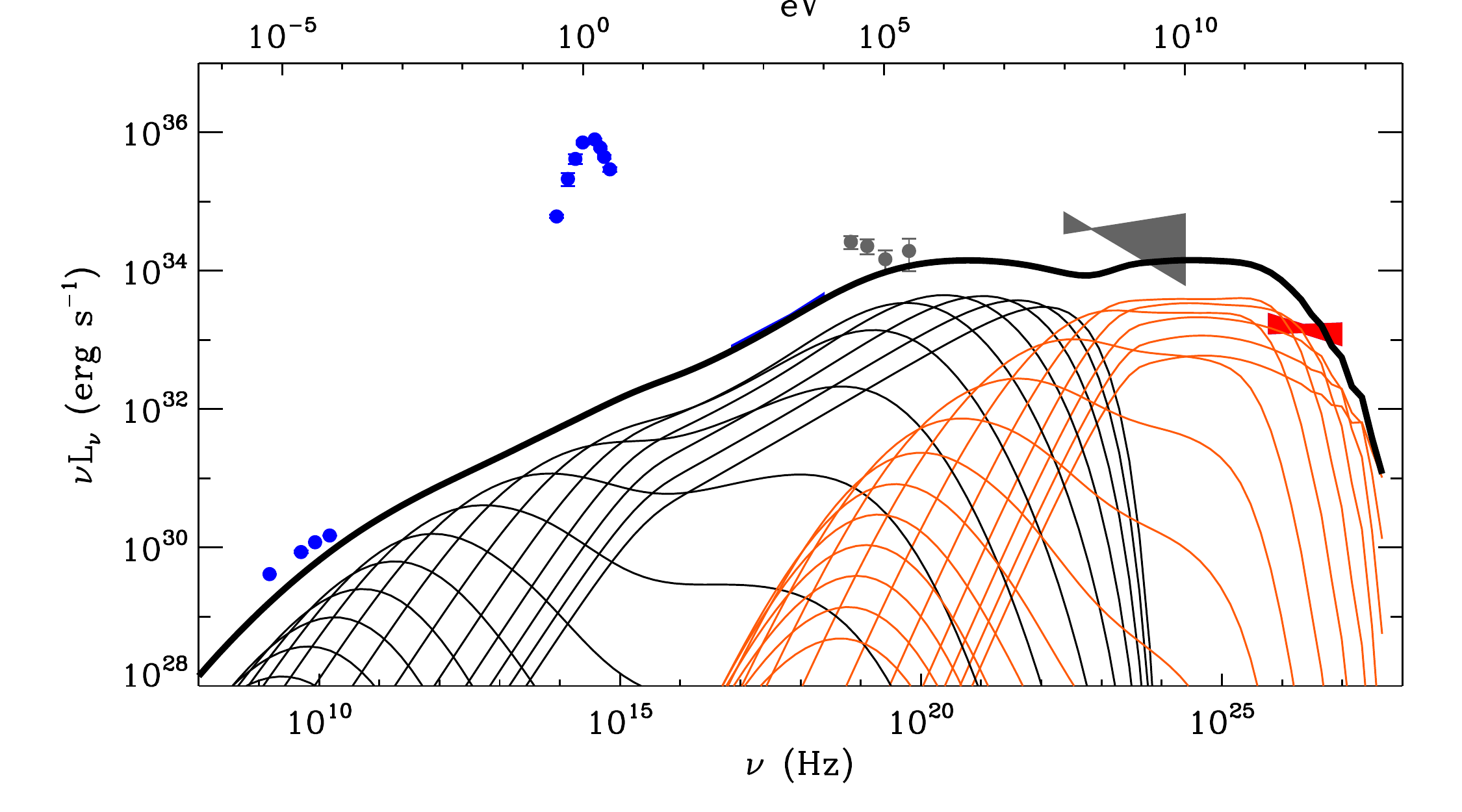}}
\resizebox{!}{4.05cm}{\includegraphics{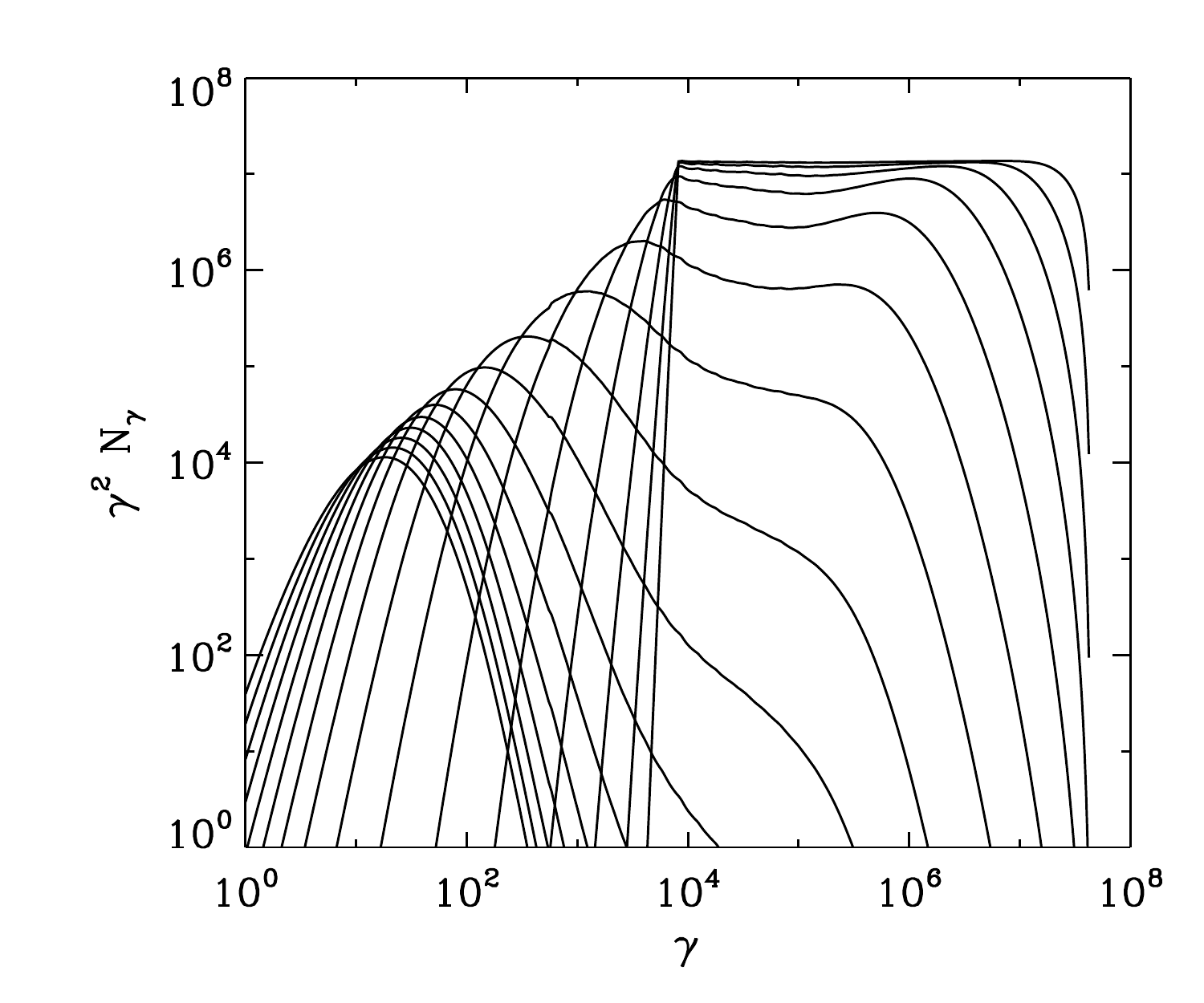}}
\caption{Left: Evolution of the emission in a pulsar wind nebula model of \ls. The shocked flow is divided into sections at various distances $z$ with the corresponding synchrotron (black) and inverse Compton (orange) spectra shown. The first section shown is for $z=R_s$, the fifth spectrum is at $z\approx 1.5 R_s$, the tenth at $z\approx 15 R_s$, the last spectrum shown is at $z\approx1000 R_s$. The total spectrum is the thick black line. Right: Corresponding evolution of the particle distribution (in units of the particle Lorentz factor $\gamma$). The initial $E^{-2}$ power law cools rapidly at high energies as a result of synchrotron emission. Adiabatic losses dominate at large distances. Figure and caption reproduced with permission from \citet{Dubus:2006lc} \copyright ESO. \label{evol}}
\end{figure}

Detailed modeling of the radio maps faces important difficulties. First, the input of particles may be inhomogeneous depending on location along the termination shock (\S\ref{conditions}), if only because the shock is much closer to the pulsar headwind than downwind. \citet{2012arXiv1212.3222Z} actually attributed the GeV and TeV emission to two different populations accelerated at different shock locations. Second, particles must followed down to low energies to model radio emission, which requires accurate knowledge of adiabatic cooling, magnetic field density, and speed along the flow. \citet{Dubus:2006lc} assumed a conical geometry with the evolution following \citet{Kennel:1984gu} (Fig.~\ref{evol}). Instabilities can mix the particles with denser wind material at the contact discontinuity \citep{Chernyakova:1999xm}, enhancing Coulomb energy losses \citep{2010MNRAS.403.1873Z} or to destruction of the spiral structure \citep{2011A&A...535A..20B,2012A&A...546A..60L,2013arXiv1301.2953B}. Finally, the structure of the flow is a complex double-armed structure with a non-zero transverse extension instead of a unidimensional Archimedean spiral. The impact of orbital motion on the relativistic flow  needs to be taken into account since radio maps show the emission is within or close to the radius where it begins to curve \citep{Dubus:2006lc,2012arXiv1209.6073M}. 

Numerical simulations can take inspiration both from models of colliding wind binaries  \citep[e.g.][]{2005xrrc.procE2.01P}, for the aspects related to stellar wind physics and orbital motion, and from models of pulsar winds interacting with their supernova remnant or the ISM \citep[e.g.][]{Gaensler:2006qi}, for the aspects related to pulsar wind physics. Simulations of gamma-ray binaries have been performed by three groups:
\begin{itemize}
\item \citet{2008MNRAS.387...63B,2012MNRAS.419.3426B} solved the axisymmetric relativistic (M)HD equations in the shocked region, finding the location of the discontinuities by iteratively solving a Riemann problem at the boundaries (Fig.~\ref{fig:bow}). They find that the shocked flow can reach extremely high Lorentz factors, up to several 100s, because the collimating wind acts as a nozzle.  They also found that the pulsar wind magnetisation or anisotropy have little impact on the overall structure, at least in the region explored close to the head. Depending on $\eta$,  the pulsar wind is completely confined or its downwind part escapes freely to infinity. Similar behaviour is found in simulations of colliding wind binaries  \citep{2011MNRAS.418.2618L}. Complete confinement maximizes the conversion of kinetic energy to thermal energy, which helps to get high radiative efficiencies (\S\ref{cwgeometry}).  
 \item \citet{Romero:2007ly,2011PASJ...63..893O,2012ApJ...750...70T} used 3D non-relativistic smooth particle hydrodynamics (SPH) to simulate the interaction of \psrb\ with the circumstellar disc of its companion. They found that a dense disc is needed to avoid strong disruption by the pulsar wind. This is consistent with optical observations that show little change in the Balmer lines during periastron passage. The high densities also imply nearly full confinement of the pulsar wind when it is nearest to the disc (Be disc ``crossings'').
 \item \citet{2012A&A...544A..59B} presented 2D (cylindrical) simulations including orbital motion using the relativistic hydro version of PLUTO, a grid-based code \citep{2012ApJS..198....7M}. Their simulations show Kelvin-Helmholtz mixing at the contact discontinuity  (Fig.~\ref{fig:numerical}). They argue that the relativistic nature of the flow is an important factor in shaping the interaction region on large scales.
  \end{itemize}
 Numerical simulations are a promising tool to address the relationship between emission and dynamics of the interaction region in gamma-ray binaries. Yet, covering the large range in radii from the interaction region to the radio emission region together with the timescale difference between the relativistic flow and the orbital motion is a very challenging computational problem.
\begin{figure}\sidecaption
\resizebox{0.42\hsize}{!}{\includegraphics*{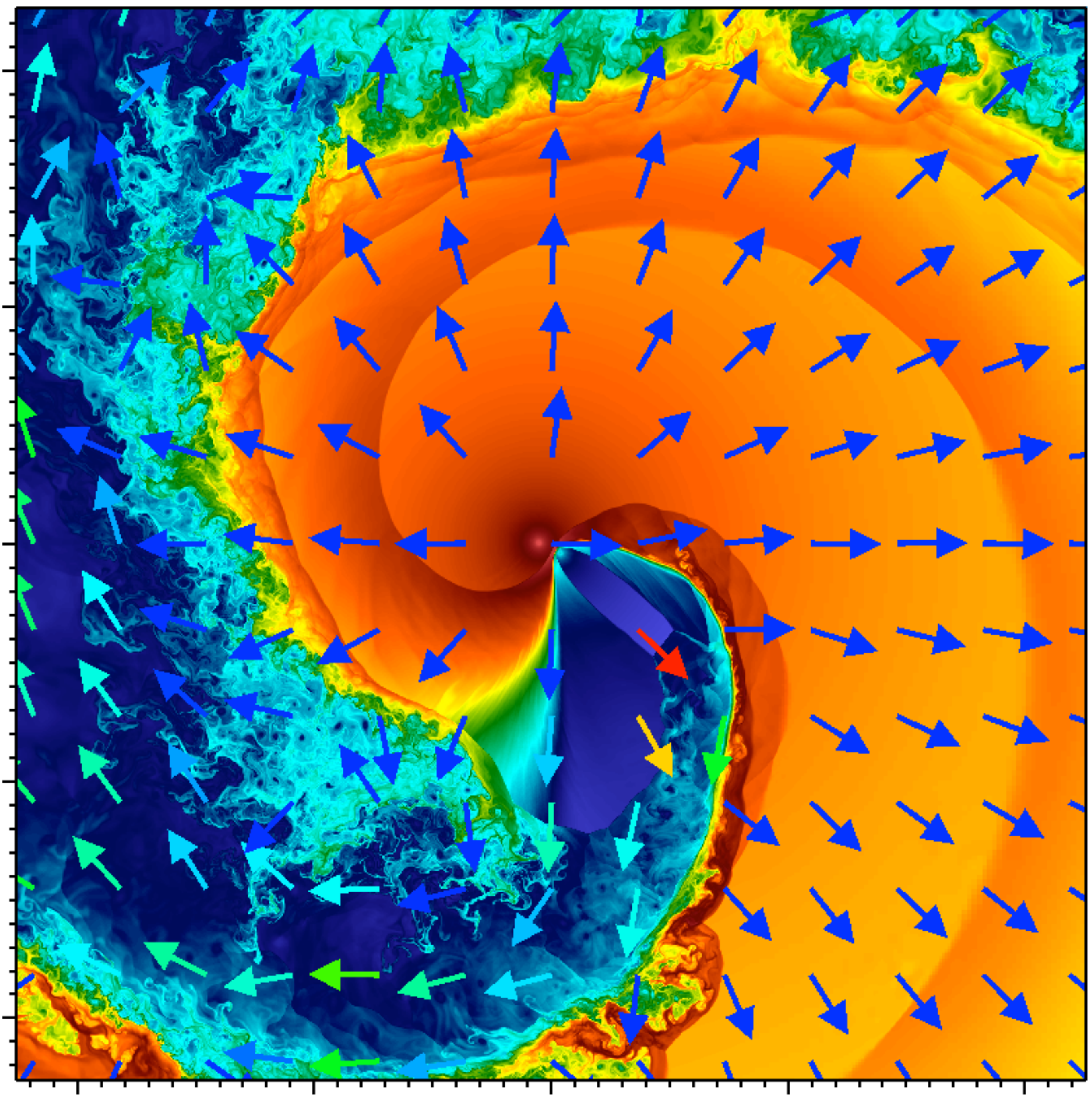}}
\caption{Relativistic hydrodynamical simulation on the scale of the orbit of a pulsar wind with $\Gamma=2$ interacting with a stellar wind ($\eta=0.3$). The color codes the density while the arrows show the velocity field (reproduced with permission from \citealt{2012A&A...544A..59B} \copyright ESO) \label{fig:numerical}}
\end{figure}

\subsubsection{The shocked stellar wind\label{signatures}}
The stellar wind is also shocked by the interaction with the pulsar wind. The post-shock temperature is high enough that X-ray emission is expected with \citep{Eichler:1993wt}
\begin{equation}
kT_s=\frac{3}{16} \mu m_H v_{\rm w}^2\approx 1.2 \left(\frac{v_{\rm w}}{1000\,{\rm km\,s}^{-1}}\right)^2 \,{\rm keV}\label{eq:thermal}
\end{equation}
for gas with $\hat{\gamma}=5/3$ and $\mu=0.62$ (ionized solar composition). In principle, the maximum that the shocked wind can radiate in X-rays is the available kinetic power from the wind
\begin{equation}
L_X\leq \frac{1}{2}f_{\rm w}\dot{M}_{\rm w} v_{\rm w}^2 = 3\times 10^{34}\, f_{\rm w}\left(\frac{\dot{M}_{\rm w}}{10^{-7}\, {\rm M}_\odot\,{\rm yr}^{-1}}\right)  \left(\frac{v_{\rm w}}{1000\,{\rm km\,s}^{-1}}\right)^2 \ {\rm erg\,s}^{-1}
\end{equation}
The fraction of the wind that is shocked is $f_{\rm w}=0.5 (1-\cos\theta_{\rm bow})$ \citep{2008ApJ...675..698T,2011ApJ...743....7Z}, about 7\% for \ls\ ($\theta_{\rm bow}\approx 30$\degr) so $L_X\leq 2\times 10^{33}$ erg\,s$^{-1}$ to compare to the observed $L_X\approx 10^{34}$ erg\,s$^{-1}$ (Tab.~\ref{mwltable}) The ratio $\chi$ of the radiative cooling timescale to the escape timescale ($\tau_{\rm esc}\sim R_{s}/v_{w}$) gives an estimate of the radiative efficiency of the shocked wind \citep{Stevens:1992on}. The maximum luminosity is reached if $\chi\ll 1$. A flow with $\chi\gg 1$ evolves adiabatically.

The shocked stellar wind in gamma-ray binaries radiates less than this upper limit because none of the X-ray spectra show evidence for bremsstrahlung emission.  \citet{2011ApJ...743....7Z} used this to derive a limit on the pulsar spindown power for \ls, since $\dot{E}$ is tied to the conditions in the shocked stellar wind via $\eta$. Using a detailed radiative cooling function, they find $L_X\propto \dot{E}^{1.4}$ and $\dot{E}\la 6\times 10^{36}$ erg\,s$^{-1}$. Particle acceleration is another way to dissipate energy at the shock besides radiative cooling. Thus, \citet{2011MNRAS.418L..49B} attributed  the HE and VHE gamma-ray emission to particles accelerated at the stellar wind and pulsar wind termination shock, respectively. However, particle acceleration at stellar wind shocks does not seem to generate much gamma-ray emission in colliding wind binaries, with only $\eta$ Car detected so far (see \S\ref{cwb}). Pulsar wind shocks are  more efficient than stellar wind shocks at converting energy into gamma rays.

\subsection{High-energy radiation and its orbital modulation\label{modulations}}

The mechanisms responsible for high-energy radiation are generic to both microquasar and pulsar wind scenarios \citep[see e.g.][]{Bosch-Ramon:2008hg}. Most models invoke synchrotron and inverse Compton from electrons or electron-positron pairs, and differ in their assumptions on the location of the particles, the flow in which they are embedded, or on the dominant radiation field. Even then, it is difficult to identify singular predictions that could provide a signature of each scenario. 

Hadronic processes have not been favoured for several reasons (but see \citealt{2004ApJ...607..949K,Bednarek:2005lp,Romero:2005go,2006JPhCS..39..408A,Chernyakova:2006cu} for models considering hadronic emission in gamma-ray binaries). First, the emission timescales are long so particles do not radiate much before leaving the system, which is a significant strain when observations suggest that a large fraction of the available power is converted to radiation. Second, while nuclei may be present in the pulsar wind, the dominant component is  likely to be pairs. Third, the proton energies required, $>100$ TeV for $pp$ and $>10$ PeV for $p\gamma$ (on stellar photons), imply Larmor radii comparable to or much greater than the shock size $R_s$. Fourth, the origin of the gamma-ray orbital modulations is not straightforward to explain using hadronic interactions.

This section presents the basic tools that have been used to model the gamma-ray emission in the context of binaries, especially the origin of the modulations present at nearly all wavelengths (\S\ref{mwl}). Modulations can be caused by changes in the physical conditions along the orbit or by the changing line-of-sight of the observer to the source of emission, i.e. because the geometry changes without any need for intrinsic variability. Models are needed to distinguish between both possibilities and provide information on the size and location of the emitter. The discussion focuses on \ls\ because of its well-established and regular behaviour at all wavelengths.

\subsubsection{Formation of the gamma-ray spectrum}
The dominant seed photon field for inverse Compton losses is usually assumed to be light from the very luminous massive star\footnote{In microquasars, the accretion disc and its corona can also contribute significantly to the ambient photon field \citep{Bosch-Ramon:2006do,Dermer:2006li}.}. The stellar radiation energy density at a distance $d=0.1$~AU from the star (the orbital separation of \ls) 
\begin{equation}
u_\star=\frac{ \sigma_{\rm SB} T_\star^4}{c} \left(\frac{R_\star}{d}\right)^2\approx 10^3 \left(\frac{T_\star}{40\,000\,{\rm K}} \right)^{4} \left(\frac{R_{\star}}{10\,{\rm R}_\odot}\right)^{2} \ {\rm erg}\,{\rm cm}^{-3} \gg u_B=\frac{B^2}{8\pi}\approx 0.04\ {\rm erg}\,{\rm cm}^{-3} 
\end{equation}
is much greater than the typical magnetic energy density at the shock (Eq.~\ref{magnetic}).

Combining Compton losses on stellar light with synchrotron losses leads to characteristic breaks at about 0.1$-$1 MeV in the synchrotron component and at about 0.1$-$1 TeV in the inverse Compton component of the SED \citep{Dubus:2006lc}. These breaks are associated to the electron energy at which cooling transits from inverse Compton to synchrotron losses \citep{Moderski:2005yd}.  Inverse Compton cooling  in the Thomson regime dominates over synchrotron because $\tau_{\rm thomson}=\tau_{\rm sync} (u_B/u_\star)$ and $u_{\star}\gg u_{B}$. The cooling timescale decreases with increasing electron energy. However, since the seed stellar photons have a typical energy $h\nu_\star \approx 2.7 kT_\star\approx 9$\,eV, electrons with energies larger than $(m_e c^2)^2 (h\nu_\star)^{-1}\approx 25$\, GeV upscatter photons in the Klein-Nishina regime. The inverse Compton timescale becomes \citep{Blumenthal:1970op}
\begin{equation}
\tau_{\rm KN}\approx40~  \left(\frac{E}{1\,{\rm TeV}} \right)  \left(\frac{d}{0.1\,{\rm AU}} \right)^2 \left[\ln \left(\frac{E}{1\,{\rm TeV}} \right)+2 \right]^{-1}~\left(\frac{40\,000\,{\rm K}}{T_\star} \right)^{2} \left(\frac{10\,{\rm R}_\odot}{R_{\star}}\right)^{2}~\mathrm{s.}
\label{eq:tkn}
\end{equation}
The inverse Compton timescale {\em increases} with electron energy above the transition to the Klein-Nishina regime at $25$\,GeV. Synchrotron emission (Eq.~\ref{eq:tsync}) takes over as the dominant radiative process for electrons with energies greater than
\begin{equation}
E_{\rm brk}\ga  3\,  \left(\frac{1\, {\rm G}}{B}\right) \left(\frac{0.1\,{\rm AU}}{d} \right) \,{\rm TeV}.
\label{hebreak}
\end{equation}
The steady injection of a power-law distribution of electrons in a zone where they cool by radiating synchrotron emission steepens the power-law index $p$ to $p+1$. The distribution is only slightly hardened if cooling is dominated by inverse Compton scattering in the Klein-Nishina regime. Therefore, the steady-state electron distribution has an index close to $p$ up to $E_{\rm brk}$, breaking to $p+1$ above $E_{\rm brk}$ (Eq.~\ref{hebreak}). The inverse Compton spectrum reflects this directly as a spectral break from a photon index $\Gamma_{\rm VHE}\approx p$ to $\Gamma_{\rm VHE}\approx p+1$  at $h\nu\approx E_{\rm brk}$ \citep[see][and their erratum]{Moderski:2005yd}. The synchrotron spectrum emitted by the same electrons has a corresponding  softening of the spectral index by $\approx 0.5$ beyond the break frequency
\begin{equation}
h\nu_{\rm sync,\ break}\approx 750\ \left(\frac{1\, {\rm G}}{B}\right) \left(\frac{0.1\,{\rm AU}}{d} \right)^2 \, \mathrm{keV}
\label{hebreak2}
\end{equation}
The spectral breaks are easily identified in Fig.~\ref{evol}, even though this is a multi-zone model (the initial $B$ was $\approx 5$~G). 

The location of spectral breaks can be used to estimate the average magnetic field in the VHE gamma-ray emission zone by applying Eq.~\ref{hebreak}-\ref{hebreak2} \citep{Dubus:2007oq}. The location of the break frequencies can change with orbital phase. For instance, the magnetic field intensity changes with the termination shock distance as $B\propto 1/R_s\propto 1/d$ for constant magnetisation $\sigma$. The magnetisation can also change with distance to the pulsar. A varying break frequency will modulate the VHE flux. The usual assumption is that the X-ray emission is related to synchrotron emission while the gamma rays are produced by inverse Compton emission. Models have also been proposed where the X-ray emission is due to inverse Compton emission \citep{2009MNRAS.397.2123C,2010MNRAS.403.1873Z}.

Adiabatic losses can modify this picture, especially for long orbital period systems like \psrb\ where the stellar radiation field and  the magnetic field at the termination shock will be lower. The expansion timescale of the shocked flow is difficult to estimate accurately without a dynamical model (e.g. using numerical simulations, \S\ref{pwn}). Radiative models have taken to parametrizing adiabatic losses, comparing spectra and lightcurves to observations in order to gain insight: see \citet{Tavani:1997wv,Kirk:1999hr,2007MNRAS.380..320K,2009ApJ...698..911U} for discussions related to \psrb, \citet{Takahashi:2008vu} for \ls, and \citet{Chernyakova:2006cu,2011A&A...527A...9Z} for \lsi. 

\subsubsection{VHE absorption by pair production\label{gg}}
\begin{figure}\sidecaption
\resizebox{0.7\hsize}{!}{\includegraphics*{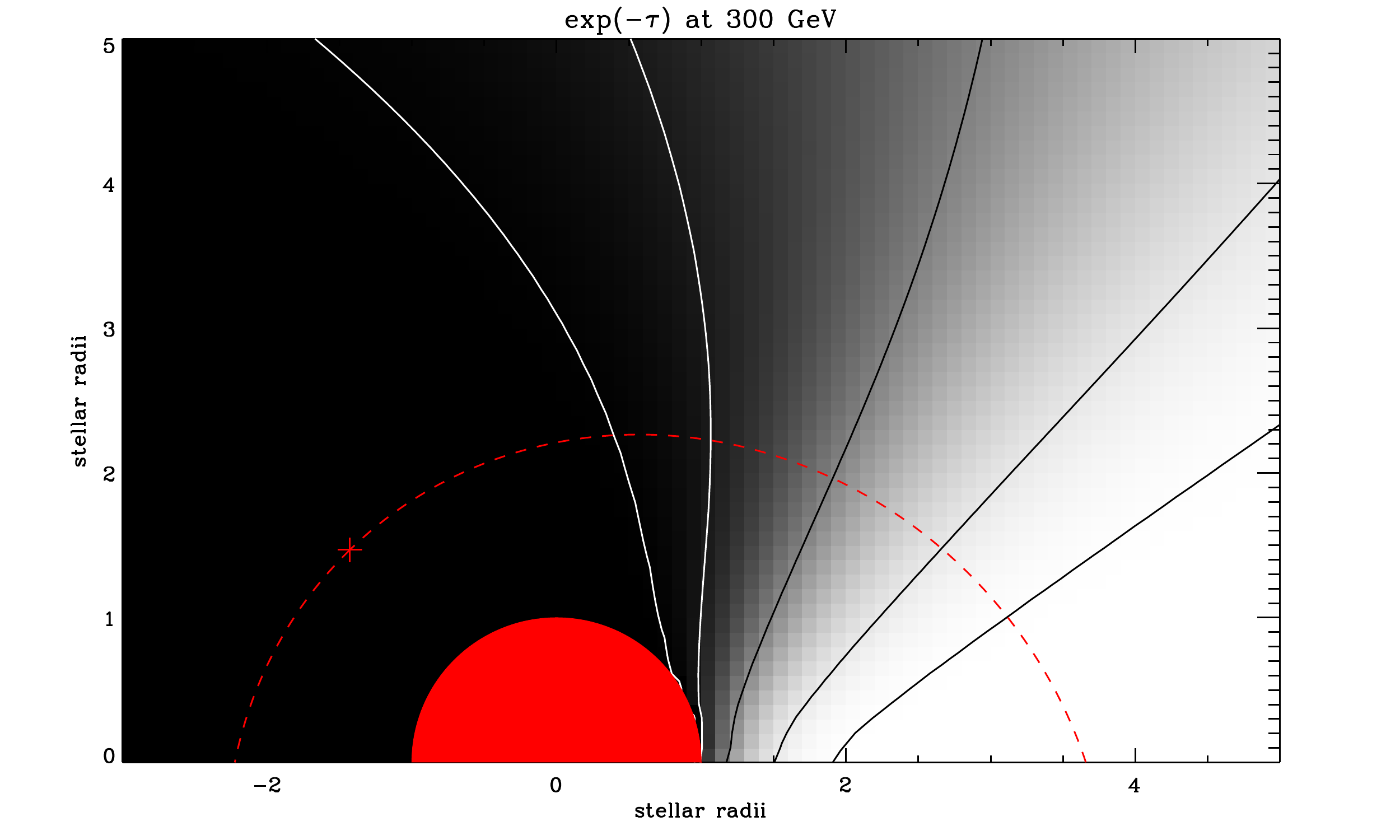}}
\caption{Absorption of 300 GeV gamma rays due to pair production with photons from the star. Greyscale codes the fraction of the flux emitted at each location that reaches an observer situated at infinity to the right. Lines from left to right correspond to 99\%, 90\%, 50\%, 10\%, and 1\% absorption.  Red dashed line is the orbit of \ls. Reproduced with permission from \citet{Dubus:2006lr} \copyright ESO.\label{fig:abs}}
\end{figure}
VHE gamma rays are energetic enough to produce a $e^+e^-$ pair by interacting with UV photons from the star. The process is akin to an absorption modulating the observed VHE gamma-ray flux \citep{1984MNRAS.211..559F,Protheroe:1993en,1995SSRv...72..593M,Bednarek:1997ph,Kirk:1999hr,Bottcher:2005zx,Dubus:2006lr}. The threshold for two photons to produce a pair, $\gamma+\gamma\rightarrow e^+ + e^-$, is
\begin{equation}
\epsilon_\star \epsilon_\gamma \ga \frac{2 m_e^2 c^4}{1-\mu}\label{threshold}
\end{equation}
The threshold and cross-section change depending on the incident angle $\psi$ between photons ($\mu\equiv \cos\psi$, with $\psi=\pi$ for head-on collisions). For \ls, the typical energy of the stellar photon is $\epsilon_\star\approx 2.7 kT_\star\approx 9$\,eV, so the threshold for pair production, realised for head-on interactions, is $\epsilon_\gamma\ga 30$\,GeV. The cross-section peaks near the threshold energy with a peak value $\sigma_{\gamma\gamma}\approx \sigma_T/4$. An estimate of the opacity is
\begin{equation}
\tau_{\gamma\gamma}=\int_{\rm l.o.s.} n_\star \sigma_{\gamma\gamma} ds \approx \frac{\sigma_T}{4} \frac{\sigma_{\rm SB} T_\star^4}{\epsilon_\star c}\left(\frac{R_\star}{d}\right)^2 d \approx 26\,\left(\frac{T_\star}{40\,000\,\rm K}\right)^3 \left(\frac{R_\star}{10\,R_\odot}\right)^2 \left(\frac{10^{12}\,\rm cm}{d}\right)
\label{eq:gammagamma}
\end{equation}
where the parameters used are appropriate for \ls. The opacity is so large that VHE emission within the system is strongly absorbed. However, the angle $\psi$ changes as the VHE source revolves around the star, leading to an observer-dependent, orbital phase-dependent absorption of the VHE emission (Fig.~\ref{fig:abs}). The absorption is minimum when the VHE photons are emitted away from the star: the interaction is tail-on and the threshold becomes very high (Eq.~\ref{threshold}). Hence, if the VHE source is close to the compact object, maximum VHE emission is observed at inferior conjunction, when the observer sees the compact object pass in front of the star.

The VHE emission from \ls\ is indeed modulated (\S\ref{vhe}), with peaks and troughs close to conjunctions as expected. However, pair production is insufficient on its own to explain the observations \citep{Aharonian:2006qw}:
\begin{enumerate}
\item Although $\tau_{\gamma\gamma}\approx40$ at $\phi_{\rm sup}=0.08$, observations with HESS still detect the source at the 11$\sigma$ level in between $\phi=0$--0.1, and at the 3.1$\sigma$ level for $\phi=0.1$--$0.2$ ; 
\item the VHE spectrum changes from hard at $\phi_{\rm inf}$ to soft at $\phi_{\rm sup}$, pivoting around 100 GeV instead of pivoting around the threshold energy of 30 GeV ;
\item the exponential cutoff in the HE domain, at a few GeV, is inconsistent with the cutoff expected from pair production at 30 GeV.
\end{enumerate}
Anisotropic inverse Compton scattering and cascade emission help with the first two items (see below). Alternatively, VHE emission arises away from the compact object, for instance in the backward-facing termination shock of the pulsar, so that VHE photons encounter fewer stellar photons along their path \citep{2008A&A...489L..21B,2012A&A...544A..59B}. The last item is taken as evidence that two different particle populations emit in the HE and VHE domain.

Absorption by pair production may also play a role in \lsi, since $\tau\approx 1$ at periastron using Eq.~\ref{eq:gammagamma}  \citep{Bednarek:2006ka,Sierpowska-Bartosik:2008wt}. However, the phasing of the VHE extrema is inconsistent with $\gamma\gamma$ absorption if the VHE source is close to the compact object. Absorption is expected to be small in binaries with wider orbits, unless the system inclination is close to edge-on and/or the orbit highly eccentric.

\subsubsection{Cascades\label{cascades}}
\begin{figure}
\centering\resizebox{4.9cm}{!}{\includegraphics{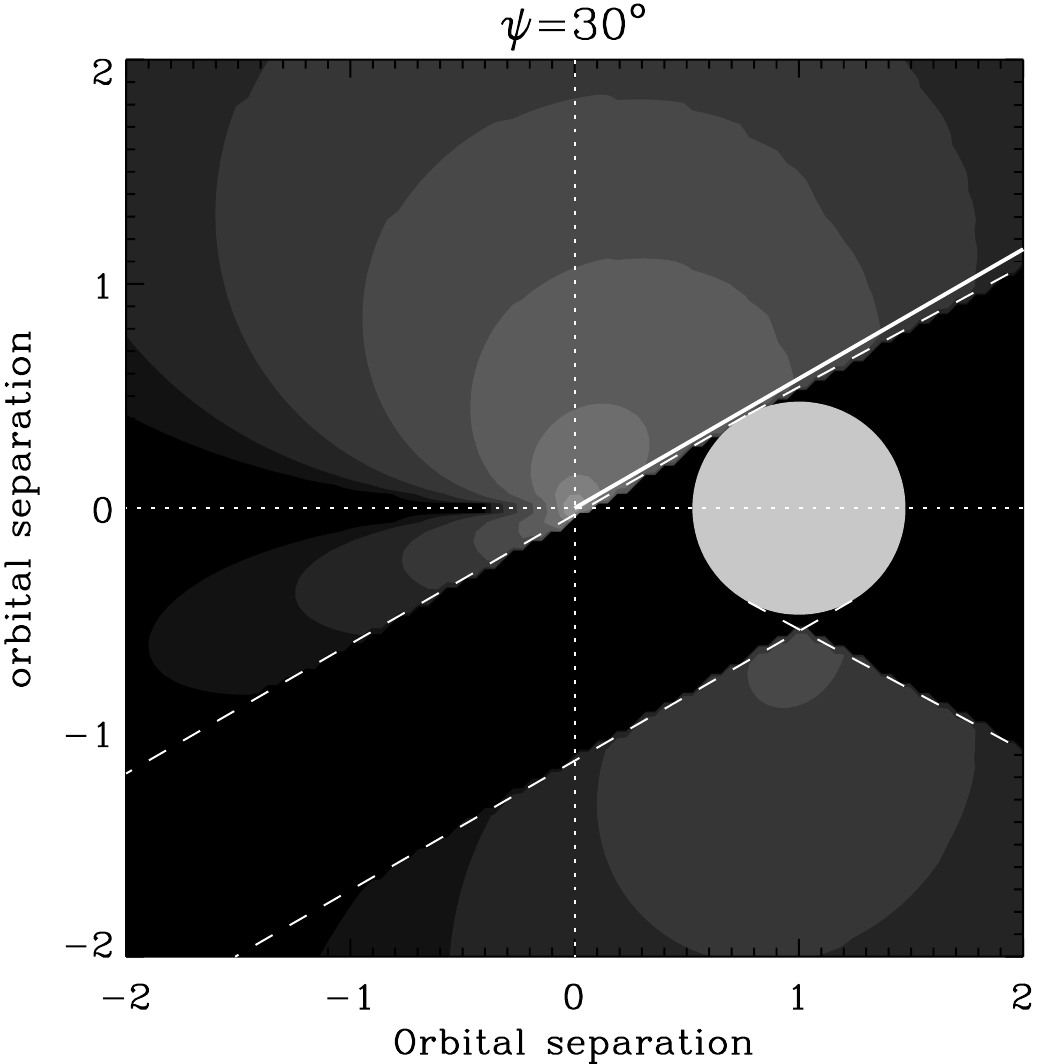}}
\centering\resizebox{6.8cm}{!}{\includegraphics{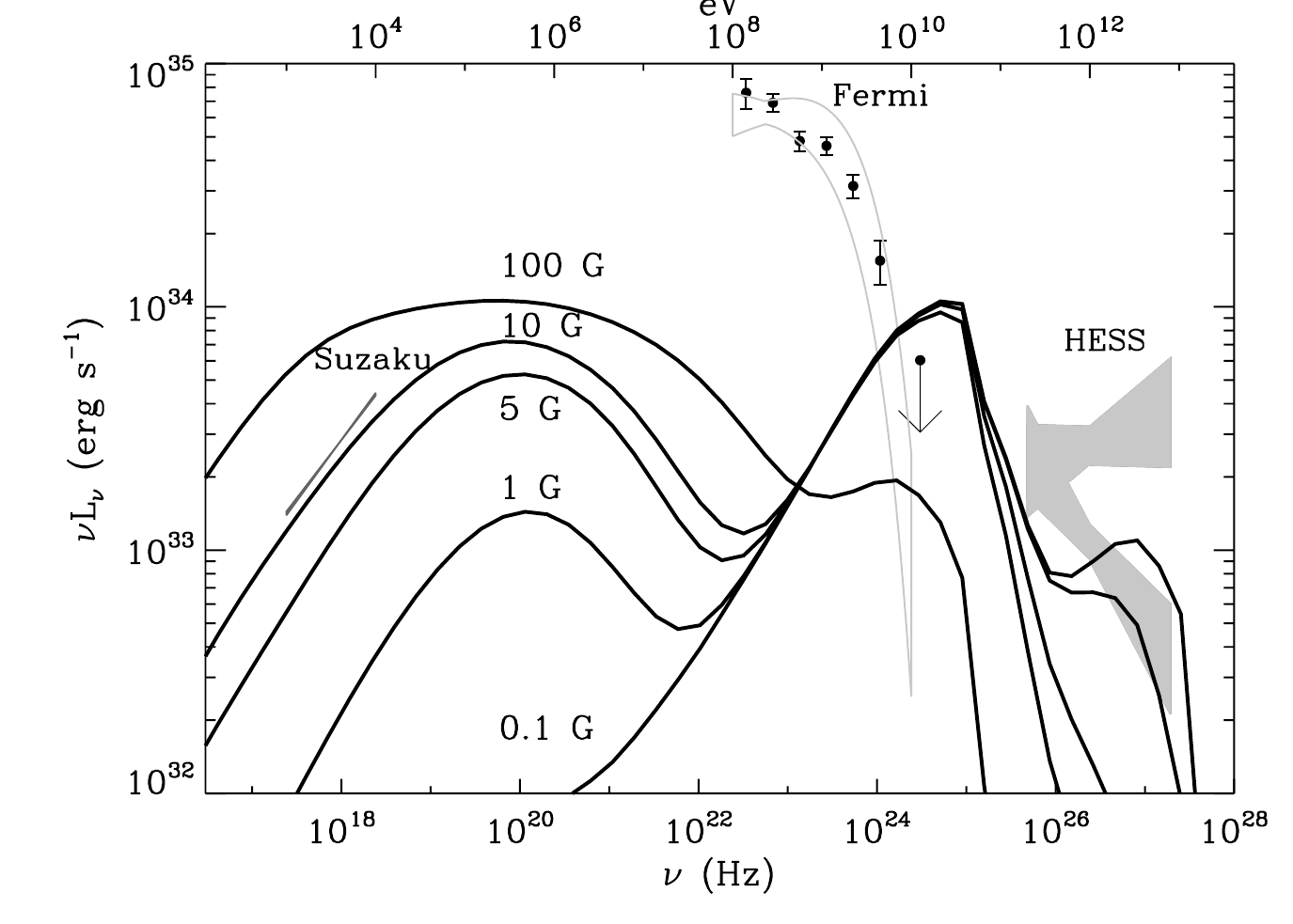}}
\caption{Left: spatial distribution of the VHE emission from the first generation of pairs in \ls, as observed when the system is at superior conjunction. Right: average spectrum of the 1st generation of pairs depending on the magnetic field intensity. Reproduced with permission from \citet{2010A&A...519A..81C} \copyright ESO. \label{fig:cascade}}
\end{figure}
The $e^{+}e^{-}$ pair created when a VHE photon interacts with a UV photon typically has one of the particles carrying away most of the energy along the initial path of the VHE photon. This particle can also upscatter stellar photons to very high energies, even if slightly lower than that of the initial VHE photon. The new VHE photon pair produces in turn, triggering a cascade that stops when the upscattered photon energy is below the pair production threshold. Cascade emission reduces the effective opacity by redistributing the absorbed flux \citep{Sierpowska:2005vy,2006JPhCS..39..408A}.

Computing cascade emission is complex because the emission and absorption processes are anisotropic and depend on location, because the pairs diffuse according to the local magnetic field properties, because there is a competition with synchrotron emission \citep{Bednarek:1997ph,Bosch-Ramon:2008vd,2010A&A...519A..81C}. Pair emission can be considered as local to the place of their creation if their Larmor radius is very small compared with the system size, 
\begin{equation}
R_{L}\ll d ~\Rightarrow~B\gg 2\times 10^{-3}\,\left(\frac{E}{1\,{\rm TeV}}\right)\left(\frac{0.1\,{\rm AU}}{d}\right)\,{\rm G}\label{rl}
\end{equation}
In this case, the pairs radiate locally the secondary gamma-ray emission (Fig.~\ref{fig:cascade}). If the magnetic field is lower, the spatial diffusion of the pairs cannot be neglected, making the problem rather intricate. In the extreme limit of no magnetic field the cascade propagates linearly, simplifying the problem, but the lightcurves and spectra do not match the observations \citep{2009A&A...507.1217C}. A very low magnetic field is unlikely: confining electrons within $R_s\approx 10^{11}$\,cm to accelerate them up to 10 TeV already requires $B\ga 0.1$\,G near the shock (Eq.~\ref{rl}). When the magnetic field is higher than $\ga 5$\,G, synchrotron starts to dominate the radiative losses of the pairs, as can be verified by equating Eq.~\ref{eq:tsync} with Eq.~\ref{eq:tkn} for a 1\,TeV electron. When $B\ga 100$\, G, the electrons with energy just above the pair product threshold (30 GeV) lose energy to synchrotron, stopping the cascade at the first generation of pairs. In all cases, the synchrotron emission from the cascade pairs cannot exceed X-ray measurements, providing an upper limit on the magnetic field intensity (Fig.~\ref{fig:cascade}).

\subsubsection{Anisotropic inverse Compton scattering\label{anis}}
\begin{figure}\sidecaption
\resizebox{0.5\hsize}{!}{\includegraphics*{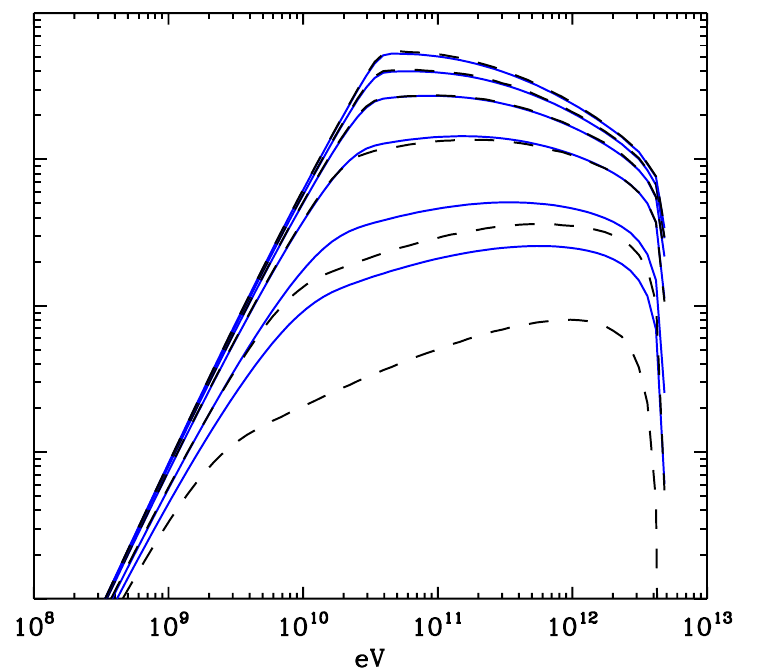}}
\caption{Dependence of the viewing angle on the inverse Compton
  spectrum (y-axis is $\nu F_\nu$ in arbitrary units).  The source of photons is a blackbody $kT_\star$=1~eV. The electrons
  are distributed according to a power-law of index $p=2$ between $10^5<\gamma_{\rm e}<10^7$. The
  spectrum is shown at viewing angles $\psi=$15\degr\ (bottom), 30\degr,
  60\degr, 90\degr, 120\degr\ and 180\degr\ (top). Solid lines take into account the finite
   size of the star for electrons at $d=2R_\star$; dashed lines assume a point
  source. Reproduced with permission from \citet{Dubus:2007oq} \copyright ESO. \label{fig:anis}}
\end{figure}
Because of the dependency of the Compton cross-section on angle between incoming and outgoing photons, an anisotropic distribution of seed photons will lead to anisotropic emission from an isotropic distribution of electrons. For instance, the total power radiated by an electron of Lorentz factor $\gamma$ upscattering photons (in the Thomson regime) from a point source of radiation energy density $U_\star$ is
\begin{equation}
P_\star= \sigma_T c U_\star (1-\beta\mu) \left[(1-\beta\mu)\gamma^2-1\right]
\label{star}
\end{equation}
where $\mu=\cos\psi$ is the angle between the incoming photon and the electron direction of motion. Therefore, the emission from electrons upscattering stellar photons will be maximum directly towards the star, corresponding to head-on collisions ($\mu=-1$), and minimum directly away from the star, when collisions are tail-on ($\mu=1$). 

Anisotropic inverse Compton scattering naturally leads to an orbital modulation of the gamma-ray lightcurve, making it particularly relevant to model gamma-ray emission from binaries (\citealt{1972NPhS..236...39J}, \citealt{Kirk:1999hr}). In practice $\mu$ is the angle between the star, the location of the electrons and the observer, the same angle that appears in Eq.~\ref{threshold}. The reason is that emission from a high-energy electron is tightly focused along its direction of motion by relativistic aberrations. Hence, the observer sees emission only from those particles whose motion is along the line-of-sight. The anisotropy also changes the emitted spectrum when scattering occurs in the Klein-Nishina regime since the transition to this regime depends on angle: the condition is $\gamma(1-\beta \mu)\epsilon_\star\ll m_e c^2$, where $\epsilon_\star$ is the energy of the incoming photon. The spectrum is typically harder away from the star because the interactions occur in the Thompson regime ($\mu=1$) instead of being softened by the Klein-Nishina drop in scattering rate with energy (see Fig.~\ref{fig:anis} and  \citealt{Dubus:2007oq} for details). 

Taking into account that $U_\star$ changes with phase, because the orbits are eccentric, the range of possible system inclinations, pair production, and anisotropic inverse Compton scattering leads to lightcurves that, despite their complexity, depend only on the geometrical configuration seen by the observer (Fig.~\ref{fig:anistau}). In principle, one can use observed gamma-ray modulations to derive a best-fitting orbit. For instance, the low amplitude GeV modulation of \fgl\ favours a low inclination and a low eccentricity \citep{2012Sci...335..189F}. 

Knowing the system parameters, the HE and VHE lightcurves can be computed and compared to observations to test whether only  geometrical effects are involved. This approach successfully predicted the anti-correlation of the HE and VHE modulations in \ls: the inverse Compton lightcurve is the same in both bands but only the VHE band is affected by  pair production. The absorption peaks at the same phases as the inverse Compton emission and is sufficiently strong to suppress the VHE peak and move it to inferior conjunction, where absorption is minimum. This is quite robust against the details of the model (\citealt{Bednarek:2007qd,Sierpowska-Bartosik:2007jo,Dubus:2007oq,Khangulyan:2007me}, \citealt{2008ApJ...672L.123N}, \citealt{2009ApJ...702..100T,2010ApJ...717...85Y}).

However, anisotropic inverse Compton emission fails to account for the phasing of the modulation peaks in \psrb\ and \lsi\ (see \S\ref{sspsrbflare} for details). Understanding why is one of the challenges that models face.
\begin{figure}
\centering\resizebox{\hsize}{!}{\includegraphics{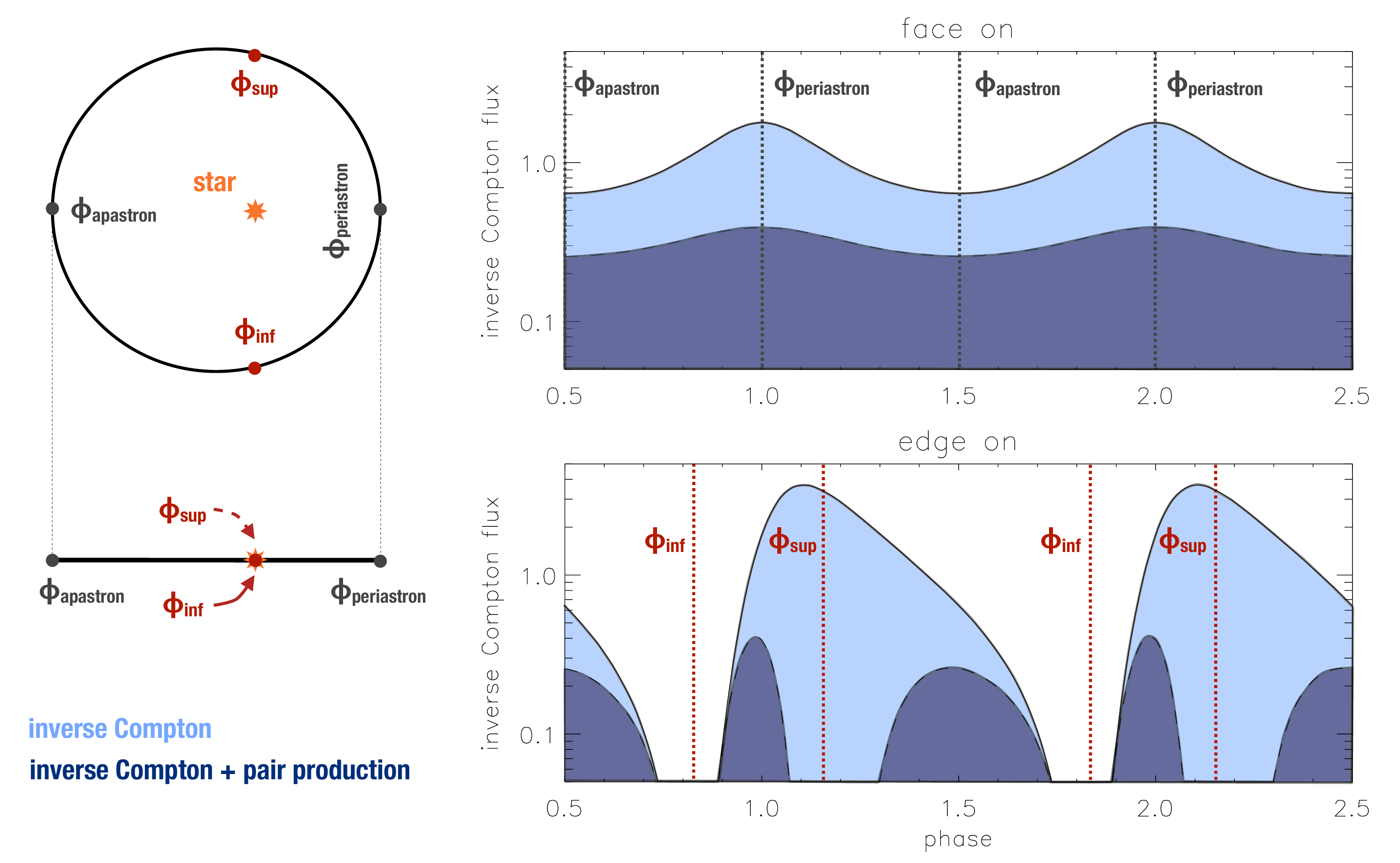}}
\caption{Orbital modulation of the anisotropic inverse Compton emission (light blue curve) for the same system seen face-on (top, $i=0$\degr) and edge-on (bottom, $i=90$\degr). The dark blue curve shows the  fraction of the gamma-ray lightcurve that remains after absorption due to pair production with stellar photons. The emitting particles are assumed to be at the location of the compact object.}
\label{fig:anistau}
\end{figure}

\subsubsection{Relativistic Doppler boosting}
Relativistic Doppler boosting can also impact the observed lightcurves. Radiating electrons advected in the shocked pulsar wind or in a relativistic jet have their synchrotron emission, isotropic in the comoving frame, boosted or deboosted in the observer frame according to the orientation of the flow. In the case of a shocked pulsar wind, if the flow is a comet tail pointing away from the star then the synchrotron flux will be higher at $\phi_{\rm inf}$ than at $\phi_{\rm sup}$. The impact on the modulation lightcurve is important even with the mildly relativistic velocities $\beta\approx1/3$ expected close to the shock.  The inverse Compton emission is also boosted but the effects are subtler. Electrons at rest in the comoving frame now see a deboosted flux of seed photons because they are flowing away from the star. The anisotropic upscattered flux must then be re-transformed to the observer frame. Doppler boosting has been proposed to account for the X-ray modulation of \ls\ \citep{2010A&A...516A..18D} and the GeV flare in \psrb\ \citep{2012ApJ...753..127K}. In the case of a microquasar, the orientation of the jet is constant on the orbital period so Doppler boosting changes the emission at all phases equally \citep{Georganopoulos:2002ci}. However, the level of emission can change significantly as the jet orientation precesses on longer timescales \citep{2002A&A...385L..10K}.

\subsection{Challenges to models\label{probe}}

The presence of pulsar winds is plausible in gamma-ray binaries (\S\ref{pw}). The interaction with the stellar wind provides good conditions for high-energy radiation (\S\ref{shock}). The radiative processes involved naturally produce orbital modulations of the lightcurve (\S\ref{modulations}). All of this provides a useful framework to make sense of the observations.  Yet, a ``standard model'' of gamma-ray binaries has yet to emerge. 

There is a consensus on some of the basic ingredients: synchrotron and inverse Compton emission of electrons are the main radiative processes, the anisotropy of inverse Compton emission and pair production plays a role in shaping the gamma-ray lightcurves that we observe, particles are accelerated very efficiently, and two populations are involved to account for the differing HE and VHE spectral components.

However, the recipes vary: principally according to where the emitting particles are assumed to be (in the pulsar wind, in the shocked pulsar wind, in a relativistic jet, close or far from the compact object, Fig.~\ref{fig:mirabel}), on how this location changes or not with orbital phase, on the strength of the magnetic field or of adiabatic cooling, on whether processes like cascade emission or Doppler boosting are incorporated. All of these choices impact the spectrum and variability of the emission; success is judged on the comparison with the observations deemed most relevant to the object or study at hand. The complexity is such that no model has succeeded yet in accurately reproducing all of the observed spectral and variability features in any single object. 

Could such a model be constructed ? One of the factors limiting consensus is the simplified description of the emission region, too often assumed to be point-like (even if not necessarily at the location of the compact object). However, X-ray emission in \ls\ is thought to be extended compared to the binary size and/or away from the compact object based on the lack of modulation X-ray absorption and the absence of eclipses \citep{Bosch-Ramon:2007fq,2011MNRAS.411..193S}. Similar arguments have been used for the VHE emission (see \S\ref{gg}). Figure~\ref{fig:geom} shows clearly that the termination shock has a significant size compared to the binary size. High-energy particles will be present throughout this region, accelerated differently depending on their location along the termination shock, evolving differently depending on streamline. Convergence between models may be helped in the future by numerical simulations, which can provide accurate estimates of the size of emission regions, adiabatic losses, wind mixing, Doppler boosts, etc.

Despite these caveats, I attempt to show that gamma-ray binaries can tell us something about pulsar physics by highlighting two of the most pressing issues that models need to confront.

\subsubsection{What is the origin of the particles emitting HE gamma rays ?\label{hespec}}
The HE gamma-ray spectral component seems to be independent of the VHE component: the HE emission is not detected in all objects (\hessj), the HE and VHE lightcurves behave differently (\lsi, \psrb), an exponential cutoff separates the HE spectrum from the VHE spectrum (Fig.~\ref{fig:exp}), the energy of this cutoff is inconsistent with expectations from pair production on stellar photons (\ls). Hence the consensus that different populations of particles are involved in each  band. A major challenge is to identify what those populations correspond to. 
\begin{figure}
\resizebox{5.5cm}{!}{\includegraphics{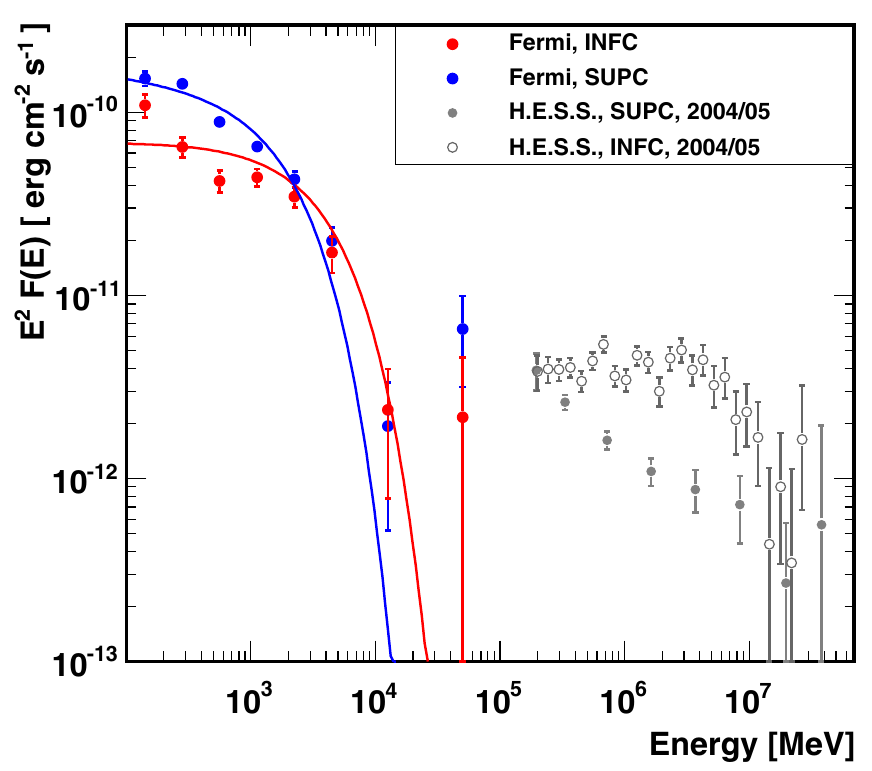}}
\resizebox{5.5cm}{!}{\includegraphics{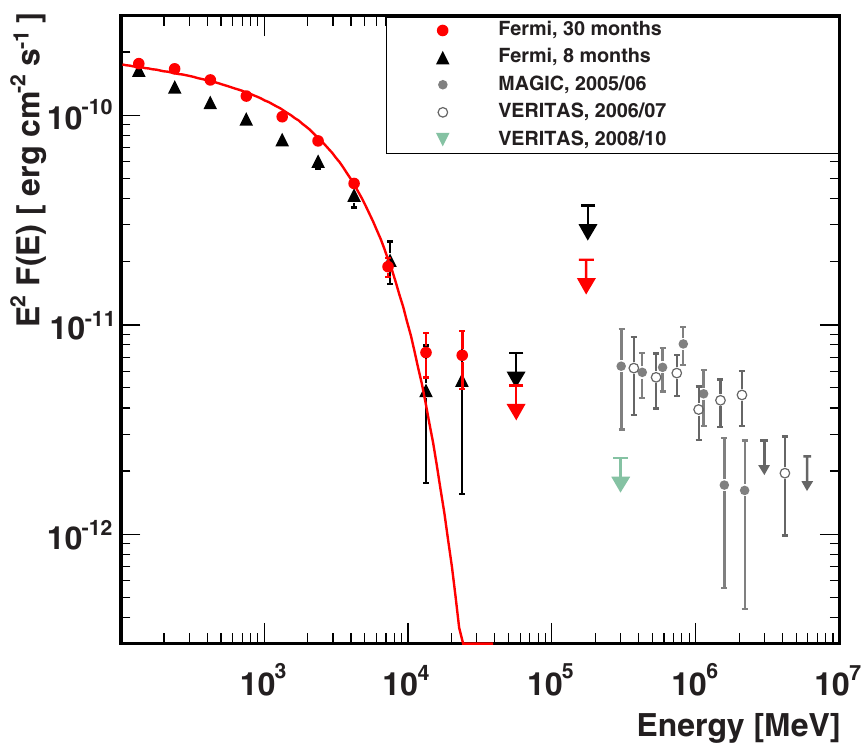}}
\caption{Spectral energy distribution in HE (0.1-100 GeV) and VHE ($>$ 100 GeV) gamma rays for \ls\ (left) and \lsi\ (right), showing the exponential cutoff separating the HE and VHE spectral components \citep[reproduced by permission of the AAS from][]{2012ApJ...749...54H}.}
\label{fig:exp}
\end{figure}

Intriguingly, the HE spectrum of gamma-ray binaries is typical of the pulsars observed by \fermi\ while the rest of the multi-wavelength emission resembles that of pulsar wind nebulae (\S\ref{indirect}). HE emission from pulsars is widely interpreted as emission from within the light cylinder by particles that are accelerated in regions (gaps) where there is a non-zero electric field, emitting curvature radiation as they travel along magnetic field lines. This ``magnetospheric'' emission is not expected to vary as the pulsar moves along its orbit in a binary. Indeed, most pulsars in binaries behave exactly like isolated pulsars (\S\ref{lmgb}). Why then is the HE component modulated on the orbital period in gamma-ray binaries ? The HE emission could be the sum of a constant magnetospheric component and a modulated inverse Compton component (e.g. \citealt{2010tsra.confE.193K}, \citealt{2011heep.conf..531T}). This possibility appears ruled out by \lsi\ since the HE modulation fraction varies strongly without the HE spectrum changing \citep{2012ApJ...749...54H}.

Alternatively, the HE emission could arise from the {unshocked} pulsar wind. In the standard picture, the pairs in the ultra-relativistic wind are cold (\S\ref{pw}). Although they do not emit synchrotron emission, these pairs can upscatter ambient photons, providing a probe into this region \citep{Bogovalov:2000ny}. The stellar radiation field in gamma-ray binaries is ideally suited to this purpose \citep{Chernyakova:1999xm,Ball:2000lr,Sierpowska:2005vy,2007MNRAS.380..320K}. Intense Compton lines are expected, peaking at an energy $\approx \Gamma mc^2$ (in the Klein-Nishina regime), where $\Gamma$ is the Lorentz factor of the wind. Applying this to \psrb, \citet{2012ApJ...752L..17K} found that its HE spectrum implies $\Gamma\approx 10^4$. However, the lines appear too narrow to reproduce the HE spectra of \ls\ and \lsi\ \citep{2008A&A...488...37C,2008APh....30..239S}.

Assuming power-laws instead of mono-energetic (cold) electrons provides a better fit to the spectrum \citep{2009arXiv0912.3722C}. Such distributions might arise as a result of magnetic reconnection in the current sheet of the striped pulsar wind (\S\ref{pw}). Little emission is expected away from the stripes where the wind is tenuous with high $\sigma$. Emission from the stripes is highly Doppler boosted in the direction of the relativistic wind: significant emission can be detected only if the line-of-sight to the pulsar crosses the current sheet. Since the stripe pattern turns with each rotation of the pulsar, its HE emission may be pulsed, providing an alternative to magnetospheric models \citep{2002A&A...388L..29K,2005ApJ...627L..37P,2011MNRAS.412.1870P}. Numerical simulations of pulsar magnetospheres support the idea that part of the gamma-ray emission arises outside the light cylinder \citep{2010ApJ...715.1282B}.  If the distribution of obliquities (the angle between pulsar magnetic and rotation axis) is random, then half of the line-of-sights to pulsars cross the stripes. Hence, HE emission associated with the stripes will not be detected in half of the VHE-selected gamma-ray binaries \citep{2011MNRAS.417..532P}, explaining why \hessj\ is not detected in HE gamma rays. A difficulty of this interpretation is to understand the identical HE spectra of gamma-ray binaries and \fermi\ pulsars: even if the particles originate in the current sheet in both cases, the emission process must differ at some level since isolated pulsars do not upscatter light from a massive stellar companion.

The conundrum has yet to be resolved: emission related to the pulsar is a tantalising interpretation of the HE gamma-ray spectrum of gamma-ray binaries but providing a consistent framework to explain the formation of this component in all high-energy pulsars is challenging. As speculated above, resolving this tension can provide new insights into pulsar wind physics. Or it may be a red herring. The peculiar HE gamma-ray spectrum could have nothing to do with pulsar emission: \cite{2011MNRAS.418L..49B} proposed that it is due to particles accelerated at the stellar wind termination shock, while \cite{2012arXiv1212.3222Z} proposed that it is due to electrons accelerated close to the apex of the pulsar wind termination shock (while the VHE emission originates further away in the termination shock). The debate is open.

\subsubsection{What is the origin of the gamma-ray flare of \psrb\ ?\label{sspsrbflare}}
A second major puzzle facing models is the HE gamma-ray flare from \psrb\ \citep{2011ApJ...736L..10T,2011ApJ...736L..11A}. A ``brightening'' was detected around periastron passage, as had been expected from computing the lightcurve for upscattering stellar photons \citep{Kirk:1999hr,2011ApJ...742...98K,2011MNRAS.417..532P}. However, the HE emission suddenly ``flared'' 30 days after periastron passage, reaching a peak flux 20$-$30 times higher than during periastron passage (Fig.~\ref{fig:psrb}). The flare started with the pulsar close to inferior conjunction, the least favourable configuration for inverse Compton scattering off stellar photons.  The Be disc also  contributes significantly to the ambient radiation field \citep{2011MNRAS.412.1721V}, mostly due to emission from the innermost disc close to the star \citep{2012MNRAS.426.3135V}.  Therefore, the configuration for scattering off Be disc photons is no more favourable. The flare started about 10 days after radio pulsations had returned from their eclipse around periastron passage, suggesting the pulsar was not interacting with the densest circumstellar material during the flare. No obvious reason stands out to explain why a HE flare started 30 days after periastron. Similar behaviour is observed from \lsi, where the HE emission peaks after periastron passage (see Fig.~\ref{fig:lsi}) despite being in an unfavourable configuration  for upscattering stellar light. 

The mysterious origin of the HE flare is all the more vexing that, at peak, the power radiated in  HE gamma-rays represented nearly 90\% of the spindown power. The flare was not accompanied by correlated or odd behaviour at other wavelengths, supporting the idea that the HE-emitting particle population is distinct from the population that radiates in X-rays or VHE.  \citet{2012ApJ...752L..17K} proposed that the flare was due to the rapid expansion of the pulsar wind, crushed by passage close to the Be disc, resulting in enhanced inverse Compton emission from the cold pulsar wind: the difficulty is the source of seed photons; \gd{a similar difficulty is faced by \cite{2013arXiv1308.4531D} where the flare is due to upscattering of X-ray photons from the pulsar wind nebula}. \citet{2011ApJ...736L..10T,2012ApJ...753..127K} proposed Doppler boosting because the orientation of the shocked flow is favourable: the difficulty is the lack of concurrent variability since all frequencies should be boosted. \gd{The flare has also been associated with the formation of a superluminal wave precursor to the termination shock \cite{2013arXiv1308.0950M}.}

Further periastron passages will confirm or infirm that the HE flare is a periodic phenomenon. If it isn't, the flare might have been caused by a  mechanism similar to the intense HE flares observed from the Crab nebula \citep{2012ApJ...749...26B}.


\section{Gamma-rays from related binary systems}

High-energy gamma-ray observations have led to the discovery of other types of binary systems besides gamma-ray binaries. Microquasars have long been on the watch list: the latest results on gamma-ray observations of X-ray (accreting) binaries are described in \S\ref{xrb}, especially Cyg X-3, the only confirmed microquasar detection. Novae  (\S\ref{novae}) and colliding wind systems  (\S\ref{cwb}) have also been detected, catching the community by surprise in the case of novae. Many of the tools developed for gamma-ray binaries also find applications in these objects. Novae and colliding winds test our understanding of diffusive shock acceleration on quite different scales than in supernovae remnants, while microquasars in gamma ray offer insights into the mechanisms linking accretion, ejection and non-thermal particle acceleration.

\subsection{X-ray binaries or microquasars\label{xrb}}
X-ray binaries are powered by accretion of matter from a stellar companion onto a black hole or neutron star. A significant fraction of the accretion energy released in X-ray binaries is channeled to launch relativistic jets, whose non-thermal emission is observed in radio \citep{Fender:2006ww}, infrared \citep{2010MNRAS.404L..21C,2010MNRAS.405.1759R}, and X-rays \citep{2002Sci...298..196C,Corbel-et-al.:2005cv}. These jets are the most striking analogy between the physics of stellar-mass black holes in X-ray binaries and supermassive black holes in Active Galactic Nuclei (AGN), hence the name ``microquasars'' for X-ray binaries with relativistic jets \citep{1992Natur.358..215M}. Similarities also exist in timing and spectral characteristics. The analogy prompted speculation that some X-ray binaries could be gamma-ray emitters, like blazars. Blazars are AGN whose relativistic jet is fortuitously aligned with the line-of-sight, boosting the jet non-thermal emission. Blazars are the dominant population of HE sources in the {\em Fermi}/LAT catalogs \citep{2012ApJS..199...31N}. With their large powers, non-thermal radio emission, and relativistic jets, microquasars have all the ingredients for gamma-ray emission \gd{\citep{1996ApJ...456L..29L}}. 

 \subsubsection{Gamma-ray emission expected from microquasars\label{expected}}
If the jet energy is in pairs, as often surmised, then the creation of the positrons must be accompanied by some gamma-ray emission. CGRO observations of X-ray binaries detected non-thermal power-law tails extending well beyond 100\,keV with a photon index $\Gamma_{LE}\approx$2.5--3  \citep{Grove:1998cc}. Cyg X-1 is the prominent example, with the power-law extending to 10~MeV in its high/soft X-ray state \citep{McConnell:2002nz}. A fainter and softer power-law component is also detected in its low/hard X-ray state (for a review of X-ray spectral states, see \citealt{Remillard:2006kg}). X-ray spectral state changes are associated with changes in the ejection regime \citep{Fender:2004hb}. Is there a link between these tails and jet formation ? Part of the available energy may be channeled into accelerating particles in the corona of the accretion disc \citep{Zdziarski:2004en,2009MNRAS.392..570M}. Synchrotron or inverse Compton emission from the compact jet has also been proposed to contribute to the X-ray flux in the low/hard state (\citealt{Markoff:2001om,Georganopoulos:2002ci}). Jet or corona emission can thus be expected to extend to HE gamma rays \citep{Romero:2003nm,Dermer:2006li,Bosch-Ramon:2006do}.  

Radio/IR observations of discrete ejections (``blobs'') in GRS 1915+105 show non-thermal emission cooling with expansion. \citet{Atoyan:1999tn} estimated a detection might be possible by integrating the HE gamma-ray emission in the day following ejection. The best evidence for particle acceleration to very high energies in accreting binaries actually comes from the observation of  X-ray emission from such blobs in the relativistic jet of two microquasars, XTE J1550-563 and H1743-322 \citep{2002Sci...298..196C,Corbel-et-al.:2005cv}. Their emission is compatible with synchrotron radiation from a power-law distribution of electrons. The maximum electron energy is $\approx 10$~TeV, assuming an equipartition magnetic field of  $\approx 500~\mu$G. The X-ray emission is detected far from the binary, on parsec scales, months after the initial ejection event. The short radiative timescale implies that particle acceleration occurs in situ: it could involve internal shocks, magnetic energy dissipation or a shock with the ISM. The particles are energetic enough to emit HE and VHE gamma-rays. Unfortunately, the expected fluxes are too faint to detect unless the magnetic field is well below equipartition with the particle energy density \citep[see Figs.~1-2 in][]{2008MNRAS.384..440X}.

Jets are now thought to be important depositaries of accretion energy, quietly feeding back energy into the ISM as high energy particles. \citet{Heinz:2002qb} estimated their contribution could reach 10\% of the Galactic cosmic ray luminosity. Radio lobes and hotspots provide diagnostics of the relativistic jet power and content in AGN. Likewise, the ISM can behave like a calorimeter for microquasar jets. Radio and optical observations reveal a bubble on parsec scales in the direction of the AU-scale jet of Cyg X-1 \citep{Gallo:2005ii,Russell:2007ju}. The power required to inflate the bubble is comparable to the bolometric luminosity of the binary. The termination shock of the jet inflating the bubble is a potential site for particle (re)acceleration and emission. \citet{Bosch-Ramon:2005tv} and \citet{2009A&A...497..325B} have considered gamma-ray emission from pion decay (when the jet interacts with a molecular cloud), bremsstrahlung, or inverse Compton. Detectability remains highly uncertain, depending on jet power, composition, duty cycle, source distance etc. 

\subsubsection{The elusive search for gamma rays from microquasars\label{xrb2}}

Microquasars have proven elusive in HE and VHE gamma rays even with the latest instrumentation. 

\paragraph{Cygnus X-1} In 2007, the MAGIC collaboration reported flaring gamma-ray emission from Cyg X-1, a binary composed of a massive star and black hole \citep{Albert:2007uw}.  A VHE source was detected for a few hours during one night of a month-long monitoring campaign, with  a  significance of 4.0$\sigma$ (3.2$\sigma$ post-trial). An origin in the bubble discussed above can be ruled out from the VHE position and variability. The flare luminosity $L(>100\,\mathrm{GeV})\approx 2\times10^{34}$\,erg\,s$^{-1}$ corresponds to a conversion in VHE gamma rays of 0.2--2\% of the total jet power ($10^{36}$ to $10^{37}$\,erg\,s$^{-1}$, \citealt{Gallo:2005ii}). The detection occurred during a very bright flare in hard X-rays, as had been observed several times prior to this episode \citep{Golenetskii:2003ps}. The X-ray flare lasted several days, with no remarkable change in flux or spectral state at the time of gamma-ray detection \citep{Malzac:2008hf}. The orbital phase of the VHE detection is close to superior conjunction of the black hole with respect to the O star companion, for which strong VHE absorption by pair production is expected \citep{2009MNRAS.394L..41Z}. Cyg X-1 was not detected during its regular hard X-ray state for which the MAGIC upper limit is $\approx 10^{33}$\,erg\,s$^{-1}$. Further observations by MAGIC, adding 40 hours to the initial 50 hours, did not re-detect the source, formally lowering the significance of the detection \citep{2009arXiv0907.1017S,2011arXiv1110.1581Z}. VERITAS also reported only upper limits, from $\approx$ 10 hours of observations \citep{2009arXiv0908.0714G}. The VHE detection of Cyg X-1 remains to be confirmed. 

In HE gamma rays, \citet{2013arXiv1305.5920M} and \citet{2013arXiv1307.3264B} report a 3 to 4$\sigma$ detection of steady gamma-ray emission from Cyg X-1 using \fermi\ observations. The source is detected only during the hard X-ray state, with a luminosity  $L({\rm\geq100\, MeV})\approx 2\times 10^{33}$\,erg\,s$^{-1}$ for a 2 kpc distance, consistent with the upper limits derived using {\em AGILE} data \citep{2010A&A...520A..67D,Sabatini:2013aa}. This is surprising because the non-thermal  power-law tail detected at MeV energies is weaker during the hard state than in the soft state  (\S\ref{expected}), so a direct extrapolation would have predicted a gamma-ray detection in the soft state. The presence of a radio jet in the hard state, absent in the soft state, could be the key ingredient. \citet{2010A&A...520A..67D} excluded the possibility of HE flares analogous to the MAGIC flare in the {\em AGILE} data, based on the extrapolation of the soft MAGIC spectrum ($\Gamma_{\rm VHE}\approx 3$) to the {\em AGILE} range. However, episodic detections of Cyg X-1 are also reported by {\em AGILE}: over a timescale $\leq 1$ day during a hard X-ray state \citep{2010ApJ...712L..10S} and over a timescale $\leq 2$ days during a hard-to-soft transition \citep{Sabatini:2013aa}. The brightest flare (4$\sigma$ post-trial significance) reached a luminosity in HE gamma rays $\approx 2\times 10^{35}$\,erg\,s$^{-1}$. These flares were not confirmed using \fermi\ data \citep{2011heep.conf..498H}. The gamma-ray detections and upper limits are constraining for models of jets, hot accretion flows, and disc coronas \citep{2013arXiv1305.5920M}. Future work is expected to clarify and confirm these detections.\\
 
\paragraph{GRS 1915+105} a famous microquasar, has a ``plateau'' state that resembles the low hard state. The plateau state is preceded and followed by a strong radio flare \citep{2002MNRAS.331..745K}. The radio jet in this state has an estimated power of 3$\times10^{38}\,\rm erg\,s^{-1}$ \citep{2000MNRAS.318L...1F}. The X-ray spectrum becomes softer before the post-plateau radio flare. HESS observed the source on several occasions for a total time of 24 hours and set an upper limit of $6\times 10^{-13}\rm\, ph\,cm^{-2}\,s^{-1}$ above 410 GeV, corresponding to a VHE luminosity of $10^{34}\rm\,erg\,s^{-1}$ at 11 kpc, so less than 0.003\% of the jet power is emitted in VHE gamma rays \citep{2009A&A...508.1135H}. This is much less than inferred in Cyg X-1 from the MAGIC detection. \\

\paragraph{Other X-ray binaries} The MAGIC and VERITAS collaborations have {published} upper limits on the HE and VHE gamma-ray flux from 1A 0535+262, a bright neutron star HMXB \citep{2011arXiv1103.3250V}, and Sco X-1, a neutron star LMXB with radio jets \citep{2011ApJ...735L...5A}. The upper limit on the ratio of VHE to X-ray flux is comparable to what was found for GRS 1915+105. In the case of 1A 0535+262, this low upper limit supports the idea that different physical processes are at work around the neutron stars of accreting HMXBs and those of gamma-ray binaries, where the ratio is close to one. In HE gamma rays, \citet{2013arXiv1307.3264B} give upper limits of $\approx 2\times 10^{-8}$~ph~cm$^{-2}$s$^{-1}$ (0.1-10~GeV) on the microquasars GRS 1915+105 and GX 339-4 using \fermi\ \gd{(see \citealt{1996ApJ...462L..67L} for earlier upper limits using {EGRET} data)}. The only detections to date are Cyg X-1 (see above) and Cyg X-3, detailed below.

\subsubsection{Cygnus X-3: observations\label{cygx3}}

Cyg X-3 is composed of a compact object in a 4.8 hours orbit with a Wolf-Rayet star \citep{1996A&A...314..521V}.  Wolf-Rayet are amongst the most extreme stars: this one has a luminosity $L_{\star}\approx 10^{39}$\,erg\,s$^{-1}$ and loses mass at the rate of 10$^{-5}$\,M$_{\odot}$\,yr$^{-1}$ as a  1000 km\,s$^{-1}$ stellar wind. There is only one such binary known in our Galaxy. The nature of the compact object remains unknown, although recent work has tended to favour a black hole \citep[][and references therein]{2012arXiv1208.5455Z}. The binary has a separation of only 4 $R_{\odot}$, the compact object comes within a stellar radius of the surface of its companion, the matter and radiation densities at this location are more than three orders-of-magnitude greater than in any other high-mass X-ray binary and six order-of-magnitude greater than in any low-mass X-ray binary. \citet{Bonnet-Bidaud:1988vw} reviewed the first two decades of work on Cyg X-3.

\begin{figure}
\centering\resizebox{11.7cm}{!}{\includegraphics{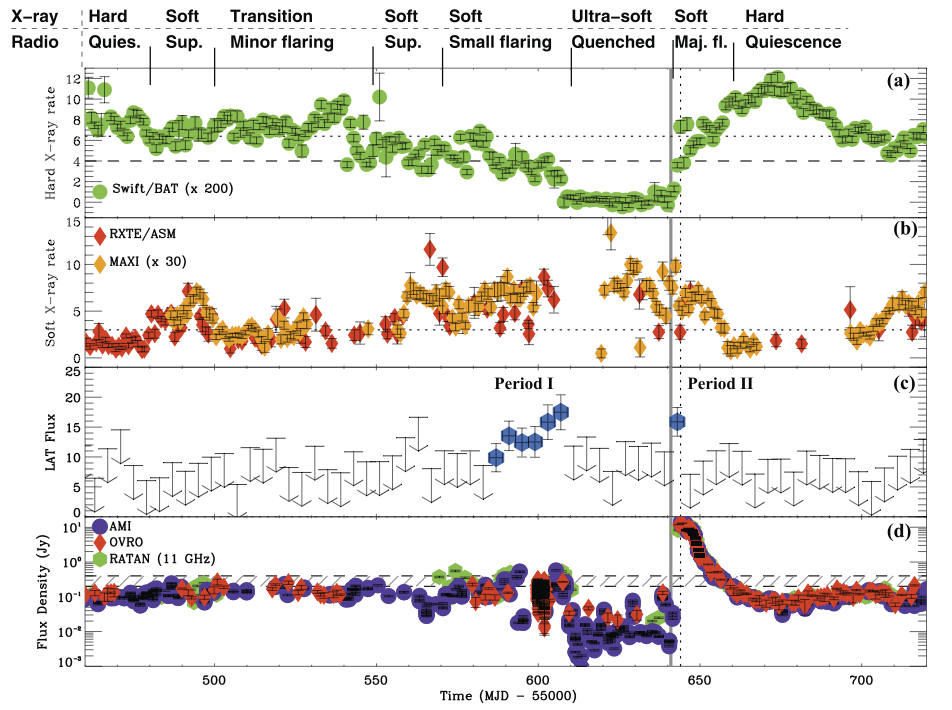}}
\caption{Multiwavelength behaviour of Cyg X-3  around the time of the 2011 major radio flare with, from top to bottom: hard X-rays, soft X-rays, HE gamma rays, radio. Figure reprinted from \citet{2012MNRAS.421.2947C}, with permission of Oxford University Press, on behalf of the RAS.}
\label{fig:radioflare}
\end{figure}
The X-ray source was discovered in 1966 by Giacconi and collaborators, rising to stardom when it was found that the system flares to become one of the brightest sources in the radio sky \citep{1972Natur.239..440G}. The flux reaches $\ga 20$ Jy during major eruptions (\citealt{1973Sci...182.1089H} gives a lively account of the discovery). These occur irregularly every year or so. The radio emission during flares was spatially resolved \citep{1983ApJ...273L..65G,1986ApJ...309..694S,1989Natur.337..234S}, making Cyg X-3 one of the first members of the class of sources that came to be known as microquasars. \cyg\ displays changes in X-ray spectral states similar to those in X-ray binaries, with the caveat that propagation in the stellar wind modifies strongly the X-ray timing and spectral properties \citep{2008MNRAS.386..593S,2010MNRAS.402..767Z}. 

\cyg\ was first claimed in gamma rays in 1973, soon after the radio detection. {\em SAS-2} reported a flux level of 4.4$\times10^{-6}$\,ph\,cm$^{-2}$\,s$^{-1}$ above 100 MeV, with the gamma-ray emission modulated on the 4.8 hr orbital period, from two observation periods of one week in 1973 \citep{Lamb:1977xw}. The flux level is comparable to that observed by {\em AGILE} and \fermi\ in the same energy band, nearly 40 years later \citep{2009Natur.462..620Ta,2009Sci...326.1512F}. However, the {\em SAS-2} gamma-ray modulation is in phase with the X-ray modulation while the \fermi\ modulation (see below) is anti-phased with the X-rays. The {\em SAS-2} detection (4.5$\sigma$) was not confirmed by either {\em COS-B} \citep{1987A&A...175..141H} or EGRET \citep{1992ApJ...401..724M}, although \citet{1997ApJ...476..842M} later reported an EGRET source at the location of \cyg.

In 2009 both {\em AGILE} and \fermi\ collaborations announced the detection of \cyg\ in HE gamma rays \citep{2009Natur.462..620Ta,2009Sci...326.1512F}. The detection is considered secure because of
\begin{itemize}
\item the high confidence of the detection ($>$20$\sigma$) at a position coincident with Cyg X-3 after careful consideration of the possible biases (gating of a strong nearby gamma-ray pulsar, improved modelling of the Galactic diffuse emission); 
\item the correlated variability of the $\gamma$-ray source with activity in radio and X-rays; 
\item the detection of a $\gamma$-ray orbital modulation at different epochs (Fig.~\ref{fig:cygx3model});
\item the independent, concurrent detections using different instruments.
\end{itemize}
The source is undetected in VHE gamma rays, even during times of HE gamma-ray emission \citep{2010ApJ...721..843A}. 

Gamma-ray detections are associated with high levels of soft X-ray emission and periods of radio activity  \citep{2009Sci...326.1512F,2011ApJ...733L..20W,2012A&A...538A..63B,2013arXiv1307.3264B}, more precisely when the following criteria are met  \citep{2012MNRAS.421.2947C,2012arXiv1207.6288P}:  the 3--5 keV {\em RXTE}/ASM count rate is above the pivotal value of 3 counts\,s$^{-1}$ marking the transition from hard to soft state, the {\em Swift}/BAT ($>$ 15 keV) count rate is  below 0.02 counts\,s$^{-1}$, the radio flux is greater than 0.2--0.4 Jy at 15 GHz. The soft X-rays are thought to arise from the accretion disc, the hard X-rays from the corona, and the radio from the jet \citep{2008MNRAS.388.1001S}. Specific conditions must be present in all these components to have gamma-ray emission. Fig.~\ref{fig:radioflare} illustrates this during the ``quenched radio state'' that precedes a major flare (Fig.~\ref{fig:radioflare}). The thin disc is thought to reach to the innermost stable orbit in this state, turning off the radio jet  \citep{2009MNRAS.392..251H}. There is a relationship between the radio emission tracing the jet on large scales, the X-ray emission tracing the accretion flow, and the gamma-ray emission tracing non-thermal processes.  All these components are simultaneously observable in \cyg.

\subsubsection{Cygnus X-3: accretion/ejection in $\gamma$-rays}
\begin{figure}
\center\includegraphics[width=10cm]{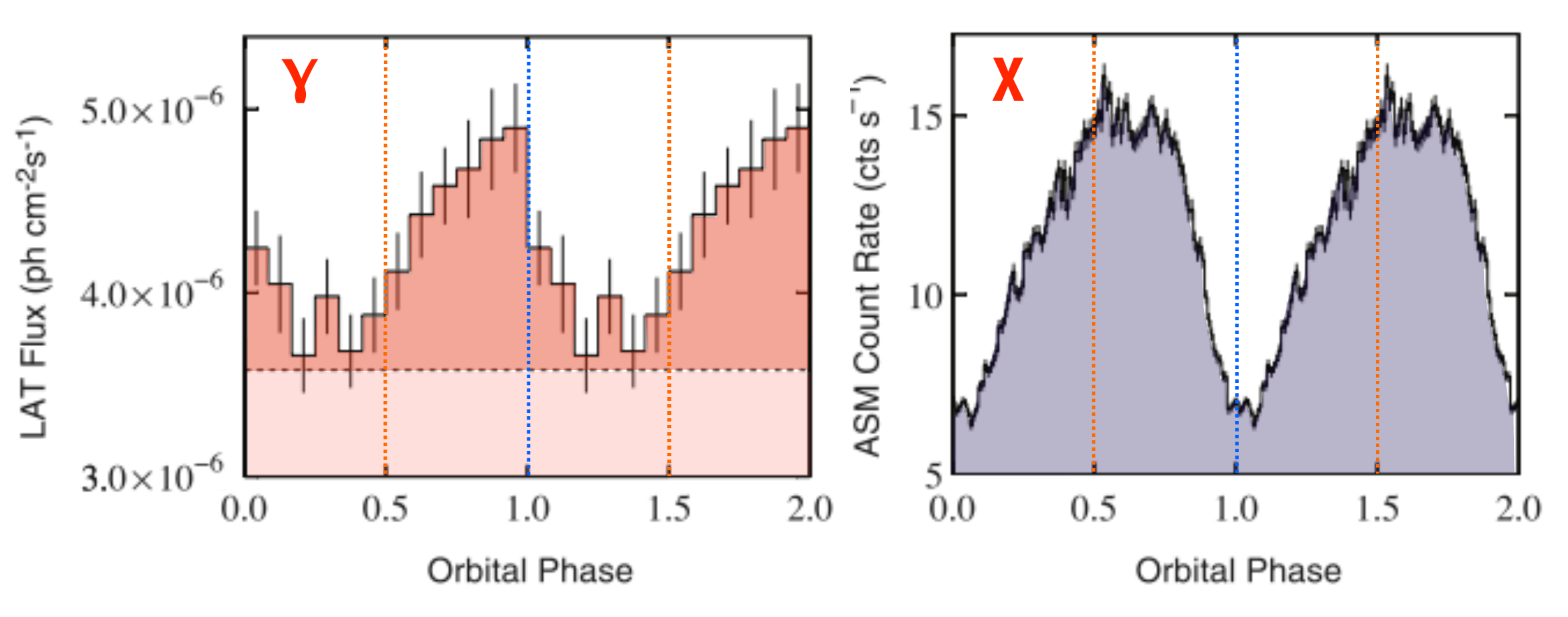}
\center\includegraphics[width=10cm]{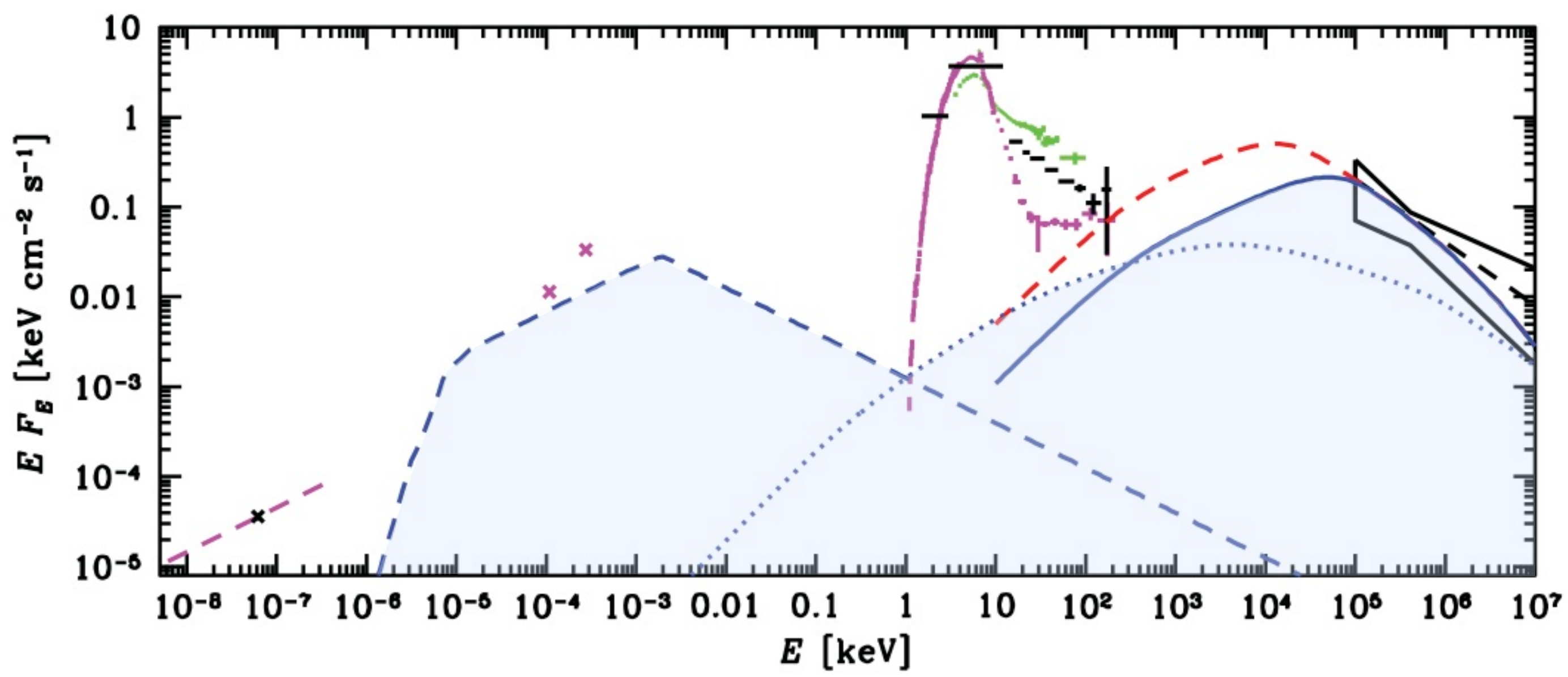}
\caption{{\em Top:} orbital modulations in gamma-rays (left) and X-rays (right). The horizontal dashed line (left panel) shows the baseline flux, including the Galactic diffuse emission, at the position of \cyg\ outside of flaring episodes. Figure taken from \citet{2009Sci...326.1512F}, reprinted with permission from AAAS. Bottom: spectral energy distribution of Cyg X-3. The shaded region corresponds to a model for the non-thermal emission (synchrotron, self-Compton, and Compton on stellar photons; figure reprinted from \citealt{2012MNRAS.421.2956Z}, with permission of Oxford University Press, on behalf of the RAS).}
\label{fig:cygx3model}
\end{figure}
The orbital modulation of the gamma-ray flux is a powerful tool to study the origin of the particle acceleration. Anisotropic inverse Compton scattering of the Wolf-Rayet photons by non-thermal electrons is a natural interpretation for the modulation  \citep{2009Sci...326.1512F}. This is supported by the anti-correlation with the X-ray modulation. The X-ray modulation is due to Thompson scattering of the X-rays ($\tau_{\rm T}\approx 1$) as they travel through the stellar wind \citep{Pringle:1974oj,2012MNRAS.426.1031Z}. Assuming the X-rays originate from the accretion disc, close to the compact object, minimum scattering (maximum X-ray flux received by the observer) occurs at inferior conjunction  and maximum scattering (minimum X-ray flux) occurs at superior conjunction. This phasing of the orbit is consistent with the interpretation of the infrared modulation as due to partial shielding of the stellar wind from the strongly ionising X-ray flux \citep{1996A&A...314..521V,1999MNRAS.308..473F}. It turns out that the gamma-ray maximum corresponds to superior conjunction (Fig.~\ref{fig:cygx3model}), just as expected from anisotropic inverse Compton scattering (\S\ref{anis}). The propagation of the gamma rays is not affected by scattering in the wind, unlike X-rays, because of the interaction occurs in the Klein-Nishina regime.

The nearly 100\% modulation fraction suggests the emission region must be localised: a large emission region would dilute the effects of anisotropic scattering. The link with the radio emission suggests the jet as the emission site. Gamma-ray emission in the accretion disc or in the corona is strongly affected by pair production with X-ray photons \citep{2011A&A...529A.120C,2012MNRAS.421..512S}. If the gamma-ray emission is localised in a small region of the jet,  \citet{2010MNRAS.404L..55D} found that the models best-fitting the modulation lightcurve involve a height in the jet of typically 1--10$\times$ the orbital separation ({\em i.e.} outside of the system) and an inclined jet of moderate speed (0.2-0.8$c$), close to our line-of-sight as already deduced from radio observations \citep{Miller-Jones:2004qg}. 

The gamma-ray emission is not a direct extrapolation of the hard power-law extending in X-rays above 30 keV: there is no direct spectral connection  (Fig.~\ref{fig:cygx3model}) and they are not modulated in phase \citep{2012MNRAS.426.1031Z}. Using the constraints from the spectral energy distribution and modulations, \citet{2012MNRAS.421.2956Z} found that the electrons are injected with a minimum Lorentz factor $\gamma\approx 300$--1000, consistent with theories of diffusive shock acceleration which have $\gamma_{\rm min}\sim m_{p}/m_{e}$, and that the jet kinetic power is comparable to the bolometric luminosity of 10$^{38}$\,\eps. 

Hence, the gamma-ray emission zone is separated from the corona and located in the jet, far from the accretion disc. Particle acceleration could be associated with the recollimation shock that forms when the jet becomes under-pressured with respect to its environment: here, the dense stellar wind from the Wolf-Rayet companion. Such shocks are seen in numerical simulations, which also show low-power jets can be disrupted by the stellar wind \citep{2010A&A...512L...4P}.  Recollimation shocks have also been invoked to explain the radio brightening of ejecta moving through (nearly) standing shocks in M87 and blazars \citep{Stawarz:2006oh,2009ApJ...699.1274B,2011ApJ...726L..13A}. The location of the shock will vary depending upon the jet and stellar wind pressure. The associated gamma-ray lightcurve will be intermittent if there is a favoured location for emission, neither too close nor too far  (a ``sweet spot'', \citealt{2012MNRAS.421.2947C}). The non-thermal particles radiate in radio as they cool down. The radio lightcurve of \cyg, with minor and major flares, has been interpreted as a succession of propagating shocks \citep{Miller-Jones:2008fk}. The formation of the recollimation shock has yet to be related to the conditions in the corona, close to the launch region, and to the radio emission further away in the jet. The tentative detection of Cyg X-1, where the stellar environment is much less stringent, may question a model designed too exclusively for \cyg. Understanding how all the pieces of the puzzle connect will provide unprecedented constraints on the dynamical links between accretion, ejection, and non-thermal processes. 

\subsection{Novae\label{novae}}
\begin{figure}
\centering
\resizebox{!}{3.75cm}{\includegraphics{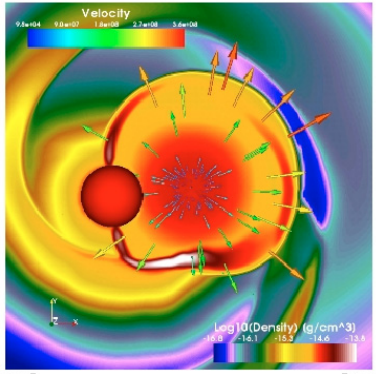}}
\resizebox{!}{3.75cm}{\includegraphics{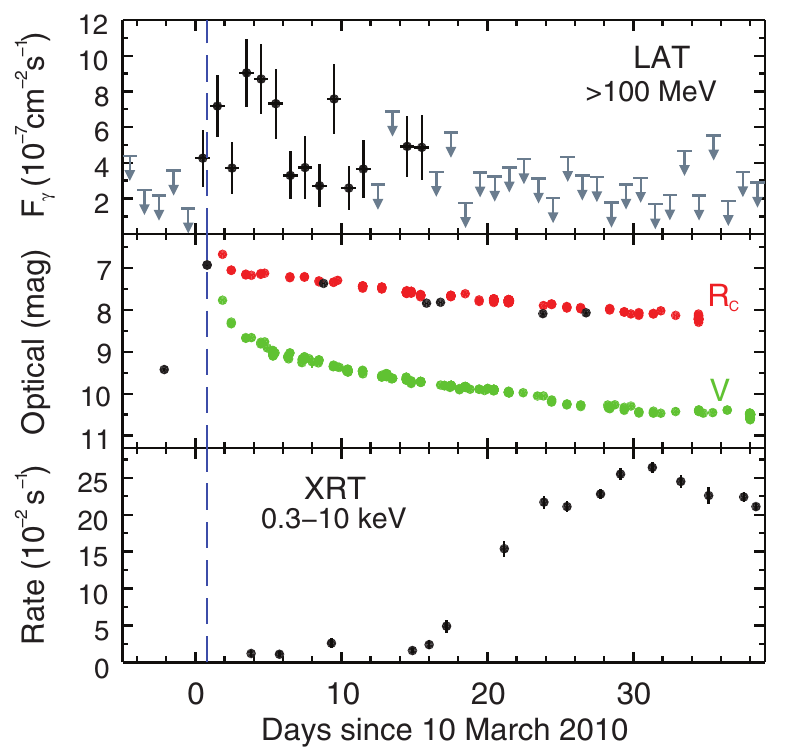}}
\resizebox{!}{3.75cm}{\includegraphics{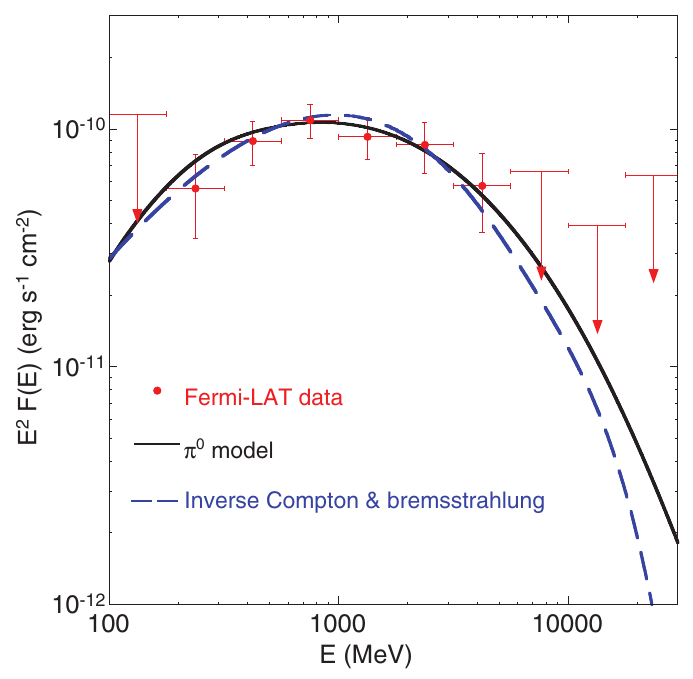}}
\caption{Gamma-ray emission from the nova V407 Cyg. {\em Left:} numerical simulation of the propagation of a shock in a symbiotic system (density map in the orbital plane, figure reproduced with permission from \citealt{2008A&A...484L...9W} \copyright ESO). {\em Middle:} gamma-ray, optical, and X-ray lightcurve of the V407 Cyg nova outburst. {\em Right}: average gamma-ray spectrum, with representative ``leptonic'' (inverse Compton) and ``hadronic'' ($\pi_{0}$ decay) models (figures taken from \citealt{2010Sci...329..817A}, reprinted with permission from AAAS). \label{fig:v407}}
\end{figure}
The discovery by the \fermi\ of gamma-ray emission associated with a symbiotic novae, V407 Cyg, was somewhat unexpected. Symbiotic novae are composed of a white dwarf in orbit around a red giant. Mass accumulates on the white dwarf through Bondi-Hoyle accretion of the stellar wind until the critical pressure for thermonuclear burning is reached. The sudden energy release forms a shock as the burning layer rapidly inflates an envelope that  reaches a size of $\approx 0.3$ AU within 0.1 day in fast novae (associated with massive white dwarfs, \citealt{2005ApJ...623..398Y}). The shock  propagates in the dense red giant wind (Fig.~\ref{fig:v407}), as observations of the symbiotic nova  RS Oph illustrated prior to the detection of V407 Cyg \citep{Sokoloski:2006uq,Rupen:2007xb}. \citet{2007ApJ...663L.101T} pointed out that diffusive shock acceleration of particles in RS Oph could explain discrepancies between the  shock speed as measured from infrared lines and from the X-ray temperature (related by $kT\propto v^{2}$ in the adiabatic limit, Eq.~\ref{eq:thermal}), a diagnostic that is also used to study the non-thermal content of shocks in supernova remnants  \citep[e.g.][]{2009Sci...325..719H}.

The discovery of gamma-ray emission from V407 Cyg with the \fermi\ confirmed particle acceleration occurs in novae \citep{2010Sci...329..817A}. V407 Cyg is a similar system to RS Oph, with a white dwarf in an estimated 40-year orbit around a Mira red giant at a distance of 2.7 kpc \citep[see][and references therein]{2011MNRAS.410L..52M,2011A&A...527A..98S,2012A&A...540A..55S}. Japanese amateur astronomers reported a nova outburst in March 2010 \citep{2010CBET.2204....2N}. The event was rapidly associated with the apparition of a new gamma-ray source at the same time and location \citep{2010ATel.2487....1C}. Gamma-ray emission was detected in the first two weeks after the optical peak  (Fig.~\ref{fig:v407}). The average gamma-ray spectrum is best-fitted by a power-law of index $\Gamma\approx 1.5$  with an exponential cutoff $E_{\rm c}\approx 2.2$\,GeV. No VHE emission was observed by the \citet{2012ApJ...754...77A}, the VHE observations starting $\approx$ one week after optical peak.

 The shock propagation depends on the direction within the system since matter is distributed anisotropically around the white dwarf. The shock slows down on a timescale $\sim$ a week as material is swept-up from the circumstellar medium. Protons and electrons accelerated at the shock lose energy via synchrotron or inverse Compton emission, $pp$ interactions leading to pion decay, and adiabatic expansion. At the same time, shock-heated material radiates thermal bremsstrahlung emission in X-rays. Radio emission appears to be due to free-free emission of the stellar wind, ionized by the nova outburst \citep{2012ApJ...761..173C}. Overall, the system behaves like a miniature supernova, with the total kinetic energy in the shock $\approx 10^{44}$\,erg, whose evolution is seen in ``real-time'' rather than through a snapshot.

The gamma-ray emission from V407 Cyg has been modeled as pion decay \citep{2010PhRvD..82l3012R,2012PhRvD..86f3011S}, while the X-ray emission has been modeled as thermal bremsstrahlung emission \citep{2012ApJ...748...43N,2012MNRAS.419.2329O}.  However, as with supernova remnants \citep[e.g.][]{2012ApJ...744...39E}, the combined study of thermal and non-thermal emission is a powerful discriminant of models. Both emissions are linked since acceleration and heating tap the same energy reservoir (the shock's kinetic energy) and the same particle reservoir (swept-up material). Because of their smaller cross-section, $pp$ interactions tend to be less efficient than inverse Compton scattering at converting particle energy to high-energy emission. Gamma-ray emission dominated by $\pi_{0}$ decay thus requires high densities in order to have a large number of protons accelerated and a high number of targets for $pp$ interactions, which also implies strong X-ray emission since a lot of material is swept-up and shock-heated. \citet{2013A&A...551A..37M} argue that this is compatible with observations only if most of the kinetic energy is channeled into particle acceleration (i.e. a ``cosmic-ray dominated shock'' with a non-thermal efficiency $>$50\%). Models where the gamma rays are predominantly due to inverse Compton emission on the red giant or nova light are less demanding in terms of energetics, with non-thermal efficiencies of 10\% and values for the fraction of particles that are accelerated compatible with those derived from supernova remnants. In all cases, the fast rise and rapid decay of the \fermi\ lightcurve  require early interaction with denser material  than expected from the stellar wind \citep{2013A&A...551A..37M}. Identical conclusions are reached by modeling the X-ray lightcurve \citep{2012MNRAS.419.2329O}. A natural candidate is denser structures arising from Bondi-Hoyle accretion onto the white dwarf prior to the outburst (\citealt{2008A&A...484L...9W}, Fig.~\ref{fig:v407}).

Since then, the \fermi\ collaboration has reported the discovery of gamma-ray emission from two other novae, Nova Mon 2012 and Nova Sco 2012 \citep{2012ATel.4284....1C,2012ATel.4310....1C}. The preliminary \fermi\ lightcurves and spectra are similar to those of V407 Cyg, with the gamma-ray emission lasting a couple of weeks and a spectrum peaking at a few GeV in $\nu F_{\nu}$ \citep{2013arXiv1304.3475C,2013arXiv1304.2427H}. Surprisingly, none are associated with a symbiotic binary. Nova Mon 2012 has a 7.2 hour orbital period, hence a main-sequence companion without a dense stellar wind \citep{2013ApJ...768L..26P,2013arXiv1303.0404S}. Any nova-generated shock quickly engulfs the system, a priori finding little matter to interact with on a timescale of weeks. What is the ejecta interacting with in those systems ? The answer could change our understanding of what happened in V407 Cyg. Classical novae are about 10$\times$ more numerous than symbiotic novae \citep{Lu:2006cu,2011MNRAS.413L..11L} so further gamma-ray detections can be expected, bringing new perspectives on nova physics as well as on  diffusive shock acceleration.

\subsection{Colliding wind binaries\label{cwb}}
\begin{figure}
\centering
\resizebox{!}{3.1cm}{\includegraphics{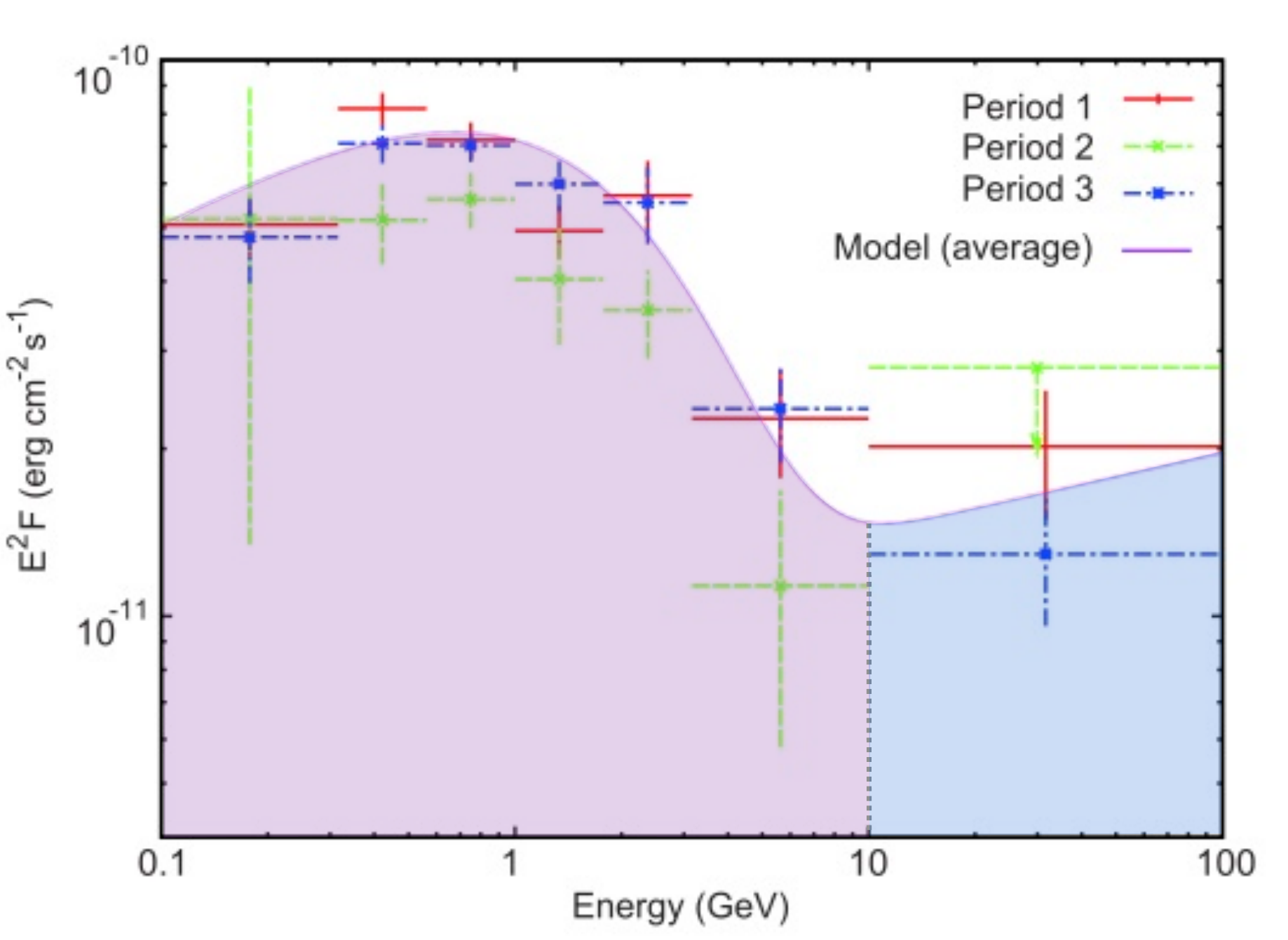}}
\resizebox{!}{3.2cm}{\includegraphics{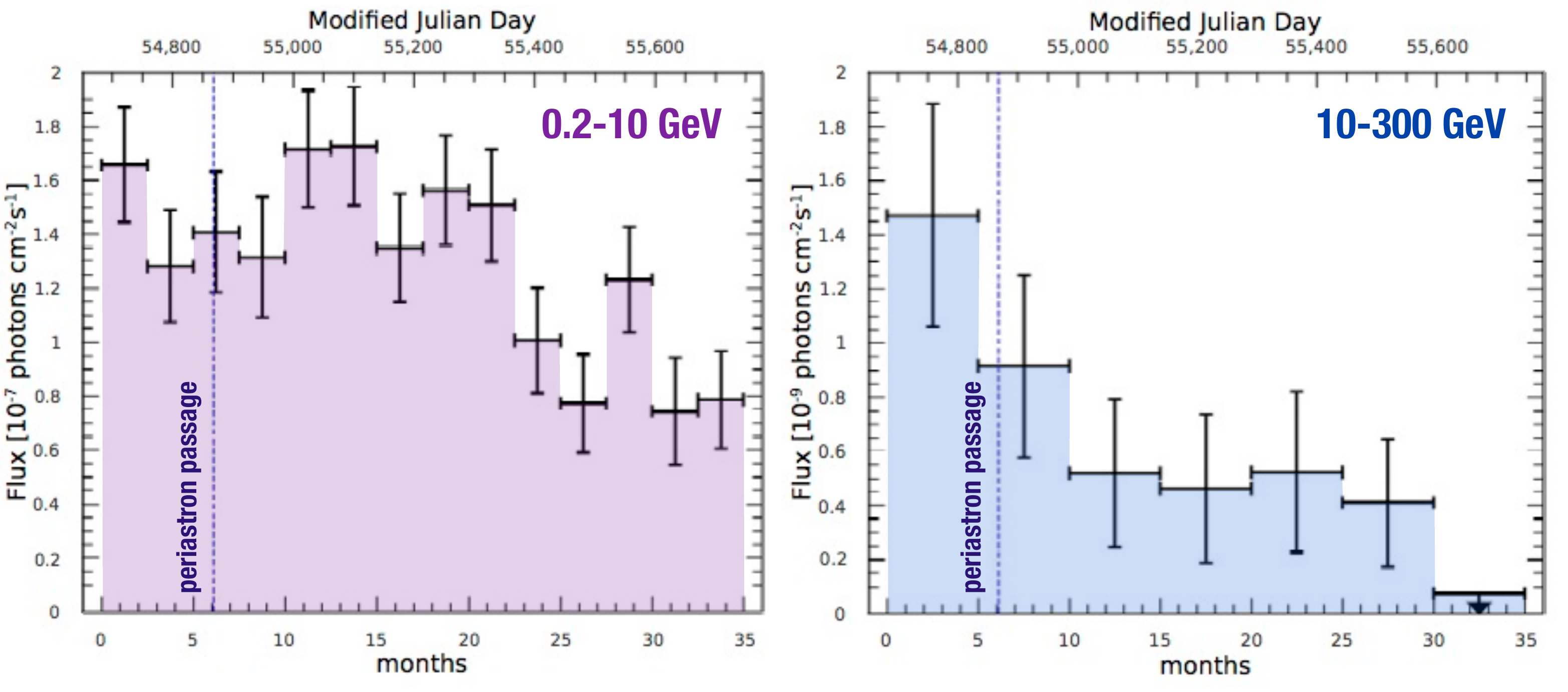}}
\caption{Spectrum (left) and lightcurves (soft and hard gamma-ray bands) of the gamma-ray source associated with the colliding wind binary $\eta$ Car. Spectrum reproduced with permission of AAS from \citet{2010ApJ...723..649A}. Lightcurves reproduced with permission from \citet{2012A&A...544A..98R} \copyright ESO. \label{fig:eta}}
\end{figure}

Colliding wind binaries (CWB) are binaries composed of two massive stars. The interaction of their stellar winds (\S\ref{wind}) generates a double-shock structure, as presented for gamma-ray binaries (Figs.~\ref{fig:mirabel}-\ref{fig:geom}). Shock heating is thought to be responsible for the X-ray emission from CWB \citep{1976PAZh....2..356C,Luo:1990mp,Usov:1992re,Stevens:1992on} ; shock compression helps obtain the high densities required to form dust and shield it from the intense UV emission from the stars \citep{1987A&A...182...91W,1991MNRAS.252...49U}. Non-thermal radio emission is also detected, attributed to diffusive shock acceleration of particles (\citealt{Eichler:1993wt}; see \citealt{2007A&ARv..14..171D} for a review \gd{and \citealt{2013arXiv1308.3149D} for a catalog}). In some CWB, the varying aspect of the shock at different orbital phases has been imaged using radio very long baseline interferometry \citep{2005ApJ...623..447D}. In others called ``pinwheel nebulae'', infrared interferometry was able to image the spiral pattern created by the rotation of the cometary ``plume'' of dust emission with the binary orbit \citep{1999Natur.398..487T}. Gamma-ray binaries are in many aspects only extreme examples of CWB (\S2-3). 

A fraction of the stellar wind kinetic power is dissipated at the shocks, where a further fraction can serve to accelerate particles and generate detectable gamma-ray emission (\citealt{2003A&A...399.1121B}, \citealt{Reimer:2006da}, \citealt{Pittard:2006wx}). Only one CWB has been detected so far in HE gamma rays. $\eta$ Car is composed of a $\approx 100\, M_\odot$ luminous blue variable in a 5.5\,year highly eccentric orbit ($e>0.9$) with an O or Wolf-Rayet type star (\citealt{1997ARA&A..35....1D,2008MNRAS.384.1649D} and references therein). The system is also rare (if not unique) amongst CWB in showing non-thermal {\em X-ray} emission: a hard tail ($\Gamma_X\approx 1.4$) is observed above 10\,keV up to at least 50\,keV, with a luminosity about $\approx 7\times10^{33} (d/\rm 2.3\,kpc)^2$\,\eps\ \citep{2004A&A...420..527V,2008A&A...477L..29L,2009PASJ...61..629S}.

A gamma-ray source positionally-coincident  with $\eta$ Car was discovered with {\em AGILE} \citep{2009ApJ...698L.142T} and \fermi\ \citep{2010ApJ...723..649A}. The HE spectrum is composed of a power-law with an exponential cutoff at $E_{\rm c}\approx1.6$\,GeV (the ``soft'' component) and a hard tail detected nearly up to 100 GeV (``hard'' component, \citealt{2010ApJ...723..649A,2011A&A...526A..57F}, Fig.~\ref{fig:eta}). This tail must cut off abruptly since observations with HESS did not detect the source in VHE gamma rays \citep{2012MNRAS.424..128H}. The flux from both spectral components has declined on a timescale of months since periastron passage, securing the identification (\citealt{2012A&A...544A..98R}, Fig.~\ref{fig:eta}). However, the reported $\la 2$\,day gamma-ray flare in the {\em AGILE} data,  which occurred 75 days before periastron passage,  is not confirmed using the \fermi\ data \citep{2010ApJ...723..649A}.

The spectrum has been interpreted as inverse Compton (soft component) and $\pi_0$ decay (hard component) emission of particles accelerated in the wind collision region \citep{2011A&A...526A..57F,2011A&A...530A..49B}. The inverse Compton emission may also connect with the hard tail of non-thermal X-ray emission. Alternatively,  the HE gamma-ray spectrum could result from the absorption of an intrinsic  power-law spectrum, due to pair production of the gamma rays with 0.1\,keV photons \citep{2012A&A...544A..98R}. Variability disfavours an origin in the large scale ($\approx 0.25$ pc) blast wave of the ``Great Eruption'' of 1843, as proposed by \citet{2010arXiv1006.2464O}. 

The gamma-ray luminosity of $\eta$ Car ($L_{\rm HE}\approx 1\times10^{35}$\,\eps) represents $\approx 0.2$\% of the kinetic power available from the winds ($\approx 6\times10^{37}$\,erg\,s$^{-1}$). There are several nearby CWB with comparable kinetic energies and non-thermal radio emission, indicating particle acceleration also occurs at some level in these systems. Yet they remain undetected in gamma rays \citep{2013A&A...555A.102W}. The key to gamma-ray emission remains to be understood, perhaps linked to how diffusive shock acceleration functions (or not) in these systems.

\section{Summary and concluding remarks}
Gamma-ray astronomy has advanced at a fast pace in the past decade thanks to the latest generation of telescopes. Binaries, suspected since the early days of gamma-ray astronomy, are now confirmed sources of HE and VHE gamma rays. A rich variety of binaries are cosmic accelerators, confirming that non-thermal processes are pervasive in astrophysical sources. At the time of writing:
\begin{itemize}
\item The five binaries detected in {\em VHE gamma rays} form a new class of systems, {\em gamma-ray binaries}, distinguished by emitting most powerfully beyond 1 MeV and by having a O or Be massive star companion. Large amplitude variability tied to the orbital period is widespread at all wavelengths. Changes in radio morphology with orbital phase are observed. The non-thermal emission in \psrb\ is powered by the spindown of a pulsar ; the same mechanism is likely to be at work in the others. In these systems, gamma-ray emission constitutes the dominant channel through which energy is radiated.
\item {\em HE gamma rays} have revealed a rich variety of binary systems: microquasars (\cyg, possibly Cyg X-1), a colliding wind binary ($\eta$ Car), and three novae (V407 Cyg, Nova Mon 2012, Nova Sco 2012). Gamma-ray emission constitutes only a small fraction of the radiation from those systems but provides unprecedented information on particle acceleration in various settings. All of these systems are variable in gamma rays. Four of the gamma-ray binaries are also detected in this energy range but one is not (\hessj). 
\item Recycled, millisecond pulsars in binaries must also be added to this roster: dozens have been detected in HE gamma rays using the \fermi \citep{2013arXiv1305.4385T}. Their emission is identical to that of isolated pulsars, their binarity irrelevant, except in the case of the black widow pulsar where there is evidence for gamma-ray emission from an intra-binary shock.
\end{itemize}
Further additions to this list can be expected as {\em AGILE} and {\em Fermi} continue to scan the HE gamma-ray sky and as the current ground-based detectors evolve towards the next generation (see \citealt{2013APh....43...81B,2012arXiv1210.3215P} for prospects with CTA, the {\em Cherenkov Telescope Array}).

What drives the high-energy emission in these {\em gamma-ray binaries and related objects} ? Where and how is power channeled to accelerate particles ? Common tools are used to address these questions: the search for modulations, the determination of orbital parameters, the use of radiative processes adapted to the case of a binary, etc. Techniques used for one type of system find use in others. Parallels can be drawn between different types of binaries: gamma-ray binaries like \psrb\ can be approached as extreme examples of colliding wind binaries such as $\eta$ Car, providing inspiration to understand the dynamics of the interaction. Gamma-ray binaries and  millisecond pulsars in orbit around low-mass companions share the same basic ingredients (a high $\dot{E}$ pulsar, a companion star) but in differing proportions, providing additional angles to understand the processes involved.

Why do some binaries emit high-energy emission and others do not ? This questions both our understanding of the physics involved in each source and the fundamental issue of particle acceleration. Gamma-ray binaries likely represent a short PWN phase in the transition towards a high-mass X-ray binary system, setting these gamma-ray detections in the general context of high-energy emission from pulsars and of binary evolution. There are dozens of microquasars, with large amounts of accretion power available, yet few have been detected. Is the extreme density and radiation environment in \cyg\ uniquely responsible for the intense gamma-ray emission ? Similarly, why are particles accelerated to high energies only in a few novae or colliding wind binaries ? Does that mean that the conditions are special in those systems, for instance because a strong shock happens to be present, or that particle acceleration is somehow stunted in the others ? 

Analogies extend to other classes of objects. The study of gamma-ray emission from microquasar jets connects to the study of relativistic jets from active galactic nuclei and gamma-ray bursts \gd{\citep{2006IJMPA..21.6015L}}. Gamma-ray binaries are another manifestation of pulsar wind nebulae. Particle acceleration in novae links up with particle acceleration in supernova remnants. Hence, binaries involve the same mechanisms that operate in major classes of sources in the gamma-ray sky, straining theories by testing them on different scales, with the advantage that different conditions and line-of-sights are accessible for the same object. Binaries are indispensable laboratories in the endeavour to build a consistent picture across objects and scales of particle acceleration and the physics of relativistic, magnetized outflows. 

\begin{acknowledgements}
I would like to thank B. Cerutti, S. Corbel, R. Dubois, S. Fromang, B. Giebels, G. Henri, A. Hill, J. Kirk, A. Lamberts, P. Martin, M. de Naurois, J. P\'etri, K. Reitberger, A. Szostek, M. Yamaguchi, and A. Zdziarski for many enlightening discussions on the topic of gamma-ray emission from binaries. I would also like to express gratitude to my HESS and \fermi\ colleagues who, together with scientists involved with other gamma-ray observatories, have brought the field forward and made this work possible. I acknowledge support by the {European Research Council}  under the European Community's 7th framework programme (contract ERC-StG-200911),  and by the Centre National d'Etudes Spatiales.
\end{acknowledgements}

\bibliographystyle{aa}       
\bibliography{../Biblio}   

\begin{thebibliography}{432}
\expandafter\ifx\csname natexlab\endcsname\relax\def\natexlab#1{#1}\fi

\bibitem[{{Agudo} {et~al.}(2011){Agudo}, {Jorstad}, {Marscher}, {Larionov},
  {G{\'o}mez}, {L{\"a}hteenm{\"a}ki}, {Gurwell}, {Smith}, {Wiesemeyer}, {Thum},
  {Heidt}, {Blinov}, {D'Arcangelo}, {Hagen-Thorn}, {Morozova}, {Nieppola},
  {Roca-Sogorb}, {Schmidt}, {Taylor}, {Tornikoski}, \&
  {Troitsky}}]{2011ApJ...726L..13A}
{Agudo}, I., {Jorstad}, S.~G., {Marscher}, A.~P., {et~al.} 2011, \apjl, 726,
  L13

\bibitem[{{Aharonian} {et~al.}(2006){Aharonian}, {Anchordoqui}, {Khangulyan},
  \& {Montaruli}}]{2006JPhCS..39..408A}
{Aharonian}, F., {Anchordoqui}, L., {Khangulyan}, D., \& {Montaruli}, T. 2006,
  Journal of Physics Conference Series, 39, 408

\bibitem[{{Aragona} {et~al.}(2010){Aragona}, {McSwain}, \& {De
  Becker}}]{2010ApJ...724..306A}
{Aragona}, C., {McSwain}, M.~V., \& {De Becker}, M. 2010, \apj, 724, 306

\bibitem[{{Aragona} {et~al.}(2009){Aragona}, {McSwain}, {Grundstrom}, {Marsh},
  {Roettenbacher}, {Hessler}, {Boyajian}, \& {Ray}}]{2009ApJ...698..514A}
{Aragona}, C., {McSwain}, M.~V., {Grundstrom}, E.~D., {et~al.} 2009, \apj, 698,
  514

\bibitem[{{Archibald} {et~al.}(2009){Archibald}, {Stairs}, {Ransom}, {Kaspi},
  {Kondratiev}, {Lorimer}, {McLaughlin}, {Boyles}, {Hessels}, {Lynch}, {van
  Leeuwen}, {Roberts}, {Jenet}, {Champion}, {Rosen}, {Barlow}, {Dunlap}, \&
  {Remillard}}]{2009Sci...324.1411A}
{Archibald}, A.~M., {Stairs}, I.~H., {Ransom}, S.~M., {et~al.} 2009, Science,
  324, 1411

\bibitem[{{Arons}(2009)}]{2009ASSL..357..373A}
{Arons}, J. 2009, in Astrophysics and Space Science Library, Vol. 357, Neutron
  Stars and Pulsars, ed. W.~{Becker} (Springer), 373

\bibitem[{{Atoyan} \& {Aharonian}(1999)}]{Atoyan:1999tn}
{Atoyan}, A.~M. \& {Aharonian}, F.~A. 1999, \mnras, 302, 253

\bibitem[{{Bai} \& {Spitkovsky}(2010)}]{2010ApJ...715.1282B}
{Bai}, X. \& {Spitkovsky}, A. 2010, \apj, 715, 1282

\bibitem[{{Ball} \& {Kirk}(2000)}]{Ball:2000lr}
{Ball}, L. \& {Kirk}, J.~G. 2000, Astroparticle Physics, 12, 335

\bibitem[{{Barkov} \& {Khangulyan}(2012)}]{2012MNRAS.421.1351B}
{Barkov}, M.~V. \& {Khangulyan}, D.~V. 2012, \mnras, 421, 1351

\bibitem[{{Barthelmy} {et~al.}(2008){Barthelmy}, {Baumgartner}, {Cummings},
  {Gehrels}, {Markwardt}, {Sakamoto}, {Godet}, {Evans}, {Osborne}, {Beardmore},
  {Kennea}, {Falcone}, {Burrows}, {Campana}, \& {de
  Pasquale}}]{2008GCN..8215....1B}
{Barthelmy}, S.~D., {Baumgartner}, W., {Cummings}, J., {et~al.} 2008, GRB
  Coordinates Network, 8215, 1

\bibitem[{{Basko} {et~al.}(1974){Basko}, {Sunyaev}, \&
  {Titarchuk}}]{Basko:1974zv}
{Basko}, M.~M., {Sunyaev}, R.~A., \& {Titarchuk}, L.~G. 1974, \aap, 31, 249

\bibitem[{{Bassa} {et~al.}(2011){Bassa}, {Brisken}, {Nelemans}, {Stairs},
  {Stappers}, \& {Kramer}}]{2011MNRAS.412L..63B}
{Bassa}, C.~G., {Brisken}, W.~F., {Nelemans}, G., {et~al.} 2011, \mnras, 412,
  L63

\bibitem[{{Bastian} {et~al.}(1988){Bastian}, {Dulk}, \&
  {Chanmugam}}]{1988ApJ...324..431B}
{Bastian}, T.~S., {Dulk}, G.~A., \& {Chanmugam}, G. 1988, \apj, 324, 431

\bibitem[{{Becker}(2009)}]{2009ASSL..357...91B}
{Becker}, W. 2009, in Astrophysics and Space Science Library, Vol. 357,
  Astrophysics and Space Science Library, ed. W.~{Becker}, 91

\bibitem[{{Bednarek}(1997)}]{Bednarek:1997ph}
{Bednarek}, W. 1997, \aap, 322, 523

\bibitem[{{Bednarek}(2005)}]{Bednarek:2005lp}
{Bednarek}, W. 2005, \apj, 631, 466

\bibitem[{{Bednarek}(2006)}]{Bednarek:2006ka}
{Bednarek}, W. 2006, \mnras, 368, 579

\bibitem[{{Bednarek}(2007)}]{Bednarek:2007qd}
{Bednarek}, W. 2007, \aap, 464, 259

\bibitem[{{Bednarek}(2009)}]{2009MNRAS.397.1420B}
{Bednarek}, W. 2009, \mnras, 397, 1420

\bibitem[{{Bednarek}(2011)}]{2011MNRAS.418L..49B}
{Bednarek}, W. 2011, \mnras, 418, L49

\bibitem[{{Bednarek}(2013)}]{2013APh....43...81B}
{Bednarek}, W. 2013, Astroparticle Physics, 43, 81

\bibitem[{{Bednarek} \& {Pabich}(2011)}]{2011A&A...530A..49B}
{Bednarek}, W. \& {Pabich}, J. 2011, \aap, 530, A49

\bibitem[{{Bednarek} \& {Sitarek}(2013)}]{2013arXiv1301.2953B}
{Bednarek}, W. \& {Sitarek}, J. 2013, \mnras, 430, 2951

\bibitem[{{Bell} {et~al.}(1995){Bell}, {Bessell}, {Stappers}, {Bailes}, \&
  {Kaspi}}]{1995ApJ...447L.117B}
{Bell}, J.~F., {Bessell}, M.~S., {Stappers}, B.~W., {Bailes}, M., \& {Kaspi},
  V.~M. 1995, \apjl, 447, L117

\bibitem[{{Beloborodov} \& {Illarionov}(2001)}]{Beloborodov:2001qx}
{Beloborodov}, A.~M. \& {Illarionov}, A.~F. 2001, \mnras, 323, 167

\bibitem[{{Benaglia} \& {Romero}(2003)}]{2003A&A...399.1121B}
{Benaglia}, P. \& {Romero}, G.~E. 2003, \aap, 399, 1121

\bibitem[{{Bhattacharyya} {et~al.}(2012){Bhattacharyya}, {Godambe}, {Bhatt},
  {Mitra}, \& {Choudhury}}]{2012MNRAS.421L...1B}
{Bhattacharyya}, S., {Godambe}, S., {Bhatt}, N., {Mitra}, A., \& {Choudhury},
  M. 2012, \mnras, 421, L1

\bibitem[{{Bignami} {et~al.}(1977){Bignami}, {Maraschi}, \&
  {Treves}}]{1977A&A....55..155B}
{Bignami}, G.~F., {Maraschi}, L., \& {Treves}, A. 1977, \aap, 55, 155

\bibitem[{{Blumenthal} \& {Gould}(1970)}]{Blumenthal:1970op}
{Blumenthal}, G.~R. \& {Gould}, R.~J. 1970, Reviews of Modern Physics, 42, 237

\bibitem[{{Bodaghee} {et~al.}(2013){Bodaghee}, {Tomsick}, {Pottschmidt},
  {Rodriguez}, {Wilms}, \& {Pooley}}]{2013arXiv1307.3264B}
{Bodaghee}, A., {Tomsick}, J.~A., {Pottschmidt}, K., {et~al.} 2013, \apj, in
  press (arXiv:1307.3264)

\bibitem[{{Bogdanov} {et~al.}(2005){Bogdanov}, {Grindlay}, \& {van den
  Berg}}]{Bogdanov:2005dt}
{Bogdanov}, S., {Grindlay}, J.~E., \& {van den Berg}, M. 2005, \apj, 630, 1029

\bibitem[{{Bogovalov} \& {Aharonian}(2000)}]{Bogovalov:2000ny}
{Bogovalov}, S.~V. \& {Aharonian}, F.~A. 2000, \mnras, 313, 504

\bibitem[{{Bogovalov} {et~al.}(2008){Bogovalov}, {Khangulyan}, {Koldoba},
  {Ustyugova}, \& {Aharonian}}]{2008MNRAS.387...63B}
{Bogovalov}, S.~V., {Khangulyan}, D., {Koldoba}, A.~V., {Ustyugova}, G.~V., \&
  {Aharonian}, F.~A. 2008, \mnras, 387, 63

\bibitem[{{Bogovalov} {et~al.}(2012){Bogovalov}, {Khangulyan}, {Koldoba},
  {Ustyugova}, \& {Aharonian}}]{2012MNRAS.419.3426B}
{Bogovalov}, S.~V., {Khangulyan}, D., {Koldoba}, A.~V., {Ustyugova}, G.~V., \&
  {Aharonian}, F.~A. 2012, \mnras, 419, 3426

\bibitem[{{Bongiorno} {et~al.}(2011){Bongiorno}, {Falcone}, {Stroh}, {Holder},
  {Skilton}, {Hinton}, {Gehrels}, \& {Grube}}]{2011ApJ...737L..11B}
{Bongiorno}, S.~D., {Falcone}, A.~D., {Stroh}, M., {et~al.} 2011, \apjl, 737,
  L11

\bibitem[{{Bonnet-Bidaud} \& {Chardin}(1988)}]{Bonnet-Bidaud:1988vw}
{Bonnet-Bidaud}, J.~M. \& {Chardin}, G. 1988, \physrep, 170, 326

\bibitem[{{Bordas} {et~al.}(2009){Bordas}, {Bosch-Ramon}, {Paredes}, \&
  {Perucho}}]{2009A&A...497..325B}
{Bordas}, P., {Bosch-Ramon}, V., {Paredes}, J.~M., \& {Perucho}, M. 2009, \aap,
  497, 325

\bibitem[{{Bordas} \& {Maier}(2012)}]{2012arXiv1212.0350B}
{Bordas}, P. \& {Maier}, G. 2012, in AIP Conf. Proc., Vol. 1505, Proceedings of
  Gamma 2012, Heidelberg, July 2012, ed. F.~A. {Aharonian}, W.~{Hofmann}, \&
  F.~M. {Rieger}, 366--369

\bibitem[{{Bosch-Ramon} {et~al.}(2005{\natexlab{a}}){Bosch-Ramon}, {Aharonian},
  \& {Paredes}}]{Bosch-Ramon:2005tv}
{Bosch-Ramon}, V., {Aharonian}, F.~A., \& {Paredes}, J.~M. 2005{\natexlab{a}},
  \aap, 432, 609

\bibitem[{{Bosch-Ramon} \& {Barkov}(2011)}]{2011A&A...535A..20B}
{Bosch-Ramon}, V. \& {Barkov}, M.~V. 2011, \aap, 535, A20

\bibitem[{{Bosch-Ramon} {et~al.}(2012){Bosch-Ramon}, {Barkov}, {Khangulyan}, \&
  {Perucho}}]{2012A&A...544A..59B}
{Bosch-Ramon}, V., {Barkov}, M.~V., {Khangulyan}, D., \& {Perucho}, M. 2012,
  \aap, 544, A59

\bibitem[{{Bosch-Ramon} \& {Khangulyan}(2009)}]{Bosch-Ramon:2008hg}
{Bosch-Ramon}, V. \& {Khangulyan}, D. 2009, International Journal of Modern
  Physics D, 18, 347

\bibitem[{{Bosch-Ramon} {et~al.}(2008{\natexlab{a}}){Bosch-Ramon},
  {Khangulyan}, \& {Aharonian}}]{Bosch-Ramon:2008vd}
{Bosch-Ramon}, V., {Khangulyan}, D., \& {Aharonian}, F.~A. 2008{\natexlab{a}},
  \aap, 482, 397

\bibitem[{{Bosch-Ramon} {et~al.}(2008{\natexlab{b}}){Bosch-Ramon},
  {Khangulyan}, \& {Aharonian}}]{2008A&A...489L..21B}
{Bosch-Ramon}, V., {Khangulyan}, D., \& {Aharonian}, F.~A. 2008{\natexlab{b}},
  \aap, 489, L21

\bibitem[{{Bosch-Ramon} {et~al.}(2007){Bosch-Ramon}, {Motch}, {Rib{\'o}},
  {Lopes de Oliveira}, {Janot-Pacheco}, {Negueruela}, {Paredes}, \&
  {Martocchia}}]{Bosch-Ramon:2007fq}
{Bosch-Ramon}, V., {Motch}, C., {Rib{\'o}}, M., {et~al.} 2007, \aap, 473, 545

\bibitem[{{Bosch-Ramon} {et~al.}(2005{\natexlab{b}}){Bosch-Ramon}, {Paredes},
  {Rib{\'o}}, {Miller}, {Reig}, \& {Mart{\'{\i}}}}]{Bosch-Ramon:2005zc}
{Bosch-Ramon}, V., {Paredes}, J.~M., {Rib{\'o}}, M., {et~al.}
  2005{\natexlab{b}}, \apj, 628, 388

\bibitem[{{Bosch-Ramon} {et~al.}(2006){Bosch-Ramon}, {Romero}, \&
  {Paredes}}]{Bosch-Ramon:2006do}
{Bosch-Ramon}, V., {Romero}, G.~E., \& {Paredes}, J.~M. 2006, \aap, 447, 263

\bibitem[{{B{\"o}ttcher} \& {Dermer}(2005)}]{Bottcher:2005zx}
{B{\"o}ttcher}, M. \& {Dermer}, C.~D. 2005, \apjl, 634, L81

\bibitem[{{Bromberg} \& {Levinson}(2009)}]{2009ApJ...699.1274B}
{Bromberg}, O. \& {Levinson}, A. 2009, \apj, 699, 1274

\bibitem[{{Buehler} {et~al.}(2012){Buehler}, {Scargle}, {Blandford}, {Baldini},
  {Baring}, {Belfiore}, {Charles}, {Chiang}, {D'Ammando}, {Dermer}, {Funk},
  {Grove}, {Harding}, {Hays}, {Kerr}, {Massaro}, {Mazziotta}, {Romani}, {Saz
  Parkinson}, {Tennant}, \& {Weisskopf}}]{2012ApJ...749...26B}
{Buehler}, R., {Scargle}, J.~D., {Blandford}, R.~D., {et~al.} 2012, \apj, 749,
  26

\bibitem[{{Bulgarelli} {et~al.}(2012){Bulgarelli}, {Tavani}, {Chen},
  {Evangelista}, {Trifoglio}, {Gianotti}, {Piano}, {Sabatini}, {Striani},
  {Pooley}, {Trushkin}, {Nizhelskij}, {McCollough}, {Koljonen}, {Hannikainen},
  {L{\"a}hteenm{\"a}ki}, {Tammi}, {Lavonen}, {Steeghs}, {Aboudan}, {Argan},
  {Barbiellini}, {Campana}, {Caraveo}, {Cattaneo}, {Cocco}, {Contessi},
  {Costa}, {D'Ammando}, {Del Monte}, {de Paris}, {Di Cocco}, {Donnarumma},
  {Feroci}, {Fiorini}, {Fuschino}, {Galli}, {Giuliani}, {Giusti}, {Labanti},
  {Lapshov}, {Lazzarotto}, {Lipari}, {Longo}, {Marisaldi}, {Mereghetti},
  {Morelli}, {Moretti}, {Morselli}, {Pacciani}, {Pellizzoni}, {Perotti},
  {Picozza}, {Pilia}, {Prest}, {Pucella}, {Rapisarda}, {Rappoldi}, {Rubini},
  {Soffitta}, {Trois}, {Vallazza}, {Vercellone}, {Vittorini}, {Zambra},
  {Zanello}, {Giommi}, {Pittori}, {Verrecchia}, {Santolamazza}, {Lucarelli},
  {Colafrancesco}, \& {Salotti}}]{2012A&A...538A..63B}
{Bulgarelli}, A., {Tavani}, M., {Chen}, A.~W., {et~al.} 2012, \aap, 538, A63

\bibitem[{{Burderi} {et~al.}(2001){Burderi}, {Possenti}, {D'Antona}, {Di
  Salvo}, {Burgay}, {Stella}, {Menna}, {Iaria}, {Campana}, \&
  {d'Amico}}]{Burderi:2001nx}
{Burderi}, L., {Possenti}, A., {D'Antona}, F., {et~al.} 2001, \apjl, 560, L71

\bibitem[{{Burrows} {et~al.}(2012){Burrows}, {Chester}, {D'Elia}, {Palmer},
  {Romano}, {Saxton}, {Sonbas}, {Stamatikos}, \&
  {Stratta}}]{2012GCN..12914...1B}
{Burrows}, D.~N., {Chester}, M.~M., {D'Elia}, V., {et~al.} 2012, GRB
  Coordinates Network, 12914, 1

\bibitem[{{Ca{\~n}ellas} {et~al.}(2012){Ca{\~n}ellas}, {Joshi}, {Paredes},
  {Ishwara-Chandra}, {Mold{\'o}n}, {Zabalza}, {Mart{\'{\i}}}, \&
  {Rib{\'o}}}]{2012A&A...543A.122C}
{Ca{\~n}ellas}, A., {Joshi}, B.~C., {Paredes}, J.~M., {et~al.} 2012, \aap, 543,
  A122

\bibitem[{{Caliandro} {et~al.}(2013){Caliandro}, {Hill}, {Torres}, {Hadasch},
  {Ray}, {Abdo}, {Hessels}, {Ridolfi}, {Possenti}, {Burgay}, {Rea}, {Tam},
  {Dubois}, {Dubus}, {Glanzman}, \& {Jogler}}]{2013arXiv1308.5234C}
{Caliandro}, G.~A., {Hill}, A.~B., {Torres}, D.~F., {et~al.} 2013, \mnras, in
  press (arXiv:1308.5234)

\bibitem[{{Caliandro} {et~al.}(2012){Caliandro}, {Torres}, \&
  {Rea}}]{2012arXiv1209.2034C}
{Caliandro}, G.~A., {Torres}, D.~F., \& {Rea}, N. 2012, \mnras, 427, 2251

\bibitem[{{Canto} {et~al.}(1996){Canto}, {Raga}, \& {Wilkin}}]{Canto:1996jj}
{Canto}, J., {Raga}, A.~C., \& {Wilkin}, F.~P. 1996, \apj, 469, 729

\bibitem[{{Casares} {et~al.}(2011){Casares}, {Corral-Santana}, {Herrero},
  {Morales}, {Mu{\~n}oz-Darias}, {Negueruela}, {Paredes}, {Ribas}, {Rib{\'o}},
  {Steeghs}, {van Spaandonk}, \& {Vilardell}}]{2011heep.conf..559C}
{Casares}, J., {Corral-Santana}, J.~M., {Herrero}, A., {et~al.} 2011, in
  High-Energy Emission from Pulsars and their Systems: Proceedings of the First
  Session of the Sant Cugat Forum on Astrophysics, Astrophysics and Space
  Science Proceedings, ed. D.~F. {Torres} \& N.~{Rea} (Springer-Verlag Berlin),
  559

\bibitem[{{Casares} {et~al.}(2005){Casares}, {Rib{\'o}}, {Ribas}, {Paredes},
  {Mart{\'{\i}}}, \& {Herrero}}]{Casares:2005gg}
{Casares}, J., {Rib{\'o}}, M., {Ribas}, I., {et~al.} 2005, \mnras, 364, 899

\bibitem[{{Casares} {et~al.}(2012){Casares}, {Rib{\'o}}, {Ribas}, {Paredes},
  {Vilardell}, \& {Negueruela}}]{2012MNRAS.421.1103C}
{Casares}, J., {Rib{\'o}}, M., {Ribas}, I., {et~al.} 2012, \mnras, 421, 1103

\bibitem[{{Casella} {et~al.}(2010){Casella}, {Maccarone}, {O'Brien}, {Fender},
  {Russell}, {van der Klis}, {Pe'Er}, {Maitra}, {Altamirano}, {Belloni},
  {Kanbach}, {Klein-Wolt}, {Mason}, {Soleri}, {Stefanescu}, {Wiersema}, \&
  {Wijnands}}]{2010MNRAS.404L..21C}
{Casella}, P., {Maccarone}, T.~J., {O'Brien}, K., {et~al.} 2010, \mnras, 404,
  L21

\bibitem[{{Cerutti} {et~al.}(2008){Cerutti}, {Dubus}, \&
  {Henri}}]{2008A&A...488...37C}
{Cerutti}, B., {Dubus}, G., \& {Henri}, G. 2008, \aap, 488, 37

\bibitem[{{Cerutti} {et~al.}(2009{\natexlab{a}}){Cerutti}, {Dubus}, \&
  {Henri}}]{2009arXiv0912.3722C}
{Cerutti}, B., {Dubus}, G., \& {Henri}, G. 2009{\natexlab{a}}, in {\em 2008
  Fermi Symposium} proceedings, eConf C091122 (arXiv:0912.3722)

\bibitem[{{Cerutti} {et~al.}(2009{\natexlab{b}}){Cerutti}, {Dubus}, \&
  {Henri}}]{2009A&A...507.1217C}
{Cerutti}, B., {Dubus}, G., \& {Henri}, G. 2009{\natexlab{b}}, \aap, 507, 1217

\bibitem[{{Cerutti} {et~al.}(2011){Cerutti}, {Dubus}, {Malzac}, {Szostek},
  {Belmont}, {Zdziarski}, \& {Henri}}]{2011A&A...529A.120C}
{Cerutti}, B., {Dubus}, G., {Malzac}, J., {et~al.} 2011, \aap, 529, A120

\bibitem[{{Cerutti} {et~al.}(2010){Cerutti}, {Malzac}, {Dubus}, \&
  {Henri}}]{2010A&A...519A..81C}
{Cerutti}, B., {Malzac}, J., {Dubus}, G., \& {Henri}, G. 2010, \aap, 519, A81+

\bibitem[{{Chardin} \& {Gerbier}(1989)}]{Chardin:1989ga}
{Chardin}, G. \& {Gerbier}, G. 1989, \aap, 210, 52

\bibitem[{{Cherepashchuk}(1976)}]{1976PAZh....2..356C}
{Cherepashchuk}, A.~M. 1976, Pis ma Astronomicheskii Zhurnal, 2, 356

\bibitem[{{Chernyakova} {et~al.}(2009){Chernyakova}, {Neronov}, {Aharonian},
  {Uchiyama}, \& {Takahashi}}]{2009MNRAS.397.2123C}
{Chernyakova}, M., {Neronov}, A., {Aharonian}, F., {Uchiyama}, Y., \&
  {Takahashi}, T. 2009, \mnras, 397, 2123

\bibitem[{{Chernyakova} {et~al.}(2006{\natexlab{a}}){Chernyakova}, {Neronov},
  {Lutovinov}, {Rodriguez}, \& {Johnston}}]{2006MNRAS.367.1201C}
{Chernyakova}, M., {Neronov}, A., {Lutovinov}, A., {Rodriguez}, J., \&
  {Johnston}, S. 2006{\natexlab{a}}, \mnras, 367, 1201

\bibitem[{{Chernyakova} {et~al.}(2012){Chernyakova}, {Neronov}, {Molkov},
  {Malyshev}, {Lutovinov}, \& {Pooley}}]{2012ApJ...747L..29C}
{Chernyakova}, M., {Neronov}, A., {Molkov}, S., {et~al.} 2012, \apjl, 747, L29

\bibitem[{{Chernyakova} {et~al.}(2006{\natexlab{b}}){Chernyakova}, {Neronov},
  \& {Walter}}]{Chernyakova:2006cu}
{Chernyakova}, M., {Neronov}, A., \& {Walter}, R. 2006{\natexlab{b}}, \mnras,
  372, 1585

\bibitem[{{Chernyakova} \& {Illarionov}(1999)}]{Chernyakova:1999xm}
{Chernyakova}, M.~A. \& {Illarionov}, A.~F. 1999, \mnras, 304, 359

\bibitem[{{Cheung}(2013)}]{2013arXiv1304.3475C}
{Cheung}, C.~C. 2013, in {\em 2012 Fermi Symposium} proceedings - eConf C121028
  (eprint arXiv:1304.3475)

\bibitem[{{Cheung} {et~al.}(2010){Cheung}, {Donato}, {Wallace}, {Corbet},
  {Dubus}, {Sokolovsky}, \& {Takahashi}}]{2010ATel.2487....1C}
{Cheung}, C.~C., {Donato}, D., {Wallace}, E., {et~al.} 2010, The Astronomer's
  Telegram, 2487, 1

\bibitem[{{Cheung} {et~al.}(2012{\natexlab{a}}){Cheung}, {Glanzman}, \&
  {Hill}}]{2012ATel.4284....1C}
{Cheung}, C.~C., {Glanzman}, T., \& {Hill}, A.~B. 2012{\natexlab{a}}, The
  Astronomer's Telegram, 4284, 1

\bibitem[{{Cheung} {et~al.}(2012{\natexlab{b}}){Cheung}, {Shore}, {De Gennaro
  Aquino}, {Charbonnel}, {Edlin}, {Hays}, {Corbet}, \&
  {Wood}}]{2012ATel.4310....1C}
{Cheung}, C.~C., {Shore}, S.~N., {De Gennaro Aquino}, I., {et~al.}
  2012{\natexlab{b}}, The Astronomer's Telegram, 4310, 1

\bibitem[{{Chomiuk} {et~al.}(2012){Chomiuk}, {Krauss}, {Rupen}, {Nelson},
  {Roy}, {Sokoloski}, {Mukai}, {Munari}, {Mioduszewski}, {Weston}, {O'Brien},
  {Eyres}, \& {Bode}}]{2012ApJ...761..173C}
{Chomiuk}, L., {Krauss}, M.~I., {Rupen}, M.~P., {et~al.} 2012, \apj, 761, 173

\bibitem[{{Corbel} {et~al.}(2012){Corbel}, {Dubus}, {Tomsick}, {Szostek},
  {Corbet}, {Miller-Jones}, {Richards}, {Pooley}, {Trushkin}, {Dubois}, {Hill},
  {Kerr}, {Max-Moerbeck}, {Readhead}, {Bodaghee}, {Tudose}, {Parent}, {Wilms},
  \& {Pottschmidt}}]{2012MNRAS.421.2947C}
{Corbel}, S., {Dubus}, G., {Tomsick}, J.~A., {et~al.} 2012, \mnras, 421, 2947

\bibitem[{{Corbel} {et~al.}(2002){Corbel}, {Fender}, {Tzioumis}, {Tomsick},
  {Orosz}, {Miller}, {Wijnands}, \& {Kaaret}}]{2002Sci...298..196C}
{Corbel}, S., {Fender}, R.~P., {Tzioumis}, A.~K., {et~al.} 2002, Science, 298,
  196

\bibitem[{{Corbel} {et~al.}(2005){Corbel}, {Kaaret}, {Fender}, {Tzioumis},
  {Tomsick}, \& {Orosz}}]{Corbel-et-al.:2005cv}
{Corbel}, S., {Kaaret}, P., {Fender}, R.~P., {et~al.} 2005, \apj, 632, 504

\bibitem[{{Corbet} {et~al.}(2011){Corbet}, {Cheung}, {Kerr}, {Dubois},
  {Donato}, {Caliandro}, {Coe}, {Edwards}, {Filipovic}, {Payne}, \&
  {Stevens}}]{2011ATel.3221....1C}
{Corbet}, R.~H.~D., {Cheung}, C.~C., {Kerr}, M., {et~al.} 2011, The
  Astronomer's Telegram, 3221, 1

\bibitem[{{Damineli} {et~al.}(2008){Damineli}, {Hillier}, {Corcoran}, {Stahl},
  {Levenhagen}, {Leister}, {Groh}, {Teodoro}, {Albacete Colombo}, {Gonzalez},
  {Arias}, {Levato}, {Grosso}, {Morrell}, {Gamen}, {Wallerstein}, \&
  {Niemela}}]{2008MNRAS.384.1649D}
{Damineli}, A., {Hillier}, D.~J., {Corcoran}, M.~F., {et~al.} 2008, \mnras,
  384, 1649

\bibitem[{{Davidson} \& {Humphreys}(1997)}]{1997ARA&A..35....1D}
{Davidson}, K. \& {Humphreys}, R.~M. 1997, \araa, 35, 1

\bibitem[{{Davidson} \& {Ostriker}(1973)}]{1973ApJ...179..585D}
{Davidson}, K. \& {Ostriker}, J.~P. 1973, \apj, 179, 585

\bibitem[{{De Becker}(2007)}]{2007A&ARv..14..171D}
{De Becker}, M. 2007, \aapr, 14, 171

\bibitem[{{De Becker} \& {Raucq}(2013)}]{2013arXiv1308.3149D}
{De Becker}, M. \& {Raucq}, F. 2013, \aap, in press (arXiv:1308.3149)

\bibitem[{{de Martino} {et~al.}(2013){de Martino}, {Belloni}, {Falanga},
  {Papitto}, {Motta}, {Pellizzoni}, {Evangelista}, {Piano}, {Masetti},
  {Bonnet-Bidaud}, {Mouchet}, {Mukai}, \& {Possenti}}]{2013A&A...550A..89D}
{de Martino}, D., {Belloni}, T., {Falanga}, M., {et~al.} 2013, \aap, 550, A89

\bibitem[{{de Martino} {et~al.}(2010){de Martino}, {Falanga}, {Bonnet-Bidaud},
  {Belloni}, {Mouchet}, {Masetti}, {Andruchow}, {Cellone}, {Mukai}, \&
  {Matt}}]{2010A&A...515A..25D}
{de Martino}, D., {Falanga}, M., {Bonnet-Bidaud}, J.-M., {et~al.} 2010, \aap,
  515, A25

\bibitem[{{Del Monte} {et~al.}(2010){Del Monte}, {Feroci}, {Evangelista},
  {Costa}, {Donnarumma}, {Lapshov}, {Lazzarotto}, {Pacciani}, {Rapisarda},
  {Soffitta}, {Argan}, {Barbiellini}, {Boffelli}, {Bulgarelli}, {Caraveo},
  {Cattaneo}, {Chen}, {D'Ammando}, {Di Cocco}, {Fuschino}, {Galli}, {Gianotti},
  {Giuliani}, {Labanti}, {Lipari}, {Longo}, {Marisaldi}, {Mereghetti},
  {Moretti}, {Morselli}, {Pellizzoni}, {Perotti}, {Piano}, {Picozza}, {Pilia},
  {Prest}, {Pucella}, {Rappoldi}, {Sabatini}, {Striani}, {Tavani}, {Trifoglio},
  {Trois}, {Vallazza}, {Vercellone}, {Vittorini}, {Zambra}, {Antonelli},
  {Cutini}, {Pittori}, {Preger}, {Santolamazza}, {Verrecchia}, {Giommi}, \&
  {Salotti}}]{2010A&A...520A..67D}
{Del Monte}, E., {Feroci}, M., {Evangelista}, Y., {et~al.} 2010, \aap, 520, A67

\bibitem[{{Dermer} \& {B{\"o}ttcher}(2006)}]{Dermer:2006li}
{Dermer}, C.~D. \& {B{\"o}ttcher}, M. 2006, \apj, 643, 1081

\bibitem[{{Dhawan} {et~al.}(2006){Dhawan}, {Mioduszewski}, \&
  {Rupen}}]{Dhawan:2006kr}
{Dhawan}, V., {Mioduszewski}, A., \& {Rupen}, M. 2006, in VI Microquasar
  Workshop: Microquasars and Beyond, Vol. MQW6 (Proceedings of Science), 52

\bibitem[{{Dougherty} {et~al.}(2005){Dougherty}, {Beasley}, {Claussen},
  {Zauderer}, \& {Bolingbroke}}]{2005ApJ...623..447D}
{Dougherty}, S.~M., {Beasley}, A.~J., {Claussen}, M.~J., {Zauderer}, B.~A., \&
  {Bolingbroke}, N.~J. 2005, \apj, 623, 447

\bibitem[{{Dubus}(2006{\natexlab{a}})}]{Dubus:2006lr}
{Dubus}, G. 2006{\natexlab{a}}, \aap, 451, 9

\bibitem[{{Dubus}(2006{\natexlab{b}})}]{Dubus:2006lc}
{Dubus}, G. 2006{\natexlab{b}}, \aap, 456, 801

\bibitem[{{Dubus}(2010)}]{2010ASPC..422...23D}
{Dubus}, G. 2010, in Astronomical Society of the Pacific Conference Series,
  Vol. 422, High Energy Phenomena in Massive Stars, ed. J.~{Mart{\'{\i}}},
  P.~L. {Luque-Escamilla}, \& J.~A. {Combi}, 23

\bibitem[{{Dubus} \& {Cerutti}(2013)}]{2013arXiv1308.4531D}
{Dubus}, G. \& {Cerutti}, B. 2013, \aap, in press (arXiv:1308.4531)

\bibitem[{{Dubus} {et~al.}(2008){Dubus}, {Cerutti}, \& {Henri}}]{Dubus:2007oq}
{Dubus}, G., {Cerutti}, B., \& {Henri}, G. 2008, \aap, 477, 691

\bibitem[{{Dubus} {et~al.}(2010{\natexlab{a}}){Dubus}, {Cerutti}, \&
  {Henri}}]{2010A&A...516A..18D}
{Dubus}, G., {Cerutti}, B., \& {Henri}, G. 2010{\natexlab{a}}, \aap, 516, A18

\bibitem[{{Dubus} {et~al.}(2010{\natexlab{b}}){Dubus}, {Cerutti}, \&
  {Henri}}]{2010MNRAS.404L..55D}
{Dubus}, G., {Cerutti}, B., \& {Henri}, G. 2010{\natexlab{b}}, \mnras, 404, L55

\bibitem[{{Dubus} \& {Giebels}(2008)}]{2008ATel.1715....1D}
{Dubus}, G. \& {Giebels}, B. 2008, The Astronomer's Telegram, 1715, 1

\bibitem[{{Durant} {et~al.}(2011){Durant}, {Kargaltsev}, {Pavlov}, {Chang}, \&
  {Garmire}}]{2011ApJ...735...58D}
{Durant}, M., {Kargaltsev}, O., {Pavlov}, G.~G., {Chang}, C., \& {Garmire},
  G.~P. 2011, \apj, 735, 58

\bibitem[{{Eichler} \& {Usov}(1993)}]{Eichler:1993wt}
{Eichler}, D. \& {Usov}, V. 1993, \apj, 402, 271

\bibitem[{{Ellison} {et~al.}(2012){Ellison}, {Slane}, {Patnaude}, \&
  {Bykov}}]{2012ApJ...744...39E}
{Ellison}, D.~C., {Slane}, P., {Patnaude}, D.~J., \& {Bykov}, A.~M. 2012, \apj,
  744, 39

\bibitem[{{{\em Fermi}/LAT collaboration} {et~al.}(2010{\natexlab{a}}){{\em
  Fermi}/LAT collaboration}, {Abdo}, {Ackermann}, {Ajello}, {Allafort},
  {Baldini}, {Ballet}, {Barbiellini}, {Bastieri}, {Bechtol}, {Bellazzini},
  {Berenji}, {Blandford}, {Bonamente}, {Borgland}, {Bouvier}, {Brandt},
  {Bregeon}, {Brez}, {Brigida}, {Bruel}, {Buehler}, {Burnett}, {Caliandro},
  {Cameron}, {Caraveo}, {Carrigan}, {Casandjian}, {Cecchi}, {{\c C}elik},
  {Chaty}, {Chekhtman}, {Cheung}, {Chiang}, {Ciprini}, {Claus}, {Cohen-Tanugi},
  {Cominsky}, {Conrad}, {Dermer}, {de Palma}, {Digel}, {Silva}, {Drell},
  {Dubois}, {Dumora}, {Favuzzi}, {Fegan}, {Ferrara}, {Frailis}, {Fukazawa},
  {Fusco}, {Gargano}, {Gehrels}, {Germani}, {Giglietto}, {Giordano}, {Godfrey},
  {Grenier}, {Grondin}, {Grove}, {Guillemot}, {Guiriec}, {Hadasch}, {Hanabata},
  {Harding}, {Hayashida}, {Hays}, {Hill}, {Horan}, {Hughes}, {Itoh}, {Jackson},
  {J{\'o}hannesson}, {Johnson}, {Johnson}, {Kamae}, {Katagiri}, {Kataoka},
  {Kerr}, {Kn{\"o}dlseder}, {Kuss}, {Lande}, {Latronico}, {Lee},
  {Lemoine-Goumard}, {Livingstone}, {Llena Garde}, {Longo}, {Loparco},
  {Lovellette}, {Lubrano}, {Makeev}, {Mazziotta}, {McEnery}, {Mehault},
  {Michelson}, {Mitthumsiri}, {Mizuno}, {Moiseev}, {Monte}, {Monzani},
  {Morselli}, {Moskalenko}, {Murgia}, {Nakamori}, {Naumann-Godo}, {Nolan},
  {Norris}, {Nuss}, {Ohsugi}, {Okumura}, {Omodei}, {Orlando}, {Ormes}, {Ozaki},
  {Panetta}, {Parent}, {Pelassa}, {Pepe}, {Pesce-Rollins}, {Piron}, {Porter},
  {Rain{\`o}}, {Rando}, {Razzano}, {Reimer}, {Reimer}, {Reposeur}, {Rodriguez},
  {Romani}, {Roth}, {Sadrozinski}, {Sander}, {Saz Parkinson}, {Scargle},
  {Sgr{\`o}}, {Siskind}, {Smith}, {Smith}, {Spandre}, {Spinelli}, {Strickman},
  {Suson}, {Takahashi}, {Takahashi}, {Tanaka}, {Thayer}, {Thayer}, {Thompson},
  {Tibaldo}, {Tibolla}, {Torres}, {Tosti}, {Tramacere}, {Uchiyama}, {Usher},
  {Vandenbroucke}, {Vasileiou}, {Vilchez}, {Vitale}, {Waite}, {Wallace},
  {Wang}, {Winer}, {Wood}, {Yang}, {Ylinen}, \&
  {Ziegler}}]{2010ApJ...723..649A}
{{\em Fermi}/LAT collaboration}, {Abdo}, A.~A., {Ackermann}, M., {et~al.}
  2010{\natexlab{a}}, \apj, 723, 649

\bibitem[{{{\em Fermi}/LAT collaboration} {et~al.}(2010{\natexlab{b}}){{\em
  Fermi}/LAT collaboration}, {Abdo}, {Ackermann}, {Ajello}, {Atwood},
  {Baldini}, {Ballet}, {Barbiellini}, {Bastieri}, {Bechtol}, {Bellazzini}, \&
  et~al.}]{2010Sci...329..817A}
{{\em Fermi}/LAT collaboration}, {Abdo}, A.~A., {Ackermann}, M., {et~al.}
  2010{\natexlab{b}}, Science, 329, 817

\bibitem[{{{\em Fermi}/LAT collaboration} {et~al.}(2009{\natexlab{a}}){{\em
  Fermi}/LAT collaboration}, {Abdo}, \& {et al.}}]{2009ApJ...701L.123A}
{{\em Fermi}/LAT collaboration}, {Abdo}, A.~A., \& {et al.} 2009{\natexlab{a}},
  \apjl, 701, L123

\bibitem[{{{\em Fermi}/LAT collaboration} {et~al.}(2009{\natexlab{b}}){{\em
  Fermi}/LAT collaboration}, {Abdo}, \& {et al.}}]{2009ApJ...706L..56A}
{{\em Fermi}/LAT collaboration}, {Abdo}, A.~A., \& {et al.} 2009{\natexlab{b}},
  \apjl, 706, L56

\bibitem[{{{\em Fermi}/LAT collaboration} {et~al.}(2009{\natexlab{c}}){{\em
  Fermi}/LAT collaboration}, {Abdo}, \& {et al.}}]{2009Sci...326.1512F}
{{\em Fermi}/LAT collaboration}, {Abdo}, A.~A., \& {et al.} 2009{\natexlab{c}},
  Science, 326, 1512

\bibitem[{{{\em Fermi}/LAT collaboration} {et~al.}(2010{\natexlab{c}}){{\em
  Fermi}/LAT collaboration}, {Abdo}, \& {et al.}}]{2010ApJS..187..460A}
{{\em Fermi}/LAT collaboration}, {Abdo}, A.~A., \& {et al.} 2010{\natexlab{c}},
  \apjs, 187, 460

\bibitem[{{{\em Fermi}/LAT collaboration} {et~al.}(2011){{\em Fermi}/LAT
  collaboration}, {Abdo}, \& {et al.}}]{2011ApJ...736L..11A}
{{\em Fermi}/LAT collaboration}, {Abdo}, A.~A., \& {et al.} 2011, \apjl, 736,
  L11

\bibitem[{{{\em Fermi}/LAT collaboration} {et~al.}(2013){{\em Fermi}/LAT
  collaboration}, {Ackermann}, {Ajello}, {Albert}, {Allafort}, {Antolini},
  {Baldini}, {Ballet}, {Barbiellini}, {Bastieri}, {Bechtol}, {Bellazzini},
  {Blandford}, {Bloom}, {Bonamente}, {Bottacini}, {Bouvier}, {Brandt},
  {Bregeon}, {Brigida}, {Bruel}, {Buehler}, {Buson}, {Caliandro}, {Cameron},
  {Caraveo}, {Cavazzuti}, {Cecchi}, {Charles}, {Chekhtman}, {Cheung}, {Chiang},
  {Chiaro}, {Ciprini}, {Claus}, {Cohen-Tanugi}, {Conrad}, {Cutini}, {Dalton},
  {D'Ammando}, {de Angelis}, {de Palma}, {Dermer}, {Di Venere}, {Drell},
  {Drlica-Wagner}, {Favuzzi}, {Fegan}, {Ferrara}, {Focke}, {Franckowiak},
  {Fukazawa}, {Funk}, {Fusco}, {Gargano}, {Gasparrini}, {Germani}, {Giglietto},
  {Giordano}, {Giroletti}, {Glanzman}, {Godfrey}, {Grenier}, {Grondin},
  {Grove}, {Guiriec}, {Hadasch}, {Hanabata}, {Harding}, {Hayashida}, {Hays},
  {Hewitt}, {Hill}, {Horan}, {Hou}, {Hughes}, {Inoue}, {Jackson}, {Jogler},
  {J{\'o}hannesson}, {Johnson}, {Kamae}, {Kataoka}, {Kawano}, {Kn{\"o}dlseder},
  {Kuss}, {Lande}, {Larsson}, {Latronico}, {Lemoine-Goumard}, {Longo},
  {Loparco}, {Lott}, {Lovellette}, {Lubrano}, {Mayer}, {Mazziotta}, {McEnery},
  {Michelson}, {Mitthumsiri}, {Mizuno}, {Monte}, {Monzani}, {Morselli},
  {Moskalenko}, {Murgia}, {Nemmen}, {Nuss}, {Ohsugi}, {Okumura}, {Omodei},
  {Orienti}, {Orlando}, {Ormes}, {Paneque}, {Panetta}, {Perkins},
  {Pesce-Rollins}, {Piron}, {Pivato}, {Porter}, {Rain{\`o}}, {Rando},
  {Razzano}, {Reimer}, {Reimer}, {Romoli}, {Roth}, {S{\'a}nchez-Conde},
  {Scargle}, {Schulz}, {Sgr{\`o}}, {Siskind}, {Spandre}, {Spinelli}, {Suson},
  {Takahashi}, {Takeuchi}, {Thayer}, {Thayer}, {Thompson}, {Tibaldo},
  {Tinivella}, {Torres}, {Tosti}, {Troja}, {Tronconi}, {Usher},
  {Vandenbroucke}, {Vasileiou}, {Vianello}, {Vitale}, {Winer}, {Wood}, {Wood},
  \& {Yang}}]{2013ApJ...771...57A}
{{\em Fermi}/LAT collaboration}, {Ackermann}, M., {Ajello}, M., {et~al.} 2013,
  \apj, 771, 57

\bibitem[{{{\em Fermi}/LAT collaboration} {et~al.}(2012{\natexlab{a}}){{\em
  Fermi}/LAT collaboration}, {Ackermann}, {Ajello}, {Allafort}, {Antolini},
  {Baldini}, {Ballet}, {Barbiellini}, {Bastieri}, {Bellazzini}, {Berenji},
  {Blandford}, {Bloom}, {Bonamente}, {Borgland}, {Bouvier}, {Brandt},
  {Bregeon}, {Brigida}, {Bruel}, {Buehler}, {Burnett}, {Buson}, {Caliandro},
  {Cameron}, {Caraveo}, {Casandjian}, {Cavazzuti}, {Cecchi}, {{\c C}elik},
  {Charles}, {Chekhtman}, {Chen}, {Cheung}, {Chiang}, {Ciprini}, {Claus},
  {Cohen-Tanugi}, {Conrad}, {Cutini}, {de Angelis}, {DeCesar}, {De Luca}, {de
  Palma}, {Dermer}, {Silva}, {Drell}, {Drlica-Wagner}, {Dubois}, {Enoto},
  {Favuzzi}, {Fegan}, {Ferrara}, {Focke}, {Fortin}, {Fukazawa}, {Funk},
  {Fusco}, {Gargano}, {Gasparrini}, {Gehrels}, {Germani}, {Giglietto},
  {Giordano}, {Giroletti}, {Glanzman}, {Godfrey}, {Grenier}, {Grondin},
  {Grove}, {Guillemot}, {Guiriec}, {Gustafsson}, {Hadasch}, {Hanabata},
  {Harding}, {Hayashida}, {Hays}, {Healey}, {Hill}, {Horan}, {Hou},
  {J{\'o}hannesson}, {Johnson}, {Johnson}, {Kamae}, {Katagiri}, {Kataoka},
  {Kerr}, {Kn{\"o}dlseder}, {Kuss}, {Lande}, {Latronico}, {Lee},
  {Lemoine-Goumard}, {Longo}, {Loparco}, {Lott}, {Lovellette}, {Lubrano},
  {Madejski}, {Mazziotta}, {McEnery}, {Mehault}, {Michelson}, {Mignani},
  {Mitthumsiri}, {Mizuno}, {Monte}, {Monzani}, {Morselli}, {Moskalenko},
  {Murgia}, {Nakamori}, {Naumann-Godo}, {Nolan}, {Norris}, {Nuss}, {Ohsugi},
  {Okumura}, {Omodei}, {Orlando}, {Ormes}, {Ozaki}, {Paneque}, {Panetta},
  {Parent}, {Pelassa}, {Pesce-Rollins}, {Pierbattista}, {Piron}, {Pivato},
  {Porter}, {Rain{\`o}}, {Rando}, {Ray}, {Razzano}, {Reimer}, {Reimer},
  {Reposeur}, {Romani}, {Sadrozinski}, {Salvetti}, {Saz Parkinson}, {Schalk},
  {Sgr{\`o}}, {Shaw}, {Siskind}, {Smith}, {Spandre}, {Spinelli}, {Suson},
  {Takahashi}, {Tanaka}, {Thayer}, {Thayer}, {Thompson}, {Tibaldo}, {Tibolla},
  {Torres}, {Tosti}, {Tramacere}, {Troja}, {Usher}, {Vandenbroucke},
  {Vasileiou}, {Vianello}, {Vilchez}, {Vitale}, {Waite}, {Wallace}, {Wang},
  {Winer}, {Wolff}, {Wood}, {Wood}, {Yang}, \& {Zimmer}}]{2012ApJ...753...83A}
{{\em Fermi}/LAT collaboration}, {Ackermann}, M., {Ajello}, M., {et~al.}
  2012{\natexlab{a}}, \apj, 753, 83

\bibitem[{{{\em Fermi}/LAT collaboration} {et~al.}(2012{\natexlab{b}}){{\em
  Fermi}/LAT collaboration}, {Ackermann}, {Ajello}, {Ballet}, {Barbiellini},
  {Bastieri}, {Belfiore}, {Bellazzini}, {Berenji}, {Blandford}, {Bloom},
  {Bonamente}, {Borgland}, {Bregeon}, {Brigida}, {Bruel}, {Buehler}, {Buson},
  {Caliandro}, {Cameron}, {Caraveo}, {Cavazzuti}, {Cecchi}, {{\c C}elik},
  {Charles}, {Chaty}, {Chekhtman}, {Cheung}, {Chiang}, {Ciprini}, {}, {Claus},
  {Cohen-Tanugi}, {Corbel}, {Corbet}, {Cutini}, {de Luca}, {den Hartog}, {de
  Palma}, {Dermer}, {Digel}, {do Couto e Silva}, {Donato}, {Drell},
  {Drlica-Wagner}, {Dubois}, {Dubus}, {Favuzzi}, {Fegan}, {Ferrara}, {Focke},
  {Fortin}, {Fukazawa}, {Funk}, {Fusco}, {Gargano}, {Gasparrini}, {Gehrels},
  {Germani}, {Giglietto}, {Giordano}, {Giroletti}, {Glanzman}, {Godfrey},
  {Grenier}, {Grove}, {Guiriec}, {Hadasch}, {Hanabata}, {Harding}, {Hayashida},
  {Hays}, {Hill}, {Hughes}, {J{\'o}hannesson}, {Johnson}, {Johnson}, {Kamae},
  {Katagiri}, {Kataoka}, {Kerr}, {Kn{\"o}dlseder}, {Kuss}, {Lande}, {Longo},
  {Loparco}, {Lovellette}, {Lubrano}, {Mazziotta}, {McEnery}, {Michelson},
  {Mitthumsiri}, {Mizuno}, {Monte}, {Monzani}, {Morselli}, {Moskalenko},
  {Murgia}, {Nakamori}, {Naumann-Godo}, {Norris}, {Nuss}, {Ohno}, {Ohsugi},
  {Okumura}, {Omodei}, {Orlando}, {Ozaki}, {Paneque}, {Parent},
  {Pesce-Rollins}, {Pierbattista}, {Piron}, {Pivato}, {Porter}, {Rain{\`o}},
  {Rando}, {Razzano}, {Reimer}, {Reimer}, {Ritz}, {Romani}, {Roth}, {Saz
  Parkinson}, {Sgr{\`o}}, {Siskind}, {Spandre}, {Spinelli}, {Suson},
  {Takahashi}, {Tanaka}, {Thayer}, {Thayer}, {Thompson}, {Tibaldo},
  {Tinivella}, {Torres}, {Tosti}, {Troja}, {Uchiyama}, {Usher},
  {Vandenbroucke}, {Vianello}, {Vitale}, {Waite}, {Winer}, {Wood}, {Wood},
  {Yang}, {Zimmer}, {Coe}, {Di Mille}, {Edwards}, {Filipovi{\'c}}, {Payne},
  {Stevens}, \& {Torres}}]{2012Sci...335..189F}
{{\em Fermi}/LAT collaboration}, {Ackermann}, M., {Ajello}, M., {et~al.}
  2012{\natexlab{b}}, Science, 335, 189

\bibitem[{{{\em Fermi}/LAT collaboration et
  al.}(2013{\natexlab{a}})}]{2013arXiv1307.6384T}
{{\em Fermi}/LAT collaboration et al.} 2013{\natexlab{a}}, \apjl, in press,
  (ArXiv:1307.6384)

\bibitem[{{{\em Fermi}/LAT collaboration et
  al.}(2013{\natexlab{b}})}]{2013arXiv1305.4385T}
{{\em Fermi}/LAT collaboration et al.} 2013{\natexlab{b}}, \apjs, in press
  (ArXiv:1305.4385)

\bibitem[{{Esposito} {et~al.}(2007){Esposito}, {Caraveo}, {Pellizzoni}, {de
  Luca}, {Gehrels}, \& {Marelli}}]{2007A&A...474..575E}
{Esposito}, P., {Caraveo}, P.~A., {Pellizzoni}, A., {et~al.} 2007, \aap, 474,
  575

\bibitem[{{Farnier} {et~al.}(2011){Farnier}, {Walter}, \&
  {Leyder}}]{2011A&A...526A..57F}
{Farnier}, C., {Walter}, R., \& {Leyder}, J.-C. 2011, \aap, 526, A57

\bibitem[{{Fender}(2006)}]{Fender:2006ww}
{Fender}, R. 2006, Cambridge Astrophysics, Vol.~39, {Jets from X-ray binaries},
  ed. W.~H.~G. {Lewin} \& M.~{van der Klis} (Compact stellar X-ray sources),
  381--419

\bibitem[{{Fender} {et~al.}(2004){Fender}, {Belloni}, \&
  {Gallo}}]{Fender:2004hb}
{Fender}, R.~P., {Belloni}, T.~M., \& {Gallo}, E. 2004, \mnras, 355, 1105

\bibitem[{{Fender} {et~al.}(1999){Fender}, {Hanson}, \&
  {Pooley}}]{1999MNRAS.308..473F}
{Fender}, R.~P., {Hanson}, M.~M., \& {Pooley}, G.~G. 1999, \mnras, 308, 473

\bibitem[{{Fender} \& {Pooley}(2000)}]{2000MNRAS.318L...1F}
{Fender}, R.~P. \& {Pooley}, G.~G. 2000, \mnras, 318, L1

\bibitem[{{Ford}(1984)}]{1984MNRAS.211..559F}
{Ford}, L.~H. 1984, \mnras, 211, 559

\bibitem[{{Frail} \& {Hjellming}(1991)}]{1991AJ....101.2126F}
{Frail}, D.~A. \& {Hjellming}, R.~M. 1991, \aj, 101, 2126

\bibitem[{{Frail} {et~al.}(1987){Frail}, {Seaquist}, \&
  {Taylor}}]{1987AJ.....93.1506F}
{Frail}, D.~A., {Seaquist}, E.~R., \& {Taylor}, A.~R. 1987, \aj, 93, 1506

\bibitem[{{Gaensler} \& {Slane}(2006)}]{Gaensler:2006qi}
{Gaensler}, B.~M. \& {Slane}, P.~O. 2006, \araa, 44, 17

\bibitem[{{Gallo} {et~al.}(2005){Gallo}, {Fender}, {Kaiser}, {Russell},
  {Morganti}, {Oosterloo}, \& {Heinz}}]{Gallo:2005ii}
{Gallo}, E., {Fender}, R., {Kaiser}, C., {et~al.} 2005, \nat, 436, 819

\bibitem[{{Geldzahler} {et~al.}(1983){Geldzahler}, {Johnston}, {Spencer},
  {Klepczynski}, {Josties}, {Angerhofer}, {Florkowski}, {McCarthy}, {Matsakis},
  \& {Hjellming}}]{1983ApJ...273L..65G}
{Geldzahler}, B.~J., {Johnston}, K.~J., {Spencer}, J.~H., {et~al.} 1983, \apjl,
  273, L65

\bibitem[{{Georganopoulos} {et~al.}(2002){Georganopoulos}, {Aharonian}, \&
  {Kirk}}]{Georganopoulos:2002ci}
{Georganopoulos}, M., {Aharonian}, F.~A., \& {Kirk}, J.~G. 2002, \aap, 388, L25

\bibitem[{{Godambe} {et~al.}(2008){Godambe}, {Bhattacharyya}, {Bhatt}, \&
  {Choudhury}}]{2008MNRAS.390L..43G}
{Godambe}, S., {Bhattacharyya}, S., {Bhatt}, N., \& {Choudhury}, M. 2008,
  \mnras, 390, L43

\bibitem[{{Golenetskii} {et~al.}(2003){Golenetskii}, {Aptekar}, {Frederiks},
  {Mazets}, {Palshin}, {Hurley}, {Cline}, \& {Stern}}]{Golenetskii:2003ps}
{Golenetskii}, S., {Aptekar}, R., {Frederiks}, D., {et~al.} 2003, \apj, 596,
  1113

\bibitem[{{Gregory}(2002)}]{2002ApJ...575..427G}
{Gregory}, P.~C. 2002, \apj, 575, 427

\bibitem[{{Gregory} \& {Kronberg}(1972)}]{1972Natur.239..440G}
{Gregory}, P.~C. \& {Kronberg}, P.~P. 1972, \nat, 239, 440

\bibitem[{{Gregory} \& {Taylor}(1978)}]{Gregory:1978or}
{Gregory}, P.~C. \& {Taylor}, A.~R. 1978, \nat, 272, 704

\bibitem[{{Gregory} {et~al.}(1979){Gregory}, {Taylor}, {Crampton}, {Hutchings},
  {Hjellming}, {Hogg}, {Hvatum}, {Gottlieb}, {Feldman}, \&
  {Kwok}}]{Gregory:1979ng}
{Gregory}, P.~C., {Taylor}, A.~R., {Crampton}, D., {et~al.} 1979, \aj, 84, 1030

\bibitem[{{Grimm} {et~al.}(2002){Grimm}, {Gilfanov}, \&
  {Sunyaev}}]{Grimm:2002xx}
{Grimm}, H.-J., {Gilfanov}, M., \& {Sunyaev}, R. 2002, \aap, 391, 923

\bibitem[{{Grove} {et~al.}(1998){Grove}, {Johnson}, {Kroeger}, {McNaron-Brown},
  {Skibo}, \& {Phlips}}]{Grove:1998cc}
{Grove}, J.~E., {Johnson}, W.~N., {Kroeger}, R.~A., {et~al.} 1998, \apj, 500,
  899

\bibitem[{{Grundstrom} {et~al.}(2007){Grundstrom}, {Caballero-Nieves}, {Gies},
  {Huang}, {McSwain}, {Rafter}, {Riddle}, {Williams}, \&
  {Wingert}}]{Grundstrom:2007kv}
{Grundstrom}, E.~D., {Caballero-Nieves}, S.~M., {Gies}, D.~R., {et~al.} 2007,
  \apj, 656, 437

\bibitem[{{Grundstrom} \& {Gies}(2006)}]{Grundstrom:2006tc}
{Grundstrom}, E.~D. \& {Gies}, D.~R. 2006, \apjl, 651, L53

\bibitem[{{Guenette}(2009)}]{2009arXiv0908.0714G}
{Guenette}, R. 2009, Proceedings of the 31st International Cosmic Ray
  Conference (ICRC), Lodz, Poland, July 2009 (arXiv:0908.0714)

\bibitem[{{Hadasch} {et~al.}(2012){Hadasch}, {Torres}, {Tanaka}, {Corbet},
  {Hill}, {Dubois}, {Dubus}, {Glanzman}, {Corbel}, {Li}, {Chen}, {Zhang},
  {Caliandro}, {Kerr}, {Richards}, {Max-Moerbeck}, {Readhead}, \&
  {Pooley}}]{2012ApJ...749...54H}
{Hadasch}, D., {Torres}, D.~F., {Tanaka}, T., {et~al.} 2012, \apj, 749, 54

\bibitem[{{Harding} \& {Gaisser}(1990)}]{Harding:1990gb}
{Harding}, A.~K. \& {Gaisser}, T.~K. 1990, \apj, 358, 561

\bibitem[{{Harrison} {et~al.}(2000){Harrison}, {Ray}, {Leahy}, {Waltman}, \&
  {Pooley}}]{Harrison:2000bk}
{Harrison}, F.~A., {Ray}, P.~S., {Leahy}, D.~A., {Waltman}, E.~B., \& {Pooley},
  G.~G. 2000, \apj, 528, 454

\bibitem[{{Hartman} {et~al.}(1999){Hartman}, {Bertsch}, {Bloom}, {Chen},
  {Deines-Jones}, {Esposito}, {Fichtel}, {Friedlander}, {Hunter}, {McDonald},
  {Sreekumar}, {Thompson}, {Jones}, {Lin}, {Michelson}, {Nolan}, {Tompkins},
  {Kanbach}, {Mayer-Hasselwander}, {M{\"u}cke}, {Pohl}, {Reimer}, {Kniffen},
  {Schneid}, {von Montigny}, {Mukherjee}, \& {Dingus}}]{Hartman:1999mn}
{Hartman}, R.~C., {Bertsch}, D.~L., {Bloom}, S.~D., {et~al.} 1999, \apjs, 123,
  79

\bibitem[{{Heinz} \& {Sunyaev}(2002)}]{Heinz:2002qb}
{Heinz}, S. \& {Sunyaev}, R. 2002, \aap, 390, 751

\bibitem[{{Helder} {et~al.}(2009){Helder}, {Vink}, {Bassa}, {Bamba}, {Bleeker},
  {Funk}, {Ghavamian}, {van der Heyden}, {Verbunt}, \&
  {Yamazaki}}]{2009Sci...325..719H}
{Helder}, E.~A., {Vink}, J., {Bassa}, C.~G., {et~al.} 2009, Science, 325, 719

\bibitem[{{Hermsen} {et~al.}(1987){Hermsen}, {Bloemen}, {Jansen}, {Bennett},
  {Buccheri}, {Mastichiadis}, {Mayer-Hasselwander}, {Strong}, {Oezel}, \&
  {Pollock}}]{1987A&A...175..141H}
{Hermsen}, W., {Bloemen}, J.~B.~G.~M., {Jansen}, F.~A., {et~al.} 1987, \aap,
  175, 141

\bibitem[{{H.E.S.S. collaboration} {et~al.}(2013){H.E.S.S. collaboration},
  {Abramowski}, {Acero}, {Aharonian}, {Akhperjanian}, {Anton}, {Balenderan},
  {Balzer}, {Barnacka}, {Becherini}, {Becker Tjus}, {Bernl{\"o}hr}, {Birsin},
  {Biteau}, {Boisson}, {Bolmont}, {Bordas}, {Brucker}, {Brun}, {Brun}, {Bulik},
  {Carrigan}, {Casanova}, {Cerruti}, {Chadwick}, {Chaves}, {Cheesebrough},
  {Colafrancesco}, {Cologna}, {Conrad}, {Couturier}, {Dalton}, {Daniel},
  {Davids}, {Degrange}, {Deil}, {deWilt}, {Dickinson}, {Djannati-Ata{\"\i}},
  {Domainko}, {Drury}, {Dubus}, {Dutson}, {Dyks}, {Dyrda}, {Egberts}, {Eger},
  {Espigat}, {Fallon}, {Farnier}, {Fegan}, {Feinstein}, {Fernandes},
  {Fernandez}, {Fiasson}, {Fontaine}, {F{\"o}rster}, {F{\"u}{\ss}ling},
  {Gajdus}, {Gallant}, {Garrigoux}, {Gast}, {Giebels}, {Glicenstein},
  {Gl{\"u}ck}, {G{\"o}ring}, {Grondin}, {Grudzi{\'n}ska}, {H{\"a}er}, {Hague},
  {Hahn}, {Hampf}, {Harris}, {Heinz}, {Heinzelmann}, {Henri}, {Hermann},
  {Hillert}, {Hinton}, {Hofmann}, {Hofverberg}, {Holler}, {Horns},
  {Jacholkowska}, {Jahn}, {Jamrozy}, {Jung}, {Kastendieck}, {Katarzy{\'n}ski},
  {Katz}, {Kaufmann}, {Kh{\'e}lifi}, {Klepser}, {Klochkov}, {Klu{\'z}niak},
  {Kneiske}, {Kolitzus}, {Komin}, {Kosack}, {Kossakowski}, {Krayzel},
  {Kr{\"u}ger}, {Lan}, {Lamanna}, {Lefaucheur}, {Lemoine-Goumard}, {Lenain},
  {Lennarz}, {Lohse}, {Lopatin}, {Lu}, {Marandon}, {Marcowith}, {Masbou},
  {Maurin}, {Maxted}, {Mayer}, {McComb}, {Medina}, {M{\'e}hault}, {Menzler},
  {Moderski}, {Mohamed}, {Moulin}, {Naumann}, {Naumann-Godo}, {de Naurois},
  {Nedbal}, {Nguyen}, {Niemiec}, {Nolan}, {Oakes}, {Ohm}, {de O{\~n}a
  Wilhelmi}, {Opitz}, {Ostrowski}, {Oya}, {Panter}, {Parsons}, {Paz Arribas},
  {Pekeur}, {Pelletier}, {Perez}, {Petrucci}, {Peyaud}, {Pita},
  {P{\"u}hlhofer}, {Punch}, {Quirrenbach}, {Raab}, {Raue}, {Reimer}, {Reimer},
  {Renaud}, {de los Reyes}, {Rieger}, {Ripken}, {Rob}, {Rosier-Lees}, {Rowell},
  {Rudak}, {Rulten}, {Sahakian}, {Sanchez}, {Santangelo}, {Schlickeiser},
  {Schulz}, {Schwanke}, {Schwarzburg}, {Schwemmer}, {Sheidaei}, {Skilton},
  {Sol}, {Spengler}, {Stawarz}, {Steenkamp}, {Stegmann}, {Stinzing}, {Stycz},
  {Sushch}, {Szostek}, {Tavernet}, {Terrier}, {Tluczykont}, {Trichard},
  {Valerius}, {van Eldik}, {Vasileiadis}, {Venter}, {Viana}, {Vincent},
  {V{\"o}lk}, {Volpe}, {Vorobiov}, {Vorster}, {Wagner}, {Ward}, {White},
  {Wierzcholska}, {Willmann}, {Wouters}, {Zacharias}, {Zajczyk}, {Zdziarski},
  {Zech}, \& {Zechlin}}]{H.E.S.S.Collaboration:2013fk}
{H.E.S.S. collaboration}, {Abramowski}, A., {Acero}, F., {et~al.} 2013, \aap,
  551, A94

\bibitem[{{H.E.S.S. collaboration} {et~al.}(2011){H.E.S.S. collaboration},
  {Abramowski}, {Acero}, {Aharonian}, {Akhperjanian}, {Anton}, {Balzer},
  {Barnacka}, {Barres de Almeida}, {Bazer-Bachi}, {Becherini}, {Becker},
  {Behera}, {Bernl{\"o}hr}, {Bochow}, {Boisson}, {Bolmont}, {Bordas}, {Borrel},
  {Brucker}, {Brun}, {Brun}, {Bulik}, {B{\"u}sching}, {Carrigan}, {Casanova},
  {Cerruti}, {Chadwick}, {Charbonnier}, {Chaves}, {Cheesebrough}, {Chounet},
  {Clapson}, {Coignet}, {Colom}, {Conrad}, {Dalton}, {Daniel}, {Davids},
  {Degrange}, {Deil}, {Dickinson}, {Djannati-Ata{\"\i}}, {Domainko}, {Drury},
  {Dubois}, {Dubus}, {Dyks}, {Dyrda}, {Egberts}, {Eger}, {Espigat}, {Fallon},
  {Farnier}, {Fegan}, {Feinstein}, {Fernandes}, {Fiasson}, {Fontaine},
  {F{\"o}rster}, {F{\"u}{\ss}ling}, {Gallant}, {Gast}, {G{\'e}rard}, {Gerbig},
  {Giebels}, {Glicenstein}, {Gl{\"u}ck}, {Goret}, {G{\"o}ring}, {H{\"a}ffner},
  {Hague}, {Hampf}, {Hauser}, {Heinz}, {Heinzelmann}, {Henri}, {Hermann},
  {Hinton}, {Hoffmann}, {Hofmann}, {Hofverberg}, {Holler}, {Horns},
  {Jacholkowska}, {de Jager}, {Jahn}, {Jamrozy}, {Jung}, {Kastendieck},
  {Katarzy{\'n}ski}, {Katz}, {Kaufmann}, {Keogh}, {Khangulyan}, {Kh{\'e}lifi},
  {Klochkov}, {Klu{\'z}niak}, {Kneiske}, {Komin}, {Kosack}, {Kossakowski},
  {Laffon}, {Lamanna}, {Lennarz}, {Lohse}, {Lopatin}, {Lu}, {Marandon},
  {Marcowith}, {Masbou}, {Maurin}, {Maxted}, {McComb}, {Medina}, {M{\'e}hault},
  {Nguyen}, {Moderski}, {Moulin}, {Naumann}, {Naumann-Godo}, {de Naurois},
  {Nedbal}, {Nekrassov}, {Nicholas}, {Niemiec}, {Nolan}, {Ohm}, {Olive}, {de
  O{\~n}a Wilhelmi}, {Opitz}, {Ostrowski}, {Panter}, {Paz Arribas},
  {Pedaletti}, {Pelletier}, {Petrucci}, {Pita}, {P{\"u}hlhofer}, {Punch},
  {Quirrenbach}, {Raue}, {Rayner}, {Reimer}, {Reimer}, {Renaud}, {de los
  Reyes}, {Rieger}, {Ripken}, {Rob}, {Rosier-Lees}, {Rowell}, {Rudak},
  {Rulten}, {Ruppel}, {Ryde}, {Sahakian}, {Santangelo}, {Schlickeiser},
  {Sch{\"o}ck}, {Sch{\"o}nwald}, {Schulz}, {Schwanke}, {Schwarzburg},
  {Schwemmer}, {Shalchi}, {Sikora}, {Skilton}, {Sol}, {Spengler}, {Stawarz},
  {Steenkamp}, {Stegmann}, {Stinzing}, {Stycz}, {Sushch}, {Szostek},
  {Tavernet}, {Terrier}, {Tibolla}, {Tluczykont}, {Valerius}, {van Eldik},
  {Vasileiadis}, {Venter}, {Vialle}, {Viana}, {Vincent}, {Vivier}, {V{\"o}lk},
  {Volpe}, {Vorobiov}, {Vorster}, {Wagner}, {Ward}, {Wierzcholska}, {Zajczyk},
  {Zdziarski}, {Zech}, {Zechlin}, {Burnett}, \& {Hill}}]{2011A&A...529A..49H}
{H.E.S.S. collaboration}, {Abramowski}, A., {Acero}, F., {et~al.} 2011, \aap,
  529, A49

\bibitem[{{H.E.S.S. collaboration} {et~al.}(2012{\natexlab{a}}){H.E.S.S.
  collaboration}, {Abramowski}, {Acero}, {Aharonian}, {Akhperjanian}, {Anton},
  {Balzer}, {Barnacka}, {Becherini}, {Becker}, {Bernl{\"o}h}, {Birsin},
  {Biteau}, {Bochow}, {Boisson}, {Bolmont}, {Bordas}, {Brucker}, {Brun},
  {Brun}, {Bulik}, {B{\"u}sching}, {Carrigan}, {Casanova}, {Cerruti},
  {Chadwick}, {Charbonnier}, {Chaves}, {Cheesebrough}, {Cologna}, {Conrad},
  {Dalton}, {Daniel}, {Davids}, {Degrange}, {Deil}, {Dickinson},
  {Djannati-Ata{\"\i}}, {Domainko}, {Drury}, {Dubus}, {Dutson}, {Dyks},
  {Dyrda}, {Egberts}, {Eger}, {Espigat}, {Fallon}, {Fegan}, {Feinstein},
  {Fernandes}, {Fiasson}, {Fontaine}, {F{\"o}rster}, {F{\"u}{\ss}ling},
  {Gallant}, {Gast}, {G{\'e}rard}, {Gerbig}, {Giebels}, {Glicenstein},
  {Gl{\"u}ck}, {G{\"o}ring}, {H{\"a}ffner}, {Hague}, {Hahn}, {Hampf}, {Harris},
  {Hauser}, {Heinz}, {Heinzelmann}, {Henri}, {Hermann}, {Hillert}, {Hinton},
  {Hofmann}, {Hofverberg}, {Holler}, {Horns}, {Jacholkowska}, {de Jager},
  {Jahn}, {Jamrozy}, {Jung}, {Kastendieck}, {Katarzy{\'n}ski}, {Katz},
  {Kaufmann}, {Keogh}, {Kh{\'e}lifi}, {Klochkov}, {Klu{\.z}niak}, {Kneiske},
  {Komin}, {Kosack}, {Kossakowski}, {Krayzel}, {Laffon}, {Lamanna}, {Lenain},
  {Lennarz}, {Lohse}, {Lopatin}, {Lu}, {Marandon}, {Marcowith}, {Masbou},
  {Maxted}, {Mayer}, {McComb}, {Medina}, {M{\'e}hault}, {Moderski}, {Mohamed},
  {Moulin}, {Naumann}, {Naumann-Godo}, {de Naurois}, {Nedbal}, {Nekrassov},
  {Nguyen}, {Nicholas}, {Niemiec}, {Nolan}, {Ohm}, {de O{\~n}a Wilhelmi},
  {Opitz}, {Ostrowski}, {Oya}, {Panter}, {Paz Arribas}, {Pekeur}, {Pelletier},
  {Perez}, {Petrucci}, {Peyaud}, {Pita}, {P{\"u}hlhofer}, {Punch},
  {Quirrenbach}, {Raue}, {Rayner}, {Reimer}, {Reimer}, {Renaud}, {de los
  Reyes}, {Rieger}, {Ripken}, {Rob}, {Rosier-Lees}, {Rowell}, {Rudak},
  {Rulten}, {Sahakian}, {Sanchez}, {Santangelo}, {Schlickeiser}, {Schulz},
  {Schwanke}, {Schwarzburg}, {Schwemmer}, {Sheidaei}, {Skilton}, {Sol},
  {Spengler}, {Stawarz}, {Steenkamp}, {Stegmann}, {Stinzing}, {Stycz},
  {Sushch}, {Szostek}, {Tavernet}, {Terrier}, {Tluczykont}, {Valerius}, {van
  Eldik}, {Vasileiadis}, {Venter}, {Viana}, {Vincent}, {V{\"o}lk}, {Volpe},
  {Vorobiov}, {Vorster}, {Wagner}, {Ward}, {White}, {Wierzcholska},
  {Zacharias}, {Zajczyk}, {Zdziarski}, {Zech}, \&
  {Zechlin}}]{2012A&A...541A...5H}
{H.E.S.S. collaboration}, {Abramowski}, A., {Acero}, F., {et~al.}
  2012{\natexlab{a}}, \aap, 541, A5

\bibitem[{{H.E.S.S. collaboration} {et~al.}(2012{\natexlab{b}}){H.E.S.S.
  collaboration}, {Abramowski}, {Acero}, {Aharonian}, {Akhperjanian}, {Anton},
  {Balzer}, {Barnacka}, {Becherini}, {Becker}, {Bernl{\"o}hr}, {Birsin},
  {Biteau}, {Bochow}, {Boisson}, {Bolmont}, {Bordas}, {Brucker}, {Brun},
  {Brun}, {Bulik}, {B{\"u}sching}, {Carrigan}, {Casanova}, {Cerruti},
  {Chadwick}, {Charbonnier}, {Chaves}, {Cheesebrough}, {Cologna}, {Conrad},
  {Dalton}, {Daniel}, {Davids}, {Degrange}, {Deil}, {Dickinson},
  {Djannati-Ata{\"\i}}, {Domainko}, {Drury}, {Dubus}, {Dutson}, {Dyks},
  {Dyrda}, {Egberts}, {Eger}, {Espigat}, {Fallon}, {Fegan}, {Feinstein},
  {Fernandes}, {Fiasson}, {Fontaine}, {F{\"o}rster}, {F{\"u}{\ss}ling},
  {Gallant}, {Garrigoux}, {Gast}, {G{\'e}rard}, {Giebels}, {Glicenstein},
  {Gl{\"u}ck}, {G{\"o}ring}, {Grondin}, {H{\"a}ffner}, {Hague}, {Hahn},
  {Hampf}, {Harris}, {Hauser}, {Heinz}, {Heinzelmann}, {Henri}, {Hermann},
  {Hillert}, {Hinton}, {Hofmann}, {Hofverberg}, {Holler}, {Horns},
  {Jacholkowska}, {Jahn}, {Jamrozy}, {Jung}, {Kastendieck}, {Katarzy{\'n}ski},
  {Katz}, {Kaufmann}, {Kh{\'e}lifi}, {Klochkov}, {Klu{\'z}niak}, {Kneiske},
  {Komin}, {Kosack}, {Kossakowski}, {Krayzel}, {Laffon}, {Lamanna}, {Lenain},
  {Lennarz}, {Lohse}, {Lopatin}, {Lu}, {Marandon}, {Marcowith}, {Masbou},
  {Maurin}, {Maxted}, {Mayer}, {McComb}, {Medina}, {M{\'e}hault}, {Moderski},
  {Mohamed}, {Moulin}, {Naumann}, {Naumann-Godo}, {de Naurois}, {Nedbal},
  {Nekrassov}, {Nguyen}, {Nicholas}, {Niemiec}, {Nolan}, {Ohm}, {de O{\~n}a
  Wilhelmi}, {Opitz}, {Ostrowski}, {Oya}, {Panter}, {Paz Arribas}, {Pekeur},
  {Pelletier}, {Perez}, {Petrucci}, {Peyaud}, {Pita}, {P{\"u}hlhofer}, {Punch},
  {Quirrenbach}, {Raue}, {Reimer}, {Reimer}, {Renaud}, {de los Reyes},
  {Rieger}, {Ripken}, {Rob}, {Rosier-Lees}, {Rowell}, {Rudak}, {Rulten},
  {Sahakian}, {Sanchez}, {Santangelo}, {Schlickeiser}, {Schulz}, {Schwanke},
  {Schwarzburg}, {Schwemmer}, {Sheidaei}, {Skilton}, {Sol}, {Spengler},
  {Stawarz}, {Steenkamp}, {Stegmann}, {Stinzing}, {Stycz}, {Sushch}, {Szostek},
  {Tavernet}, {Terrier}, {Tluczykont}, {Valerius}, {van Eldik}, {Vasileiadis},
  {Venter}, {Viana}, {Vincent}, {V{\"o}lk}, {Volpe}, {Vorobiov}, {Vorster},
  {Wagner}, {Ward}, {White}, {Wierzcholska}, {Zacharias}, {Zajczyk},
  {Zdziarski}, {Zech}, {Zechlin}, \& {Montmerle}}]{2012MNRAS.424..128H}
{H.E.S.S. collaboration}, {Abramowski}, A., {Acero}, F., {et~al.}
  2012{\natexlab{b}}, \mnras, 424, 128

\bibitem[{{H.E.S.S. collaboration} {et~al.}(2009{\natexlab{a}}){H.E.S.S.
  collaboration}, {Acero}, {Aharonian}, {Akhperjanian}, {Anton}, {Barres de
  Almeida}, {Bazer-Bachi}, {Becherini}, {Behera}, {Bernl{\"o}hr}, {Bochow},
  {Boisson}, {Bolmont}, {Borrel}, {Brucker}, {Brun}, {Brun}, {Bulik},
  {B{\"u}sching}, {Boutelier}, {Chadwick}, {Charbonnier}, {Chaves},
  {Cheesebrough}, {Conrad}, {Chounet}, {Clapson}, {Coignet}, {Dalton},
  {Daniel}, {Davids}, {Degrange}, {Deil}, {Dickinson}, {Djannati-Ata{\"\i}},
  {Domainko}, {Drury}, {Dubois}, {Dubus}, {Dyks}, {Dyrda}, {Egberts}, {Eger},
  {Espigat}, {Fallon}, {Farnier}, {Fegan}, {Feinstein}, {Fiasson},
  {F{\"o}rster}, {Fontaine}, {F{\"u}{\ss}ling}, {Gabici}, {Gallant},
  {G{\'e}rard}, {Gerbig}, {Giebels}, {Glicenstein}, {Gl{\"u}ck}, {Goret},
  {G{\"o}ring}, {Hauser}, {Heinz}, {Heinzelmann}, {Henri}, {Hermann}, {Hinton},
  {Hoffmann}, {Hofmann}, {Hofverberg}, {Holleran}, {Hoppe}, {Horns},
  {Jacholkowska}, {de Jager}, {Jahn}, {Jung}, {Katarzy{\'n}ski}, {Katz},
  {Kaufmann}, {Kerschhaggl}, {Khangulyan}, {Kh{\'e}lifi}, {Keogh}, {Klochkov},
  {Klu{\'z}niak}, {Kneiske}, {Komin}, {Kosack}, {Kossakowski}, {Lamanna},
  {Lenain}, {Lohse}, {Marandon}, {Marcowith}, {Masbou}, {Maurin}, {McComb},
  {Medina}, {M{\'e}hault}, {Moderski}, {Moulin}, {Naumann-Godo}, {de Naurois},
  {Nedbal}, {Nekrassov}, {Nicholas}, {Niemiec}, {Nolan}, {Ohm}, {Olive}, {de
  O{\~n}a Wilhelmi}, {Orford}, {Ostrowski}, {Panter}, {Paz Arribas},
  {Pedaletti}, {Pelletier}, {Petrucci}, {Pita}, {P{\"u}hlhofer}, {Punch},
  {Quirrenbach}, {Raubenheimer}, {Raue}, {Rayner}, {Reimer}, {Renaud}, {de Los
  Reyes}, {Rieger}, {Ripken}, {Rob}, {Rosier-Lees}, {Rowell}, {Rudak},
  {Rulten}, {Ruppel}, {Ryde}, {Sahakian}, {Santangelo}, {Schlickeiser},
  {Sch{\"o}ck}, {Sch{\"o}nwald}, {Schwanke}, {Schwarzburg}, {Schwemmer},
  {Shalchi}, {Sushch}, {Sikora}, {Skilton}, {Sol}, {Stawarz}, {Steenkamp},
  {Stegmann}, {Stinzing}, {Superina}, {Szostek}, {Tam}, {Tavernet}, {Terrier},
  {Tibolla}, {Tluczykont}, {van Eldik}, {Vasileiadis}, {Venter}, {Venter},
  {Vialle}, {Vincent}, {Vivier}, {V{\"o}lk}, {Volpe}, {Vorobiov}, {Wagner},
  {Ward}, {Zdziarski}, \& {Zech}}]{2009A&A...508.1135H}
{H.E.S.S. collaboration}, {Acero}, F., {Aharonian}, F.~A., {et~al.}
  2009{\natexlab{a}}, \aap, 508, 1135

\bibitem[{{H.E.S.S. collaboration} {et~al.}(2009{\natexlab{b}}){H.E.S.S.
  collaboration}, {Aharonian}, {Akhperjanian}, {Anton}, {Barres de Almeida},
  {Bazer-Bachi}, {Becherini}, {Behera}, {Bernl{\"o}hr}, {Bochow}, {Boisson},
  {Bolmont}, {Borrel}, {Brucker}, {Brun}, {Brun}, {B{\"u}hler}, {Bulik},
  {B{\"u}sching}, {Boutelier}, {Chadwick}, {Charbonnier}, {Chaves},
  {Cheesebrough}, {Chounet}, {Clapson}, {Coignet}, {Dalton}, {Daniel},
  {Davids}, {Degrange}, {Deil}, {Dickinson}, {Djannati-Ata{\"\i}}, {Domainko},
  {O'C.~Drury}, {Dubois}, {Dubus}, {Dyks}, {Dyrda}, {Egberts},
  {Emmanoulopoulos}, {Espigat}, {Farnier}, {Feinstein}, {Fiasson},
  {F{\"o}rster}, {Fontaine}, {F{\"u}{\ss}ling}, {Gabici}, {Gallant},
  {G{\'e}rard}, {Gerbig}, {Giebels}, {Glicenstein}, {Gl{\"u}ck}, {Goret},
  {G{\"o}ring}, {Hauser}, {Hauser}, {Heinz}, {Heinzelmann}, {Henri}, {Hermann},
  {Hinton}, {Hoffmann}, {Hofmann}, {Holleran}, {Hoppe}, {Horns},
  {Jacholkowska}, {de Jager}, {Jahn}, {Jung}, {Katarzy{\'n}ski}, {Katz},
  {Kaufmann}, {Kerschhaggl}, {Khangulyan}, {Kh{\'e}lifi}, {Keogh}, {Klochkov},
  {Klu{\'z}niak}, {Kneiske}, {Komin}, {Kosack}, {Kossakowski}, {Lamanna},
  {Lenain}, {Lohse}, {Marandon}, {Martineau-Huynh}, {Marcowith}, {Masbou},
  {Maurin}, {McComb}, {Medina}, {Moderski}, {Moulin}, {Naumann-Godo}, {de
  Naurois}, {Nedbal}, {Nekrassov}, {Nicholas}, {Niemiec}, {Nolan}, {Ohm},
  {Olive}, {de O{\~n}a Wilhelmi}, {Orford}, {Ostrowski}, {Panter}, {Paz
  Arribas}, {Pedaletti}, {Pelletier}, {Petrucci}, {Pita}, {P{\"u}hlhofer},
  {Punch}, {Quirrenbach}, {Raubenheimer}, {Raue}, {Rayner}, {Renaud}, {Rieger},
  {Ripken}, {Rob}, {Rosier-Lees}, {Rowell}, {Rudak}, {Rulten}, {Ruppel},
  {Sahakian}, {Santangelo}, {Schlickeiser}, {Sch{\"o}ck}, {Schwanke},
  {Schwarzburg}, {Schwemmer}, {Shalchi}, {Sikora}, {Skilton}, {Sol},
  {Spangler}, {Stawarz}, {Steenkamp}, {Stegmann}, {Stinzing}, {Superina},
  {Szostek}, {Tam}, {Tavernet}, {Terrier}, {Tibolla}, {Tluczykont}, {van
  Eldik}, {Vasileiadis}, {Venter}, {Venter}, {Vialle}, {Vincent}, {Vivier},
  {V{\"o}lk}, {Volpe}, {Wagner}, {Ward}, {Zdziarski}, \&
  {Zech}}]{2009A&A...507..389A}
{H.E.S.S. collaboration}, {Aharonian}, F.~A., {Akhperjanian}, A.~G., {et~al.}
  2009{\natexlab{b}}, \aap, 507, 389

\bibitem[{{H.E.S.S. collaboration} {et~al.}(2005{\natexlab{a}}){H.E.S.S.
  collaboration}, {Aharonian}, {Akhperjanian}, {Aye}, {Bazer-Bachi},
  {Beilicke}, {Benbow}, {Berge}, {Berghaus}, {Bernl{\"o}hr}, {Boisson}, {Bolz},
  {Borrel}, {Braun}, {Breitling}, {Brown}, {Gordo}, {Chadwick}, {Chounet},
  {Cornils}, {Costamante}, {Degrange}, {Dickinson}, {Djannati-Ata{\"\i}},
  {Drury}, {Dubus}, {Emmanoulopoulos}, {Espigat}, {Feinstein}, {Fleury},
  {Fontaine}, {Fuchs}, {Funk}, {Gallant}, {Giebels}, {Gillessen},
  {Glicenstein}, {Goret}, {Hadjichristidis}, {Hauser}, {Heinzelmann}, {Henri},
  {Hermann}, {Hinton}, {Hofmann}, {Holleran}, {Horns}, {Jacholkowska}, {de
  Jager}, {Kh{\'e}lifi}, {Komin}, {Konopelko}, {Latham}, {Le Gallou},
  {Lemi{\`e}re}, {Lemoine-Goumard}, {Leroy}, {Lohse}, {Marcowith}, {Martin},
  {Martineau-Huynh}, {Masterson}, {McComb}, {de Naurois}, {Nolan}, {Noutsos},
  {Orford}, {Osborne}, {Ouchrif}, {Panter}, {Pelletier}, {Pita},
  {P{\"u}hlhofer}, {Punch}, {Raubenheimer}, {Raue}, {Raux}, {Rayner}, {Reimer},
  {Reimer}, {Ripken}, {Rob}, {Rolland}, {Rowell}, {Sahakian}, {Saug{\'e}},
  {Schlenker}, {Schlickeiser}, {Schuster}, {Schwanke}, {Siewert}, {Sol},
  {Spangler}, {Steenkamp}, {Stegmann}, {Tavernet}, {Terrier}, {Th{\'e}oret},
  {Tluczykont}, {Vasileiadis}, {Venter}, {Vincent}, {V{\"o}lk}, \&
  {Wagner}}]{Aharonian:2005nj}
{H.E.S.S. collaboration}, {Aharonian}, F.~A., {Akhperjanian}, A.~G., {et~al.}
  2005{\natexlab{a}}, Science, 309, 746

\bibitem[{{H.E.S.S. collaboration} {et~al.}(2005{\natexlab{b}}){H.E.S.S.
  collaboration}, {Aharonian}, {Akhperjanian}, {Aye}, {Bazer-Bachi},
  {Beilicke}, {Benbow}, {Berge}, {Berghaus}, {Bernl{\"o}hr}, {Boisson}, {Bolz},
  {Braun}, {Breitling}, {Brown}, {Bussons Gordo}, {Chadwick}, {Chounet},
  {Cornils}, {Costamante}, {Degrange}, {Djannati-Ata{\"\i}}, {O'C.~Drury},
  {Dubus}, {Emmanoulopoulos}, {Espigat}, {Feinstein}, {Fleury}, {Fontaine},
  {Fuchs}, {Funk}, {Gallant}, {Giebels}, {Gillessen}, {Glicenstein}, {Goret},
  {Hadjichristidis}, {Hauser}, {Heinzelmann}, {Henri}, {Hermann}, {Hinton},
  {Hofmann}, {Holleran}, {Horns}, {de Jager}, {Johnston}, {Kh{\'e}lifi},
  {Kirk}, {Komin}, {Konopelko}, {Latham}, {Le Gallou}, {Lemi{\`e}re},
  {Lemoine-Goumard}, {Leroy}, {Martineau-Huynh}, {Lohse}, {Marcowith},
  {Masterson}, {McComb}, {de Naurois}, {Nolan}, {Noutsos}, {Orford}, {Osborne},
  {Ouchrif}, {Panter}, {Pelletier}, {Pita}, {P{\"u}hlhofer}, {Punch},
  {Raubenheimer}, {Raue}, {Raux}, {Rayner}, {Redondo}, {Reimer}, {Reimer},
  {Ripken}, {Rob}, {Rolland}, {Rowell}, {Sahakian}, {Saug{\'e}}, {Schlenker},
  {Schlickeiser}, {Schuster}, {Schwanke}, {Siewert}, {Skj{\ae}raasen}, {Sol},
  {Steenkamp}, {Stegmann}, {Tavernet}, {Terrier}, {Th{\'e}oret}, {Tluczykont},
  {Vasileiadis}, {Venter}, {Vincent}, {V{\"o}lk}, \&
  {Wagner}}]{Aharonian:2005br}
{H.E.S.S. collaboration}, {Aharonian}, F.~A., {Akhperjanian}, A.~G., {et~al.}
  2005{\natexlab{b}}, \aap, 442, 1

\bibitem[{{H.E.S.S. collaboration} {et~al.}(2007){H.E.S.S. collaboration},
  {Aharonian}, {Akhperjanian}, {Bazer-Bachi}, {Behera}, {Beilicke}, {Benbow},
  {Berge}, {Bernl{\"o}hr}, {Boisson}, {Bolz}, {Borrel}, {Braun}, {Brion},
  {Brown}, {B{\"u}hler}, {B{\"u}sching}, {Boutelier}, {Carrigan}, {Chadwick},
  {Chounet}, {Coignet}, {Cornils}, {Costamante}, {Degrange}, {Dickinson},
  {Djannati-Ata{\"\i}}, {Domainko}, {O'C.~Drury}, {Dubus}, {Egberts},
  {Emmanoulopoulos}, {Espigat}, {Farnier}, {Feinstein}, {Fiasson},
  {F{\"o}rster}, {Fontaine}, {Funk}, {Funk}, {F{\"u}{\ss}ling}, {Gallant},
  {Giebels}, {Glicenstein}, {Gl{\"u}ck}, {Goret}, {Hadjichristidis}, {Hauser},
  {Hauser}, {Heinzelmann}, {Henri}, {Hermann}, {Hinton}, {Hoffmann}, {Hofmann},
  {Holleran}, {Hoppe}, {Horns}, {Jacholkowska}, {de Jager}, {Kendziorra},
  {Kerschhaggl}, {Kh{\'e}lifi}, {Komin}, {Kosack}, {Lamanna}, {Latham}, {Le
  Gallou}, {Lemi{\`e}re}, {Lemoine-Goumard}, {Lohse}, {Martin},
  {Martineau-Huynh}, {Marcowith}, {Masterson}, {Maurin}, {McComb}, {Moulin},
  {de Naurois}, {Nedbal}, {Nolan}, {Noutsos}, {Olive}, {Orford}, {Osborne},
  {Panter}, {Pedaletti}, {Pelletier}, {Petrucci}, {Pita}, {P{\"u}hlhofer},
  {Punch}, {Ranchon}, {Raubenheimer}, {Raue}, {Rayner}, {Reimer}, {Ripken},
  {Rob}, {Rolland}, {Rosier-Lees}, {Rowell}, {Ruppel}, {Sahakian},
  {Santangelo}, {Saug{\'e}}, {Schlenker}, {Schlickeiser}, {Schr{\"o}der},
  {Schwanke}, {Schwarzburg}, {Schwemmer}, {Shalchi}, {Sol}, {Spangler},
  {Steenkamp}, {Stegmann}, {Superina}, {Tam}, {Tavernet}, {Terrier},
  {Tluczykont}, {van Eldik}, {Vasileiadis}, {Venter}, {Vialle}, {Vincent},
  {V{\"o}lk}, {Wagner}, {Ward}, {Moriguchi}, \& {Fukui}}]{2007A&A...469L...1A}
{H.E.S.S. collaboration}, {Aharonian}, F.~A., {Akhperjanian}, A.~G., {et~al.}
  2007, \aap, 469, L1

\bibitem[{{H.E.S.S. collaboration} {et~al.}(2006{\natexlab{a}}){H.E.S.S.
  collaboration}, {Aharonian}, {Akhperjanian}, {Bazer-Bachi}, {Beilicke},
  {Benbow}, {Berge}, {Bernl{\"o}hr}, {Boisson}, {Bolz}, {Borrel}, {Braun},
  {Breitling}, {Brown}, {B{\"u}hler}, {B{\"u}sching}, {Carrigan}, {Chadwick},
  {Chounet}, {Cornils}, {Costamante}, {Degrange}, {Dickinson},
  {Djannati-Ata{\"\i}}, {O'C.~Drury}, {Dubus}, {Egberts}, {Emmanoulopoulos},
  {Espigat}, {Feinstein}, {Ferrero}, {Fiasson}, {Fontaine}, {Funk}, {Funk},
  {Gallant}, {Giebels}, {Glicenstein}, {Goret}, {Hadjichristidis}, {Hauser},
  {Hauser}, {Heinzelmann}, {Henri}, {Hermann}, {Hinton}, {Hofmann}, {Holleran},
  {Horns}, {Jacholkowska}, {de Jager}, {Kh{\'e}lifi}, {Komin}, {Konopelko},
  {Kosack}, {Latham}, {Le Gallou}, {Lemi{\`e}re}, {Lemoine-Goumard}, {Lohse},
  {Martin}, {Martineau-Huynh}, {Marcowith}, {Masterson}, {McComb}, {de
  Naurois}, {Nedbal}, {Nolan}, {Noutsos}, {Orford}, {Osborne}, {Ouchrif},
  {Panter}, {Pelletier}, {Pita}, {P{\"u}hlhofer}, {Punch}, {Raubenheimer},
  {Raue}, {Rayner}, {Reimer}, {Reimer}, {Ripken}, {Rob}, {Rolland}, {Rowell},
  {Sahakian}, {Saug{\'e}}, {Schlenker}, {Schlickeiser}, {Schwanke}, {Sol},
  {Spangler}, {Spanier}, {Steenkamp}, {Stegmann}, {Superina}, {Tavernet},
  {Terrier}, {Th{\'e}oret}, {Tluczykont}, {van Eldik}, {Vasileiadis}, {Venter},
  {Vincent}, {V{\"o}lk}, {Wagner}, \& {Ward}}]{2006A&A...457..899A}
{H.E.S.S. collaboration}, {Aharonian}, F.~A., {Akhperjanian}, A.~G., {et~al.}
  2006{\natexlab{a}}, \aap, 457, 899

\bibitem[{{H.E.S.S. collaboration} {et~al.}(2006{\natexlab{b}}){H.E.S.S.
  collaboration}, {Aharonian}, {Akhperjanian}, {Bazer-Bachi}, {Beilicke},
  {Benbow}, {Berge}, {Bernl{\"o}hr}, {Boisson}, {Bolz}, {Borrel}, {Braun},
  {Brown}, {B{\"u}hler}, {B{\"u}sching}, {Carrigan}, {Chadwick}, {Chounet},
  {Cornils}, {Costamante}, {Degrange}, {Dickinson}, {Djannati-Ata{\"\i}},
  {O'C.~Drury}, {Dubus}, {Egberts}, {Emmanoulopoulos}, {Espigat}, {Feinstein},
  {Ferrero}, {Fiasson}, {Fontaine}, {Funk}, {Funk}, {F{\"u}{\ss}ling},
  {Gallant}, {Giebels}, {Glicenstein}, {Goret}, {Hadjichristidis}, {Hauser},
  {Hauser}, {Heinzelmann}, {Henri}, {Hermann}, {Hinton}, {Hoffmann}, {Hofmann},
  {Holleran}, {Horns}, {Jacholkowska}, {de Jager}, {Kendziorra}, {Kh{\'e}lifi},
  {Komin}, {Konopelko}, {Kosack}, {Latham}, {Le Gallou}, {Lemi{\`e}re},
  {Lemoine-Goumard}, {Lohse}, {Martin}, {Martineau-Huynh}, {Marcowith},
  {Masterson}, {Maurin}, {McComb}, {Moulin}, {de Naurois}, {Nedbal}, {Nolan},
  {Noutsos}, {Orford}, {Osborne}, {Ouchrif}, {Panter}, {Pelletier}, {Pita},
  {P{\"u}hlhofer}, {Punch}, {Raubenheimer}, {Raue}, {Rayner}, {Reimer},
  {Reimer}, {Ripken}, {Rob}, {Rolland}, {Rowell}, {Sahakian}, {Santangelo},
  {Saug{\'e}}, {Schlenker}, {Schlickeiser}, {Schr{\"o}der}, {Schwanke},
  {Schwarzburg}, {Shalchi}, {Sol}, {Spangler}, {Spanier}, {Steenkamp},
  {Stegmann}, {Superina}, {Tavernet}, {Terrier}, {Tluczykont}, {van Eldik},
  {Vasileiadis}, {Venter}, {Vincent}, {V{\"o}lk}, {Wagner}, \&
  {Ward}}]{Aharonian:2006qw}
{H.E.S.S. collaboration}, {Aharonian}, F.~A., {Akhperjanian}, A.~G., {et~al.}
  2006{\natexlab{b}}, \aap, 460, 743

\bibitem[{{Hill}(2013)}]{2013arXiv1304.2427H}
{Hill}, A.~B. 2013, in {\em 2012 Fermi Symposium} proceedings - eConf C121028
  (eprint arXiv:1304.2427)

\bibitem[{{Hill} {et~al.}(2011{\natexlab{a}}){Hill}, {Dubois}, {Torres}, \&
  {Fermi LAT Collaboration}}]{2011heep.conf..498H}
{Hill}, A.~B., {Dubois}, R., {Torres}, D.~F., \& {Fermi LAT Collaboration}.
  2011{\natexlab{a}}, in High-Energy Emission from Pulsars and their Systems:
  Proceedings of the First Session of the Sant Cugat Forum on Astrophysics,
  Astrophysics and Space Science Proceedings, ed. D.~F. {Torres} \& N.~{Rea}
  (Springer-Verlag Berlin), 498

\bibitem[{{Hill} {et~al.}(2011{\natexlab{b}}){Hill}, {Szostek}, {Corbel},
  {Camilo}, {Corbet}, {Dubois}, {Dubus}, {Edwards}, {Ferrara}, {Kerr},
  {Koerding}, {Kozie{\l}}, \& {Stawarz}}]{2011MNRAS.415..235H}
{Hill}, A.~B., {Szostek}, A., {Corbel}, S., {et~al.} 2011{\natexlab{b}},
  \mnras, 415, 235

\bibitem[{{Hinton} {et~al.}(2009){Hinton}, {Skilton}, {Funk}, {Brucker},
  {Aharonian}, {Dubus}, {Fiasson}, {Gallant}, {Hofmann}, {Marcowith}, \&
  {Reimer}}]{Hinton:2008eg}
{Hinton}, J.~A., {Skilton}, J.~L., {Funk}, S., {et~al.} 2009, \apjl, 690, L101

\bibitem[{{Hirayama} {et~al.}(1999){Hirayama}, {Cominsky}, {Kaspi}, {Nagase},
  {Tavani}, {Kawai}, \& {Grove}}]{1999ApJ...521..718H}
{Hirayama}, M., {Cominsky}, L.~R., {Kaspi}, V.~M., {et~al.} 1999, \apj, 521,
  718

\bibitem[{{Hjalmarsdotter} {et~al.}(2009){Hjalmarsdotter}, {Zdziarski},
  {Szostek}, \& {Hannikainen}}]{2009MNRAS.392..251H}
{Hjalmarsdotter}, L., {Zdziarski}, A.~A., {Szostek}, A., \& {Hannikainen},
  D.~C. 2009, \mnras, 392, 251

\bibitem[{{Hjellming}(1973)}]{1973Sci...182.1089H}
{Hjellming}, R.~M. 1973, Science, 182, 1089

\bibitem[{{Hoffmann} {et~al.}(2009){Hoffmann}, {Klochkov}, {Santangelo},
  {Horns}, {Segreto}, {Staubert}, \& {Puehlhofer}}]{Hoffmann:2008ys}
{Hoffmann}, A.~D., {Klochkov}, D., {Santangelo}, A., {et~al.} 2009, \aap, 494,
  L37

\bibitem[{{Howarth}(1983)}]{1983MNRAS.203..801H}
{Howarth}, I.~D. 1983, \mnras, 203, 801

\bibitem[{{Iben} {et~al.}(1995){Iben}, {Tutukov}, \& {Yungelson}}]{Iben:1995vu}
{Iben}, I.~J., {Tutukov}, A.~V., \& {Yungelson}, L.~R. 1995, \apjs, 100, 217

\bibitem[{{Illarionov} \& {Sunyaev}(1975)}]{Illarionov:1975yk}
{Illarionov}, A.~F. \& {Sunyaev}, R.~A. 1975, \aap, 39, 185

\bibitem[{{Jackson}(1972)}]{1972NPhS..236...39J}
{Jackson}, J.~C. 1972, Nature Physical Science, 236, 39

\bibitem[{{Johnston} {et~al.}(2005){Johnston}, {Ball}, {Wang}, \&
  {Manchester}}]{Johnston:2005jq}
{Johnston}, S., {Ball}, L., {Wang}, N., \& {Manchester}, R.~N. 2005, \mnras,
  358, 1069

\bibitem[{{Johnston} {et~al.}(1992){Johnston}, {Manchester}, {Lyne}, {Bailes},
  {Kaspi}, {Qiao}, \& {D'Amico}}]{1992ApJ...387L..37J}
{Johnston}, S., {Manchester}, R.~N., {Lyne}, A.~G., {et~al.} 1992, \apjl, 387,
  L37

\bibitem[{{Johnston} {et~al.}(1994){Johnston}, {Manchester}, {Lyne},
  {Nicastro}, \& {Spyromilio}}]{1994MNRAS.268..430J}
{Johnston}, S., {Manchester}, R.~N., {Lyne}, A.~G., {Nicastro}, L., \&
  {Spyromilio}, J. 1994, \mnras, 268, 430

\bibitem[{{Kaaret} {et~al.}(1999){Kaaret}, {Piraino}, {Halpern}, \&
  {Eracleous}}]{1999ApJ...523..197K}
{Kaaret}, P., {Piraino}, S., {Halpern}, J., \& {Eracleous}, M. 1999, \apj, 523,
  197

\bibitem[{{Kapala} {et~al.}(2010){Kapala}, {Bulik}, {Rudak}, {Dubus}, \&
  {Lyczek}}]{2010tsra.confE.193K}
{Kapala}, M., {Bulik}, T., {Rudak}, B., {Dubus}, G., \& {Lyczek}, M. 2010, in
  25th Texas Symposium on Relativistic Astrophysics, Vol. Texas 2010
  (Proceedings of Science), 193

\bibitem[{{Kargaltsev} \& {Pavlov}(2010)}]{2010AIPC.1248...25K}
{Kargaltsev}, O. \& {Pavlov}, G.~G. 2010, X-ray Astronomy 2009; Present Status,
  Multi-Wavelength Approach and Future Perspectives, AIP Conference
  Proceedings, 1248, 25

\bibitem[{{Kaspi} {et~al.}(1996{\natexlab{a}}){Kaspi}, {Bailes}, {Manchester},
  {Stappers}, \& {Bell}}]{1996Natur.381..584K}
{Kaspi}, V.~M., {Bailes}, M., {Manchester}, R.~N., {Stappers}, B.~W., \&
  {Bell}, J.~F. 1996{\natexlab{a}}, \nat, 381, 584

\bibitem[{{Kaspi} {et~al.}(1994){Kaspi}, {Johnston}, {Bell}, {Manchester},
  {Bailes}, {Bessell}, {Lyne}, \& {D'Amico}}]{1994ApJ...423L..43K}
{Kaspi}, V.~M., {Johnston}, S., {Bell}, J.~F., {et~al.} 1994, \apjl, 423, L43

\bibitem[{{Kaspi} {et~al.}(1996{\natexlab{b}}){Kaspi}, {Tauris}, \&
  {Manchester}}]{Kaspi:1996id}
{Kaspi}, V.~M., {Tauris}, T.~M., \& {Manchester}, R.~N. 1996{\natexlab{b}},
  \apj, 459, 717

\bibitem[{{Kaufman Bernad{\'o}} {et~al.}(2002){Kaufman Bernad{\'o}}, {Romero},
  \& {Mirabel}}]{2002A&A...385L..10K}
{Kaufman Bernad{\'o}}, M.~M., {Romero}, G.~E., \& {Mirabel}, I.~F. 2002, \aap,
  385, L10

\bibitem[{{Kawachi} {et~al.}(2004){Kawachi}, {Naito}, {Patterson}, {Edwards},
  {Asahara}, {Bicknell}, {Clay}, {Enomoto}, {Gunji}, {Hara}, {Hara}, {Hattori},
  {Hayashi}, {Hayashi}, {Itoh}, {Kabuki}, {Kajino}, {Katagiri}, {Kifune},
  {Ksenofontov}, {Kubo}, {Kushida}, {Matsubara}, {Mizumoto}, {Mori}, {Moro},
  {Muraishi}, {Muraki}, {Nakase}, {Nishida}, {Nishijima}, {Ohishi}, {Okumura},
  {Protheroe}, {Sakurazawa}, {Swaby}, {Tanimori}, {Tokanai}, {Tsuchiya},
  {Tsunoo}, {Uchida}, {Watanabe}, {Watanabe}, {Yanagita}, {Yoshida}, \&
  {Yoshikoshi}}]{2004ApJ...607..949K}
{Kawachi}, A., {Naito}, T., {Patterson}, J.~R., {et~al.} 2004, \apj, 607, 949

\bibitem[{{Kennel} \& {Coroniti}(1984{\natexlab{a}})}]{Kennel:1984pd}
{Kennel}, C.~F. \& {Coroniti}, F.~V. 1984{\natexlab{a}}, \apj, 283, 694

\bibitem[{{Kennel} \& {Coroniti}(1984{\natexlab{b}})}]{Kennel:1984gu}
{Kennel}, C.~F. \& {Coroniti}, F.~V. 1984{\natexlab{b}}, \apj, 283, 710

\bibitem[{{Khangulyan} {et~al.}(2011){Khangulyan}, {Aharonian}, {Bogovalov}, \&
  {Rib{\'o}}}]{2011ApJ...742...98K}
{Khangulyan}, D., {Aharonian}, F.~A., {Bogovalov}, S.~V., \& {Rib{\'o}}, M.
  2011, \apj, 742, 98

\bibitem[{{Khangulyan} {et~al.}(2012){Khangulyan}, {Aharonian}, {Bogovalov}, \&
  {Rib{\'o}}}]{2012ApJ...752L..17K}
{Khangulyan}, D., {Aharonian}, F.~A., {Bogovalov}, S.~V., \& {Rib{\'o}}, M.
  2012, \apjl, 752, L17

\bibitem[{{Khangulyan} {et~al.}(2008){Khangulyan}, {Aharonian}, \&
  {Bosch-Ramon}}]{Khangulyan:2007me}
{Khangulyan}, D., {Aharonian}, F.~A., \& {Bosch-Ramon}, V. 2008, \mnras, 383,
  467

\bibitem[{{Khangulyan} {et~al.}(2007){Khangulyan}, {Hnatic}, {Aharonian}, \&
  {Bogovalov}}]{2007MNRAS.380..320K}
{Khangulyan}, D., {Hnatic}, S., {Aharonian}, F., \& {Bogovalov}, S. 2007,
  \mnras, 380, 320

\bibitem[{{Kijak} {et~al.}(2011){Kijak}, {Dembska}, {Lewandowski}, {Melikidze},
  \& {Sendyk}}]{2011MNRAS.418L.114K}
{Kijak}, J., {Dembska}, M., {Lewandowski}, W., {Melikidze}, G., \& {Sendyk}, M.
  2011, \mnras, 418, L114

\bibitem[{{Kirk} {et~al.}(1999){Kirk}, {Ball}, \& {Skjaeraasen}}]{Kirk:1999hr}
{Kirk}, J.~G., {Ball}, L., \& {Skjaeraasen}, O. 1999, Astroparticle Physics,
  10, 31

\bibitem[{{Kirk} {et~al.}(2009){Kirk}, {Lyubarsky}, \& {P\'etri}}]{Kirk:2007wh}
{Kirk}, J.~G., {Lyubarsky}, Y., \& {P\'etri}, J. 2009, in Astrophysics and
  Space Science Library, Vol. 357, Neutron Stars and Pulsars, ed. W.~{Becker}
  (Springer), 421

\bibitem[{{Kirk} {et~al.}(2002){Kirk}, {Skj{\ae}raasen}, \&
  {Gallant}}]{2002A&A...388L..29K}
{Kirk}, J.~G., {Skj{\ae}raasen}, O., \& {Gallant}, Y.~A. 2002, \aap, 388, L29

\bibitem[{{Kishishita} {et~al.}(2009){Kishishita}, {Tanaka}, {Uchiyama}, \&
  {Takahashi}}]{2009ApJ...697L...1K}
{Kishishita}, T., {Tanaka}, T., {Uchiyama}, Y., \& {Takahashi}, T. 2009, \apjl,
  697, L1

\bibitem[{{Klein-Wolt} {et~al.}(2002){Klein-Wolt}, {Fender}, {Pooley},
  {Belloni}, {Migliari}, {Morgan}, \& {van der Klis}}]{2002MNRAS.331..745K}
{Klein-Wolt}, M., {Fender}, R.~P., {Pooley}, G.~G., {et~al.} 2002, \mnras, 331,
  745

\bibitem[{{Kochanek}(1993)}]{1993ApJ...406..638K}
{Kochanek}, C.~S. 1993, \apj, 406, 638

\bibitem[{{Kong} {et~al.}(2012){Kong}, {Cheng}, \&
  {Huang}}]{2012ApJ...753..127K}
{Kong}, S.~W., {Cheng}, K.~S., \& {Huang}, Y.~F. 2012, \apj, 753, 127

\bibitem[{{Lamb} {et~al.}(1977){Lamb}, {Fichtel}, {Hartman}, {Kniffen}, \&
  {Thompson}}]{Lamb:1977xw}
{Lamb}, R.~C., {Fichtel}, C.~E., {Hartman}, R.~C., {Kniffen}, D.~A., \&
  {Thompson}, D.~J. 1977, \apjl, 212, L63

\bibitem[{{Lamberts} {et~al.}(2012){Lamberts}, {Dubus}, {Lesur}, \&
  {Fromang}}]{2012A&A...546A..60L}
{Lamberts}, A., {Dubus}, G., {Lesur}, G., \& {Fromang}, S. 2012, \aap, 546, A60

\bibitem[{{Lamberts} {et~al.}(2011){Lamberts}, {Fromang}, \&
  {Dubus}}]{2011MNRAS.418.2618L}
{Lamberts}, A., {Fromang}, S., \& {Dubus}, G. 2011, \mnras, 418, 2618

\bibitem[{{Lang} {et~al.}(1998){Lang}, {Buckley}, {Carter-Lewis}, {Catanese},
  {Cawley}, {Colombo}, {Connaughton}, {Fegan}, {Finley}, {Gaidos},
  {Gillanders}, {Hillas}, {Kertzman}, {Krennrich}, {Lessard}, {Moriarty},
  {Quinn}, {Rose}, {Sembroski}, \& {Weekes}}]{Lang:1998rz}
{Lang}, M.~J., {Buckley}, J.~H., {Carter-Lewis}, D.~A., {et~al.} 1998,
  Astroparticle Physics, 9, 203

\bibitem[{{Lebedev} \& {Myasnikov}(1990)}]{1990FlDy...25..629L}
{Lebedev}, M.~G. \& {Myasnikov}, A.~V. 1990, Fluid Dynamics, 25, 629

\bibitem[{{Lemoine} \& {Pelletier}(2011)}]{2011CRPhy..12..234L}
{Lemoine}, M. \& {Pelletier}, G. 2011, Comptes Rendus Physique, 12, 234

\bibitem[{{Levinson}(2006)}]{2006IJMPA..21.6015L}
{Levinson}, A. 2006, International Journal of Modern Physics A, 21, 6015

\bibitem[{{Levinson} \& {Blandford}(1996)}]{1996ApJ...456L..29L}
{Levinson}, A. \& {Blandford}, R. 1996, \apjl, 456, L29

\bibitem[{{Levinson} \& {Mattox}(1996)}]{1996ApJ...462L..67L}
{Levinson}, A. \& {Mattox}, J.~R. 1996, \apjl, 462, L67

\bibitem[{{Leyder} {et~al.}(2008){Leyder}, {Walter}, \&
  {Rauw}}]{2008A&A...477L..29L}
{Leyder}, J.-C., {Walter}, R., \& {Rauw}, G. 2008, \aap, 477, L29

\bibitem[{{Li} {et~al.}(2011{\natexlab{a}}){Li}, {Torres}, {Chen}, {G{\"o}tz},
  {Rea}, {Zhang}, {Caliandro}, \& {Wang}}]{2011ApJ...738L..31L}
{Li}, J., {Torres}, D.~F., {Chen}, Y., {et~al.} 2011{\natexlab{a}}, \apjl, 738,
  L31

\bibitem[{{Li} {et~al.}(2011{\natexlab{b}}){Li}, {Torres}, {Zhang}, {Chen},
  {Hadasch}, {Ray}, {Kretschmar}, {Rea}, \& {Wang}}]{2011ApJ...733...89L}
{Li}, J., {Torres}, D.~F., {Zhang}, S., {et~al.} 2011{\natexlab{b}}, \apj, 733,
  89

\bibitem[{{Li} {et~al.}(2012){Li}, {Torres}, {Zhang}, {Hadasch}, {Rea},
  {Caliandro}, {Chen}, \& {Wang}}]{2012ApJ...744L..13L}
{Li}, J., {Torres}, D.~F., {Zhang}, S., {et~al.} 2012, \apjl, 744, L13

\bibitem[{{Liu} {et~al.}(2006){Liu}, {van Paradijs}, \& {van den
  Heuvel}}]{Liu:2006kk}
{Liu}, Q.~Z., {van Paradijs}, J., \& {van den Heuvel}, E.~P.~J. 2006, \aap,
  455, 1165

\bibitem[{{Lorimer}(2005)}]{Lorimer:2005ic}
{Lorimer}, D.~R. 2005, Living Reviews in Relativity, 8, 7

\bibitem[{{Lorimer} {et~al.}(2006){Lorimer}, {Faulkner}, {Lyne}, {Manchester},
  {Kramer}, {McLaughlin}, {Hobbs}, {Possenti}, {Stairs}, {Camilo}, {Burgay},
  {D'Amico}, {Corongiu}, \& {Crawford}}]{2006MNRAS.372..777L}
{Lorimer}, D.~R., {Faulkner}, A.~J., {Lyne}, A.~G., {et~al.} 2006, \mnras, 372,
  777

\bibitem[{{L{\"u}} {et~al.}(2006){L{\"u}}, {Yungelson}, \& {Han}}]{Lu:2006cu}
{L{\"u}}, G., {Yungelson}, L., \& {Han}, Z. 2006, \mnras, 372, 1389

\bibitem[{{L{\"u}} {et~al.}(2011){L{\"u}}, {Zhu}, {Wang}, {Huo}, \&
  {Yang}}]{2011MNRAS.413L..11L}
{L{\"u}}, G., {Zhu}, C., {Wang}, Z., {Huo}, W., \& {Yang}, Y. 2011, \mnras,
  413, L11

\bibitem[{{Lucarelli} {et~al.}(2010){Lucarelli}, {Verrecchia}, {Striani},
  {Pittori}, {Tavani}, {Vercellone}, {Bulgarelli}, {Gianotti}, {Trifoglio},
  {Chen}, {Giuliani}, {Mereghetti}, {Caraveo}, {Perotti}, {Donnarumma},
  {D'Ammando}, {Del Monte}, {Evangelista}, {Feroci}, {Lazzarotto}, {Pacciani},
  {Soffitta}, {Costa}, {Lapshov}, {Rapisarda}, {Argan}, {Piano}, {Pucella},
  {Sabatini}, {Trois}, {Vittorini}, {Fuschino}, {Galli}, {Labanti},
  {Marisaldi}, {Di Cocco}, {Pellizzoni}, {Pilia}, {Barbiellini}, {Longo},
  {Moretti}, {Vallazza}, {Morselli}, {Picozza}, {Prest}, {Lipari}, {Zanello},
  {Cattaneo}, {Rappoldi}, {Santolamazza}, {Colafrancesco}, {Giommi}, \&
  {Salotti}}]{2010ATel.2761....1L}
{Lucarelli}, F., {Verrecchia}, F., {Striani}, E., {et~al.} 2010, The
  Astronomer's Telegram, 2761, 1

\bibitem[{{Luo} {et~al.}(1990){Luo}, {McCray}, \& {Mac Low}}]{Luo:1990mp}
{Luo}, D., {McCray}, R., \& {Mac Low}, M.-M. 1990, \apj, 362, 267

\bibitem[{{Lyubarsky}(2003)}]{2003MNRAS.345..153L}
{Lyubarsky}, Y.~E. 2003, \mnras, 345, 153

\bibitem[{{Madsen} {et~al.}(2012){Madsen}, {Stairs}, {Kramer}, {Camilo},
  {Hobbs}, {Janssen}, {Lyne}, {Manchester}, {Possenti}, \&
  {Stappers}}]{2012MNRAS.425.2378M}
{Madsen}, E.~C., {Stairs}, I.~H., {Kramer}, M., {et~al.} 2012, \mnras, 425,
  2378

\bibitem[{{MAGIC collaboration} {et~al.}(2009{\natexlab{a}}){MAGIC
  collaboration}, {Albert}, {Aliu}, {Anderhub}, {Antonelli}, {Antoranz},
  {Backes}, {Baixeras}, {Barrio}, {Bartko}, {Bastieri}, {Becker}, {Bednarek},
  {Berger}, {Bernardini}, {Bigongiari}, {Biland}, {Bock}, {Bonnoli}, {Bordas},
  {Bosch-Ramon}, {Bretz}, {Britvitch}, {Camara}, {Carmona}, {Chilingarian},
  {Commichau}, {Contreras}, {Cortina}, {Costado}, {Covino}, {Curtef}, {Dazzi},
  {DeAngelis}, {DeCea del Pozo}, {de los Reyes}, {DeLotto}, {DeMaria},
  {DeSabata}, {Delgado Mendez}, {Dominguez}, {Dorner}, {Doro}, {Errando},
  {Fagiolini}, {Ferenc}, {Fern{\'a}ndez}, {Firpo}, {Fonseca}, {Font},
  {Galante}, {Garc{\'{\i}}a L{\'o}pez}, {Garczarczyk}, {Gaug}, {Goebel},
  {Hayashida}, {Herrero}, {H{\"o}hne}, {Hose}, {Hsu}, {Huber}, {Jogler},
  {Kranich}, {La Barbera}, {Laille}, {Leonardo}, {Lindfors}, {Lombardi},
  {Longo}, {L{\'o}pez}, {Lorenz}, {Majumdar}, {Maneva}, {Mankuzhiyil},
  {Mannheim}, {Maraschi}, {Mariotti}, {Mart{\'{\i}}nez}, {Mazin}, {Meucci},
  {Meyer}, {Miranda}, {Mirzoyan}, {Mizobuchi}, {Moles}, {Moralejo}, {Nieto},
  {Nilsson}, {Ninkovic}, {Otte}, {Oya}, {Panniello}, {Paoletti}, {Paredes},
  {Pasanen}, {Pascoli}, {Pauss}, {Pegna}, {Perez-Torres}, {Persic}, {Peruzzo},
  {Piccioli}, {Prada}, {Prandini}, {Puchades}, {Raymers}, {Rhode}, {Rib{\'o}},
  {Rico}, {Rissi}, {Robert}, {R{\"u}gamer}, {Saggion}, {Saito}, {Salvati},
  {Sanchez-Conde}, {Sartori}, {Satalecka}, {Scalzotto}, {Scapin}, {Schmitt},
  {Schweizer}, {Shayduk}, {Shinozaki}, {Shore}, {Sidro}, {Sierpowska-Bartosik},
  {Sillanp{\"a}{\"a}}, {Sobczynska}, {Spanier}, {Stamerra}, {Stark}, {Takalo},
  {Tavecchio}, {Temnikov}, {Tescaro}, {Teshima}, {Tluczykont}, {Torres},
  {Turini}, {Vankov}, {Venturini}, {Vitale}, {Wagner}, {Wittek}, {Zabalza},
  {Zandanel}, {Zanin}, \& {Zapatero}}]{2009ApJ...693..303A}
{MAGIC collaboration}, {Albert}, J., {Aliu}, E., {et~al.} 2009{\natexlab{a}},
  \apj, 693, 303

\bibitem[{{MAGIC collaboration} {et~al.}(2006){MAGIC collaboration}, {Albert},
  {Aliu}, {Anderhub}, {Antoranz}, {Armada}, {Asensio}, {Baixeras}, {Barrio},
  {Bartelt}, {Bartko}, {Bastieri}, {Bavikadi}, {Bednarek}, {Berger},
  {Bigongiari}, {Biland}, {Bisesi}, {Bock}, {Bordas}, {Bosch-Ramon}, {Bretz},
  {Britvitch}, {Camara}, {Carmona}, {Chilingarian}, {Ciprini}, {Coarasa},
  {Commichau}, {Contreras}, {Cortina}, {Curtef}, {Danielyan}, {Dazzi}, {De
  Angelis}, {de los Reyes}, {De Lotto}, {Domingo-Santamar{\'{\i}}a}, {Dorner},
  {Doro}, {Errando}, {Fagiolini}, {Ferenc}, {Fern{\'a}ndez}, {Firpo}, {Flix},
  {Fonseca}, {Font}, {Fuchs}, {Galante}, {Garczarczyk}, {Gaug}, {Giller},
  {Goebel}, {Hakobyan}, {Hayashida}, {Hengstebeck}, {H{\"o}hne}, {Hose}, {Hsu},
  {Isar}, {Jacon}, {Kalekin}, {Kosyra}, {Kranich}, {Laatiaoui}, {Laille},
  {Lenisa}, {Liebing}, {Lindfors}, {Lombardi}, {Longo}, {L{\'o}pez},
  {L{\'o}pez}, {Lorenz}, {Lucarelli}, {Majumdar}, {Maneva}, {Mannheim},
  {Mansutti}, {Mariotti}, {Mart{\'{\i}}nez}, {Mase}, {Mazin}, {Merck},
  {Meucci}, {Meyer}, {Miranda}, {Mirzoyan}, {Mizobuchi}, {Moralejo}, {Nilsson},
  {O{\~n}a-Wilhelmi}, {Ordu{\~n}a}, {Otte}, {Oya}, {Paneque}, {Paoletti},
  {Paredes}, {Pasanen}, {Pascoli}, {Pauss}, {Pavel}, {Pegna}, {Persic},
  {Peruzzo}, {Piccioli}, {Poller}, {Pooley}, {Prandini}, {Raymers}, {Rhode},
  {Rib{\'o}}, {Rico}, {Riegel}, {Rissi}, {Robert}, {Romero}, {R{\"u}gamer},
  {Saggion}, {S{\'a}nchez}, {Sartori}, {Scalzotto}, {Scapin}, {Schmitt},
  {Schweizer}, {Shayduk}, {Shinozaki}, {Shore}, {Sidro}, {Sillanp{\"a}{\"a}},
  {Sobczynska}, {Stamerra}, {Stark}, {Takalo}, {Temnikov}, {Tescaro},
  {Teshima}, {Tonello}, {Torres}, {Torres}, {Turini}, {Vankov}, {Vitale},
  {Wagner}, {Wibig}, {Wittek}, {Zanin}, \& {Zapatero}}]{Albert:2006wi}
{MAGIC collaboration}, {Albert}, J., {Aliu}, E., {et~al.} 2006, Science, 312,
  1771

\bibitem[{{MAGIC collaboration} {et~al.}(2007){MAGIC collaboration}, {Albert},
  {Aliu}, {Anderhub}, {Antoranz}, {Armada}, {Baixeras}, {Barrio}, {Bartko},
  {Bastieri}, {Becker}, {Bednarek}, {Berger}, {Bigongiari}, {Biland}, {Bock},
  {Bordas}, {Bosch-Ramon}, {Bretz}, {Britvitch}, {Camara}, {Carmona},
  {Chilingarian}, {Coarasa}, {Commichau}, {Contreras}, {Cortina}, {Costado},
  {Curtef}, {Danielyan}, {Dazzi}, {De Angelis}, {Delgado}, {de los Reyes}, {De
  Lotto}, {Domingo-Santamar{\'{\i}}a}, {Dorner}, {Doro}, {Errando},
  {Fagiolini}, {Ferenc}, {Fern{\'a}ndez}, {Firpo}, {Flix}, {Fonseca}, {Font},
  {Fuchs}, {Galante}, {Garc{\'{\i}}a-L{\'o}pez}, {Garczarczyk}, {Gaug},
  {Giller}, {Goebel}, {Hakobyan}, {Hayashida}, {Hengstebeck}, {Herrero},
  {H{\"o}hne}, {Hose}, {Hsu}, {Jacon}, {Jogler}, {Kosyra}, {Kranich},
  {Kritzer}, {Laille}, {Lindfors}, {Lombardi}, {Longo}, {L{\'o}pez},
  {L{\'o}pez}, {Lorenz}, {Majumdar}, {Maneva}, {Mannheim}, {Mansutti},
  {Mariotti}, {Mart{\'{\i}}nez}, {Mazin}, {Merck}, {Meucci}, {Meyer},
  {Miranda}, {Mirzoyan}, {Mizobuchi}, {Moralejo}, {Nieto}, {Nilsson},
  {Ninkovic}, {O{\~n}a-Wilhelmi}, {Otte}, {Oya}, {Panniello}, {Paoletti},
  {Paredes}, {Pasanen}, {Pascoli}, {Pauss}, {Pegna}, {Persic}, {Peruzzo},
  {Piccioli}, {Prandini}, {Puchades}, {Raymers}, {Rhode}, {Rib{\'o}}, {Rico},
  {Rissi}, {Robert}, {R{\"u}gamer}, {Saggion}, {Saito}, {S{\'a}nchez},
  {Sartori}, {Scalzotto}, {Scapin}, {Schmitt}, {Schweizer}, {Shayduk},
  {Shinozaki}, {Shore}, {Sidro}, {Sillanp{\"a}{\"a}}, {Sobczynska}, {Stamerra},
  {Stark}, {Takalo}, {Temnikov}, {Tescaro}, {Teshima}, {Torres}, {Turini},
  {Vankov}, {Vitale}, {Wagner}, {Wibig}, {Wittek}, {Zandanel}, {Zanin}, \&
  {Zapatero}}]{Albert:2007uw}
{MAGIC collaboration}, {Albert}, J., {Aliu}, E., {et~al.} 2007, \apjl, 665, L51

\bibitem[{{MAGIC collaboration} {et~al.}(2008){MAGIC collaboration}, {Albert},
  {Aliu}, {Anderhub}, {Antoranz}, {Backes}, {Baixeras}, {Barrio}, {Bartko},
  {Bastieri}, {Becker}, {Bednarek}, {Berger}, {Bigongiari}, {Biland}, {Bock},
  {Bonnoli}, {Bordas}, {Bosch-Ramon}, {Bretz}, {Britvitch}, {Camara},
  {Carmona}, {Chilingarian}, {Commichau}, {Contreras}, {Cortina}, {Costado},
  {Curtef}, {Dazzi}, {De Angelis}, {de los Reyes}, {De Lotto}, {De Maria}, {De
  Sabata}, {Delgado Mendez}, {Dorner}, {Doro}, {Errando}, {Fagiolini},
  {Ferenc}, {Fern{\'a}ndez}, {Firpo}, {Fonseca}, {Font}, {Galante},
  {Garc{\'{\i}}a L{\'o}pez}, {Garczarczyk}, {Gaug}, {Goebel}, {Hayashida},
  {Herrero}, {H{\"o}hne}, {Hose}, {Hsu}, {Huber}, {Jogler}, {Kosyra},
  {Kranich}, {Laille}, {Leonardo}, {Lindfors}, {Lombardi}, {Longo},
  {L{\'o}pez}, {Lorenz}, {Majumdar}, {Maneva}, {Mankuzhiyil}, {Mannheim},
  {Mariotti}, {Mart{\'{\i}}nez}, {Mazin}, {Merck}, {Meucci}, {Meyer},
  {Miranda}, {Mirzoyan}, {Mizobuchi}, {Moles}, {Moralejo}, {Nieto}, {Nilsson},
  {Ninkovic}, {O{\~n}a-Wilhelmi}, {Otte}, {Oya}, {Panniello}, {Paoletti},
  {Paredes}, {Pasanen}, {Pascoli}, {Pauss}, {Pegna}, {P{\'e}rez-Torres},
  {Persic}, {Peruzzo}, {Piccioli}, {Prada}, {Prandini}, {Puchades}, {Raymers},
  {Rhode}, {Rib{\'o}}, {Rico}, {Rissi}, {Robert}, {R{\"u}gamer}, {Saggion},
  {Saito}, {S{\'a}nchez}, {S{\'a}nchez-Conde}, {Sartori}, {Scalzotto},
  {Scapin}, {Schmitt}, {Schweizer}, {Shayduk}, {Shinozaki}, {Shore}, {Sidro},
  {Sillanp{\"a}{\"a}}, {Sobczynska}, {Spanier}, {Stamerra}, {Stark}, {Takalo},
  {Temnikov}, {Tescaro}, {Teshima}, {Torres}, {Turini}, {Vankov}, {Venturini},
  {Vitale}, {Wagner}, {Wittek}, {Zandanel}, {Zanin}, {Zapatero}, {Guerrero},
  {Alberdi}, {Paragi}, {Muxlow}, \& {Diamond}}]{Albert:2008zs}
{MAGIC collaboration}, {Albert}, J., {Aliu}, E., {et~al.} 2008, \apj, 684, 1351

\bibitem[{{MAGIC collaboration} {et~al.}(2012{\natexlab{a}}){MAGIC
  collaboration}, {Aleksi{\'c}}, {Alvarez}, {Antonelli}, {Antoranz}, {Asensio},
  {Backes}, {Barres de Almeida}, {Barrio}, {Bastieri}, {Becerra Gonz{\'a}lez},
  {Bednarek}, {Berger}, {Bernardini}, {Biland}, {Blanch}, {Bock}, {Boller},
  {Bonnoli}, {Borla Tridon}, {Bosch-Ramon}, {Bretz}, {Ca{\~n}ellas}, {Carmona},
  {Carosi}, {Colin}, {Colombo}, {Contreras}, {Cortina}, {Cossio}, {Covino}, {Da
  Vela}, {Dazzi}, {De Angelis}, {De Caneva}, {De Cea del Pozo}, {De Lotto},
  {Delgado Mendez}, {Diago Ortega}, {Doert}, {Dom{\'{\i}}nguez}, {Dominis
  Prester}, {Dorner}, {Doro}, {Eisenacher}, {Elsaesser}, {Ferenc}, {Fonseca},
  {Font}, {Fruck}, {Garc{\'{\i}}a L{\'o}pez}, {Garczarczyk}, {Garrido Terrats},
  {Giavitto}, {Godinovi{\'c}}, {Gonz{\'a}lez Mu{\~n}oz}, {Gozzini}, {Hadasch},
  {H{\"a}fner}, {Herrero}, {Hildebrand}, {Hose}, {Hrupec}, {Huber},
  {Jankowski}, {Jogler}, {Kadenius}, {Kellermann}, {Klepser},
  {Kr{\"a}henb{\"u}hl}, {Krause}, {La Barbera}, {Lelas}, {Leonardo},
  {Lewandowska}, {Lindfors}, {Lombardi}, {L{\'o}pez}, {L{\'o}pez-Coto},
  {L{\'o}pez-Oramas}, {Lorenz}, {Makariev}, {Maneva}, {Mankuzhiyil},
  {Mannheim}, {Maraschi}, {Mariotti}, {Mart{\'{\i}}nez}, {Mazin}, {Meucci},
  {Miranda}, {Mirzoyan}, {Mold{\'o}n}, {Moralejo}, {Munar-Adrover},
  {Niedzwiecki}, {Nieto}, {Nilsson}, {Nowak}, {Orito}, {Paiano}, {Paneque},
  {Paoletti}, {Pardo}, {Paredes}, {Partini}, {Perez-Torres}, {Persic}, {Pilia},
  {Pochon}, {Prada}, {Prada Moroni}, {Prandini}, {Puerto Gimenez}, {Puljak},
  {Reichardt}, {Reinthal}, {Rhode}, {Rib{\'o}}, {Rico}, {R{\"u}gamer},
  {Saggion}, {Saito}, {Saito}, {Salvati}, {Satalecka}, {Scalzotto}, {Scapin},
  {Schultz}, {Schweizer}, {Shore}, {Sillanp{\"a}{\"a}}, {Sitarek}, {Snidaric},
  {Sobczynska}, {Spanier}, {Spiro}, {Stamatescu}, {Stamerra}, {Steinke},
  {Storz}, {Strah}, {Sun}, {Suri{\'c}}, {Takalo}, {Takami}, {Tavecchio},
  {Temnikov}, {Terzi{\'c}}, {Tescaro}, {Teshima}, {Tibolla}, {Torres},
  {Treves}, {Uellenbeck}, {Vogler}, {Wagner}, {Weitzel}, {Zabalza}, {Zandanel},
  \& {Zanin}}]{2012ApJ...754L..10A}
{MAGIC collaboration}, {Aleksi{\'c}}, J., {Alvarez}, E.~A., {et~al.}
  2012{\natexlab{a}}, \apjl, 754, L10

\bibitem[{{MAGIC collaboration} {et~al.}(2012{\natexlab{b}}){MAGIC
  collaboration}, {Aleksi{\'c}}, {Alvarez}, {Antonelli}, {Antoranz}, {Asensio},
  {Backes}, {Barrio}, {Bastieri}, {Becerra Gonz{\'a}lez}, {Bednarek},
  {Berdyugin}, {Berger}, {Bernardini}, {Biland}, {Blanch}, {Bock}, {Boller},
  {Bonnoli}, {Borla Tridon}, {Bosch-Ramon}, {Braun}, {Bretz}, {Ca{\~n}ellas},
  {Carmona}, {Carosi}, {Colin}, {Colombo}, {Contreras}, {Cortina}, {Cossio},
  {Covino}, {Dazzi}, {De Angelis}, {De Caneva}, {De Cea del Pozo}, {De Lotto},
  {Delgado Mendez}, {Diago Ortega}, {Doert}, {Dom{\'{\i}}nguez}, {Dominis
  Prester}, {Dorner}, {Doro}, {Elsaesser}, {Ferenc}, {Fonseca}, {Font},
  {Fruck}, {Garc{\'{\i}}a L{\'o}pez}, {Garczarczyk}, {Garrido}, {Giavitto},
  {Godinovi{\'c}}, {Hadasch}, {H{\"a}fner}, {Herrero}, {Hildebrand},
  {H{\"o}hne-M{\"o}nch}, {Hose}, {Hrupec}, {Huber}, {Jogler}, {Kellermann},
  {Klepser}, {Kr{\"a}henb{\"u}hl}, {Krause}, {La Barbera}, {Lelas}, {Leonardo},
  {Lindfors}, {Lombardi}, {L{\'o}pez}, {L{\'o}pez}, {Lorenz}, {Makariev},
  {Maneva}, {Mankuzhiyil}, {Mannheim}, {Maraschi}, {Mariotti},
  {Mart{\'{\i}}nez}, {Mazin}, {Meucci}, {Miranda}, {Mirzoyan}, {Miyamoto},
  {Mold{\'o}n}, {Moralejo}, {Munar-Adrover}, {Nieto}, {Nilsson}, {Orito},
  {Oya}, {Paneque}, {Paoletti}, {Pardo}, {Paredes}, {Partini}, {Pasanen},
  {Pauss}, {Perez-Torres}, {Persic}, {Peruzzo}, {Pilia}, {Pochon}, {Prada},
  {Prada Moroni}, {Prandini}, {Puljak}, {Reichardt}, {Reinthal}, {Rhode},
  {Rib{\'o}}, {Rico}, {R{\"u}gamer}, {Saggion}, {Saito}, {Saito}, {Salvati},
  {Satalecka}, {Scalzotto}, {Scapin}, {Schultz}, {Schweizer}, {Shayduk},
  {Shore}, {Sillanp{\"a}{\"a}}, {Sitarek}, {Sobczynska}, {Spanier}, {Spiro},
  {Stamerra}, {Steinke}, {Storz}, {Strah}, {Suri{\'c}}, {Takalo}, {Takami},
  {Tavecchio}, {Temnikov}, {Terzi{\'c}}, {Tescaro}, {Teshima}, {Tibolla},
  {Torres}, {Treves}, {Uellenbeck}, {Vankov}, {Vogler}, {Wagner}, {Weitzel},
  {Zabalza}, {Zandanel}, \& {Zanin}}]{2012ApJ...746...80A}
{MAGIC collaboration}, {Aleksi{\'c}}, J., {Alvarez}, E.~A., {et~al.}
  2012{\natexlab{b}}, \apj, 746, 80

\bibitem[{{MAGIC collaboration} {et~al.}(2011){MAGIC collaboration},
  {Aleksi{\'c}}, {Alvarez}, {Antonelli}, {Antoranz}, {Asensio}, {Backes},
  {Barrio}, {Bastieri}, {Becerra Gonz{\'a}lez}, {Bednarek}, {Berdyugin},
  {Berger}, {Bernardini}, {Biland}, {Blanch}, {Bock}, {Boller}, {Bonnoli},
  {Bordas}, {Borla Tridon}, {Bosch-Ramon}, {Braun}, {Bretz}, {Ca{\~n}ellas},
  {Carmona}, {Carosi}, {Colin}, {Colombo}, {Contreras}, {Cortina}, {Cossio},
  {Covino}, {Dazzi}, {De Angelis}, {De Cea del Pozo}, {De Lotto}, {Delgado
  Mendez}, {Diago Ortega}, {Doert}, {Dom{\'{\i}}nguez}, {Dominis Prester},
  {Dorner}, {Doro}, {Elsaesser}, {Ferenc}, {Fonseca}, {Font}, {Fruck},
  {Garc{\'{\i}}a L{\'o}pez}, {Garczarczyk}, {Garrido}, {Giavitto},
  {Godinovi{\'c}}, {Hadasch}, {H{\"a}fner}, {Herrero}, {Hildebrand},
  {H{\"o}hne-M{\"o}nch}, {Hose}, {Hrupec}, {Huber}, {Jogler}, {Klepser},
  {Kr{\"a}henb{\"u}hl}, {Krause}, {La Barbera}, {Lelas}, {Leonardo},
  {Lindfors}, {Lombardi}, {L{\'o}pez}, {Lorenz}, {Makariev}, {Maneva},
  {Mankuzhiyil}, {Mannheim}, {Maraschi}, {Mariotti}, {Mart{\'{\i}}nez},
  {Mazin}, {Meucci}, {Miranda}, {Mirzoyan}, {Miyamoto}, {Mold{\'o}n},
  {Moralejo}, {Munar-Adrover}, {Nieto}, {Nilsson}, {Orito}, {Oya}, {Paneque},
  {Paoletti}, {Pardo}, {Paredes}, {Partini}, {Pasanen}, {Pauss},
  {Perez-Torres}, {Persic}, {Peruzzo}, {Pilia}, {Pochon}, {Prada}, {Prada
  Moroni}, {Prandini}, {Puljak}, {Reichardt}, {Reinthal}, {Rhode}, {Rib{\'o}},
  {Rico}, {R{\"u}gamer}, {Saggion}, {Saito}, {Saito}, {Salvati}, {Satalecka},
  {Scalzotto}, {Scapin}, {Schultz}, {Schweizer}, {Shayduk}, {Shore},
  {Sillanp{\"a}{\"a}}, {Sitarek}, {Sobczynska}, {Spanier}, {Spiro}, {Stamerra},
  {Steinke}, {Storz}, {Strah}, {Suri{\'c}}, {Takalo}, {Takami}, {Tavecchio},
  {Temnikov}, {Terzi{\'c}}, {Tescaro}, {Teshima}, {Thom}, {Tibolla}, {Torres},
  {Treves}, {Vankov}, {Vogler}, {Wagner}, {Weitzel}, {Zabalza}, {Zandanel}, \&
  {Zanin}}]{2011ApJ...735L...5A}
{MAGIC collaboration}, {Aleksi{\'c}}, J., {Alvarez}, E.~A., {et~al.} 2011,
  \apjl, 735, L5

\bibitem[{{MAGIC collaboration} {et~al.}(2010){MAGIC collaboration},
  {Aleksi{\'c}}, {Antonelli}, {Antoranz}, {Backes}, {Baixeras}, {Barrio},
  {Bastieri}, {Becerra Gonz{\'a}lez}, {Bednarek}, {Berdyugin}, {Berger},
  {Bernardini}, {Biland}, {Blanch}, {Bock}, {Boller}, {Bonnoli}, {Bordas},
  {Borla Tridon}, {Bosch-Ramon}, {Bose}, {Braun}, {Bretz}, {Britzger},
  {Camara}, {Carmona}, {Carosi}, {Colin}, {Contreras}, {Cortina}, {Costado},
  {Covino}, {Dazzi}, {De Angelis}, {De Cea del Pozo}, {De Lotto}, {De Maria},
  {De Sabata}, {Delgado Mendez}, {Doert}, {Dom{\'{\i}}nguez}, {Dominis
  Prester}, {Dorner}, {Doro}, {Elsaesser}, {Errando}, {Ferenc}, {Fonseca},
  {Font}, {Garc{\'{\i}}a L{\'o}pez}, {Garczarczyk}, {Gaug}, {Godinovic},
  {G{\"o}ebel}, {Hadasch}, {Herrero}, {Hildebrand}, {H{\"o}hne-M{\"o}nch},
  {Hose}, {Hrupec}, {Hsu}, {Jogler}, {Klepser}, {Kr{\"a}henb{\"u}hl},
  {Kranich}, {La Barbera}, {Laille}, {Leonardo}, {Lindfors}, {Lombardi},
  {Longo}, {L{\'o}pez}, {Lorenz}, {Majumdar}, {Maneva}, {Mankuzhiyil},
  {Mannheim}, {Maraschi}, {Mariotti}, {Mart{\'{\i}}nez}, {Mazin}, {Meucci},
  {Miranda}, {Mirzoyan}, {Miyamoto}, {Mold{\'o}n}, {Moles}, {Moralejo},
  {Nieto}, {Nilsson}, {Ninkovic}, {Orito}, {Oya}, {Paiano}, {Paoletti},
  {Paredes}, {Partini}, {Pasanen}, {Pascoli}, {Pauss}, {Pegna}, {Perez-Torres},
  {Persic}, {Peruzzo}, {Prada}, {Prandini}, {Puchades}, {Puljak}, {Reichardt},
  {Rhode}, {Rib{\'o}}, {Rico}, {Rissi}, {R{\"u}gamer}, {Saggion}, {Saito},
  {Saito}, {Salvati}, {S{\'a}nchez-Conde}, {Satalecka}, {Scalzotto}, {Scapin},
  {Schultz}, {Schweizer}, {Shayduk}, {Shore}, {Sierpowska-Bartosik},
  {Sillanp{\"a}{\"a}}, {Sitarek}, {Sobczynska}, {Spanier}, {Spiro}, {Stamerra},
  {Steinke}, {Struebig}, {Suric}, {Takalo}, {Tavecchio}, {Temnikov}, {Terzic},
  {Tescaro}, {Teshima}, {Torres}, {Vankov}, {Wagner}, {Weitzel}, {Zabalza},
  {Zandanel}, {Zanin}, \& {The MAGIC Collaboration}}]{2010ApJ...721..843A}
{MAGIC collaboration}, {Aleksi{\'c}}, J., {Antonelli}, L.~A., {et~al.} 2010,
  \apj, 721, 843

\bibitem[{{MAGIC collaboration} {et~al.}(2009{\natexlab{b}}){MAGIC
  collaboration}, {Anderhub}, {Antonelli}, {Antoranz}, {Backes}, {Baixeras},
  {Balestra}, {Barrio}, {Bastieri}, {Becerra Gonz{\'a}lez}, {Becker},
  {Bednarek}, {Berger}, {Bernardini}, {Biland}, {Blanch Bigas}, {Bock},
  {Bonnoli}, {Bordas}, {Borla Tridon}, {Bosch-Ramon}, {Bose}, {Braun}, {Bretz},
  {Britzger}, {Camara}, {Carmona}, {Carosi}, {Colin}, {Commichau}, {Contreras},
  {Cortina}, {Costado}, {Covino}, {Dazzi}, {De Angelis}, {de Cea del Pozo}, {De
  los Reyes}, {De Lotto}, {De Maria}, {De Sabata}, {Delgado Mendez},
  {Dom{\'{\i}}nguez}, {Dominis Prester}, {Dorner}, {Doro}, {Elsaesser},
  {Errando}, {Ferenc}, {Fern{\'a}ndez}, {Firpo}, {Fonseca}, {Font}, {Galante},
  {Garc{\'{\i}}a L{\'o}pez}, {Garczarczyk}, {Gaug}, {Godinovic}, {Goebel},
  {Hadasch}, {Herrero}, {Hildebrand}, {H{\"o}hne-M{\"o}nch}, {Hose}, {Hrupec},
  {Hsu}, {Jogler}, {Klepser}, {Kranich}, {La Barbera}, {Laille}, {Leonardo},
  {Lindfors}, {Lombardi}, {Longo}, {L{\'o}pez}, {Lorenz}, {Majumdar}, {Maneva},
  {Mankuzhiyil}, {Mannheim}, {Maraschi}, {Mariotti}, {Mart{\'{\i}}nez},
  {Mazin}, {Meucci}, {Miranda}, {Mirzoyan}, {Miyamoto}, {Mold{\'o}n}, {Moles},
  {Moralejo}, {Nieto}, {Nilsson}, {Ninkovic}, {Orito}, {Oya}, {Paoletti},
  {Paredes}, {Pasanen}, {Pascoli}, {Pauss}, {Pegna}, {Perez-Torres}, {Persic},
  {Peruzzo}, {Prada}, {Prandini}, {Puchades}, {Puljak}, {Reichardt}, {Rhode},
  {Rib{\'o}}, {Rico}, {Rissi}, {Robert}, {R{\"u}gamer}, {Saggion}, {Saito},
  {Salvati}, {S{\'a}nchez-Conde}, {Satalecka}, {Scalzotto}, {Scapin},
  {Schweizer}, {Shayduk}, {Shore}, {Sidro}, {Sierpowska-Bartosik},
  {Sillanp{\"a}{\"a}}, {Sitarek}, {Sobczynska}, {Spanier}, {Spiro}, {Stamerra},
  {Stark}, {Suric}, {Takalo}, {Tavecchio}, {Temnikov}, {Tescaro}, {Teshima},
  {Torres}, {Turini}, {Vankov}, {Wagner}, {Zabalza}, {Zandanel}, {Zanin},
  {Zapatero}, {The MAGIC Collaboration}, {Falcone}, {Vetere}, {Gehrels},
  {Trushkin}, {Dhawan}, \& {Reig}}]{2009ApJ...706L..27A}
{MAGIC collaboration}, {Anderhub}, H., {Antonelli}, L.~A., {et~al.}
  2009{\natexlab{b}}, \apjl, 706, L27

\bibitem[{{Malyshev} {et~al.}(2013){Malyshev}, {Zdziarski}, \&
  {Chernyakova}}]{2013arXiv1305.5920M}
{Malyshev}, D., {Zdziarski}, A.~A., \& {Chernyakova}, M. 2013, \mnras, 434,
  2380

\bibitem[{{Malzac} \& {Belmont}(2009)}]{2009MNRAS.392..570M}
{Malzac}, J. \& {Belmont}, R. 2009, \mnras, 392, 570

\bibitem[{{Malzac} {et~al.}(2008){Malzac}, {Lubi{\'n}ski}, {Zdziarski},
  {Cadolle Bel}, {T{\"u}rler}, \& {Laurent}}]{Malzac:2008hf}
{Malzac}, J., {Lubi{\'n}ski}, P., {Zdziarski}, A.~A., {et~al.} 2008, \aap, 492,
  527

\bibitem[{{Manchester} {et~al.}(2005){Manchester}, {Hobbs}, {Teoh}, \&
  {Hobbs}}]{2005AJ....129.1993M}
{Manchester}, R.~N., {Hobbs}, G.~B., {Teoh}, A., \& {Hobbs}, M. 2005, \aj, 129,
  1993

\bibitem[{{Manchester} {et~al.}(1995){Manchester}, {Johnston}, {Lyne},
  {D'Amico}, {Bailes}, \& {Nicastro}}]{Manchester:1995ck}
{Manchester}, R.~N., {Johnston}, S., {Lyne}, A.~G., {et~al.} 1995, \apjl, 445,
  L137

\bibitem[{{Manchester} \& {Taylor}(1977)}]{1977puls.book.....M}
{Manchester}, R.~N. \& {Taylor}, J.~H. 1977, Pulsars (San Francisco, CA (USA):
  W.~H.~Freeman, 281 p.)

\bibitem[{{Maraschi} \& {Treves}(1981)}]{Maraschi:1981nj}
{Maraschi}, L. \& {Treves}, A. 1981, \mnras, 194, 1P

\bibitem[{{Markoff} {et~al.}(2001){Markoff}, {Falcke}, \&
  {Fender}}]{Markoff:2001om}
{Markoff}, S., {Falcke}, H., \& {Fender}, R. 2001, \aap, 372, L25

\bibitem[{{Mart{\'{\i}}} \& {Paredes}(1995)}]{1995AA...298..151M}
{Mart{\'{\i}}}, J. \& {Paredes}, J.~M. 1995, \aap, 298, 151

\bibitem[{{Mart{\'{\i}}} {et~al.}(1998){Mart{\'{\i}}}, {Peracaula}, {Paredes},
  {Massi}, \& {Estalella}}]{Marti:1998mf}
{Mart{\'{\i}}}, J., {Peracaula}, M., {Paredes}, J.~M., {Massi}, M., \&
  {Estalella}, R. 1998, \aap, 329, 951

\bibitem[{{Martin} \& {Dubus}(2013)}]{2013A&A...551A..37M}
{Martin}, P. \& {Dubus}, G. 2013, \aap, 551, A37

\bibitem[{{Martocchia} {et~al.}(2005){Martocchia}, {Motch}, \&
  {Negueruela}}]{Martocchia:2005kq}
{Martocchia}, A., {Motch}, C., \& {Negueruela}, I. 2005, \aap, 430, 245

\bibitem[{{Massi} \& {Kaufman Bernad{\'o}}(2009)}]{2009ApJ...702.1179M}
{Massi}, M. \& {Kaufman Bernad{\'o}}, M. 2009, \apj, 702, 1179

\bibitem[{{Massi} {et~al.}(2004){Massi}, {Rib{\'o}}, {Paredes}, {Garrington},
  {Peracaula}, \& {Mart{\'{\i}}}}]{Massi:2004pf}
{Massi}, M., {Rib{\'o}}, M., {Paredes}, J.~M., {et~al.} 2004, \aap, 414, L1

\bibitem[{{Massi} {et~al.}(2001){Massi}, {Rib{\'o}}, {Paredes}, {Peracaula}, \&
  {Estalella}}]{Massi:2001jg}
{Massi}, M., {Rib{\'o}}, M., {Paredes}, J.~M., {Peracaula}, M., \& {Estalella},
  R. 2001, \aap, 376, 217

\bibitem[{{Mattana} {et~al.}(2009){Mattana}, {Falanga}, {G{\"o}tz}, {Terrier},
  {Esposito}, {Pellizzoni}, {De Luca}, {Marandon}, {Goldwurm}, \&
  {Caraveo}}]{2009ApJ...694...12M}
{Mattana}, F., {Falanga}, M., {G{\"o}tz}, D., {et~al.} 2009, \apj, 694, 12

\bibitem[{{McConnell} {et~al.}(2002){McConnell}, {Zdziarski}, {Bennett},
  {Bloemen}, {Collmar}, {Hermsen}, {Kuiper}, {Paciesas}, {Phlips}, {Poutanen},
  {Ryan}, {Sch{\"o}nfelder}, {Steinle}, \& {Strong}}]{McConnell:2002nz}
{McConnell}, M.~L., {Zdziarski}, A.~A., {Bennett}, K., {et~al.} 2002, \apj,
  572, 984

\bibitem[{{McLaughlin}(2004)}]{2004xmm..prop..229M}
{McLaughlin}, M. 2004, XMM-Newton Proposal 03071701

\bibitem[{{McSwain} \& {Gies}(2002)}]{McSwain:2002hp}
{McSwain}, M.~V. \& {Gies}, D.~R. 2002, \apjl, 568, L27

\bibitem[{{McSwain} {et~al.}(2004){McSwain}, {Gies}, {Huang}, {Wiita},
  {Wingert}, \& {Kaper}}]{2004ApJ...600..927M}
{McSwain}, M.~V., {Gies}, D.~R., {Huang}, W., {et~al.} 2004, \apj, 600, 927

\bibitem[{{McSwain} {et~al.}(2010){McSwain}, {Grundstrom}, {Gies}, \&
  {Ray}}]{2010ApJ...724..379M}
{McSwain}, M.~V., {Grundstrom}, E.~D., {Gies}, D.~R., \& {Ray}, P.~S. 2010,
  \apj, 724, 379

\bibitem[{{McSwain} {et~al.}(2011){McSwain}, {Ray}, {Ransom}, {Roberts},
  {Dougherty}, \& {Pooley}}]{2011ApJ...738..105M}
{McSwain}, M.~V., {Ray}, P.~S., {Ransom}, S.~M., {et~al.} 2011, \apj, 738, 105

\bibitem[{{Melatos} {et~al.}(1995){Melatos}, {Johnston}, \&
  {Melrose}}]{1995MNRAS.275..381M}
{Melatos}, A., {Johnston}, S., \& {Melrose}, D.~B. 1995, \mnras, 275, 381

\bibitem[{{Mendelson} \& {Mazeh}(1989)}]{1989MNRAS.239..733M}
{Mendelson}, H. \& {Mazeh}, T. 1989, \mnras, 239, 733

\bibitem[{{Meurs} \& {van den Heuvel}(1989)}]{Meurs:1989zp}
{Meurs}, E.~J.~A. \& {van den Heuvel}, E.~P.~J. 1989, \aap, 226, 88

\bibitem[{{Michelson} {et~al.}(1992){Michelson}, {Bertsch}, {Chiang},
  {Fichtel}, {Hartman}, {Hunter}, {Kanbach}, {Kniffen}, {Kwok}, {Lin},
  {Mattox}, {Mayer-Hasselwander}, {von Montigny}, {Nolan}, {Pinkau},
  {Rothermel}, {Schneid}, {Sommer}, {Sreekumar}, \&
  {Thompson}}]{1992ApJ...401..724M}
{Michelson}, P.~F., {Bertsch}, D.~L., {Chiang}, J., {et~al.} 1992, \apj, 401,
  724

\bibitem[{{Mignone} {et~al.}(2012){Mignone}, {Zanni}, {Tzeferacos}, {van
  Straalen}, {Colella}, \& {Bodo}}]{2012ApJS..198....7M}
{Mignone}, A., {Zanni}, C., {Tzeferacos}, P., {et~al.} 2012, \apjs, 198, 7

\bibitem[{{Miller-Jones} {et~al.}(2004){Miller-Jones}, {Blundell}, {Rupen},
  {Mioduszewski}, {Duffy}, \& {Beasley}}]{Miller-Jones:2004qg}
{Miller-Jones}, J.~C.~A., {Blundell}, K.~M., {Rupen}, M.~P., {et~al.} 2004,
  \apj, 600, 368

\bibitem[{{Miller-Jones} {et~al.}(2009){Miller-Jones}, {Rupen}, {T{\"u}rler},
  {Lindfors}, {Blundell}, \& {Pooley}}]{Miller-Jones:2008fk}
{Miller-Jones}, J.~C.~A., {Rupen}, M.~P., {T{\"u}rler}, M., {et~al.} 2009,
  \mnras, 394, 309

\bibitem[{{Mirabel}(2006)}]{2006Sci...312.1759M}
{Mirabel}, I.~F. 2006, Science, 312, 1759

\bibitem[{{Mirabel} {et~al.}(1992){Mirabel}, {Rodriguez}, {Cordier}, {Paul}, \&
  {Lebrun}}]{1992Natur.358..215M}
{Mirabel}, I.~F., {Rodriguez}, L.~F., {Cordier}, B., {Paul}, J., \& {Lebrun},
  F. 1992, \nat, 358, 215

\bibitem[{{Mochol} \& {Kirk}(2013)}]{2013arXiv1308.0950M}
{Mochol}, I. \& {Kirk}, J.~G. 2013, ApJ in press, (arXiv:1308.0950)

\bibitem[{{Moderski} {et~al.}(2005){Moderski}, {Sikora}, {Coppi}, \&
  {Aharonian}}]{Moderski:2005yd}
{Moderski}, R., {Sikora}, M., {Coppi}, P.~S., \& {Aharonian}, F. 2005, \mnras,
  363, 954

\bibitem[{{Mold{\'o}n} {et~al.}(2011{\natexlab{a}}){Mold{\'o}n}, {Johnston},
  {Rib{\'o}}, {Paredes}, \& {Deller}}]{2011ApJ...732L..10M}
{Mold{\'o}n}, J., {Johnston}, S., {Rib{\'o}}, M., {Paredes}, J.~M., \&
  {Deller}, A.~T. 2011{\natexlab{a}}, \apjl, 732, L10

\bibitem[{{Mold{\'o}n} {et~al.}(2011{\natexlab{b}}){Mold{\'o}n}, {Rib{\'o}}, \&
  {Paredes}}]{2011A&A...533L...7M}
{Mold{\'o}n}, J., {Rib{\'o}}, M., \& {Paredes}, J.~M. 2011{\natexlab{b}}, \aap,
  533, L7

\bibitem[{{Mold{\'o}n} {et~al.}(2012{\natexlab{a}}){Mold{\'o}n}, {Rib{\'o}}, \&
  {Paredes}}]{2012arXiv1209.6073M}
{Mold{\'o}n}, J., {Rib{\'o}}, M., \& {Paredes}, J.~M. 2012{\natexlab{a}}, \aap,
  548, A103

\bibitem[{{Mold{\'o}n} {et~al.}(2012{\natexlab{b}}){Mold{\'o}n}, {Rib{\'o}},
  {Paredes}, {Brisken}, {Dhawan}, {Kramer}, {Lyne}, \&
  {Stappers}}]{2012A&A...543A..26M}
{Mold{\'o}n}, J., {Rib{\'o}}, M., {Paredes}, J.~M., {et~al.}
  2012{\natexlab{b}}, \aap, 543, A26

\bibitem[{{Mori} {et~al.}(1997){Mori}, {et al.}, \& {}}]{1997ApJ...476..842M}
{Mori}, M., {et al.}, \& {}. 1997, \apj, 476, 842

\bibitem[{{Moskalenko}(1995)}]{1995SSRv...72..593M}
{Moskalenko}, I.~V. 1995, Space Science Reviews, 72, 593

\bibitem[{{Motch} {et~al.}(1997){Motch}, {Haberl}, {Dennerl}, {Pakull}, \&
  {Janot-Pacheco}}]{Motch:1997qk}
{Motch}, C., {Haberl}, F., {Dennerl}, K., {Pakull}, M., \& {Janot-Pacheco}, E.
  1997, \aap, 323, 853

\bibitem[{{Mu{\~n}oz-Arjonilla} {et~al.}(2009){Mu{\~n}oz-Arjonilla},
  {Mart{\'{\i}}}, {Combi}, {Luque-Escamilla}, {S{\'a}nchez-Sutil}, {Zabalza},
  \& {Paredes}}]{2009A&A...497..457M}
{Mu{\~n}oz-Arjonilla}, A.~J., {Mart{\'{\i}}}, J., {Combi}, J.~A., {et~al.}
  2009, \aap, 497, 457

\bibitem[{{Munari} {et~al.}(2011){Munari}, {Joshi}, {Ashok}, {Banerjee},
  {Valisa}, {Milani}, {Siviero}, {Dallaporta}, \&
  {Castellani}}]{2011MNRAS.410L..52M}
{Munari}, U., {Joshi}, V.~H., {Ashok}, N.~M., {et~al.} 2011, \mnras, 410, L52

\bibitem[{{Napoli} {et~al.}(2011){Napoli}, {McSwain}, {Boyer}, \&
  {Roettenbacher}}]{2011PASP..123.1262N}
{Napoli}, V.~J., {McSwain}, M.~V., {Boyer}, A.~N.~M., \& {Roettenbacher}, R.~M.
  2011, \pasp, 123, 1262

\bibitem[{{Negueruela} {et~al.}(2011){Negueruela}, {Rib{\'o}}, {Herrero},
  {Lorenzo}, {Khangulyan}, \& {Aharonian}}]{2011ApJ...732L..11N}
{Negueruela}, I., {Rib{\'o}}, M., {Herrero}, A., {et~al.} 2011, \apjl, 732, L11

\bibitem[{{Nelson} {et~al.}(2012){Nelson}, {Donato}, {Mukai}, {Sokoloski}, \&
  {Chomiuk}}]{2012ApJ...748...43N}
{Nelson}, T., {Donato}, D., {Mukai}, K., {Sokoloski}, J., \& {Chomiuk}, L.
  2012, \apj, 748, 43

\bibitem[{{Neronov} \& {Chernyakova}(2008)}]{2008ApJ...672L.123N}
{Neronov}, A. \& {Chernyakova}, M. 2008, \apjl, 672, L123

\bibitem[{{Neronov} {et~al.}(2012){Neronov}, {Malyshev}, {Chernyakova}, \&
  {Lutovinov}}]{2012A&A...543L...9N}
{Neronov}, A., {Malyshev}, D., {Chernyakova}, M., \& {Lutovinov}, A. 2012,
  \aap, 543, L9

\bibitem[{{Nishiyama} {et~al.}(2010){Nishiyama}, {Kabashima}, {Guido},
  {Sostero}, \& {Schmeer}}]{2010CBET.2204....2N}
{Nishiyama}, K., {Kabashima}, F., {Guido}, E., {Sostero}, G., \& {Schmeer}, P.
  2010, Central Bureau Electronic Telegrams, 2204, 2

\bibitem[{{Nolan} {et~al.}(2012){Nolan}, {Abdo}, {Ackermann}, {Ajello},
  {Allafort}, {Antolini}, {Atwood}, {Axelsson}, {Baldini}, {Ballet}, \&
  et~al.}]{2012ApJS..199...31N}
{Nolan}, P.~L., {Abdo}, A.~A., {Ackermann}, M., {et~al.} 2012, \apjs, 199, 31

\bibitem[{{Nolan} {et~al.}(2003){Nolan}, {Tompkins}, {Grenier}, \&
  {Michelson}}]{Nolan:2003sn}
{Nolan}, P.~L., {Tompkins}, W.~F., {Grenier}, I.~A., \& {Michelson}, P.~F.
  2003, \apj, 597, 615

\bibitem[{{Ohm} {et~al.}(2010){Ohm}, {Hinton}, \&
  {Domainko}}]{2010arXiv1006.2464O}
{Ohm}, S., {Hinton}, J.~A., \& {Domainko}, W. 2010, \apjl, 718, L161

\bibitem[{{Okazaki} {et~al.}(2011){Okazaki}, {Nagataki}, {Naito}, {Kawachi},
  {Hayasaki}, {Owocki}, \& {Takata}}]{2011PASJ...63..893O}
{Okazaki}, A.~T., {Nagataki}, S., {Naito}, T., {et~al.} 2011, \pasj, 63, 893

\bibitem[{{Ong}(2011)}]{2011ATel.3153....1O}
{Ong}, R.~A. 2011, The Astronomer's Telegram, 3153, 1

\bibitem[{{Orlando} \& {Drake}(2012)}]{2012MNRAS.419.2329O}
{Orlando}, S. \& {Drake}, J.~J. 2012, \mnras, 419, 2329

\bibitem[{{Page} {et~al.}(2013){Page}, {Osborne}, {Wagner}, {Beardmore},
  {Shore}, {Starrfield}, \& {Woodward}}]{2013ApJ...768L..26P}
{Page}, K.~L., {Osborne}, J.~P., {Wagner}, R.~M., {et~al.} 2013, \apjl, 768,
  L26

\bibitem[{{Papitto} {et~al.}(2013){Papitto}, {Ferrigno}, {Bozzo}, {Rea},
  {Pavan}, {Campana}, {Romano}, {Burderi}, {Di Salvo}, {Riggio}, {Torres},
  {Falanga}, {Hessels}, {Burgay}, {Sarkissian}, {Wieringa}, {Filipovi{\'c}}, \&
  {Wong}}]{2013arXiv1305.3884P}
{Papitto}, A., {Ferrigno}, C., {Bozzo}, E., {et~al.} 2013, preprint, submitted
  (ArXiv:1305.3884)

\bibitem[{{Paredes} {et~al.}(2013){Paredes}, {Bednarek}, {Bordas},
  {Bosch-Ramon}, {De Cea del Pozo}, {Dubus}, {Funk}, {Hadasch}, {Khangulyan},
  {Markoff}, {Moldon}, {Munar-Adrover}, {Nagataki}, {Naito}, {de Naurois},
  {Pedaletti}, {Reimer}, {Ribo}, {Szostek}, {Terada}, {Torres}, {Zabalza},
  {Zdziarski}, \& {CTA Consortium}}]{2012arXiv1210.3215P}
{Paredes}, J.~M., {Bednarek}, W., {Bordas}, P., {et~al.} 2013, Astroparticle
  Physics, 43, 301

\bibitem[{{Paredes} {et~al.}(1997){Paredes}, {Marti}, {Peracaula}, \&
  {Ribo}}]{1997A&A...320L..25P}
{Paredes}, J.~M., {Marti}, J., {Peracaula}, M., \& {Ribo}, M. 1997, \aap, 320,
  L25

\bibitem[{{Paredes} {et~al.}(2000){Paredes}, {Mart{\'{\i}}}, {Rib{\'o}}, \&
  {Massi}}]{2000Sci...288.2340P}
{Paredes}, J.~M., {Mart{\'{\i}}}, J., {Rib{\'o}}, M., \& {Massi}, M. 2000,
  Science, 288, 2340

\bibitem[{{Paredes} {et~al.}(1994){Paredes}, {Marziani}, {Marti}, {Fabregat},
  {Coe}, {Everall}, {Figueras}, {Jordi}, {Norton}, {Prince}, {Reglero},
  {Roche}, {Torra}, {Unger}, \& {Zamanov}}]{1994A&A...288..519P}
{Paredes}, J.~M., {Marziani}, P., {Marti}, J., {et~al.} 1994, \aap, 288, 519

\bibitem[{{Paredes} {et~al.}(2007){Paredes}, {Rib{\'o}}, {Bosch-Ramon}, {West},
  {Butt}, {Torres}, \& {Mart{\'{\i}}}}]{2007ApJ...664L..39P}
{Paredes}, J.~M., {Rib{\'o}}, M., {Bosch-Ramon}, V., {et~al.} 2007, \apjl, 664,
  L39

\bibitem[{{Paredes} {et~al.}(2002){Paredes}, {Rib{\'o}}, {Ros}, {Mart{\'{\i}}},
  \& {Massi}}]{2002A&A...393L..99P}
{Paredes}, J.~M., {Rib{\'o}}, M., {Ros}, E., {Mart{\'{\i}}}, J., \& {Massi}, M.
  2002, \aap, 393, L99

\bibitem[{{Pavlov} {et~al.}(2011){Pavlov}, {Chang}, \&
  {Kargaltsev}}]{2011ApJ...730....2P}
{Pavlov}, G.~G., {Chang}, C., \& {Kargaltsev}, O. 2011, \apj, 730, 2

\bibitem[{{Perucho} {et~al.}(2010){Perucho}, {Bosch-Ramon}, \&
  {Khangulyan}}]{2010A&A...512L...4P}
{Perucho}, M., {Bosch-Ramon}, V., \& {Khangulyan}, D. 2010, \aap, 512, L4

\bibitem[{{P{\'e}tri}(2011)}]{2011MNRAS.412.1870P}
{P{\'e}tri}, J. 2011, \mnras, 412, 1870

\bibitem[{{P{\'e}tri}(2012)}]{2012MNRAS.424..605P}
{P{\'e}tri}, J. 2012, \mnras, 424, 605

\bibitem[{{P{\'e}tri} \& {Dubus}(2011)}]{2011MNRAS.417..532P}
{P{\'e}tri}, J. \& {Dubus}, G. 2011, \mnras, 417, 532

\bibitem[{{P{\'e}tri} \& {Kirk}(2005)}]{2005ApJ...627L..37P}
{P{\'e}tri}, J. \& {Kirk}, J.~G. 2005, \apjl, 627, L37

\bibitem[{{Phinney} {et~al.}(1988){Phinney}, {Evans}, {Blandford}, \&
  {Kulkarni}}]{1988Natur.333..832P}
{Phinney}, E.~S., {Evans}, C.~R., {Blandford}, R.~D., \& {Kulkarni}, S.~R.
  1988, \nat, 333, 832

\bibitem[{{Piano} {et~al.}(2012){Piano}, {Tavani}, {Vittorini}, {Trois},
  {Giuliani}, {Bulgarelli}, {Evangelista}, {Coppi}, {Del Monte}, {Sabatini},
  {Striani}, {Donnarumma}, {Hannikainen}, {Koljonen}, {McCollough}, {Pooley},
  {Trushkin}, {Zanin}, {Barbiellini}, {Cardillo}, {Cattaneo}, {Chen},
  {Colafrancesco}, {Feroci}, {Fuschino}, {Giusti}, {Longo}, {Morselli},
  {Pellizzoni}, {Pittori}, {Pucella}, {Rapisarda}, {Rappoldi}, {Soffitta},
  {Trifoglio}, {Vercellone}, \& {Verrecchia}}]{2012arXiv1207.6288P}
{Piano}, G., {Tavani}, M., {Vittorini}, V., {et~al.} 2012, \aap, 545, A110

\bibitem[{{Pittard} \& {Dougherty}(2006)}]{Pittard:2006wx}
{Pittard}, J.~M. \& {Dougherty}, S.~M. 2006, \mnras, 372, 801

\bibitem[{{Pittard} {et~al.}(2005){Pittard}, {Dougherty}, {Coker}, \&
  {Corcoran}}]{2005xrrc.procE2.01P}
{Pittard}, J.~M., {Dougherty}, S.~M., {Coker}, R.~F., \& {Corcoran}, M.~F.
  2005, in X-Ray and Radio Connections, Published electronically by NRAO, Held
  3-6 February, 2004 in Santa Fe, New Mexico, USA, (E2.01), ed.
  {L.~O.~Sjouwerman \& K.~K.~Dyer}

\bibitem[{Pletsch {et~al.}(2012)Pletsch, Guillemot, Fehrmann, Allen, Kramer,
  Aulbert, Ackermann, Ajello, de~Angelis, Atwood, Baldini, Ballet, Barbiellini,
  Bastieri, Bechtol, Bellazzini, Borgland, Bottacini, Brandt, Bregeon, Brigida,
  Bruel, Buehler, Buson, Caliandro, Cameron, Caraveo, Casandjian, Cecchi, {\c
  C}elik, Charles, Chaves, Cheung, Chiang, Ciprini, Claus, Cohen-Tanugi,
  Conrad, Cutini, D'Ammando, Dermer, Digel, Drell, Drlica-Wagner, Dubois,
  Dumora, Favuzzi, Ferrara, Franckowiak, Fukazawa, Fusco, Gargano, Gehrels,
  Germani, Giglietto, Giordano, Giroletti, Godfrey, Grenier, Grondin, Grove,
  Guiriec, Hadasch, Hanabata, Harding, den Hartog, Hayashida, Hays, Hill, Hou,
  Hughes, J{\'o}hannesson, Jackson, Jogler, Johnson, Johnson, Kataoka, Kerr,
  Kn{\"o}dlseder, Kuss, Lande, Larsson, Latronico, Lemoine-Goumard, Longo,
  Loparco, Lovellette, Lubrano, Massaro, Mayer, Mazziotta, McEnery, Mehault,
  Michelson, Mitthumsiri, Mizuno, Monzani, Morselli, Moskalenko, Murgia,
  Nakamori, Nemmen, Nuss, Ohno, Ohsugi, Omodei, Orienti, Orlando, de~Palma,
  Paneque, Perkins, Piron, Pivato, Porter, Rain{\`o}, Rando, Ray, Razzano,
  Reimer, Reimer, Reposeur, Ritz, Romani, Romoli, Sanchez, Parkinson, Schulz,
  Sgr{\`o}, do~Couto~e Silva, Siskind, Smith, Spandre, Spinelli, Suson,
  Takahashi, Tanaka, Thayer, Thayer, Thompson, Tibaldo, Tinivella, Troja,
  Usher, Vandenbroucke, Vasileiou, Vianello, Vitale, Waite, Winer, Wood, Wood,
  Yang, \& Zimmer}]{Pletsch25102012}
Pletsch, H.~J., Guillemot, L., Fehrmann, H., {et~al.} 2012, Science, 338, 1314

\bibitem[{{Portegies Zwart} \& {Verbunt}(1996)}]{Portegies-Zwart:1996ky}
{Portegies Zwart}, S.~F. \& {Verbunt}, F. 1996, \aap, 309, 179

\bibitem[{Portegies~Zwart \& Yungelson(1998)}]{Portegies-Zwart:1998ic}
Portegies~Zwart, S.~F. \& Yungelson, L.~R. 1998, \aap, 332, 173

\bibitem[{{Porter} \& {Rivinius}(2003)}]{2003PASP..115.1153P}
{Porter}, J.~M. \& {Rivinius}, T. 2003, \pasp, 115, 1153

\bibitem[{{Postnov} \& {Yungelson}(2006)}]{2006LRR.....9....6P}
{Postnov}, K.~A. \& {Yungelson}, L.~R. 2006, Living Reviews in Relativity, 9, 6

\bibitem[{{Pringle}(1974)}]{Pringle:1974oj}
{Pringle}, J.~E. 1974, \nat, 247, 21

\bibitem[{{Pringle} \& {Rees}(1972)}]{Pringle:1972kx}
{Pringle}, J.~E. \& {Rees}, M.~J. 1972, \aap, 21, 1

\bibitem[{{Protheroe} \& {Stanev}(1993)}]{Protheroe:1993en}
{Protheroe}, R.~J. \& {Stanev}, T. 1993, \mnras, 264, 191

\bibitem[{{Rajoelimanana} {et~al.}(2011){Rajoelimanana}, {Charles}, \&
  {Udalski}}]{2011MNRAS.413.1600R}
{Rajoelimanana}, A.~F., {Charles}, P.~A., \& {Udalski}, A. 2011, \mnras, 413,
  1600

\bibitem[{{Ray} {et~al.}(1997){Ray}, {Foster}, {Waltman}, {Tavani}, \&
  {Ghigo}}]{1997ApJ...491..381R}
{Ray}, P.~S., {Foster}, R.~S., {Waltman}, E.~B., {Tavani}, M., \& {Ghigo},
  F.~D. 1997, \apj, 491, 381

\bibitem[{{Razzaque} {et~al.}(2010){Razzaque}, {Jean}, \&
  {Mena}}]{2010PhRvD..82l3012R}
{Razzaque}, S., {Jean}, P., \& {Mena}, O. 2010, \prd, 82, 123012

\bibitem[{{Rea} \& {Torres}(2011)}]{2011ApJ...737L..12R}
{Rea}, N. \& {Torres}, D.~F. 2011, \apjl, 737, L12

\bibitem[{{Rea} {et~al.}(2011){Rea}, {Torres}, {Caliandro}, {Hadasch}, {van der
  Klis}, {Jonker}, {M{\'e}ndez}, \&
  {Sierpowska-Bartosik}}]{2011MNRAS.416.1514R}
{Rea}, N., {Torres}, D.~F., {Caliandro}, G.~A., {et~al.} 2011, \mnras, 416,
  1514

\bibitem[{{Rea} {et~al.}(2010){Rea}, {Torres}, {van der Klis}, {Jonker},
  {M{\'e}ndez}, \& {Sierpowska-Bartosik}}]{2010MNRAS.405.2206R}
{Rea}, N., {Torres}, D.~F., {van der Klis}, M., {et~al.} 2010, \mnras, 405,
  2206

\bibitem[{{Rees} \& {Gunn}(1974)}]{Rees:1974xr}
{Rees}, M.~J. \& {Gunn}, J.~E. 1974, \mnras, 167, 1

\bibitem[{{Reimer} {et~al.}(2006){Reimer}, {Pohl}, \& {Reimer}}]{Reimer:2006da}
{Reimer}, A., {Pohl}, M., \& {Reimer}, O. 2006, \apj, 644, 1118

\bibitem[{{Reimer} \& {Iyudin}(2003)}]{Reimer:2003nx}
{Reimer}, O. \& {Iyudin}, A. 2003, in Proceedings of the 28th International
  Cosmic Ray Conference (Tsukuba, Japan), ed. T.~Kajita, Y.~Asaoka, A.~Kawachi,
  Y.~Matsubara, \& M.~Sasaki, Vol.~4 (Tokyo, Japan: Universal Academy Press),
  2341

\bibitem[{{Reitberger} {et~al.}(2012){Reitberger}, {Reimer}, {Reimer},
  {Werner}, {Egberts}, \& {Takahashi}}]{2012A&A...544A..98R}
{Reitberger}, K., {Reimer}, O., {Reimer}, A., {et~al.} 2012, \aap, 544, A98

\bibitem[{{Remillard} \& {McClintock}(2006)}]{Remillard:2006kg}
{Remillard}, R.~A. \& {McClintock}, J.~E. 2006, \araa, 44, 49

\bibitem[{{Renaud}(2009)}]{2009arXiv0905.1287R}
{Renaud}, M. 2009, Proceedings of 44th Recontres de Moriond 2009 - Very High
  Energy Phenomena in the Universe (arXiv:0905.1287)

\bibitem[{{Rib{\'o}} {et~al.}(2008){Rib{\'o}}, {Paredes}, {Mold{\'o}n},
  {Mart{\'{\i}}}, \& {Massi}}]{2008A&A...481...17R}
{Rib{\'o}}, M., {Paredes}, J.~M., {Mold{\'o}n}, J., {Mart{\'{\i}}}, J., \&
  {Massi}, M. 2008, \aap, 481, 17

\bibitem[{{Rib{\'o}} {et~al.}(2002){Rib{\'o}}, {Paredes}, {Romero}, {Benaglia},
  {Mart{\'{\i}}}, {Fors}, \& {Garc{\'{\i}}a-S{\'a}nchez}}]{2002A&A...384..954R}
{Rib{\'o}}, M., {Paredes}, J.~M., {Romero}, G.~E., {et~al.} 2002, \aap, 384,
  954

\bibitem[{{Rib{\'o}} {et~al.}(1999){Rib{\'o}}, {Reig}, {Mart{\'{\i}}}, \&
  {Paredes}}]{Ribo:1999qo}
{Rib{\'o}}, M., {Reig}, P., {Mart{\'{\i}}}, J., \& {Paredes}, J.~M. 1999, \aap,
  347, 518

\bibitem[{Roberts(2012)}]{Roberts:2013aa}
Roberts, M. S.~E. 2012, Proceedings of the International Astronomical Union, 8,
  127

\bibitem[{{Romero} {et~al.}(2005){Romero}, {Christiansen}, \&
  {Orellana}}]{Romero:2005go}
{Romero}, G.~E., {Christiansen}, H.~R., \& {Orellana}, M. 2005, \apj, 632, 1093

\bibitem[{{Romero} {et~al.}(2007){Romero}, {Okazaki}, {Orellana}, \&
  {Owocki}}]{Romero:2007ly}
{Romero}, G.~E., {Okazaki}, A.~T., {Orellana}, M., \& {Owocki}, S.~P. 2007,
  \aap, 474, 15

\bibitem[{{Romero} {et~al.}(2003){Romero}, {Torres}, {Kaufman Bernad{\'o}}, \&
  {Mirabel}}]{Romero:2003nm}
{Romero}, G.~E., {Torres}, D.~F., {Kaufman Bernad{\'o}}, M.~M., \& {Mirabel},
  I.~F. 2003, \aap, 410, L1

\bibitem[{{Rupen} {et~al.}(2008){Rupen}, {Mioduszewski}, \&
  {Sokoloski}}]{Rupen:2007xb}
{Rupen}, M.~P., {Mioduszewski}, A.~J., \& {Sokoloski}, J.~L. 2008, \apj, 688,
  559

\bibitem[{{Russell} {et~al.}(2007){Russell}, {Fender}, {Gallo}, \&
  {Kaiser}}]{Russell:2007ju}
{Russell}, D.~M., {Fender}, R.~P., {Gallo}, E., \& {Kaiser}, C.~R. 2007,
  \mnras, 376, 1341

\bibitem[{{Russell} {et~al.}(2010){Russell}, {Maitra}, {Dunn}, \&
  {Markoff}}]{2010MNRAS.405.1759R}
{Russell}, D.~M., {Maitra}, D., {Dunn}, R.~J.~H., \& {Markoff}, S. 2010,
  \mnras, 405, 1759

\bibitem[{{Rybicki} \& {Lightman}(1979)}]{Rybicki:1979za}
{Rybicki}, G.~B. \& {Lightman}, A.~P. 1979, {Radiative processes in
  astrophysics} (New York, Wiley-Interscience, 1979.~393 p.)

\bibitem[{{Sabatini} {et~al.}(2013){Sabatini}, {Tavani}, {Coppi}, {Pooley},
  {Del Santo}, {Campana}, {Chen}, {Evangelista}, {Piano}, {Bulgarelli},
  {Cattaneo}, {Colafrancesco}, {Del Monte}, {Giuliani}, {Giusti}, {Longo},
  {Morselli}, {Pellizzoni}, {Pilia}, {Striani}, {Trifoglio}, \&
  {Vercellone}}]{Sabatini:2013aa}
{Sabatini}, S., {Tavani}, M., {Coppi}, P., {et~al.} 2013, \apj, 766, 83

\bibitem[{{Sabatini} {et~al.}(2010){Sabatini}, {Tavani}, {Striani},
  {Bulgarelli}, {Vittorini}, {Piano}, {Del Monte}, {Feroci}, {de Pasquale},
  {Trifoglio}, {Gianotti}, {Argan}, {Barbiellini}, {Caraveo}, {Cattaneo},
  {Chen}, {D'Ammando}, {Costa}, {De Paris}, {Di Cocco}, {Donnarumma},
  {Evangelista}, {Ferrari}, {Fiorini}, {Fuschino}, {Galli}, {Giuliani},
  {Giusti}, {Labanti}, {Lazzarotto}, {Lipari}, {Longo}, {Marisaldi},
  {Mereghetti}, {Morelli}, {Moretti}, {Morselli}, {Pacciani}, {Pellizzoni},
  {Perotti}, {Picozza}, {Pilia}, {Pucella}, {Prest}, {Rapisarda}, {Rappoldi},
  {Rubini}, {Scalise}, {Soffitta}, {Trois}, {Vallazza}, {Vercellone}, {Zambra},
  {Zanello}, {Pittori}, {Verrecchia}, {Santolamazza}, {Giommi},
  {Colafrancesco}, {Antonelli}, \& {Salotti}}]{2010ApJ...712L..10S}
{Sabatini}, S., {Tavani}, M., {Striani}, E., {et~al.} 2010, \apjl, 712, L10

\bibitem[{{Saito} {et~al.}(2009){Saito}, {Zanin}, {Bordas}, {Bosh-Ramon},
  {Jogler}, {Paredes}, {Ribo}, {Rissi}, {Rico}, {Torres}, \& {for the MAGIC
  Collaboration}}]{2009arXiv0907.1017S}
{Saito}, T.~Y., {Zanin}, R., {Bordas}, P., {et~al.} 2009, Proceedings of the
  31st International Cosmic Ray Conference (ICRC), Lodz, Poland, July 2009
  (arXiv:0907.1017)

\bibitem[{{Sarty} {et~al.}(2011){Sarty}, {Szalai}, {Kiss}, {Matthews}, {Wu},
  {Kuschnig}, {Guenther}, {Moffat}, {Rucinski}, {Sasselov}, {Weiss}, {Huziak},
  {Johnston}, {Phillips}, \& {Ashley}}]{2011MNRAS.411.1293S}
{Sarty}, G.~E., {Szalai}, T., {Kiss}, L.~L., {et~al.} 2011, \mnras, 411, 1293

\bibitem[{{Sekiguchi} {et~al.}(2009){Sekiguchi}, {Tsujimoto}, {Kitamoto},
  {Ishida}, {Hamaguchi}, {Mori}, \& {Tsuboi}}]{2009PASJ...61..629S}
{Sekiguchi}, A., {Tsujimoto}, M., {Kitamoto}, S., {et~al.} 2009, \pasj, 61, 629

\bibitem[{{Shore} {et~al.}(2013){Shore}, {De Gennaro Aquino}, {Schwarz},
  {Augusteijn}, {Cheung}, {Walter}, \& {Starrfield}}]{2013arXiv1303.0404S}
{Shore}, S.~N., {De Gennaro Aquino}, I., {Schwarz}, G.~J., {et~al.} 2013, \aap,
  553, A123

\bibitem[{{Shore} {et~al.}(2012){Shore}, {Wahlgren}, {Augusteijn}, {Liimets},
  {Koubsky}, {{\v S}lechta}, \& {Votruba}}]{2012A&A...540A..55S}
{Shore}, S.~N., {Wahlgren}, G.~M., {Augusteijn}, T., {et~al.} 2012, \aap, 540,
  A55

\bibitem[{{Shore} {et~al.}(2011){Shore}, {Wahlgren}, {Augusteijn}, {Liimets},
  {Page}, {Osborne}, {Beardmore}, {Koubsky}, {{\v S}lechta}, \&
  {Votruba}}]{2011A&A...527A..98S}
{Shore}, S.~N., {Wahlgren}, G.~M., {Augusteijn}, T., {et~al.} 2011, \aap, 527,
  A98

\bibitem[{{Shvartsman}(1971)}]{Shvartsman:1971rw}
{Shvartsman}, V.~F. 1971, Soviet Astronomy, 15, 342

\bibitem[{{Sidoli} {et~al.}(2006){Sidoli}, {Pellizzoni}, {Vercellone},
  {Moroni}, {Mereghetti}, \& {Tavani}}]{Sidoli:2006an}
{Sidoli}, L., {Pellizzoni}, A., {Vercellone}, S., {et~al.} 2006, \aap, 459, 901

\bibitem[{{Sidro} {et~al.}(2008){Sidro}, {Cortina}, {Mauche}, \& {et
  al.}}]{2008ICRC....2..715S}
{Sidro}, N., {Cortina}, J., {Mauche}, C.~W., \& {et al.} 2008, in Proceedings
  of the 30th ICRC, July 3 - 11, 2007, M{\'e}rida, Yucat{\'a}n, Mexico, ed.
  {Caballero, R.}, {D'Olivo, J. C.}, {Medina-Tanco, G.}, {Nellen, L.},
  {S{\'a}nchez, F. A.}, \& {Vald{\'e}s-Galicia, J. F.}, Vol.~2 (Universidad
  Nacional Aut{\'o}noma de M{\'e}xico), 715--718

\bibitem[{{Sierpowska} \& {Bednarek}(2005)}]{Sierpowska:2005vy}
{Sierpowska}, A. \& {Bednarek}, W. 2005, \mnras, 356, 711

\bibitem[{{Sierpowska-Bartosik} \& {Torres}(2007)}]{Sierpowska-Bartosik:2007jo}
{Sierpowska-Bartosik}, A. \& {Torres}, D.~F. 2007, \apjl, 671, L145

\bibitem[{{Sierpowska-Bartosik} \& {Torres}(2008)}]{2008APh....30..239S}
{Sierpowska-Bartosik}, A. \& {Torres}, D.~F. 2008, Astroparticle Physics, 30,
  239

\bibitem[{{Sierpowska-Bartosik} \& {Torres}(2009)}]{Sierpowska-Bartosik:2008wt}
{Sierpowska-Bartosik}, A. \& {Torres}, D.~F. 2009, \apj, 693, 1462

\bibitem[{{Sironi} \& {Spitkovsky}(2009)}]{2009ApJ...698.1523S}
{Sironi}, L. \& {Spitkovsky}, A. 2009, \apj, 698, 1523

\bibitem[{{Sironi} \& {Spitkovsky}(2011)}]{2011ApJ...741...39S}
{Sironi}, L. \& {Spitkovsky}, A. 2011, \apj, 741, 39

\bibitem[{{Sitarek} \& {Bednarek}(2012{\natexlab{a}})}]{2012MNRAS.421..512S}
{Sitarek}, J. \& {Bednarek}, W. 2012{\natexlab{a}}, \mnras, 421, 512

\bibitem[{{Sitarek} \& {Bednarek}(2012{\natexlab{b}})}]{2012PhRvD..86f3011S}
{Sitarek}, J. \& {Bednarek}, W. 2012{\natexlab{b}}, \prd, 86, 063011

\bibitem[{{Skilton} {et~al.}(2009){Skilton}, {Pandey-Pommier}, {Hinton},
  {Cheung}, {Aharonian}, {Brucker}, {Dubus}, {Fiasson}, {Funk}, {Gallant},
  {Marcowith}, \& {Reimer}}]{2009MNRAS.399..317S}
{Skilton}, J.~L., {Pandey-Pommier}, M., {Hinton}, J.~A., {et~al.} 2009, \mnras,
  399, 317

\bibitem[{{Smith} {et~al.}(2009){Smith}, {Kaaret}, {Holder}, {Falcone},
  {Maier}, {Pandel}, \& {Stroh}}]{2009ApJ...693.1621S}
{Smith}, A., {Kaaret}, P., {Holder}, J., {et~al.} 2009, \apj, 693, 1621

\bibitem[{{Sokoloski} {et~al.}(2006){Sokoloski}, {Luna}, {Mukai}, \&
  {Kenyon}}]{Sokoloski:2006uq}
{Sokoloski}, J.~L., {Luna}, G.~J.~M., {Mukai}, K., \& {Kenyon}, S.~J. 2006,
  \nat, 442, 276

\bibitem[{{Spencer} {et~al.}(1986){Spencer}, {Swinney}, {Johnston}, \&
  {Hjellming}}]{1986ApJ...309..694S}
{Spencer}, R.~E., {Swinney}, R.~W., {Johnston}, K.~J., \& {Hjellming}, R.~M.
  1986, \apj, 309, 694

\bibitem[{{Spitkovsky}(2006)}]{Spitkovsky:2006mq}
{Spitkovsky}, A. 2006, \apjl, 648, L51

\bibitem[{{Spitkovsky}(2008)}]{2008ApJ...682L...5S}
{Spitkovsky}, A. 2008, \apjl, 682, L5

\bibitem[{{Stairs} {et~al.}(2001){Stairs}, {Manchester}, {Lyne}, {Kaspi},
  {Camilo}, {Bell}, {D'Amico}, {Kramer}, {Crawford}, {Morris}, {Possenti},
  {McKay}, {Lumsden}, {Tacconi-Garman}, {Cannon}, {Hambly}, \&
  {Wood}}]{2001MNRAS.325..979S}
{Stairs}, I.~H., {Manchester}, R.~N., {Lyne}, A.~G., {et~al.} 2001, \mnras,
  325, 979

\bibitem[{{Stawarz} {et~al.}(2006){Stawarz}, {Aharonian}, {Kataoka},
  {Ostrowski}, {Siemiginowska}, \& {Sikora}}]{Stawarz:2006oh}
{Stawarz}, {\L}., {Aharonian}, F., {Kataoka}, J., {et~al.} 2006, \mnras, 370,
  981

\bibitem[{{Stella} {et~al.}(1986){Stella}, {White}, \&
  {Rosner}}]{Stella:1986qd}
{Stella}, L., {White}, N.~E., \& {Rosner}, R. 1986, \apj, 308, 669

\bibitem[{{Stevens} {et~al.}(1992){Stevens}, {Blondin}, \&
  {Pollock}}]{Stevens:1992on}
{Stevens}, I.~R., {Blondin}, J.~M., \& {Pollock}, A.~M.~T. 1992, \apj, 386, 265

\bibitem[{{Strickman} {et~al.}(1998){Strickman}, {Tavani}, {Coe}, {Steele},
  {Fabregat}, {Marti}, {Paredes}, \& {Ray}}]{1998ApJ...497..419S}
{Strickman}, M.~S., {Tavani}, M., {Coe}, M.~J., {et~al.} 1998, \apj, 497, 419

\bibitem[{{Strom} {et~al.}(1989){Strom}, {van Paradijs}, \& {van der
  Klis}}]{1989Natur.337..234S}
{Strom}, R.~G., {van Paradijs}, J., \& {van der Klis}, M. 1989, \nat, 337, 234

\bibitem[{{Strong} {et~al.}(2001){Strong}, {Collmar}, {Bennett}, {Bloemen},
  {Diehl}, {Hermsen}, {Iyudin}, {Mayer-Hasselwander}, {Ryan}, \&
  {Sch{\"o}nfelder}}]{2001AIPC..587...21S}
{Strong}, A.~W., {Collmar}, W., {Bennett}, K., {et~al.} 2001, in American
  Institute of Physics Conference Series, Vol. 587, Gamma 2001: Gamma-Ray
  Astrophysics, ed. S.~{Ritz}, N.~{Gehrels}, \& C.~R. {Shrader}, 21--25

\bibitem[{{Szostek} \& {Dubus}(2011)}]{2011MNRAS.411..193S}
{Szostek}, A. \& {Dubus}, G. 2011, \mnras, 411, 193

\bibitem[{{Szostek} {et~al.}(2012){Szostek}, {Dubus}, \&
  {McSwain}}]{2012MNRAS.420.3521S}
{Szostek}, A., {Dubus}, G., \& {McSwain}, M.~V. 2012, \mnras, 420, 3521

\bibitem[{{Szostek} \& {Zdziarski}(2008)}]{2008MNRAS.386..593S}
{Szostek}, A. \& {Zdziarski}, A.~A. 2008, \mnras, 386, 593

\bibitem[{{Szostek} {et~al.}(2008){Szostek}, {Zdziarski}, \&
  {McCollough}}]{2008MNRAS.388.1001S}
{Szostek}, A., {Zdziarski}, A.~A., \& {McCollough}, M.~L. 2008, \mnras, 388,
  1001

\bibitem[{{Takahashi} {et~al.}(2009){Takahashi}, {Kishishita}, {Uchiyama},
  {Tanaka}, {Yamaoka}, {Khangulyan}, {Aharonian}, {Bosch-Ramon}, \&
  {Hinton}}]{Takahashi:2008vu}
{Takahashi}, T., {Kishishita}, T., {Uchiyama}, Y., {et~al.} 2009, \apj, 697,
  592

\bibitem[{{Takata} {et~al.}(2012){Takata}, {Okazaki}, {Nagataki}, {Naito},
  {Kawachi}, {Lee}, {Mori}, {Hayasaki}, {Yamaguchi}, \&
  {Owocki}}]{2012ApJ...750...70T}
{Takata}, J., {Okazaki}, A.~T., {Nagataki}, S., {et~al.} 2012, \apj, 750, 70

\bibitem[{{Takata} \& {Taam}(2009)}]{2009ApJ...702..100T}
{Takata}, J. \& {Taam}, R.~E. 2009, \apj, 702, 100

\bibitem[{{Tam} {et~al.}(2011){Tam}, {Huang}, {Takata}, {Hui}, {Kong}, \&
  {Cheng}}]{2011ApJ...736L..10T}
{Tam}, P.~H.~T., {Huang}, R.~H.~H., {Takata}, J., {et~al.} 2011, \apjl, 736,
  L10

\bibitem[{{Tam} {et~al.}(2010){Tam}, {Hui}, {Huang}, {Kong}, {Takata}, {Lin},
  {Yang}, {Cheng}, \& {Taam}}]{2010ApJ...724L.207T}
{Tam}, P.~H.~T., {Hui}, C.~Y., {Huang}, R.~H.~H., {et~al.} 2010, \apjl, 724,
  L207

\bibitem[{{Tatischeff} \& {Hernanz}(2007)}]{2007ApJ...663L.101T}
{Tatischeff}, V. \& {Hernanz}, M. 2007, \apjl, 663, L101

\bibitem[{{Tavani} \& {Arons}(1997)}]{Tavani:1997wv}
{Tavani}, M. \& {Arons}, J. 1997, \apj, 477, 439

\bibitem[{{Tavani} {et~al.}(1994){Tavani}, {Arons}, \& {Kaspi}}]{Tavani:1994qu}
{Tavani}, M., {Arons}, J., \& {Kaspi}, V.~M. 1994, \apjl, 433, L37

\bibitem[{{Tavani} {et~al.}(2009{\natexlab{a}}){Tavani}, {Bulgarelli}, {Piano},
  {Sabatini}, {Striani}, {Evangelista}, {Trois}, {Pooley}, {Trushkin},
  {Nizhelskij}, {McCollough}, {Koljonen}, {Pucella}, {Giuliani}, {Chen},
  {Costa}, {Vittorini}, {Trifoglio}, {Gianotti}, {Argan}, {Barbiellini},
  {Caraveo}, {Cattaneo}, {Cocco}, {Contessi}, {D'Ammando}, {Monte}, {de Paris},
  {di Cocco}, {di Persio}, {Donnarumma}, {Feroci}, {Ferrari}, {Fuschino},
  {Galli}, {Labanti}, {Lapshov}, {Lazzarotto}, {Lipari}, {Longo}, {Mattaini},
  {Marisaldi}, {Mastropietro}, {Mauri}, {Mereghetti}, {Morelli}, {Morselli},
  {Pacciani}, {Pellizzoni}, {Perotti}, {Picozza}, {Pilia}, {Prest},
  {Rapisarda}, {Rappoldi}, {Rossi}, {Rubini}, {Scalise}, {Soffitta},
  {Vallazza}, {Vercellone}, {Zambra}, {Zanello}, {Pittori}, {Verrecchia},
  {Giommi}, {Colafrancesco}, {Santolamazza}, {Antonelli}, \&
  {Salotti}}]{2009Natur.462..620Ta}
{Tavani}, M., {Bulgarelli}, A., {Piano}, G., {et~al.} 2009{\natexlab{a}}, \nat,
  462, 620

\bibitem[{{Tavani} {et~al.}(1996{\natexlab{a}}){Tavani}, {Grove}, {Purcell},
  {Hermsen}, {Kuiper}, {Kaaret}, {Ford}, {Wilson}, {Finger}, {Harmon}, {Zhang},
  {Mattox}, {Thompson}, \& {Arons}}]{1996A&AS..120C.221T}
{Tavani}, M., {Grove}, J.~E., {Purcell}, W., {et~al.} 1996{\natexlab{a}},
  \aaps, 120, C221

\bibitem[{{Tavani} {et~al.}(1996{\natexlab{b}}){Tavani}, {Hermsen}, {van Dijk},
  {Strickman}, {Zhang}, {Foster}, {Ray}, {Mattox}, {Ulmer}, {Purcell}, \&
  {Coe}}]{1996A&AS..120C.243T}
{Tavani}, M., {Hermsen}, W., {van Dijk}, R., {et~al.} 1996{\natexlab{b}},
  \aaps, 120, C243

\bibitem[{{Tavani} {et~al.}(1998){Tavani}, {Kniffen}, {Mattox}, {Paredes}, \&
  {Foster}}]{Tavani:1998qi}
{Tavani}, M., {Kniffen}, D., {Mattox}, J.~R., {Paredes}, J.~M., \& {Foster}, R.
  1998, \apjl, 497, L89

\bibitem[{{Tavani} {et~al.}(2010){Tavani}, {Lucarelli}, {Pittori},
  {Verrecchia}, {Bulgarelli}, {Gianotti}, {Trifoglio}, {Striani}, {Sabatini},
  {Piano}, {Argan}, {Trois}, {de Paris}, {Vittorini}, {Costa}, {Donnarumma},
  {Feroci}, {Pacciani}, {Del Monte}, {Lazzarotto}, {Soffitta}, {Evangelista},
  {Lapshov}, {Chen}, {Giuliani}, {Marisaldi}, {Di Cocco}, {Labanti},
  {Fuschino}, {Galli}, {Caraveo}, {Mereghetti}, {Perotti}, {Pucella},
  {Rapisarda}, {Pellizzoni}, {Pilia}, {Barbiellini}, {Longo}, {Picozza},
  {Morselli}, {Prest}, {Lipari}, {Zanello}, {Vercellone}, {D'Ammando},
  {Cattaneo}, {Rappoldi}, {Giommi}, {Santolamazza}, {Colafrancesco}, \&
  {Salotti}}]{2010ATel.2772....1T}
{Tavani}, M., {Lucarelli}, F., {Pittori}, C., {et~al.} 2010, The Astronomer's
  Telegram, 2772, 1

\bibitem[{{Tavani} {et~al.}(1997){Tavani}, {Mukherjee}, {Mattox}, {Halpern},
  {Thompson}, {Kanbach}, {Hermsen}, {Zhang}, \& {Foster}}]{1997ApJ...479L.109T}
{Tavani}, M., {Mukherjee}, R., {Mattox}, J.~R., {et~al.} 1997, \apjl, 479, L109

\bibitem[{{Tavani} {et~al.}(2009{\natexlab{b}}){Tavani}, {Sabatini}, {Pian},
  {Bulgarelli}, {Caraveo}, {Viotti}, {Corcoran}, {Giuliani}, {Pittori},
  {Verrecchia}, {Vercellone}, {Mereghetti}, {Argan}, {Barbiellini}, {Boffelli},
  {Cattaneo}, {Chen}, {Cocco}, {D'Ammando}, {Costa}, {DeParis}, {Del Monte},
  {Di Cocco}, {Donnarumma}, {Evangelista}, {Ferrari}, {Feroci}, {Fiorini},
  {Froysland}, {Fuschino}, {Galli}, {Gianotti}, {Labanti}, {Lapshov},
  {Lazzarotto}, {Lipari}, {Longo}, {Marisaldi}, {Mastropietro}, {Morelli},
  {Moretti}, {Morselli}, {Pacciani}, {Pellizzoni}, {Perotti}, {Piano},
  {Picozza}, {Pilia}, {Porrovecchio}, {Pucella}, {Prest}, {Rapisarda},
  {Rappoldi}, {Rubini}, {Soffitta}, {Trifoglio}, {Trois}, {Vallazza},
  {Vittorini}, {Zambra}, {Zanello}, {Santolamazza}, {Giommi}, {Colafrancesco},
  {Antonelli}, \& {Salotti}}]{2009ApJ...698L.142T}
{Tavani}, M., {Sabatini}, S., {Pian}, E., {et~al.} 2009{\natexlab{b}}, \apjl,
  698, L142

\bibitem[{{Taylor} \& {Gregory}(1982)}]{1982ApJ...255..210T}
{Taylor}, A.~R. \& {Gregory}, P.~C. 1982, \apj, 255, 210

\bibitem[{{Torres}(2011)}]{2011heep.conf..531T}
{Torres}, D.~F. 2011, in High-Energy Emission from Pulsars and their Systems:
  Proceedings of the First Session of the Sant Cugat Forum on Astrophysics,
  Astrophysics and Space Science Proceedings, ed. D.~F. {Torres} \& N.~{Rea}
  (Springer-Verlag Berlin), 531

\bibitem[{{Torres} {et~al.}(2012){Torres}, {Rea}, {Esposito}, {Li}, {Chen}, \&
  {Zhang}}]{2012ApJ...744..106T}
{Torres}, D.~F., {Rea}, N., {Esposito}, P., {et~al.} 2012, \apj, 744, 106

\bibitem[{{Torres} {et~al.}(2010){Torres}, {Zhang}, {Li}, {Rea}, {Caliandro},
  {Hadasch}, {Chen}, {Wang}, \& {Ray}}]{2010ApJ...719L.104T}
{Torres}, D.~F., {Zhang}, S., {Li}, J., {et~al.} 2010, \apjl, 719, L104

\bibitem[{{Tuthill} {et~al.}(1999){Tuthill}, {Monnier}, \&
  {Danchi}}]{1999Natur.398..487T}
{Tuthill}, P.~G., {Monnier}, J.~D., \& {Danchi}, W.~C. 1999, \nat, 398, 487

\bibitem[{{Tuthill} {et~al.}(2008){Tuthill}, {Monnier}, {Lawrance}, {Danchi},
  {Owocki}, \& {Gayley}}]{2008ApJ...675..698T}
{Tuthill}, P.~G., {Monnier}, J.~D., {Lawrance}, N., {et~al.} 2008, \apj, 675,
  698

\bibitem[{{Uchiyama} {et~al.}(2009){Uchiyama}, {Tanaka}, {Takahashi}, {Mori},
  \& {Nakazawa}}]{2009ApJ...698..911U}
{Uchiyama}, Y., {Tanaka}, T., {Takahashi}, T., {Mori}, K., \& {Nakazawa}, K.
  2009, \apj, 698, 911

\bibitem[{{Usov}(1991)}]{1991MNRAS.252...49U}
{Usov}, V.~V. 1991, \mnras, 252, 49

\bibitem[{{Usov}(1992)}]{Usov:1992re}
{Usov}, V.~V. 1992, \apj, 389, 635

\bibitem[{{Uzdensky} {et~al.}(2011){Uzdensky}, {Cerutti}, \&
  {Begelman}}]{2011ApJ...737L..40U}
{Uzdensky}, D.~A., {Cerutti}, B., \& {Begelman}, M.~C. 2011, \apjl, 737, L40

\bibitem[{{van Dijk} {et~al.}(1996){van Dijk}, {Bennett}, {Bloemen}, {Collmar},
  {Connors}, {Diehl}, {Hermsen}, {Lichti}, {McConnell}, {Much}, {Schoenfelder},
  {Steinle}, {Strong}, \& {Tavani}}]{1996A&A...315..485V}
{van Dijk}, R., {Bennett}, K., {Bloemen}, H., {et~al.} 1996, \aap, 315, 485

\bibitem[{{van Kerkwijk} {et~al.}(1996){van Kerkwijk}, {Geballe}, {King}, {van
  der Klis}, \& {van Paradijs}}]{1996A&A...314..521V}
{van Kerkwijk}, M.~H., {Geballe}, T.~R., {King}, D.~L., {van der Klis}, M., \&
  {van Paradijs}, J. 1996, \aap, 314, 521

\bibitem[{{van Soelen} \& {Meintjes}(2011)}]{2011MNRAS.412.1721V}
{van Soelen}, B. \& {Meintjes}, P.~J. 2011, \mnras, 412, 1721

\bibitem[{{van Soelen} {et~al.}(2012){van Soelen}, {Meintjes}, {Odendaal}, \&
  {Townsend}}]{2012MNRAS.426.3135V}
{van Soelen}, B., {Meintjes}, P.~J., {Odendaal}, A., \& {Townsend}, L.~J. 2012,
  \mnras, 426, 3135

\bibitem[{{VERITAS collaboration} {et~al.}(2011{\natexlab{a}}){VERITAS
  collaboration}, {Acciari}, {Aliu}, {Araya}, {Arlen}, {Aune}, {Beilicke},
  {Benbow}, {Bradbury}, {Buckley}, {Bugaev}, {Byrum}, {Cannon}, {Cesarini},
  {Ciupik}, {Collins-Hughes}, {Cui}, {Dickherber}, {Duke}, {Falcone}, {Finley},
  {Fortson}, {Furniss}, {Galante}, {Gall}, {Godambe}, {Griffin}, {Guenette},
  {Gyuk}, {Hanna}, {Holder}, {Hughes}, {Hui}, {Humensky}, {Imran}, {Kaaret},
  {Kertzman}, {Krawczynski}, {Krennrich}, {Madhavan}, {Maier}, {Majumdar},
  {McArthur}, {Moriarty}, {Ong}, {Otte}, {Pandel}, {Park}, {Perkins}, {Pohl},
  {Prokoph}, {Quinn}, {Ragan}, {Reyes}, {Reynolds}, {Roache}, {Rose}, {Saxon},
  {Sembroski}, {{\c S}ent{\"u}rk}, {Smith}, {Te{\v s}i{\'c}}, {Theiling},
  {Thibadeau}, {Varlotta}, {Vincent}, {Vivier}, {Wakely}, {Ward}, {Weekes},
  {Weinstein}, {Weisgarber}, {Weng}, {Williams}, {Wood}, \&
  {Zitzer}}]{2011arXiv1103.3250V}
{VERITAS collaboration}, {Acciari}, V.~A., {Aliu}, E., {et~al.}
  2011{\natexlab{a}}, \apj, 733, 96

\bibitem[{{VERITAS collaboration} {et~al.}(2011{\natexlab{b}}){VERITAS
  collaboration}, {Acciari}, {Aliu}, {Arlen}, {Aune}, {Beilicke}, {Benbow},
  {Bradbury}, {Buckley}, {Bugaev}, {Byrum}, {Cannon}, {Cesarini}, {Ciupik},
  {Collins-Hughes}, {Connolly}, {Cui}, {Dickherber}, {Duke}, {Errando},
  {Falcone}, {Finley}, {Finnegan}, {Fortson}, {Furniss}, {Galante}, {Gall},
  {Gillanders}, {Godambe}, {Griffin}, {Grube}, {Guenette}, {Gyuk}, {Hanna},
  {Holder}, {Hughes}, {Hui}, {Humensky}, {Kaaret}, {Karlsson}, {Kertzman},
  {Kieda}, {Krawczynski}, {Krennrich}, {Lang}, {LeBohec}, {Maier}, {Majumdar},
  {McArthur}, {McCann}, {Moriarty}, {Mukherjee}, {Ong}, {Orr}, {Otte}, {Park},
  {Perkins}, {Pohl}, {Prokoph}, {Quinn}, {Ragan}, {Reyes}, {Reynolds},
  {Roache}, {Rose}, {Ruppel}, {Saxon}, {Schroedter}, {Sembroski}, {Senturk},
  {Smith}, {Staszak}, {Te{\v s}i{\'c}}, {Theiling}, {Thibadeau}, {Tsurusaki},
  {Varlotta}, {Vassiliev}, {Vincent}, {Vivier}, {Wakely}, {Ward}, {Weekes},
  {Weinstein}, {Weisgarber}, {Williams}, \& {Zitzer}}]{2011ApJ...738....3A}
{VERITAS collaboration}, {Acciari}, V.~A., {Aliu}, E., {et~al.}
  2011{\natexlab{b}}, \apj, 738, 3

\bibitem[{{VERITAS collaboration} {et~al.}(2009{\natexlab{a}}){VERITAS
  collaboration}, {Acciari}, {Aliu}, {Arlen}, {Bautista}, {Beilicke}, {Benbow},
  {B{\"o}ttcher}, {Bradbury}, {Bugaev}, {Butt}, {Butt}, {Byrum}, {Cannon},
  {Cesarini}, {Chow}, {Ciupik}, {Cogan}, {Colin}, {Cui}, {Daniel},
  {Dickherber}, {Ergin}, {Falcone}, {Fegan}, {Finley}, {Fortin}, {Fortson},
  {Furniss}, {Gall}, {Gillanders}, {Grube}, {Guenette}, {Gyuk}, {Hanna},
  {Hays}, {Holder}, {Horan}, {Hui}, {Humensky}, {Kaaret}, {Karlsson}, {Kieda},
  {Kildea}, {Konopelko}, {Krawczynski}, {Krennrich}, {Lang}, {LeBohec},
  {Maier}, {McCann}, {McCutcheon}, {Millis}, {Moriarty}, {Mukherjee}, {Nagai},
  {Ong}, {Otte}, {Pandel}, {Perkins}, {Perkins}, {Pohl}, {Quinn}, {Ragan},
  {Reyes}, {Reynolds}, {Roache}, {Joachim Rose}, {Schroedter}, {Sembroski},
  {Smith}, {Steele}, {Stroh}, {Swordy}, {Theiling}, {Toner}, {Varlotta},
  {Vassiliev}, {Wagner}, {Wakely}, {Ward}, {Weekes}, {Weinstein}, {White},
  {Williams}, {Wissel}, {Wood}, \& {Zitzer}}]{2009ApJ...700.1034A}
{VERITAS collaboration}, {Acciari}, V.~A., {Aliu}, E., {et~al.}
  2009{\natexlab{a}}, \apj, 700, 1034

\bibitem[{{VERITAS collaboration} {et~al.}(2009{\natexlab{b}}){VERITAS
  collaboration}, {Acciari}, {Aliu}, {Arlen}, {Beilicke}, {Benbow}, {Boltuch},
  {Bradbury}, {Buckley}, {Bugaev}, {Byrum}, {Cannon}, {Cesarini}, {Cesarini},
  {Chow}, {Ciupik}, {Cogan}, {Dickherber}, {Duke}, {Ergin}, {Falcone}, {Fegan},
  {Finley}, {Finnegan}, {Fortin}, {Fortson}, {Furniss}, {Gibbs}, {Gillanders},
  {Grube}, {Guenette}, {Gyuk}, {Hanna}, {Holder}, {Horan}, {Hui}, {Humensky},
  {Imran}, {Kaaret}, {Karlsson}, {Kertzman}, {Kieda}, {Kildea}, {Konopelko},
  {Krawczynski}, {Krennrich}, {Lang}, {LeBohec}, {LeBohec}, {Maier}, {McCann},
  {McCutcheon}, {Millis}, {Millis}, {Moriarty}, {Mukherjee}, {Ong}, {Otte},
  {Pandel}, {Perkins}, {Petry}, {Pohl}, {Quinn}, {Ragan}, {Reyes}, {Reynolds},
  {Rose}, {Schroedter}, {Sembroski}, {Smith}, {Steele}, {Swordy}, {Theiling},
  {Toner}, {Varlotta}, {Vassiliev}, {Vincent}, {Wagner}, {Wakely}, {Ward},
  {Weekes}, {Weinstein}, {Weisgarber}, {Williams}, {Wissel}, \&
  {Wood}}]{2009ApJ...698L..94A}
{VERITAS collaboration}, {Acciari}, V.~A., {Aliu}, E., {et~al.}
  2009{\natexlab{b}}, \apjl, 698, L94

\bibitem[{{VERITAS collaboration} {et~al.}(2008){VERITAS collaboration},
  {Acciari}, {Beilicke}, {Blaylock}, {Bradbury}, {Buckley}, {Bugaev}, {Butt},
  {Byrum}, {Celik}, {Cesarini}, {Ciupik}, {Chow}, {Cogan}, {Colin}, {Cui},
  {Daniel}, {Duke}, {Ergin}, {Falcone}, {Fegan}, {Finley}, {Fortin}, {Fortson},
  {Gall}, {Gibbs}, {Gillanders}, {Grube}, {Guenette}, {Hanna}, {Hays},
  {Holder}, {Horan}, {Hughes}, {Hui}, {Humensky}, {Kaaret}, {Kieda}, {Kildea},
  {Konopelko}, {Krawczynski}, {Krennrich}, {Lang}, {LeBohec}, {Lee}, {Maier},
  {McCann}, {McCutcheon}, {Millis}, {Moriarty}, {Mukherjee}, {Nagai}, {Ong},
  {Pandel}, {Perkins}, {Pizlo}, {Pohl}, {Quinn}, {Ragan}, {Reynolds}, {Rose},
  {Schroedter}, {Sembroski}, {Smith}, {Steele}, {Swordy}, {Toner}, {Valcarcel},
  {Vassiliev}, {Wagner}, {Wakely}, {Ward}, {Weekes}, {Weinstein}, {White},
  {Williams}, {Wissel}, {Wood}, \& {Zitzer}}]{Acciari:2008vf}
{VERITAS collaboration}, {Acciari}, V.~A., {Beilicke}, M., {et~al.} 2008, \apj,
  679, 1427

\bibitem[{{VERITAS collaboration} {et~al.}(2012){VERITAS collaboration},
  {Aliu}, {Archambault}, {Arlen}, {Aune}, {Beilicke}, {Benbow}, {Bouvier},
  {Bradbury}, {Buckley}, {Bugaev}, {Byrum}, {Cannon}, {Cesarini}, {Ciupik},
  {Collins-Hughes}, {Connolly}, {Cui}, {Decerprit}, {Dickherber}, {Duke},
  {Dumm}, {Dwarkadas}, {Errando}, {Falcone}, {Feng}, {Finley}, {Finnegan},
  {Fortson}, {Furniss}, {Galante}, {Gall}, {Godambe}, {Griffin}, {Grube},
  {Gyuk}, {Hanna}, {Holder}, {Huan}, {Hughes}, {Humensky}, {Kaaret},
  {Karlsson}, {Kertzman}, {Khassen}, {Kieda}, {Krawczynski}, {Krennrich},
  {Lang}, {Lee}, {Maier}, {Majumdar}, {McArthur}, {McCann}, {Millis},
  {Moriarty}, {Mukherjee}, {Nu{\~n}ez}, {Ong}, {Orr}, {Otte}, {Pandel}, {Park},
  {Perkins}, {Pohl}, {Prokoph}, {Quinn}, {Ragan}, {Reyes}, {Reynolds},
  {Roache}, {Rose}, {Ruppel}, {Saxon}, {Schroedter}, {Sembroski}, {Skole},
  {Smith}, {Staszak}, {Telezhinsky}, {Te{\v s}i{\'c}}, {Theiling}, {Thibadeau},
  {Tsurusaki}, {Tyler}, {Varlotta}, {Vincent}, {Vivier}, {Wakely}, {Ward},
  {Weekes}, {Weinstein}, {Weisgarber}, {Welsing}, {Williams}, \&
  {Zitzer}}]{2012ApJ...754...77A}
{VERITAS collaboration}, {Aliu}, E., {Archambault}, S., {et~al.} 2012, \apj,
  754, 77

\bibitem[{{Vestrand} {et~al.}(1997){Vestrand}, {Sreekumar}, \&
  {Mori}}]{1997ApJ...483L..49V}
{Vestrand}, W.~T., {Sreekumar}, P., \& {Mori}, M. 1997, \apjl, 483, L49

\bibitem[{{Viotti} {et~al.}(2004){Viotti}, {Antonelli}, {Rossi}, \&
  {Rebecchi}}]{2004A&A...420..527V}
{Viotti}, R.~F., {Antonelli}, L.~A., {Rossi}, C., \& {Rebecchi}, S. 2004, \aap,
  420, 527

\bibitem[{{Walder} {et~al.}(2008){Walder}, {Folini}, \&
  {Shore}}]{2008A&A...484L...9W}
{Walder}, R., {Folini}, D., \& {Shore}, S.~N. 2008, \aap, 484, L9

\bibitem[{{Wang} {et~al.}(2004){Wang}, {Johnston}, \&
  {Manchester}}]{Wang:2004gt}
{Wang}, N., {Johnston}, S., \& {Manchester}, R.~N. 2004, \mnras, 351, 599

\bibitem[{{Waters}(1986)}]{1986A&A...162..121W}
{Waters}, L.~B.~F.~M. 1986, \aap, 162, 121

\bibitem[{{Waters} {et~al.}(1988){Waters}, {van den Heuvel}, {Taylor},
  {Habets}, \& {Persi}}]{1988A&A...198..200W}
{Waters}, L.~B.~F.~M., {van den Heuvel}, E.~P.~J., {Taylor}, A.~R., {Habets},
  G.~M.~H.~J., \& {Persi}, P. 1988, \aap, 198, 200

\bibitem[{{Weekes}(1992)}]{Weekes:1992lf}
{Weekes}, T.~C. 1992, Space Science Reviews, 59, 315

\bibitem[{{Werner} {et~al.}(2013){Werner}, {Reimer}, {Reimer}, \&
  {Egberts}}]{2013A&A...555A.102W}
{Werner}, M., {Reimer}, O., {Reimer}, A., \& {Egberts}, K. 2013, \aap, 555,
  A102

\bibitem[{{Wex} {et~al.}(1998){Wex}, {Johnston}, {Manchester}, {Lyne},
  {Stappers}, \& {Bailes}}]{1998MNRAS.298..997W}
{Wex}, N., {Johnston}, S., {Manchester}, R.~N., {et~al.} 1998, \mnras, 298, 997

\bibitem[{White {et~al.}(1995)White, Nagase, \& Parmar}]{White:1995ld}
White, N.~E., Nagase, F., \& Parmar, A.~N. 1995, in X-ray binaries, ed.
  W.~H.~G. Lewin, J.~{van Paradijs}, \& E.~{van den Heuvel} (Cambridge
  University Press), 1--49

\bibitem[{{Williams} {et~al.}(2011){Williams}, {Tomsick}, {Bodaghee}, {Bower},
  {Pooley}, {Pottschmidt}, {Rodriguez}, {Wilms}, {Migliari}, \&
  {Trushkin}}]{2011ApJ...733L..20W}
{Williams}, P.~K.~G., {Tomsick}, J.~A., {Bodaghee}, A., {et~al.} 2011, \apjl,
  733, L20

\bibitem[{{Williams} {et~al.}(1987){Williams}, {van der Hucht}, \&
  {The}}]{1987A&A...182...91W}
{Williams}, P.~M., {van der Hucht}, K.~A., \& {The}, P.~S. 1987, \aap, 182, 91

\bibitem[{{Williams} {et~al.}(2010){Williams}, {Gies}, {Matson}, {Touhami},
  {Grundstrom}, {Huang}, \& {McSwain}}]{2010ApJ...723L..93W}
{Williams}, S.~J., {Gies}, D.~R., {Matson}, R.~A., {et~al.} 2010, \apjl, 723,
  L93

\bibitem[{{Wilson-Hodge} {et~al.}(2011){Wilson-Hodge}, {Cherry}, {Case},
  {Baumgartner}, {Beklen}, {Narayana Bhat}, {Briggs}, {Camero-Arranz},
  {Chaplin}, {Connaughton}, {Finger}, {Gehrels}, {Greiner}, {Jahoda}, {Jenke},
  {Kippen}, {Kouveliotou}, {Krimm}, {Kuulkers}, {Lund}, {Meegan}, {Natalucci},
  {Paciesas}, {Preece}, {Rodi}, {Shaposhnikov}, {Skinner}, {Swartz}, {von
  Kienlin}, {Diehl}, \& {Zhang}}]{2011ApJ...727L..40W}
{Wilson-Hodge}, C.~A., {Cherry}, M.~L., {Case}, G.~L., {et~al.} 2011, \apjl,
  727, L40

\bibitem[{{Wu} {et~al.}(2012){Wu}, {Takata}, {Cheng}, {Huang}, {Hui}, {Kong},
  {Tam}, \& {Wu}}]{2012ApJ...761..181W}
{Wu}, E.~M.~H., {Takata}, J., {Cheng}, K.~S., {et~al.} 2012, \apj, 761, 181

\bibitem[{{Xue} {et~al.}(2008){Xue}, {Wu}, \& {Cui}}]{2008MNRAS.384..440X}
{Xue}, Y., {Wu}, X.-B., \& {Cui}, W. 2008, \mnras, 384, 440

\bibitem[{{Yamaguchi} \& {Takahara}(2010)}]{2010ApJ...717...85Y}
{Yamaguchi}, M.~S. \& {Takahara}, F. 2010, \apj, 717, 85

\bibitem[{{Yaron} {et~al.}(2005){Yaron}, {Prialnik}, {Shara}, \&
  {Kovetz}}]{2005ApJ...623..398Y}
{Yaron}, O., {Prialnik}, D., {Shara}, M.~M., \& {Kovetz}, A. 2005, \apj, 623,
  398

\bibitem[{{Zabalza} {et~al.}(2013){Zabalza}, {Bosch-Ramon}, {Aharonian}, \&
  {Khangulyan}}]{2012arXiv1212.3222Z}
{Zabalza}, V., {Bosch-Ramon}, V., {Aharonian}, F., \& {Khangulyan}, D. 2013,
  \aap, 551, A17

\bibitem[{{Zabalza} {et~al.}(2011{\natexlab{a}}){Zabalza}, {Bosch-Ramon}, \&
  {Paredes}}]{2011ApJ...743....7Z}
{Zabalza}, V., {Bosch-Ramon}, V., \& {Paredes}, J.~M. 2011{\natexlab{a}}, \apj,
  743, 7

\bibitem[{{Zabalza} {et~al.}(2011{\natexlab{b}}){Zabalza}, {Paredes}, \&
  {Bosch-Ramon}}]{2011A&A...527A...9Z}
{Zabalza}, V., {Paredes}, J.~M., \& {Bosch-Ramon}, V. 2011{\natexlab{b}}, \aap,
  527, A9

\bibitem[{{Zaitseva} \& {Borisov}(2003)}]{2003AstL...29..188Z}
{Zaitseva}, G.~V. \& {Borisov}, G.~V. 2003, Astronomy Letters, 29, 188

\bibitem[{{Zamanov} {et~al.}(1999){Zamanov}, {Mart{\'{\i}}}, {Paredes},
  {Fabregat}, {Rib{\'o}}, \& {Tarasov}}]{Zamanov:1999of}
{Zamanov}, R.~K., {Mart{\'{\i}}}, J., {Paredes}, J.~M., {et~al.} 1999, \aap,
  351, 543

\bibitem[{{Zanin} {et~al.}(2011){Zanin}, {Sayto}, {Zabalza}, {Bordas},
  {Jogler}, {Cortina}, {Paredes}, {Rib{\'o}}, \& {Javier Rico for the MAGIC
  collaboration}}]{2011arXiv1110.1581Z}
{Zanin}, R., {Sayto}, T., {Zabalza}, V., {et~al.} 2011, Proceedings of the 32nd
  International Cosmic Ray Conference (ICRC), Beijing, China, August 2011
  (arXiv:1110.1581)

\bibitem[{{Zdziarski} \& {Gierli{\'n}ski}(2004)}]{Zdziarski:2004en}
{Zdziarski}, A.~A. \& {Gierli{\'n}ski}, M. 2004, Progress of Theoretical
  Physics Supplement, 155, 99

\bibitem[{{Zdziarski} {et~al.}(2012{\natexlab{a}}){Zdziarski}, {Maitra},
  {Frankowski}, {Skinner}, \& {Misra}}]{2012MNRAS.426.1031Z}
{Zdziarski}, A.~A., {Maitra}, C., {Frankowski}, A., {Skinner}, G.~K., \&
  {Misra}, R. 2012{\natexlab{a}}, \mnras, 426, 1031

\bibitem[{{Zdziarski} {et~al.}(2009){Zdziarski}, {Malzac}, \&
  {Bednarek}}]{2009MNRAS.394L..41Z}
{Zdziarski}, A.~A., {Malzac}, J., \& {Bednarek}, W. 2009, \mnras, 394, L41

\bibitem[{{Zdziarski} {et~al.}(2013){Zdziarski}, {Mikolajewska}, \&
  {Belczynski}}]{2012arXiv1208.5455Z}
{Zdziarski}, A.~A., {Mikolajewska}, J., \& {Belczynski}, K. 2013, \mnras, 429,
  L104

\bibitem[{{Zdziarski} {et~al.}(2010{\natexlab{a}}){Zdziarski}, {Misra}, \&
  {Gierli{\'n}ski}}]{2010MNRAS.402..767Z}
{Zdziarski}, A.~A., {Misra}, R., \& {Gierli{\'n}ski}, M. 2010{\natexlab{a}},
  \mnras, 402, 767

\bibitem[{{Zdziarski} {et~al.}(2010{\natexlab{b}}){Zdziarski}, {Neronov}, \&
  {Chernyakova}}]{2010MNRAS.403.1873Z}
{Zdziarski}, A.~A., {Neronov}, A., \& {Chernyakova}, M. 2010{\natexlab{b}},
  \mnras, 403, 1873

\bibitem[{{Zdziarski} {et~al.}(2012{\natexlab{b}}){Zdziarski}, {Sikora},
  {Dubus}, {Yuan}, {Cerutti}, \& {Ogorza{\l}ek}}]{2012MNRAS.421.2956Z}
{Zdziarski}, A.~A., {Sikora}, M., {Dubus}, G., {et~al.} 2012{\natexlab{b}},
  \mnras, 421, 2956

\end{thebibliography}
\end{document}